\documentclass{article}

\PassOptionsToPackage{dvips}{graphicx}
\usepackage{graphicx}            % to include images
\usepackage{pslatex}      % to use PostScript fonts

\usepackage{mathptm}    % to use math fonts
\usepackage{mathptmx}   % to use math fonts

\usepackage{color}

\usepackage{soul,xcolor}

\usepackage[left=3cm,right=3cm,top=2cm,bottom=2cm]{geometry}
\usepackage{xspace}
\usepackage{amsmath, amssymb}
\usepackage{subfigure}
\usepackage{array}
\usepackage{tabularx}
\usepackage{fancyhdr}
\usepackage{mathtools}
\usepackage{isotope}
\usepackage{physics}
\usepackage[capitalise]{cleveref} %  use \cref instead of \ref to automatically add "Fig." "Table " "Equation " etc. \Cref at beginnig of a sentence
\usepackage{authblk}

\def\urltilda{\kern -.15em\lower .7ex\hbox{\~{}}\kern .04em}
\def\urldot{\kern -.10em.\kern -.10em}
\def\urlhttp{http\kern -.10em\lower -.1ex\hbox{:}\kern -.12em\lower 0ex\hbox{/}\kern -.18em\lower 0ex\hbox{/}}

\def\sp{\kern 0.08333em}

\newcommand{\mueg}{$\mu^{+} \rightarrow e^{+} \gamma$\xspace}

\newcommand{\metoee}{$\mu^{-}e^{-} \rightarrow e^{-}e^{-}$\xspace}

\newcommand{\muec}{$\mu^{-}N \rightarrow e^{-}N$\xspace}
\newcommand{\mue}{$\mu$$-$$e$\xspace}

\newcommand{\mupc}{$\mu^{-}$$-$$e^{+}$\xspace}

\newcommand{\ethane}{C$_2$H$_6$~}
\newcommand{\Deg}{\(^{\circ} \) \kern-.4em \ }

\newcommand{\kgf}{\ensuremath{\mathrm{kg}_F}\xspace}

\IfFileExists{upgreek.sty}%
  { \usepackage{upgreek}%
    \newcommand{\micro} {\ensuremath{\upmu}}%
  }%
  { \IfFileExists{textgreek.sty}%
    { \usepackage{textgreek}%
      \newcommand{\micro} {\text{\textmu}}%
    }%
    {\newcommand{\micro} {\ensuremath{\mu}}%
    }%
  }%

\newcommand{\captionfonts}{\small}

\makeatletter  % Allow the use of @ in command names
\long\def\@makecaption#1#2{%
  \vskip\abovecaptionskip
  \sbox\@tempboxa{{\captionfonts #1: #2}}%
  \ifdim \wd\@tempboxa >\hsize
    {\captionfonts #1: #2\par}
  \else
    \hbox to\hsize{\hfil\box\@tempboxa\hfil}%
  \fi
  \vskip\belowcaptionskip}
\makeatother   % Cancel the effect of \makeatletter

\begin{document}
\title{COMET Phase-I Technical Design Report}
\author[10]{\textbf{The COMET Collaboration:}  R.~Abramishvili}
\author[10,16]{G.~Adamov}
\author[3,29]{R.~R.~Akhmetshin}
\author[22]{A.~Allin}
\author[4]{J.~C.~Ang\'elique}
\author[1]{V.~Anishchik}
\author[30]{M.~Aoki}
\author[14]{D.~Aznabayev}
\author[10]{I.~Bagaturia}
\author[4]{G.~Ban}
% ...
\affil[1]{Belarusian State University (BSU), Minsk, Belarus}
% ...
\affil[3]{Budker Institute of Nuclear Physics (BINP), Novosibirsk, Russia}
\affil[4]{Laboratoire de Physique Corpusculaire (LPC), Caen, France}
% ...
\affil[10]{Georgian Technical University (GTU), Tbilisi, Georgia}
% ...
\affil[14]{Institute of Nuclear Physics, (INP), Almaty, Kazakhstan}
% ...
\affil[16]{Joint Institute for Nuclear Research (JINR), Dubna, Russia}
% ...
\affil[22]{P.~N.~Lebedev Physical Institute of the Russian Academy of Sciences, Moscow, Russia}
% ...
\affil[29]{Novosibirsk State University (NSU), Novosibirsk, Russia}
\affil[30]{Osaka University, Osaka, Japan}
% ...

%%%% To generate auto affiliation numbers please use \author{}\affil{} command

%\author{Insert first author name here}
%\affil{Insert first author address here \email{xxxx@xxxx.ac.jp}}
%
%\author{Insert second author name here}
%\affil{Insert second author address here}
%
%\author{Insert third author name here}
%\author[3]{Insert fourth author name here} %%% Use optional bracket [3] to change the respective address
%\affil{Insert third author address here}
%
%\author{Insert last author name here\thanks{These authors contributed equally to this work}}
%\affil{Insert last author address here}

%%% To include the collaborator name... Please use the command "\collaborator"
%%% For example: \collaborator{ATLAS Collaboration}
\maketitle
\clearpage

\begin{abstract}%
The Technical Design for the COMET Phase-I experiment is presented in this paper.
COMET is an experiment at J-PARC, Japan, 
which will search for neutrinoless conversion of muons into electrons in the field of an 
aluminium
nucleus (\mue conversion, \muec); a lepton flavor violating process. 
The experimental sensitivity goal for this process in  the Phase-I experiment is $3.1\times10^{-15}$, 
or 90~\% upper limit of branching ratio of $7\times 10^{-15}$,
which is a factor of 100 improvement  over the existing limit.
The expected number of background events is 0.032.
To achieve the target sensitivity
and background level,
the 3.2\,kW 8\,GeV proton beam from J-PARC will be used. 
Two types of detectors, CyDet and StrECAL,  
will be used for detecting the \mue conversion events, 
and for measuring the beam-related background events in view of the Phase-II experiment, respectively. 
Results from simulation on signal and background estimations are also described. % }
\end{abstract}
\clearpage

%\subjectindex{C01, C08, C30, H10, H50}

\tableofcontents
\clearpage

\section{Introduction}

Despite successfully predicting and allowing to understand 
the phenomena in the particle physics such as, most notably, the prediction and discovery of Higgs boson,
the Standard Model (SM) cannot provide the ultimate description of Nature: it lacks a viable dark matter candidate, offers no explanation to the observed matter-antimatter asymmetry of the Universe, and does not account for neutrino oscillation phenomena. There are also 
theoretical difficulties such as 
the
Hierarchy problem or number of parameters in 
the
SM, 
further suggesting the need for physics beyond the SM (BSM).

The observation of neutrino oscillations 
implies
that neutrinos are massive, and that
individual lepton flavours are not conserved.
This contradicts
the original SM formulation, in which neutrinos are massless by
construction, and an (accidental) symmetry leads to the conservation
of total and individual lepton numbers. Such departure from the SM paradigm also indicates that numerous other
processes
that are forbidden in the SM
might indeed occur in Nature. In particular, the violation of flavour conservation in the neutral lepton sector opens the door to the interesting possibility of charged lepton flavour violation (CLFV).
In
addition to constituting a discovery of New Physics,
the observation of a CLFV transition could provide crucial information on the nature of the BSM physics at work.
In the presence of New
Physics, one of the most interesting CLFV processes which can occur
is the transition of a muon to an electron in the presence of a nucleus \muec. The aim of the COMET experiment~\cite{cometproposal,cdr,phase-i-loi} is to search for the \muec process. 

COMET
is an international collaboration and will take place at the Japan Proton Accelerator Research Complex (J-PARC) in Tokai, Japan. 
The
experiment will be conducted in two phases. Phase-I will employ a simplified detector and will be used to investigate the beam and backgrounds whilst aiming 
at
a sensitivity 2 orders of magnitude better than the current limit. 
Phase-II will use the information gained in Phase-I, a much more intense beam and a more complex detection system to achieve a further two orders of magnitude of sensitivity. A third phase, PRISM (Phase-Rotated Intense Slow Muons)~\cite{Pasternak:2010zz,JPARC-LOI-L25}, is being investigated and could potentially provide a further factor of 100 improvement.

\section{Charged Lepton Flavour Violation and Muon to Electron Conversion}

In the 
minimal extension of the SM, Dirac masses for neutrinos are incorporated and total lepton number remains a good symmetry, but leptonic mixings are possible including CLFV. 
Although
allowed at the loop level (mediated by massive neutrinos and $W^\pm$ bosons), CLFV processes such as
radiative decays ($\ell_i \to \ell_j \gamma$) yield extremely small rates, being suppressed by 
$(m_\nu/M_W)^2$, where $m_\nu$ and $M_W$ are the masses of neutrino and W boson, respectively. As an example 
in the SM extended with massive Dirac neutrinos, 
the predicted rate for a $\mu \to e \gamma$ transition is
~\cite{Petcov:1976ff,Bilenky:1977du,MARCIANO1977303,PhysRevLett.38.937}
\begin{equation}\label{eq:lfv.exp.BR.SMmnu}
\text{BR}(\mu \to e \gamma) = \frac{3 \alpha}{32 \pi} 
\left|\sum_{j=1}^3 U_{ej}U_{\mu j}^* \frac{m_{\nu_j}^2}{M_W^2} \right|^2 
\simeq  \mathcal{O}(10^{-55}\textrm{--}10^{-54}),
\end{equation}
where best-fit values for neutrino data ($m_{\nu_j}$ for neutrino mass, and $U_{ij}$ for the element of PMNS(Pontecorvo-Maki-Nakagawa-Sakata) neutrino mixing  matrix) 
were used.

The observation of a CLFV signal would thus require a more ambitious extension of the SM. Many appealing 
New Physics 
models---motivated by explaining the 
observational and theoretical issues of
the
SM
---not only allow for CLFV, but predict rates that could be within current and future experimental sensitivity.
Muons are one of the best laboratories to look for CLFV, and several processes can be studied, all associated with the conversion of muon to electron flavour. In addition to the radiative and three-body decays, $\mu \to e \gamma$ and $\mu \to 3e$, flavour violating muon conversions can occur in the presence of nuclear Coulomb interactions. This is the case of  \muec, a process which yields a monoenergetic electron and hence an excellent experimental signature.

\subsection{ \muec conversion }\label{sc:whatis}

One of the most important muon CLFV processes is the coherent neutrinoless
conversion of muons to electrons (\mue conversion). When a negatively charged
muon is stopped in matter, a muonic atom is formed and,
after cascading down the
energy levels, the muon becomes bound in the $1s$
ground state. It will then normally either decay in orbit
($\mu^{-} \rightarrow e^{-}\nu_{\mu}
\overline{\nu}_{e}$) or be captured by the nucleus  $\mu^{-} N(A,Z) \rightarrow \nu_{\mu} N(A,Z-1)$.
However, BSM processes can lead to
 neutrinoless muon capture.
\begin{equation}
\mu^{-} + N(A,Z) \rightarrow e^{-} + N(A,Z).
\end{equation}
This violates the conservation of individual
lepton flavours, $L_{e}$ and $L_{\mu}$, but conserves
the total lepton number, $L_\text{total}$.

The branching or conversion ratio 
\footnote{
As \mue conversion is not a decay process, it is correct to call its probability as conversion ratio or CR. 
However, as it is widely called as branching ratio in various literature, we call it as branching ratio or BR through out this paper.
}
of \mue conversion is defined as
\begin{equation}
{\rm BR}(\mu^{-}N \rightarrow e^{-}N)  \equiv {\Gamma(\mu^{-}N \rightarrow e^{-} N)
\over \Gamma(\mu^{-}  N \rightarrow {\rm capture} )},
\end{equation}
in which $\Gamma$ is the decay width. The time distribution of \mue conversion
follows the lifetime of a muonic atom, which depends on the type of nucleus. For aluminium it is 864 ns.

\begin{figure}[htb!]
\begin{center}
 \includegraphics[width=0.4\textwidth]{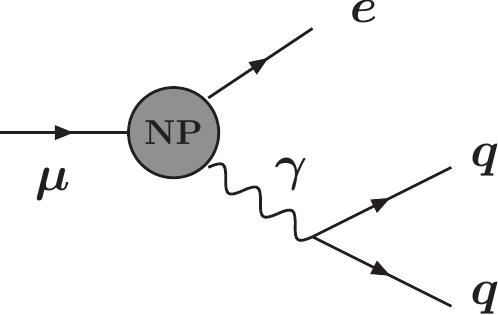}
\hspace*{0.1\textwidth}
 \includegraphics[width=0.4\textwidth]{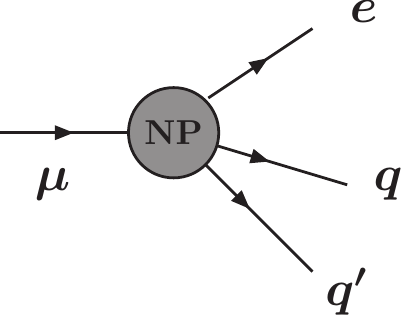}
\caption{Schematic description of the two (tree level) effective
  contributions to \mue conversion: on the left (right) panel,
  the photonic (four-fermion/contact) interaction. The shaded circles denote a BSM flavour violating interaction.}
\label{fig:NewPhysics}
\end{center}
\end{figure}

\paragraph{Photonic and non-photonic contributions}\label{sc:photonic}
As schematically depicted in the panels of \cref{fig:NewPhysics}, two distinct contributions can give rise to  \mue conversion: the photonic
(electromagnetic) dipole contribution, responsible for
$\mu \to e \gamma$ decays and the non-photonic (contact) interaction,
which does not contribute to radiative CLFV muon decays. While for the
former the photon is absorbed by the capturing nucleus,
for the latter CLFV is due to the exchange of
heavy virtual particles that couple
to the quark system\footnote{To calculate the rate
of \mue conversion, proper treatments from the quark level to the
nucleon level, and to the nucleus level have to be
made~\cite{Kosmas:1999cw}.}.
This is in contrast to the $\mu \rightarrow e\gamma$
process which is only sensitive to electromagnetic
dipole interactions. Thus if CLFV transitions are observed, a comparison
between the results from dedicated experiments (for example MEG and
COMET) can  be a powerful discriminator between CLFV extensions of
the SM. With solely BSM dipole interactions~\cite{Marciano} the
rate of the $\mu \rightarrow e\gamma$ process is typically
$\mathcal{O}(200-400)$ times that of the
neutrinoless \mue conversion process
~\cite{Czarnecki:1997pa}. The ratio between the long-distance photonic contribution to the \mue conversion process
and the rate of $\mu^+ \to e^+ \gamma$ decays can be parametrized
by
\begin{eqnarray}
{\text{BR}(\mu^{+}\rightarrow e^{+}\gamma)
\over \text{BR}(\mu^{-}N\rightarrow e^{-}N)}
= {96 \pi^3\alpha \over G^2_{F}m_{\mu}^4}
{1\over 3\times 10^{12}  B(A,Z)} %\cr
\sim {428 \over B(A,Z)}.
\end{eqnarray}
where $B(A,Z)$ represents the rate dependence on the
target nucleus with mass and atomic numbers ($A$ and $Z$).
This has been
calculated based on various approximations, e.g. using 
$B(A=27,Z=13)=1.1$
~\cite{Czarnecki:1998iz}, one obtains a
BR(\mueg)/BR($\mu N\rightarrow eN$) 
of 389 for \isotope[27]{Al}.

A wide variety of New Physics models mediated by (pseudo)scalar, (axial) vector, or tensor
currents can give rise to
short-distance (non-photonic) CLFV interactions.
Following~\cite{Cirigliano:2009bz}, the width for  muon to electron
conversion (for a target $N$) can be written as
\begin{align}
  \Gamma(\mu^{-} N \rightarrow e^{-} N) = \frac{m_\mu^5}{4 \Lambda^4}
& \left|
e  C_L^D  D_N  + 4 \sum_{h=p,n} \left\{
G_F m_\mu m_h S_N^{(h)}
\left(\sum_{q=u,d,s}
\frac{C^{SLL}_{qq} +C^{SLR}_{qq}}{m_\mu m_q G_F} f^{(q)}_{Sh}
+ \tilde{C}^L_{gg} f_{Gh} \right) \right. \right.
\nonumber \\
& \left. \left.
+ V_N^{(h)} \left(
\sum_{q=u,d,s} ( C^{VRL}_{qq} +C^{VRR}_{qq} ) f^{(q)}_{Vh}
\right) 
\right\}
\right|^2 + (L \leftrightarrow R).
\end{align}
where $C$ denote the Wilson coefficients evaluated
at the nucleon scale (for
example, $ C^D$ is responsible for the dipole interactions at the
origin of $\mu \to e \gamma$), $f^{(q)}_n$ is the nucleon form factors,
and the quantities $D_N$, $S_N$ and $V_N$ are target-dependent.

\begin{figure}[htb!]
\begin{center}
\includegraphics[width=0.8\textwidth]{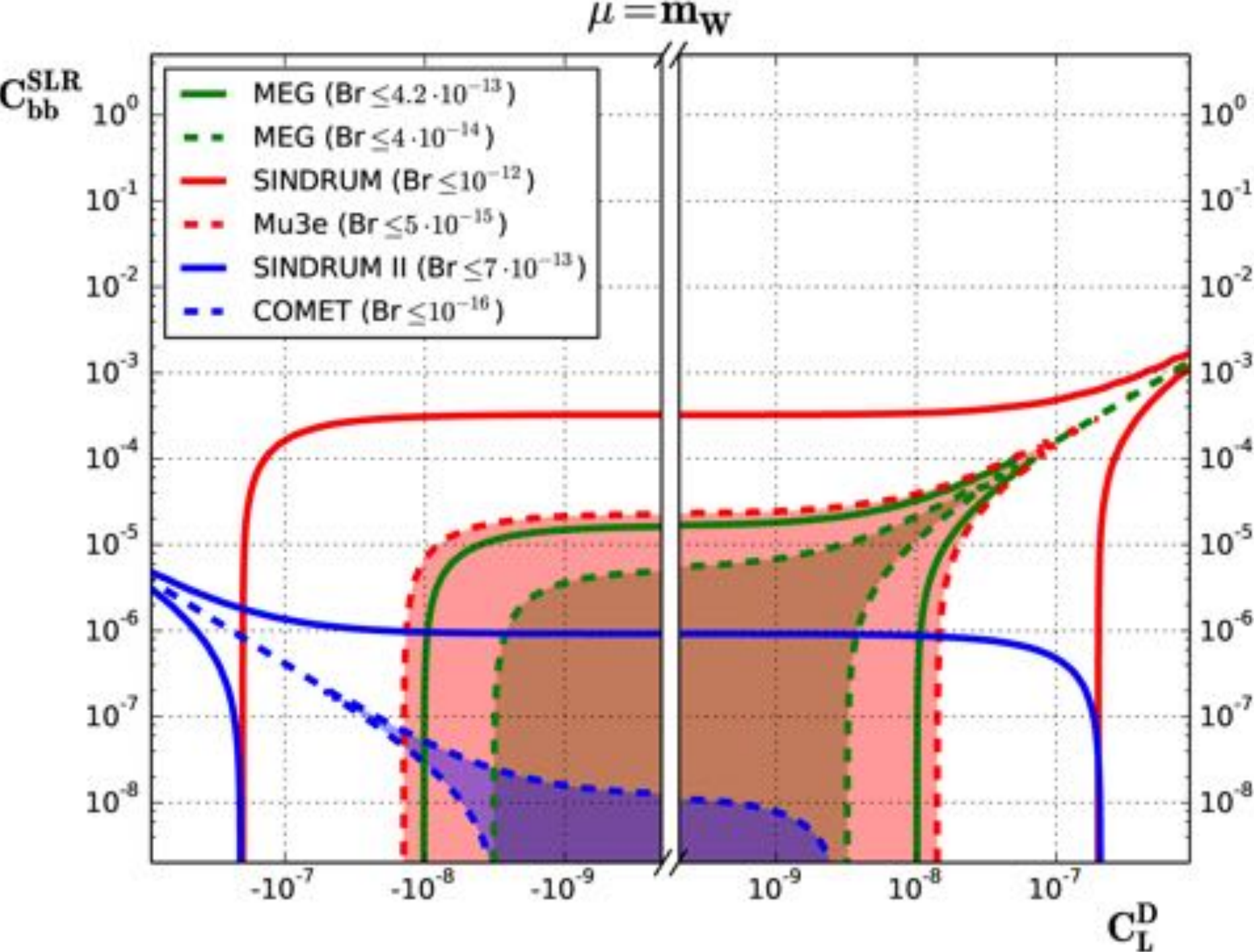}
\caption{Allowed regions in the $C_L^D - C_{bb}^{S LR}$ parameter space,
from $\mu \to e \gamma$ (green), $\mu \to 3 e$ (red) and $\mu \to e$
  conversion (blue), for current experimental bounds (full lines) and future
sensitivities (dashed lines).
From~\cite{Crivellin:2017rmk}.}
\label{fig:effective.RGE}
\end{center}
\end{figure}

As discussed in~\cite{Crivellin:2017rmk}, while the current experimental
bounds on $\mu \to e \gamma$ are clearly powerful in constraining the
dipole operators (and indirectly other scalar and tensor operators, via
mixing effects), neutrinoless muon to electron conversion is
the most sensitive observable to explore operators involving quarks; it
also appears to be the best to study (most) vector
interactions and, as a result of the cleaner experimental conditions, it may even eventually prove
more sensitive to the dipole operators than $\mu \to e \gamma$.   For the four-fermion operators, those involving $b$, $c$ or $s$ quarks can lead to significant contributions and thus to important
constraints due to
the renormalisation effects, whilst
 the
three-body decay, $\mu \to 3e$, is the most powerful observable to
explore and constrain the four-fermion operators with $\mu eee$ flavour
structure. An example of the comparative constraining power of
different CLFV processes on pairs of effective couplings can be seen in
\cref{fig:effective.RGE} (obtained under several simplifying
assumptions).

\bigskip

\begin{figure}[htb!]
\begin{center}
\includegraphics[width=0.8\textwidth]{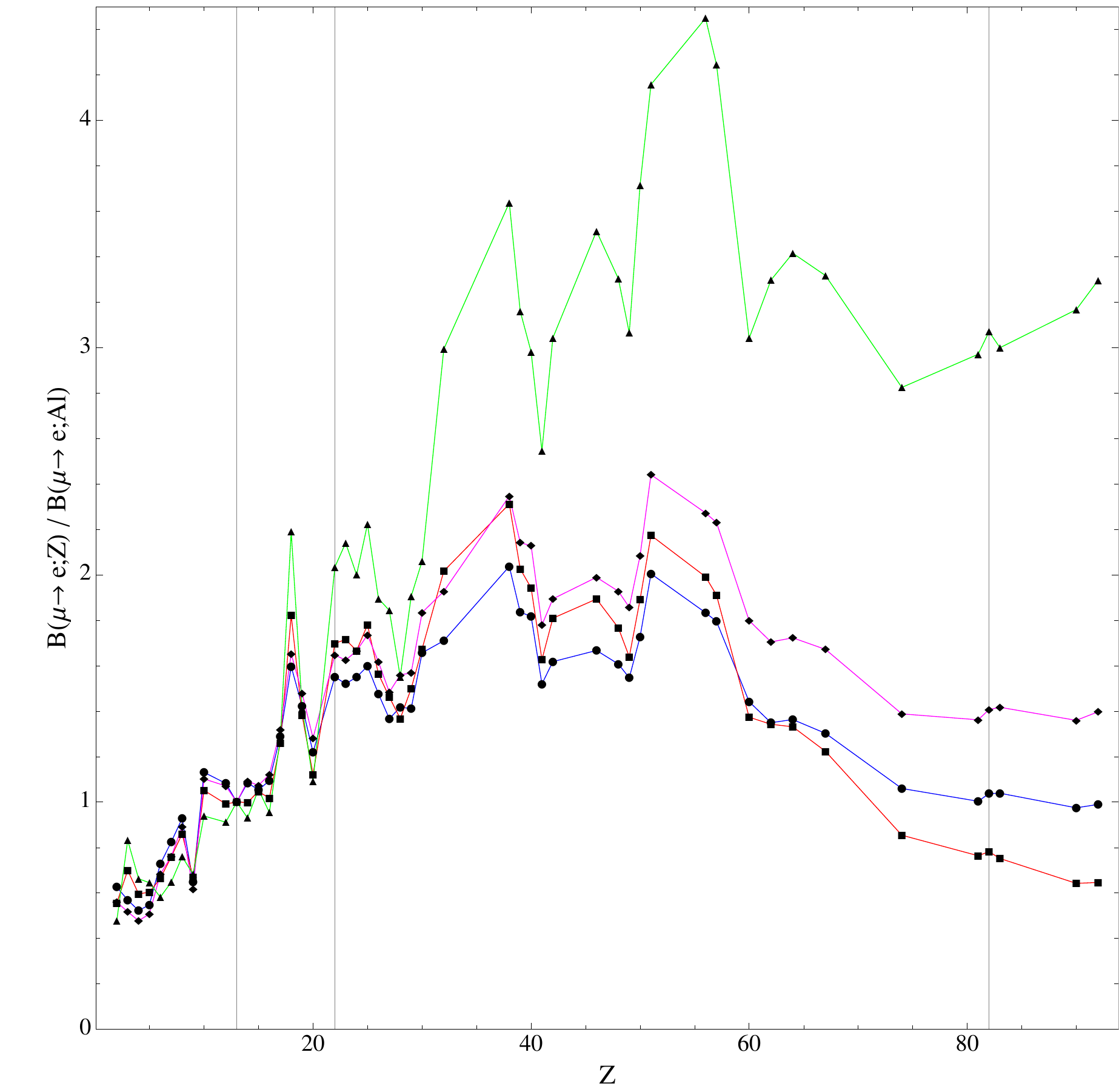}
\caption{Target dependence of the muon to electron conversion rate
in different single-operator dominance models. The different lines
correspond to the \mue conversion rate for a given atomic number $Z$,
normalised to the rate in Aluminium ($Z$ = 13), as a function of
the atomic number, for four theoretical models: dipole interaction
(blue), scalar interaction (red), vector interaction with photons
(magenta), vector interaction with $Z^0$ bosons (green). The vertical
lines correspond to $Z$ = 13 (Al), $Z$ = 22 (Ti), and $Z$ = 82 (Pb).
Taken from~\cite{Cirigliano:2009bz}.}
\label{fig:muonstoppingtarget}
\end{center}
\end{figure}

\paragraph{Dependence on muon-stopping target material}
The rates of coherent \mue conversion  for general
effective CLFV interactions (such as dipole, scalar and vector
interactions) have been calculated for various
nuclei~\cite{Kitano:2002mt,Cirigliano:2009bz} taking
into account relativistic wave functions as well as
the proton and neutron distributions
(with associated ambiguities). The results, shown in \cref{fig:muonstoppingtarget}, indicate that the
branching ratios for \mue conversion increase for light nuclei up to the
atomic number of $Z\ \sim 30$, remain large for the region of $Z = 30 - 60$,
and then decrease for heavy nuclei of $Z>60$.
It should be noted that the different physics models estimate different estimations on the \mue conversion rates, which also depend on the target materials. Therefore, experiments with different target materials and comparing the results are important to have a hint on the CLFV model.

\paragraph{Spin-dependent and spin-independent contributions}
CLFV tensor and axial-vector four-fermion operators could
also contribute to \mue conversion; these
couple to the spin of
nucleons, and can therefore mediate a spin-dependent
\mue conversion~\cite{Cirigliano:2017azj,Davidson:2017nrp}.
As these have a different atomic number dependence than the spin independent terms this could in principle be used to investigate the operators responsible.

\subsection{CLFV Models and \mue Conversion}
\begin{figure}[h]
\begin{center}
\includegraphics[clip, width=0.4\textwidth]{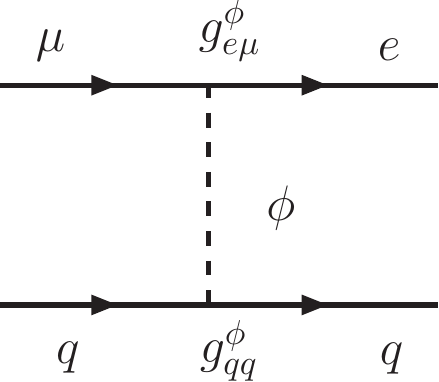}
\end{center}
\caption{
Schematic representation of \muec conversion at the tree-level
($t$-channel exchange), mediated by a scalar or vector
boson ($\phi$).
}
\label{fig:mueconv_tree_phi}
\end{figure}

BSM models can lead to
significant CLFV contributions via the introduction of new sources of
flavour violation (corrections to SM vertices, or new flavour
violating interactions) and/or new currents. The size of these contributions depends upon the New Physics model; however, the generic features for \muec conversion can be illustrated with the following (new) interaction terms in the BSM Lagrangian:
\begin{eqnarray}
& \mathcal{L}_1 \,\sim \, g^\phi_{e \mu}\,
\bar\mu \,e \,\phi \,+\, g^\phi_{q q} \, \bar{q}\,q\, \phi \,,
\label{eq:lfvcoupling1concrete} \\
& \mathcal{L}_2 \,\sim \, h_{\mu \psi \phi}\,\bar\mu \,\psi\, \phi\,+
\, h_{e \psi \phi}\,\bar{e}\,\psi\, \phi\,.
\label{eq:lfvcoupling2}
\end{eqnarray}
In Eq.~(\ref{eq:lfvcoupling1concrete}),
$\phi$ generically refers to a scalar or vector
boson (under the assumption of appropriate
Lorentz contractions), while $g^\phi_{ff}$ denotes its couplings to fermions,
which must be non-diagonal  for charged leptons. In this case
contributions to CLFV  can occur both at  tree level, as
 depicted in \cref{fig:mueconv_tree_phi}
($t$-channel), or  at higher orders. 

\begin{figure}[h]
\begin{center}
\includegraphics[clip, width=0.4\textwidth]{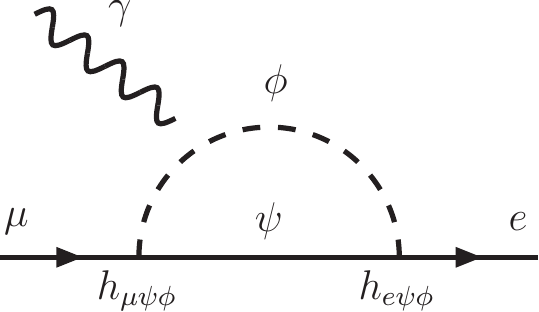}
\hspace*{0.1\textwidth}
\includegraphics[clip, width=0.4\textwidth]{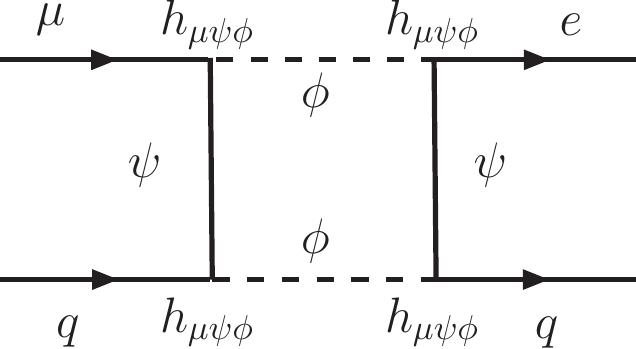}
\end{center}
\caption{
Schematic representation of photonic contributions to \mue flavour
transitions (left panel), and of box-diagram contribution to \muec
conversion (right panel), in the presence of new fermions ($\psi$)
and new vector/scalar bosons ($\phi$).
}
\label{fig:mueconv_phipsi_box}
\end{figure}

The possibility referred to in Eq.~(\ref{eq:lfvcoupling2}) requires new fermions $\psi$ in addition to the vector/scalar bosons.
The combination $\psi\phi$ must carry lepton flavour, and
the coupling $h_{\ell \psi\phi}$ must be non-diagonal.
 In BSM
constructions, at least one of $\psi$ and $\phi$ is a new particle,
and often both are (as in models in which some symmetry ensures
that new particles are
always pair produced). A diagrammatic view of the new contributions can be
found in \cref{fig:mueconv_phipsi_box}.

\paragraph{Tree-level
 contributions to \mue conversion }

\begin{figure}[t!]
\begin{center}
\includegraphics[clip, width=0.8\textwidth]{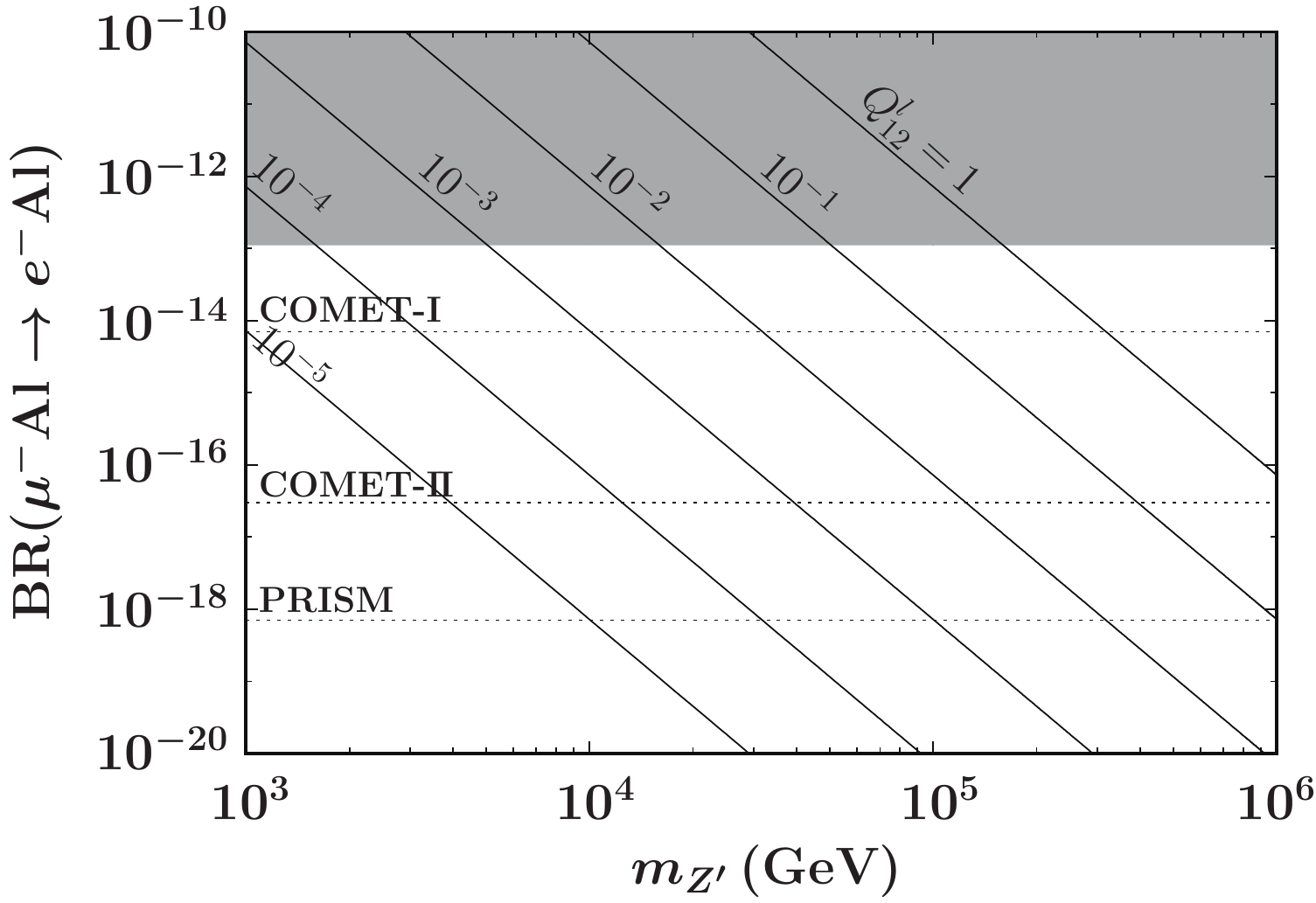}
\end{center}
\caption{The rate of neutrinoless muon to electron conversion as a function of the $Z'$ mass. Diagonal lines denote regimes for the CLFV coupling $Q_{12}^{\ell}$ ($Z'\bar{\ell}_{i} \ell_{i}$). The grey region is excluded by
the SINDRUM experiment. COMET Phases I and II are denoted by ``COMET-I'' and ``COMET-II'' here.
}
\label{Fig:BRZprime}
\end{figure}

Many BSM models
predict the existence of a massive neutral vector boson,
often called a $Z^{\prime}$.
One example is
 the
Sequential Standard Model (SSM), in which the
interaction of the $Z^\prime$ and ordinary matter inherit the
structure of the SM interactions. CLFV arises from  new,
flavour violating couplings of the vector boson to leptons $Z'
\bar{\ell}_{i} \ell_{i}$, whose strength is parametrised by
$Q_{ij}^{\ell}$ couplings~\cite{Murakami:2001cs}.
The LHC experiments have searched for neutral heavy particles decaying
into dileptons of different flavour $e^{\pm} \mu^{\mp}$, and have placed
a lower limit on the mass of a SSM $Z^\prime$, $m_{Z^\prime} \gtrsim
3.01~$TeV (at 95\% C.L.), for $Q_{12}^{\ell} =
1$~\cite{TheATLAScollaboration:2015ucd}.
\Cref{Fig:BRZprime} displays the rate of neutrinoless muon to electron
conversion as a function of the SSM $Z^\prime$ mass, for different
regimes of the CLFV coupling $Q_{12}^{\ell}$. This clearly
shows that  \muec conversion offers a sensitivity to New Physics
scales well beyond the reach of the LHC.

\medskip
Another case is  the type III realisation of the seesaw
mechanism, in which the SM content is extended by two or more
generations of fermion triplets, $\Sigma$~\cite{Foot:1988aq, Ma:1998dn}.
 The charged states of the
triplet mix with charged leptons  leading to lepton flavour violating
couplings to the $Z^0$ boson.
This induces significant differences in
the contributions to CLFV observables. Radiative
decays  remain a one-loop transition whilst $\ell_i \to 3 \ell_j$
decays and neutrinoless \mue conversion can occur at
tree-level. For the \mue system one finds~\cite{Abada:2008ea},
\begin{equation}
\text{BR}(\mu \to e \gamma) =
1.3  \times 10^{-3}  \times  \text{BR}(\mu \to 3e) =
3.1  \times 10^{-4}  \times   \text{BR}(\mu^{-} \text{Ti} \to e^{-} \text{Ti}),
\end{equation}
in striking difference with other seesaw realisations.

\begin{figure}[t!]
\begin{center}
\includegraphics[clip, width=0.8\textwidth]{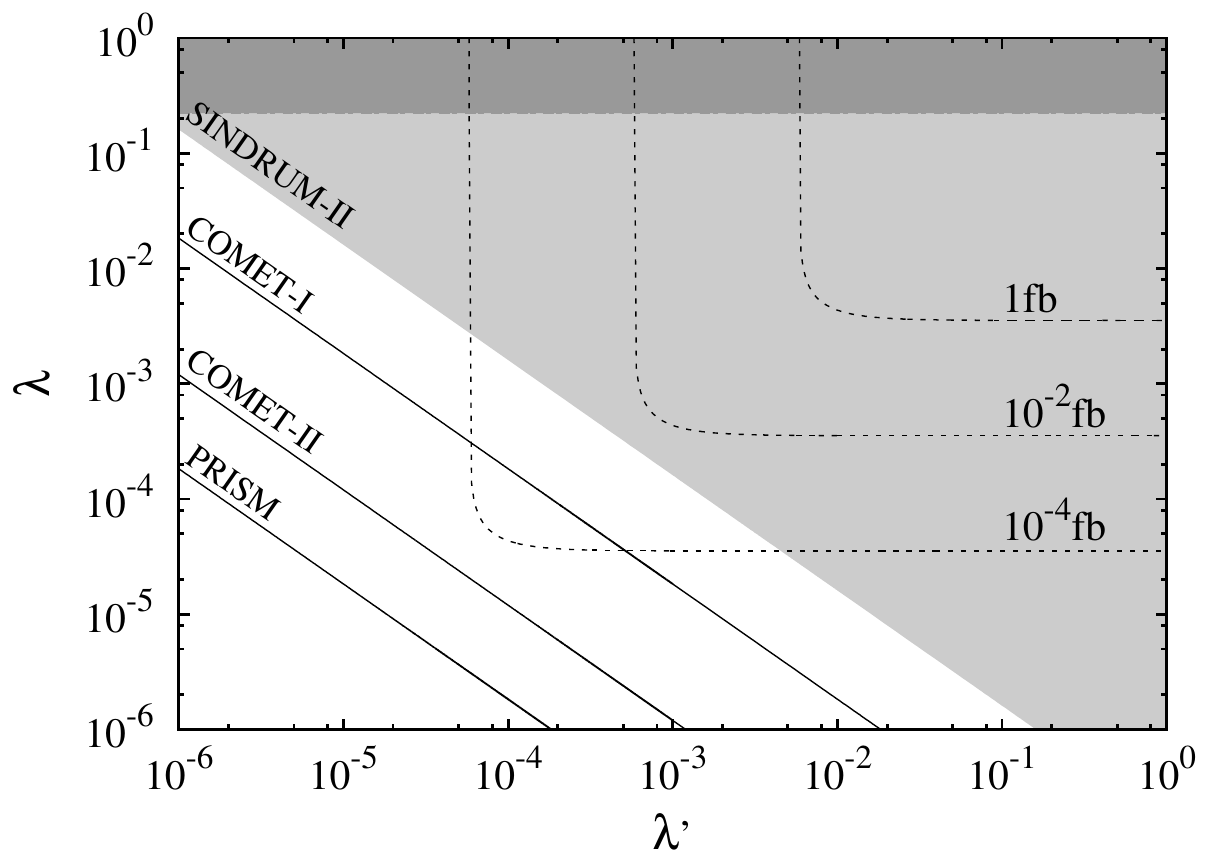}
\end{center}
\caption{Exclusion contours in 
$\text{BR}(\mu^{-} \text{Al} \to e^{-} \text{Al})$ 
and $\sigma (pp \to \mu^{-} e^{+})$ in
an RPV scenario, for $m_{\tilde{\nu}} = 1\,\text{TeV}$,
and $\lambda = \lambda_{312} = \lambda_{321}
= - \lambda_{132} = - \lambda_{231}$.
The shaded regions denote limits on
$\lambda$ (dark shaded band) and the combination $(\lambda, \lambda')$
(light shaded region) arising from CLFV
muonium-antimuonium oscillation searches and from SINDRUM bounds on
\muec conversion, respectively. COMET Phases I and II are denoted by ``COMET-I'' and ``COMET-II'' here.
Taken from~\cite{Sato:2014ita}.}
\label{Fig:RPV}
\end{figure}

SUSY models which violate R-parity can lead to sizable
CLFV rates giving tree-level contributions to \muec conversion mediated by
($t$-channel) scalar exchange. In this case the scalar neutrino $\tilde{\nu}_{iL}$
plays the role of $\phi$ in Eq.~(\ref{eq:lfvcoupling1concrete}).
LHC searches for different flavour dileptons have set limits on the
$\tilde{\nu}_{\tau}$ mass, $m_{\tilde{\nu}_{\tau}} > 1.0\,\text{TeV}$
for $\lambda_{132} = \lambda_{231} = \lambda'_{311} = 0.01$, and
$m_{\tilde{\nu}_{\tau}} > 3.3\,\text{TeV}$ for $\lambda_{132} =
\lambda_{231} = \lambda'_{311} = 0.2$~\cite{CMS:2016dfe} (with $1,2,3$
denoting $e,\mu$ and $\tau$ flavours).
\Cref{Fig:RPV} shows the contour of 
$\text{BR} (\mu^{-} \text{Al} \to e^{-} \text{Al})$ 
and $\sigma (pp \to
\mu^{-} e^{+})$ for $m_{\tilde{\nu}} = 1\,\text{TeV}$.
The LHC limit excludes only a part of the upper-right corner, leading
to much looser constraints than those imposed by the SINDRUM limit on
\muec conversion. In fact, a synergy of different muon CLFV processes,
muon $g-2$ and direct LHC search results leads to powerful constraints on the
RPV model parameters~\cite{Sato:2014ita}.

BSM contributions to
\muec conversion can also be mediated by the exchange of a scalar particle in the $s$-channel. This corresponds to interactions of the type
$h_{\ell q \phi}\,\bar\ell \,q\, \phi$, as given in
Eq.~(\ref{eq:lfvcoupling2}) and can be realised if the scalar
mediator carries both hadron and lepton numbers, as is the case of
(scalar) leptoquarks models~\cite{Dorsner:2016wpm}.

\paragraph{Loop contributions to \mue conversion}

Several higher-order processes including photon, $Z$ and Higgs
penguins and  boxes mediated by fermions and vector/scalar
bosons, with both SM and new exotic particles, can induce \muec.
Although contributions can occur at two
loop-order or even higher levels,
most BSM constructions induce CLFV observables already at the loop level.

\begin{figure}[h]
\begin{center}
\includegraphics[clip, width=0.4\textwidth]{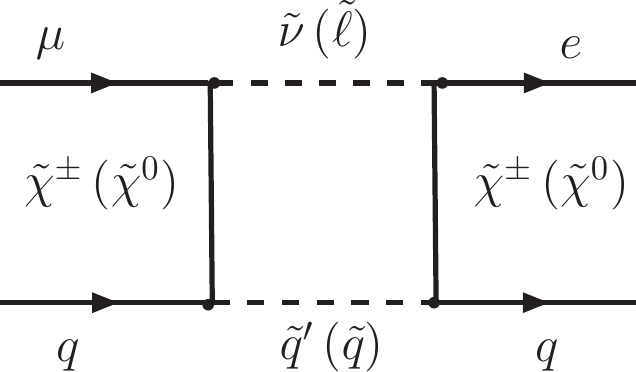}
\hspace*{0.1\textwidth}
\includegraphics[clip, width=0.4\textwidth]{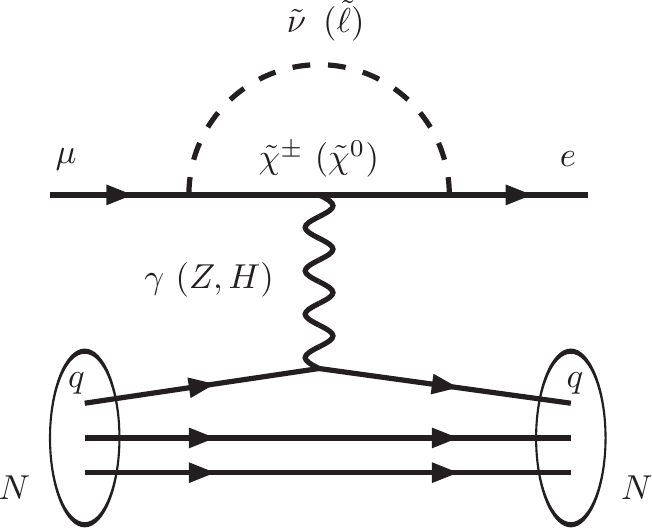}
\end{center}
\caption{
Diagrams for \muec conversion via
box (left) and penguin (right) diagrams, arising from SUSY R-parity conserving interactions, $\chi$ denotes a neutralino or chargino, $\tilde{\nu}$ $(\tilde{\ell})$
represents a scalar lepton (neutral or charged) and
$\tilde{q}$ stands for a scalar quark.
}
\label{fig:SUSYexample}
\end{figure}

From the interaction terms  in Eq.~(\ref{eq:lfvcoupling2}), when both $\psi$ and $\phi$ are new particles, CLFV
transitions are only  realised  by higher order
processes: loops are constructed by connecting $\mu$ to $e$
via exotic fermion ($\psi$) and boson ($\phi$) closed lines.
The
addition of interactions with quarks such as $\bar{q}\psi\phi$ allows
for box diagram contributions to \muec conversion.
An illustration is provided by SUSY models with R-parity conservation, with an example of the contributions to CLFV observables depicted in \cref{fig:SUSYexample}.

General SUSY models (as is the case of the MSSM) do not offer  an
explanation for neutrino oscillation, and are thus a good
illustrative example of having CLFV completely decorrelated from
flavour violation in the {\it neutral} lepton sector.
An estimate of generic
SUSY contributions to radiative lepton decays
(e.g. $\mu \to e \gamma$) arising from loops mediated by
charged sleptons (sneutrinos) and neutralinos (charginos), leads to
the approximate expression
\begin{equation}
\text{BR}(\mu \to e \gamma)\, \sim \,
\frac{\alpha}{4 \pi} \left(\frac{M_W}{M_\text{SUSY}}\right)^4
\, \sin^2 \theta_{\tilde e \tilde \mu } \,
\left(\frac{\Delta m^2_{\tilde \ell}}{M^2_\text{SUSY}}\right)^2\,,
\end{equation}
in which $M_\text{SUSY}$ denotes the SUSY
breaking scale, $\theta_{\tilde e \tilde \mu }$
and $\Delta m^2_{\tilde \ell}$ corresponding to the
slepton mixing angle and mass square difference.

Similarly with the tree-level case, 
in a regime in which
 virtual photon exchange proves to be the dominant
contribution to \mue conversion (e.g.~\cite{Hisano:2001qz,Arganda:2007jw},
and the comparative study of~\cite{Abada:2014kba}),
one also recovers a relation
between the latter and the radiative decays,
\begin{equation}\label{eq:models.lfv:susy:CLFV:relations.3}
\text{BR}(\mu^{-}N\to e^{-}N)\, \approx \, \mathcal{O}(\alpha) \, \times \,
\text{BR}(\mu \to e \gamma)\,.
\end{equation}

\subsection{Other Possible BSM Processes}

\subsubsection{Lepton number-violating $\mu^{-}N \rightarrow e^{+}N^{(\prime)}$ conversion}
If the New Physics responsible for CLFV also includes a source of Lepton Number Violation (LNV; in
general associated with the presence of BSM Majorana states), then the
muonic atom can  undergo both a CLFV and LNV
transition~\cite{Littenberg:2000fg},
\begin{equation}
\mu^{-} + N(A,Z) \rightarrow e^{+} + N^{\prime}(A,Z-2)\,,
\end{equation}
Having different
initial and final state nuclei  precludes the coherent
enhancement of the transition amplitude---which implies that it will
not be augmented in large $Z$ atoms. The experimental signal is less
clean than that of the coherent conversion; the emitted
positron is no longer monoenergetic and there are more sources of
background~\cite{kamal1979kamal, zee1980theory, vergados1981study, vergados1982study,leontaris1983study,  KOSMAS1994397, missimer1994muonic, babu1995new, domin2004nuclear,  PhysRevD.96.075027,PhysRevD.95.055009, GEIB2017157, PhysRevD.95.115010}.

\subsubsection{CLFV muonic atom decay $\mu^-e^-\to e^-e^-$}
In the presence of new physics, there is another CLFV process that
can occur in a muonic atom, the Coulomb enhanced decay into a pair of
electrons~\cite{Koike:2010xr},
\begin{equation}
\mu^-\, +\, e^-\, \to e^-\, +\, e^-\,,
\end{equation}
in which the initial fermions are the muon and the atomic $1s$ electron. As with the neutrinoless
conversion, this can be induced by dipole
and contact interactions. Experimentally this has several advantages  compared with  $\mu \to e \gamma$, or $\mu \to 3 e$. Compared with $\mu \to e \gamma$ the CLFV muonic atom decay is sensitive to both contact and
dipole interactions and its measurement is easier as no photon
detection is involved; when compared to $\mu \to 3 e$, the
new observable has a larger phase space and a cleaner experimental
signature, consisting of back-to-back electrons with a well defined
energy ($\sim m_\mu/2$).

\begin{figure}[htb!]
\begin{center}
\hspace*{10mm}
\includegraphics[height=0.8\textwidth,angle=270]{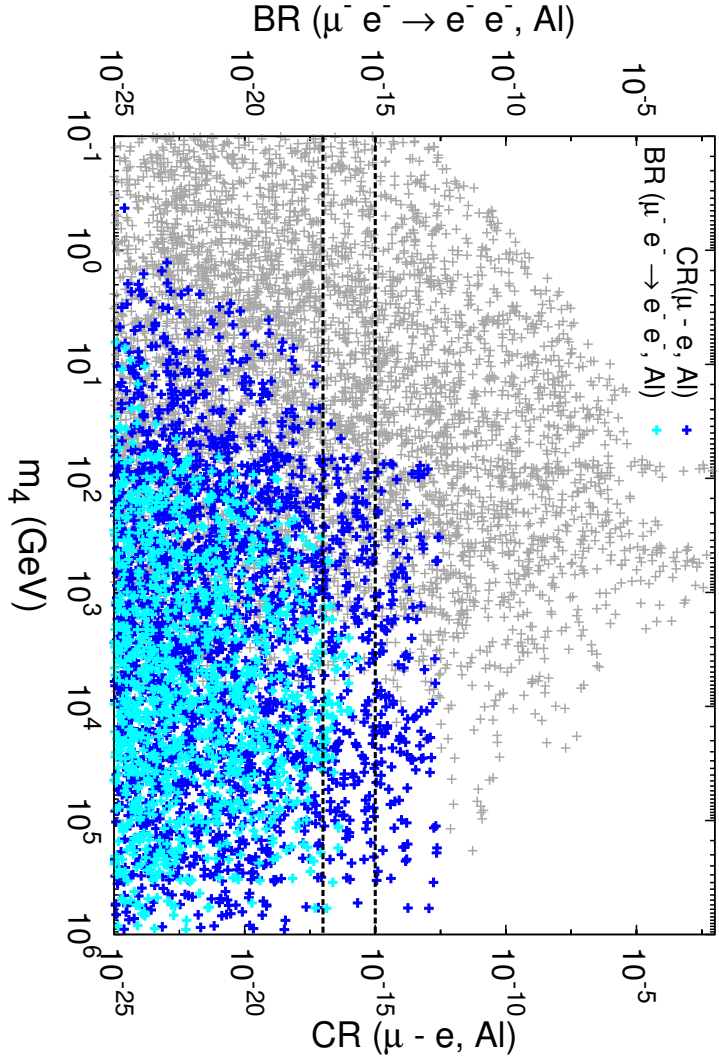}
\caption{ \mue
conversion (blue) and $\mu e \to e e$ (cyan) for Aluminium targets,
as a function of the sterile fermion mass ($m_4$), in a ``3+1'' toy
model; horizontal dashed
lines denote COMET's Phase-I and II sensitivities~\cite{Abada:2015oba}.
}
\label{fig:seesaw:clfv}
\end{center}
\end{figure}

A phenomenological study of this observable was carried out
in~\cite{Abada:2015oba}, in the framework of the SM extended by
sterile fermions (ad-hoc extensions and a (3,3) Inverse Seesaw
realisation); the comparative prospects for the COMET experiment are
displayed in \cref{fig:seesaw:clfv}.

\subsection{Experimental Aspects of \muec}
The event signature of coherent neutrinoless \mue conversion in a muonic atom
is the emission of a mono-energetic single electron in a defined time interval. The  energy of the signal electron ($E_{\mu e}$) is given by
\begin{equation}
E_{\mu e} = m_{\mu} - B_{\mu} - E_\text{recoil}
\end{equation}
where $m_{\mu}$ is the muon mass, $B_{\mu}$ is the binding energy of
the 1$s$-state muonic atom, and $E_\text{recoil}$ denotes the nuclear recoil energy
which is small. For aluminium  $E_{\mu e}$ = 104.97~MeV and the lifetime of the muonic atom is 864 ns.

This makes neutrinoless
\mue conversion  very attractive experimentally.
 Firstly, the $e^{-}$ energy of about 105~MeV is well above the
end-point energy of the muon decay spectrum ($\sim
52.8$~MeV). Secondly, since the event signature is a mono-energetic
electron, no coincidence measurement is required. Thirdly, the long lifetime means backgrounds associated with the beam flash can be eliminated. Thus the search for
this process has the potential to improve sensitivity by using a high
muon rate without suffering from accidental background events. Backgrounds are discussed in more detail in \cref{ch:SensitivityBackgrounds}.

\section{The COMET Experiment}

COMET stands for COherent Muon to Electron Transition and the
experiment seeks to measure the neutrinoless, coherent
transition of a muon to an electron  (\mue conversion) in the field of an aluminium
nucleus.
The  experiment will be carried out using a two-staged approach. 
\begin{figure}[tb!]
 \begin{center}
 \includegraphics[width=\textwidth]{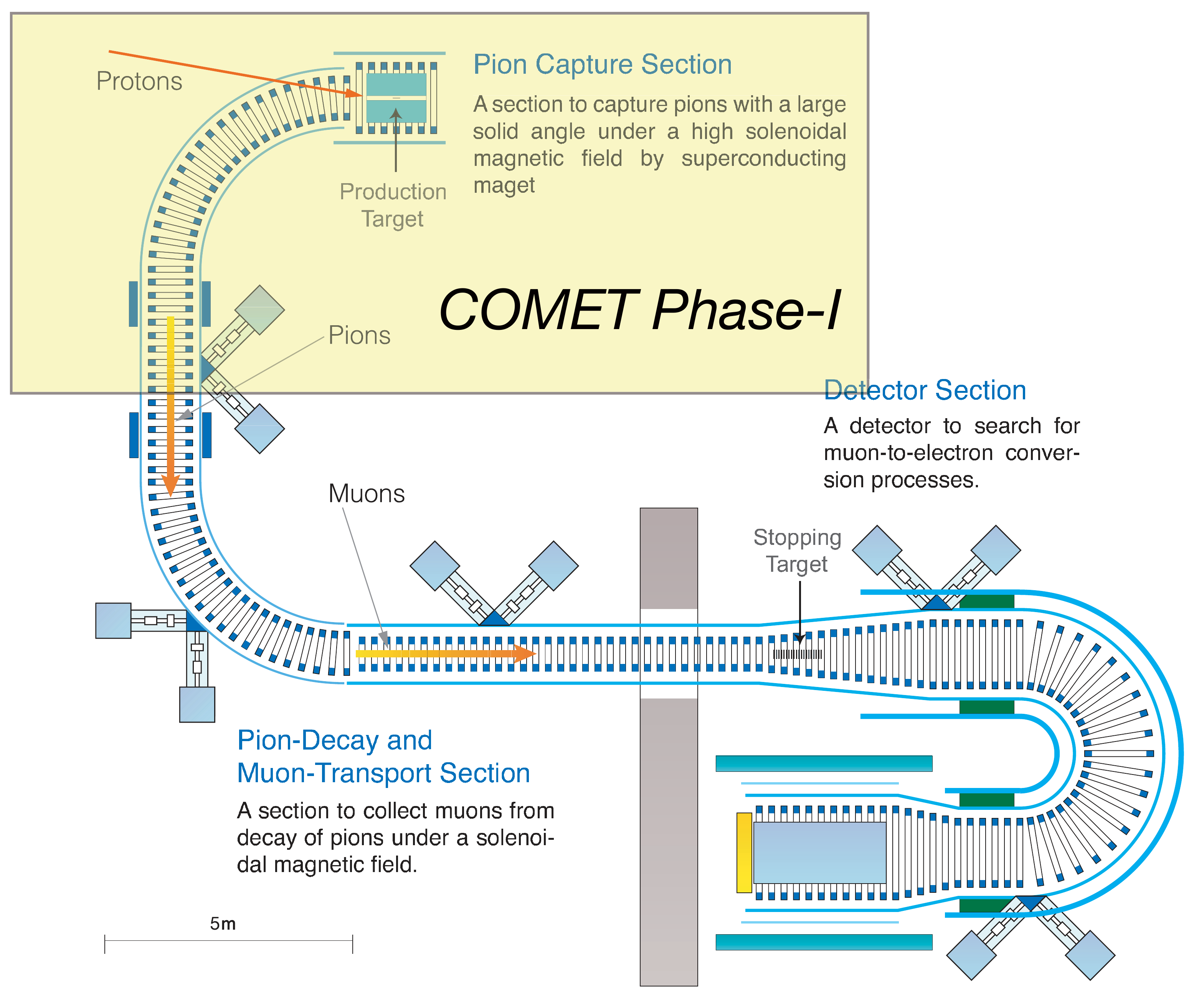}
 \end{center}
 \caption{%
 Schematic layout of COMET (Phase-II) and COMET Phase-I (not to scale).}
 \label{fig:comet-layout}
\end{figure}

The COMET Phase-I aims at a single event sensitivity (SES) of $3.1 \times 10^{-15}$,  roughly a factor 100 better than the current experimental limit.
The goal of the full experiment is a SES of $2.6 \times 10^{-17}$, which we refer to as Phase-II.
This ultimate sensitivity goal is a factor of about 10,000 better than the current experimental limit of $B(\mu^- + \mbox{Au}\rightarrow e^- + \mbox{Au}) \le 7 \times 10^{-13}$ from SINDRUM-II at PSI~\cite{Bertl:2006up}. 

A schematic layout of the COMET experiment is shown in \cref{fig:comet-layout}.  The experiment will be carried out in the Nuclear and Particle Physics Experimental Hall (NP Hall) at J-PARC using a bunched 8~GeV proton beam that is slow-extracted from the J-PARC main ring.
Muons for the COMET experiment will be generated from the decay of pions produced by collisions of the 8~GeV proton beam on a production target.
The yield of low-momentum muons transported to the experimental area is enhanced using a superconducting pion-capture solenoid surrounding the proton target in the pion-capture section shown in \cref{fig:comet-layout}. Muons are momentum- and charge-selected using curved superconducting solenoids in the muon-transport section, before being stopped in an aluminium target. The signal electrons from the muon stopping target are then  transported by additional curved solenoids to the main detector, a straw-tube tracker and  electron calorimeter, called the StrECAL detector.

\subsection{COMET Phase-I}
\begin{figure}
 \begin{center}
 \includegraphics[width=\textwidth]{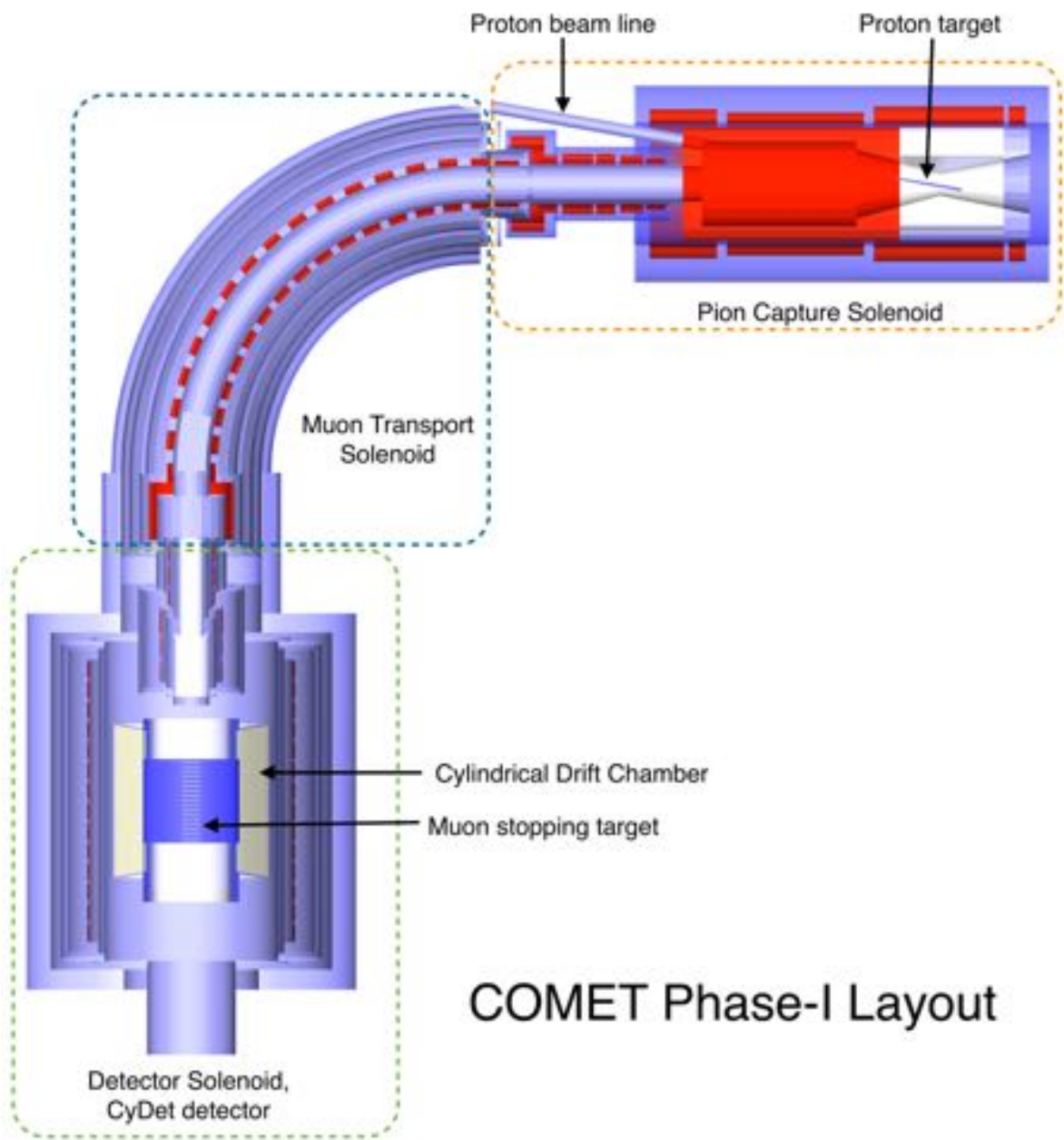}
 \end{center}
 \caption{%
 Schematic layout of COMET Phase-I.
 }
 \label{fig:CometPhaseILayout0}
\end{figure}

The COMET Phase-I will have the pion-capture and the muon-transport sections up to the end of the first 90$^{\circ}$ bend of the full experiment. The  muons will then be stopped in the aluminium target at the centre of a cylindrical drift chamber in a 1T magnetic field.  A schematic layout of the COMET Phase-I setup is shown in \cref{fig:CometPhaseILayout0} and  an illustration of how COMET Phase-I relates to Phase-II indicated in \cref{fig:comet-layout}.
For COMET Phase-I,
the primary detector for the neutrinoless \mue
conversion signals consists of a   cylindrical drift chamber and a set of trigger
hodoscope counters, referred to as the CyDet detector.
The experimental setup for Phase-I will be augmented with
prototypes of the Phase-II StrECAL detector. As well as
providing valuable experience with the detectors, the StrECAL and CyDet detectors
will be used to
characterise the beam and measure backgrounds  to ensure that the Phase-II single event
sensitivity of $2.6\times 10^{-17}$ can be realised~\cite{Krikler:2016gdq}.

For Phase-I
 a total number of protons on target (POT) of $3.2 \times 10^{19}$ is planned which will provide around  $1.5 \times 10^{16}$ muons  stopped in the target. This will enable  the design goal of COMET Phase-I to be achieved; a single event sensitivity  which, in
the absence of a signal, translates to a 90\% confidence level
branching ratio limit of $7 \times 10^{-15}$. This is a factor of about 100 better than
the current limit on gold from SINDRUM-II~\cite{Bertl:2006up}.
The important experimental parameters of COMET Phase-I and II, and the SINDRUM-II experiment are compared in \cref{tab:expcomparison}. 

\begin{table}[htb!]
\begin{center}
\caption{
Comparison of experimental plans of COMET Phase-I and II, and the SINDRUM-II experiment. 
} \label{tab:expcomparison}
\begin{tabular}{lccc} \hline\hline
Experiment    & SINDRUM-II      & COMET Phase-I   & COMET Phase-II  \\
\hline
Location      & PSI (Switzerland) & J-PARC (Japan) & J-PARC (Japan)  \\
Proton energy & 590 MeV         & 8 GeV           & 8 GeV                    \\
Proton beam power &             & 3.2 kW          & 56 kW               \\
N(proton)     &                 & $3.2 \times 10^{19}$ & $6.8\times 10^{20}$  \\
N(stopped muon) & $4.37 \times 10^{13}$ & $1.5 \times 10^{16}$  & $1.1 \times 10^{18}$    \\
Transport solenoid shape &  Linear & Half C-shape    & Full C-shape      \\
Muon target material
              &       Au        & Al              & Al                          \\
Sensitivity (90 \% C.L.)
              & $7 \times 10^{-13}$ & $7 \times 10^{-15}$ & $2.6\times 10^{-17}$  \\
Total DAQ time &   81 days      & $\sim$150 days        &  $\sim$180 days         \\
\hline\hline
\end{tabular}
\end{center}
\end{table}

\subsection{COMET Requirements}

In order to obtain the desired improvement in sensitivity, the
experiment requires an intense muon source, coming from a pulsed
proton beam with high inter-bunch extinction factor.

\paragraph{Highly intense muon source}
To achieve an experimental sensitivity better than $10^{-16}$, $\mathcal{O}(10^{18})$ muons are needed.
Two methods are adopted
to increase the muon beam intensity.
One is to use a high-power proton beam from J-PARC,
the other is to use a highly efficient pion collection system. The latter is achieved by surrounding the proton target with a $5\,\mathrm{T}$ superconducting solenoid.
 The principle of this pion-capture system has
been experimentally demonstrated at the MuSIC (Muon Science Innovative beam Channel) facility at Research Centre for Nuclear Physics (RCNP), Osaka University~\cite{Tomono:2018zkl}.

\paragraph{Proton beam pulsing with high proton extinction}
There are several potential sources of electron background events in
the signal energy region,
one of which is prompt beam-related background events.  In order to
suppress the occurrence of such background events, a pulsed
proton beam will be employed, where proton leakage between the pulses is tightly
controlled.
As a muon in an aluminium muonic atom has a lifetime of the order of 1~\micro{}s, a pulsed beam can be used to eliminate prompt beam background events by performing measurements in a delayed time window, provided that the beam pulses are shorter than this lifetime and the spacing between them is comparable or longer.
Stringent requirements
on the beam extinction, defined as the number of leakage protons with respect to
the number of protons in a beam pulse, are necessary.  Tuning of the proton beam in
the accelerator ring, as well as making use of additional extinction-improving techniques (such as modifying the timing of kicker magnets), will also be done.

\paragraph{Curved solenoids for charge and momentum selection}
High momentum  muons can  produce electron background events
in the energy region of 100~MeV, and therefore
must be eliminated. This is achieved by transporting the pion/muon beam through a system of curved superconducting solenoids.
As they pass through the curved solenoid,
 the centres of the helical motion of the charged particles  drift perpendicularly to
the plane in which their paths are curved,
with  the magnitude of the drift  proportional to their momentum. To compensate for this a
 dipole field parallel to the drift direction will  be applied for a given reference  momentum to keep the centres of
the helical trajectories in the bending plane. Hence, with suitably placed collimators,
 high momentum and positively charged particles can be eliminated.
Since the muon momentum dispersion is proportional to a total bending
angle, the COMET C-shape beam line  produces a larger separation of the
muon tracks as a function of momentum and hence an improved momentum
selection. In COMET Phase-II, additional curved solenoids will be used in a C-shaped electron transport
system between the muon stopping target and the electron spectrometer to eliminate low-momentum backgrounds to the electron signal.

\subsection{The Phase-I Programme}

The purpose of COMET Phase-I is two-fold. The first is to make
background measurements for COMET Phase-II, and the second is a search
for \mue conversion at an intermediate sensitivity.
COMET Phase-I serves several roles that are highly
complementary to the Phase-II experiment. 
It provides a working experience of many of the
components to be used in Phase-II and enables a direct measurement of backgrounds.  
Significantly it will also produce
competitive physics results,
both of the \mue conversion process 
and of other processes that COMET Phase-II cannot investigate.

\paragraph{Background measurements}Currently, background levels must be estimated by
extrapolating the existing data over several orders of magnitude.
Phase-I will be used to
obtain data-driven estimates of backgrounds, and hence inform the detailed design of COMET Phase-II.
Using a shorter $90^\circ$ muon-transport solenoid in Phase-I enables the investigation of
the secondary beam in the kinematic region that will be used in Phase-II. In Phase-I the StrECAL detector will be placed at the
downstream end of the muon-transport beam line and will be
dedicated to  background measurements, in particular

\begin{itemize}
\item Direct measurement of the inter-bunch extinction factor.
\item Direct measurement of unwanted secondary particles in the
  beam line such as pions, neutrons, antiprotons, photons and electrons.
\item Direct measurement of background processes that have not been
measured at the required accuracy, such as muon decays
  in orbit and radiative muon capture.
\end{itemize}

\paragraph{Search for \mue conversion}
Even in this partial configuration,  COMET Phase-I will  conduct a world-leading measurement of \mue conversion using the CyDet detector located inside a $1~\mathrm{T}$ solenoid magnet surrounding the muon stopping target.
This cylindrical geometry is necessary, since the curved electron transport solenoid will not be deployed in Phase-I and thus a planar type detector such as the StrECAL detector
would suffer from backgrounds caused by beam related particles.

\paragraph{Other searches} In contrast to COMET Phase-II, the CyDet detector
surrounds the muon stopping target directly in Phase-I, and can observe both
positive and negative particles from the muon stopping target.  This allows for a search for the
lepton-number-violating process \mbox{$\mu^{-} N \rightarrow e^{+} N'$
} (\mupc conversion) concurrently with the \muec search. The
anticipated experimental sensitivity for \mupc conversion could be
similar to \muec conversion, although a detailed estimation has not
yet been performed.
In addition, the Cylindrical Drift Chamber will have a relatively large
geometrical coverage, and thereby a coincidence measurement with a
large solid angle is achievable.  This allows a search for \mbox{\metoee}
conversion in a muonic atom, which is an as-yet unmeasured
process. Using a lower intensity beam, $< 10^{7} \mathrm{muon/s}$, a
measurement of \mbox{\metoee} could be carried out with the CyDet detector.

\subsection{Backgrounds}

While the signal of \muec is 105~MeV mono-energetic electron, 
there are several potential sources of electron background events in the
energy region around 100~MeV, which can be grouped into three
categories as follows:  intrinsic physics backgrounds
which come from muons stopped in the  target;
beam-related backgrounds which are caused by both muons and
other particles in  the  muon beam; other miscellaneous
backgrounds due to cosmic-rays, fake tracking events etc.

\begin{figure}[htb!]
\begin{center}
\includegraphics[width=0.8\textwidth, angle=0]{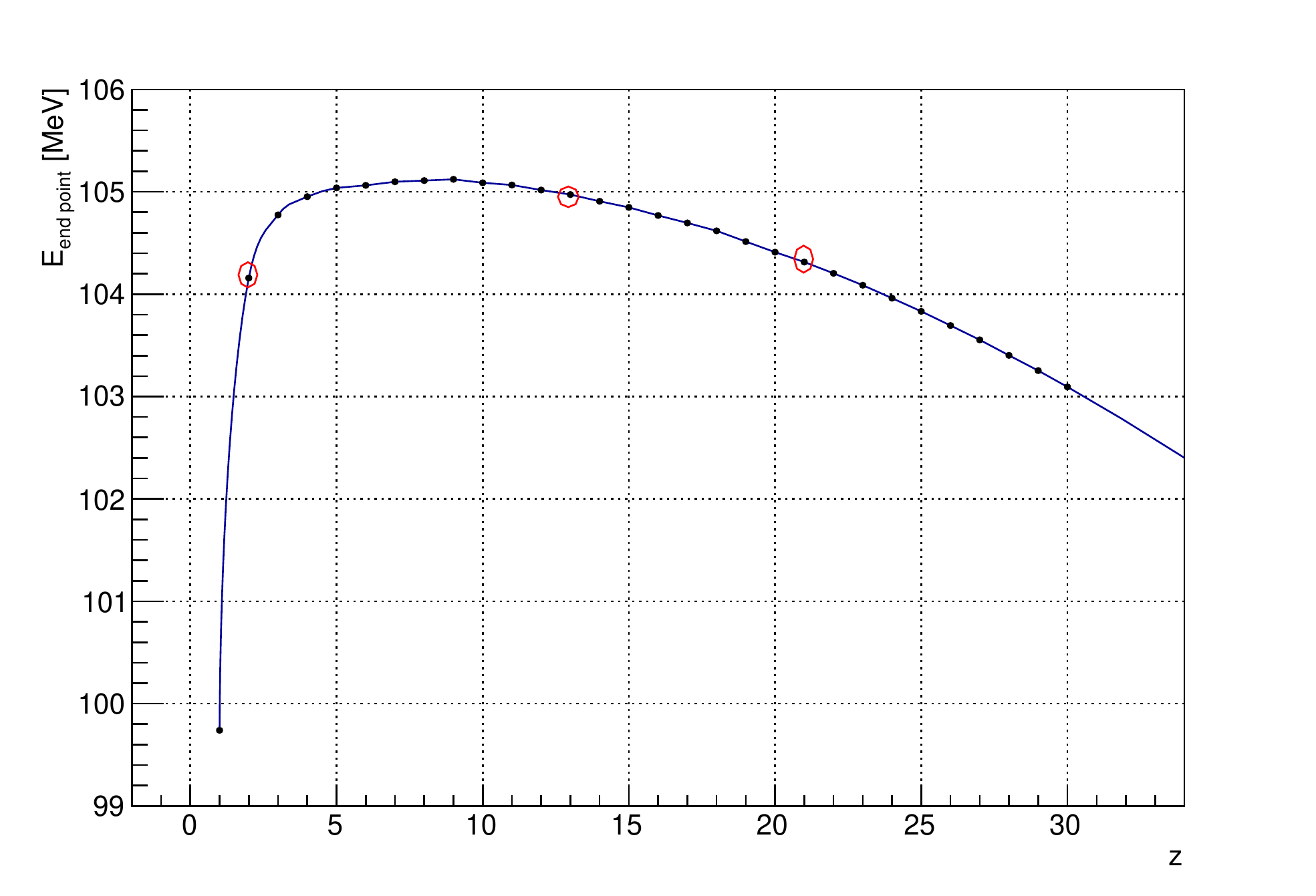}
\end{center}
\caption{DIO endpoint energy as a function of atomic number. The red circles highlight the atomic numbers corresponding to He, Al, and Ti (from left to right).}
\label{fig:dioendenergy} \end {figure}

\paragraph{Intrinsic physics backgrounds}
The major intrinsic physics background  is
muon decay in orbit (DIO) in the muonic atom. For this  the $e^{-}$ endpoint energy can extend to the energy of the \mue
conversion signal when the kinematics correspond to the limit of producing the neutrinos at rest and with the system recoiling against the nucleus.
The DIO endpoint energy depends on the element as shown in \cref{fig:dioendenergy}. Hence with an aluminium stopping target it is important to avoid materials
whose DIO end-point energy
is higher than aluminium, i.e., materials from $Z=5$ to $Z=12$, such as carbon and
nitrogen. However,
helium ($Z=2$) can be used.

The energy distribution of
DIO falls steeply toward its endpoint
as the fifth power of ($E_{\mu e}-E_{e}$),
where $E_{\mu e}$ and $E_{e}$ are the energy of the signal
electron and that of DIO electrons, respectively~\cite{czarnecki11, Czarnecki:2014cxa,
Szafron:2015mxa,Szafron:2015kja}.  The momentum
resolution of the $e^{-}$ detector is crucial  to eliminate this background; for a resolution better than 0.2\%,
the contribution from DIO occurs at a level  below $10^{-16}$.

Another prominent background process is radiative muon capture (RMC), given by
\begin{equation}
\mu^{-} + N(A,Z) \rightarrow \nu_{\mu} + N(A,Z-1) + \gamma,
\end{equation}
followed by internal and/or external asymmetric $e^{+}e^{-}$ conversion
of the photon ($\gamma \rightarrow e^{+}e^{-})$.  The kinematic
end-point ($E^{\rm end}_\text{RMC}$) of radiative muon capture is given by
\begin{equation}
E^{\rm end}_\text{RMC} \sim m_{\mu} - B_{\mu} - \Delta_{Z-1}\,,
\end{equation}
where $\Delta_{Z-1}$ is the difference in nuclear binding energy of
the two nuclei.  Other
intrinsic physics backgrounds could result from particle emission (such as
protons and neutrons)  after nuclear muon capture.

\paragraph{Beam-related backgrounds}
Beam-related background events may originate from muons, pions or
electrons in the beam. Muon decays in flight may create electrons in the
energy range of 100~MeV if the muon momentum is greater than
75~MeV/$c$. Pions in the beam may also produce background events by
radiative pion capture (RPC)
\begin{equation}
\pi^{-} + N(A,Z) \rightarrow N(A,Z-1) + \gamma\,,
\end{equation}
followed by internal and external asymmetric $e^{+}e^{-}$ conversion. There are also
electrons arising directly in the secondary beam  from the
production target. To eliminate the backgrounds from pions and
electrons, the purity of the beam (after transport) is highly important.

\paragraph{Other backgrounds}
 Cosmic ray backgrounds must be eliminated by shielding  and detecting and vetoing the signals.
 % subsection
\section{Producing the Muon Beam} % Was A Chapter Not A Section

 COMET  requires negatively-charged low-energy muons which
can be easily stopped in a  thin target to   efficiently produce muonic atoms.  Muons of appropriate momentum originate from the decay of low-energy pions produced
in the backward direction by an incident pulsed proton beam from the
J-PARC main ring onto a carbon target in the Nuclear and Experimental Hall.  The beam must be pulsed as the lifetime of the muonic atom is a critical factor for isolating the signal. High-energy pions must also be eliminated as they can potentially cause background events.

The Phase-I beam line consists of a section for pion production and capture (pion capture section), a section of muon transport (muon transport section) and a bridging section from the muon beam line to the detector (bridge section).  At the `downstream' end of the muon beam line is the aluminium target in the Detector Solenoid (DS). A schematic layout of the COMET Phase-I muon beam line is shown in \cref{fig:CometPhaseILayout0} 
and the top figure of \cref{fig:capture}.

\label{ch:ProtonAccelerators}

\subsection{The Proton Beam}

The
proton beam  pulse width must be much less than  the
gap between pulses and significantly shorter than the lifetime of a muonic atom in
aluminium, which is 864~ns. It is
critical that an  extremely high extinction rate, better than $10^{-10}$,
between pulses be achieved. A proton beam of 8 GeV is employed with
pulses of 100~ns duration, separated by 
at least 1.17~\micro{}s.
The beam energy is chosen to be 8 GeV, which is sufficiently high to produce an
adequate number of muons but low enough to minimise antiproton production,
which could lead to unwelcome population of particles in the signal
time window.

In the J-PARC LINAC, a chopper with a very fast rise time (10~ns) is
required to ensure that the Rapid Cycling Synchrotron (RCS) can be filled with high efficiency and
with the appropriate gaps between bunches. Inefficiencies could result
in stray protons between the bunches and this  needs to be minimised
in order to avoid placing unachievable demands on the extinction system.
The RCS will accept 400~MeV protons from the LINAC and accelerate them
to 3~GeV. Four sets of acceleration are performed in the RCS with
two bunches for each Main Ring (MR) acceleration cycle. 

\begin{figure}[htb!]
\begin{center}
\includegraphics[width=0.8\textwidth]{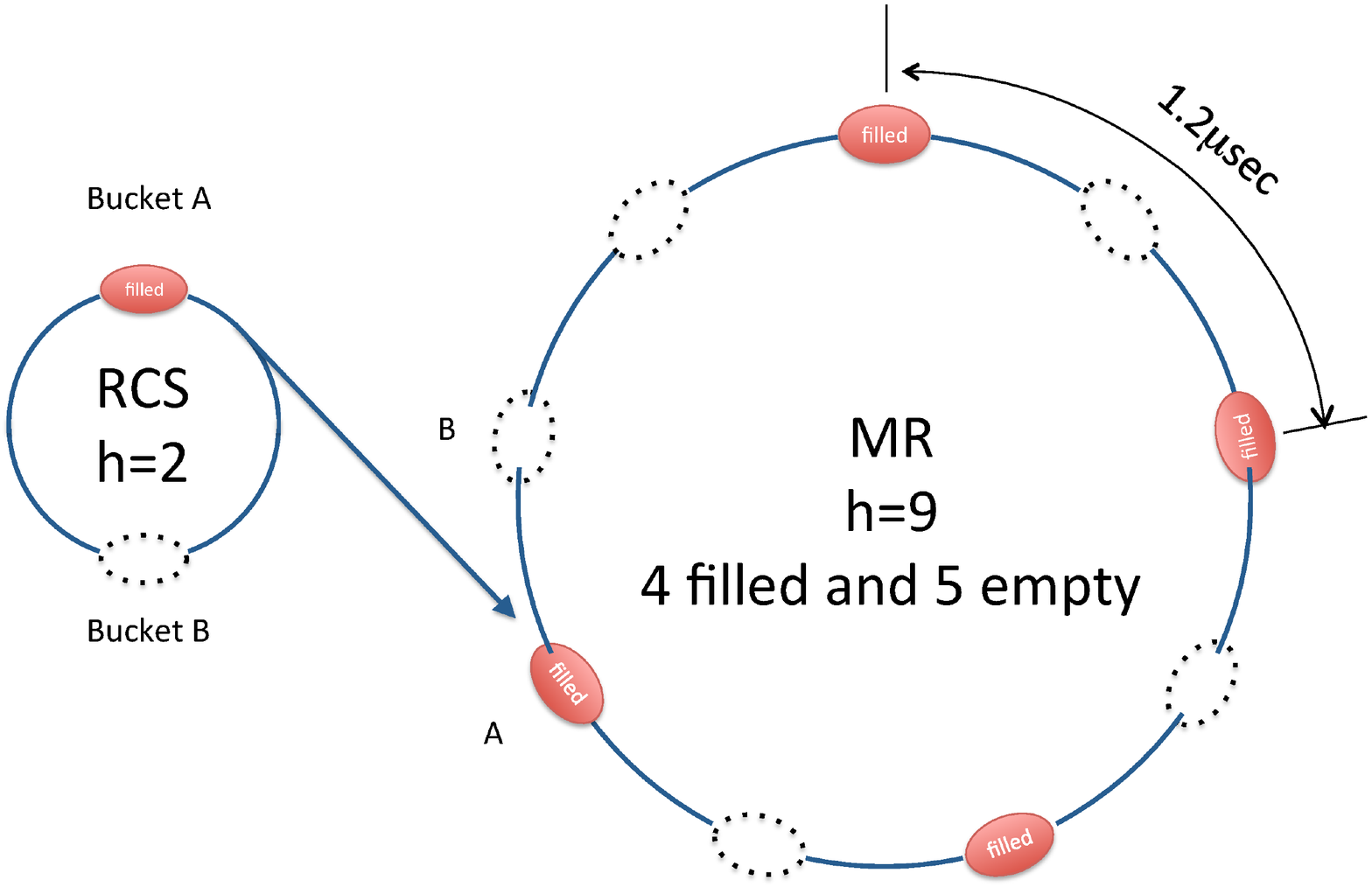}
\caption{The COMET bunch structure in the RCS and MR where four
  buckets are filled producing 100~ns proton bunches separated by at least 1.17~\micro{}s. }
\label{fig:RCS-MR-9}
\end{center}
\end{figure}

A 1.17~\micro{}s  pulsed beam structure is achieved by filling 
only four out of the nine MR buckets for MR operation at a harmonic 
number of nine. 
The four filled buckets are distributed around the ring in such a way 
that an empty bucket exists between the filled buckets. 
\footnote{
Also, a 1.75~\micro{}s pulsed beam structure is possible by filling only three out of the nine MR buckets. In this case the three filled buckets are distributed around the ring in such a way that two empty buckets exist between filled buckets. 
}
A schematic showing the four bucket structure is presented in \cref{fig:RCS-MR-9}.

Beam injection from the RCS into the MR using kicker magnets is a critical aspect for COMET due to the inter-bunch extinction requirements. A dedicated injection method, ``Single Bunch Kicking'', is realised by shifting the injection kicker excitation timing by 600~ns such that any particles remaining in empty buckets are not injected into the MR. A preliminary test in 2012 showed this to be effective at improving the extinction significantly and that the extinction level could be maintained through acceleration and extraction if the RF acceleration voltage was raised above its nominal value.

Slow extraction for COMET will be similar to that of the 30~GeV beam
into the NP Hall, but needs to be modified so that the beam is extracted at the lower energy of 8~GeV, with the bunch structure retained.
The major change for this ``bunched slow
extraction'', in contrast to the normal slow extraction, is that the
RF voltages need to be maintained and not turned off during
extraction. 
The required extinction factor can be achieved by increasing the RF voltage, however
there is a trade-off with the heat load on the cavity.

\subsubsection{Acceleration test}\label{sec:accelerationtest}
A series of proton beam acceleration tests were conducted in May 2014.
Every second acceleration bucket of the MR were filled with 3~GeV protons
from the RCS and accelerated to 8~GeV before extraction to the abort line.
Protons corresponding to the 3.2~kW operation were accelerated
to measure various beam parameters. The accelerator configuration
was then optimised for COMET operation, in order to minimise beam loss.

Systematic studies of the proton beam extinction factor were
carried out with an extinction monitor installed in the MR abort
line.
The monitor is sensitive to single protons while covering a large
dynamic range by using a plastic scintillator with four photomultipliers
with different light attenuators.
The beam extinction factor was studied by counting
the number of protons scattered off a pulse after extracting
whole beam bunches to the MR abort line with fast-extraction kickers.
During the flat-top period, where beam extraction is usually conducted,
the acceleration RF voltage was kept on to study the extinction factor dependence
on the RF voltage.

\begin{figure}[bt]
 \begin{center}
 \includegraphics[width=0.8\textwidth]{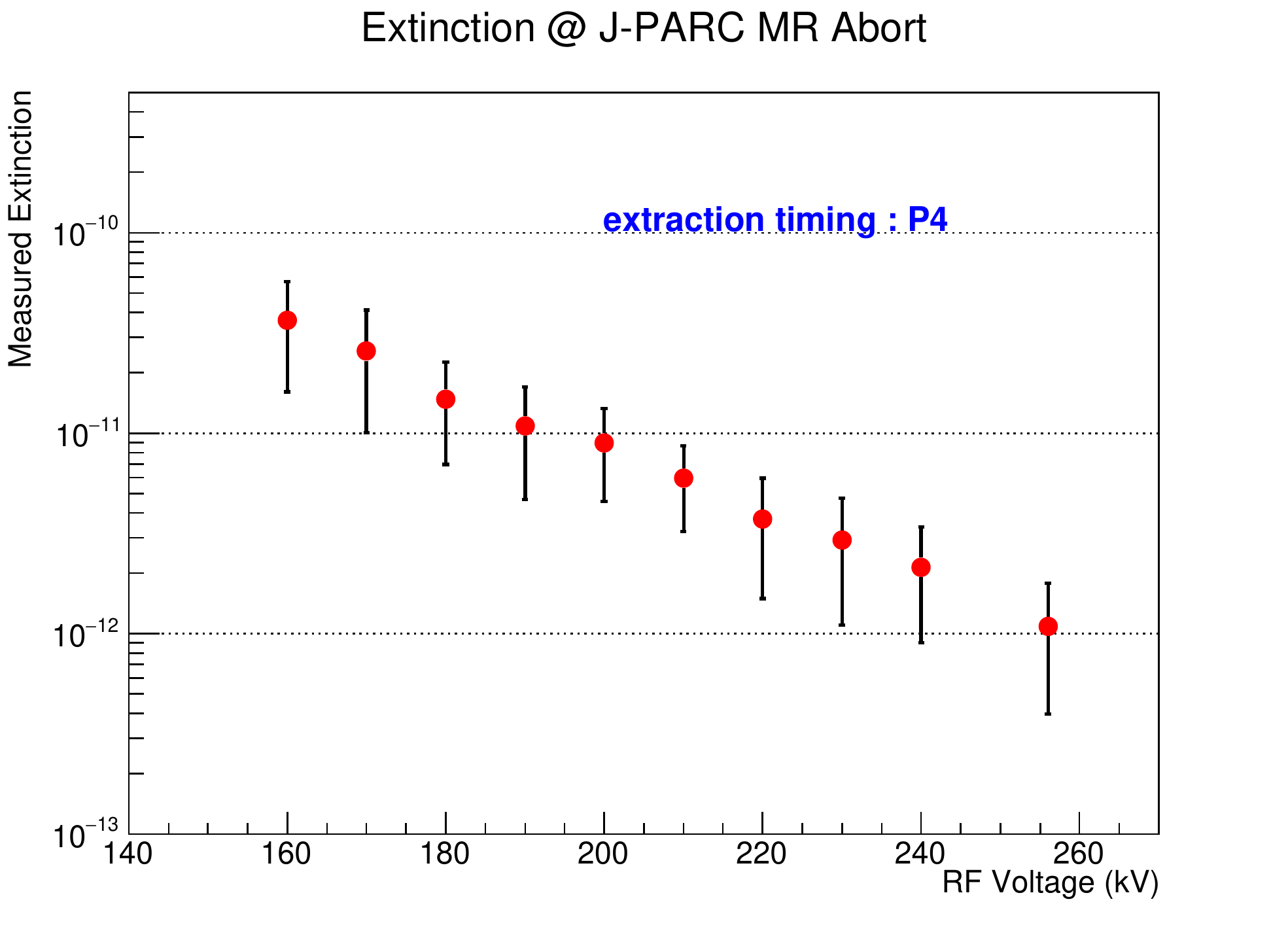}
 \end{center}
 \caption{Extinction levels measured at the MR abort line with single bucket filling with
the number of protons equivalent to that of 3.2~kW operation, as a function of the applied RF voltage during beam
circulation after acceleration. }
 \label{fig:ExtinctionvsRF}
\end{figure}

Results of extinction factor study are shown in \cref{fig:ExtinctionvsRF}.
When the RF voltage was reduced less than ~100~kV, it was observed that accelerated particles start to be scattered
along the ring into the gaps between the bunches, resulting deterioration of the extinction factor as large as $10^{-10}$.
The extinction could be improved to as low as 10$^{-12}$ by applying an RF
voltage of 255kV. This is sufficiently small for the COMET
experiment, so the voltage will be optimised for long-term operations in
order to keep the RF cavity temperatures stable within the
capabilities of the water cooling system. 

\subsubsection{Proton beam line}
\label{sec:proton-beam line}

\begin{figure}[htb!]
\centering
\includegraphics[width=\textwidth]{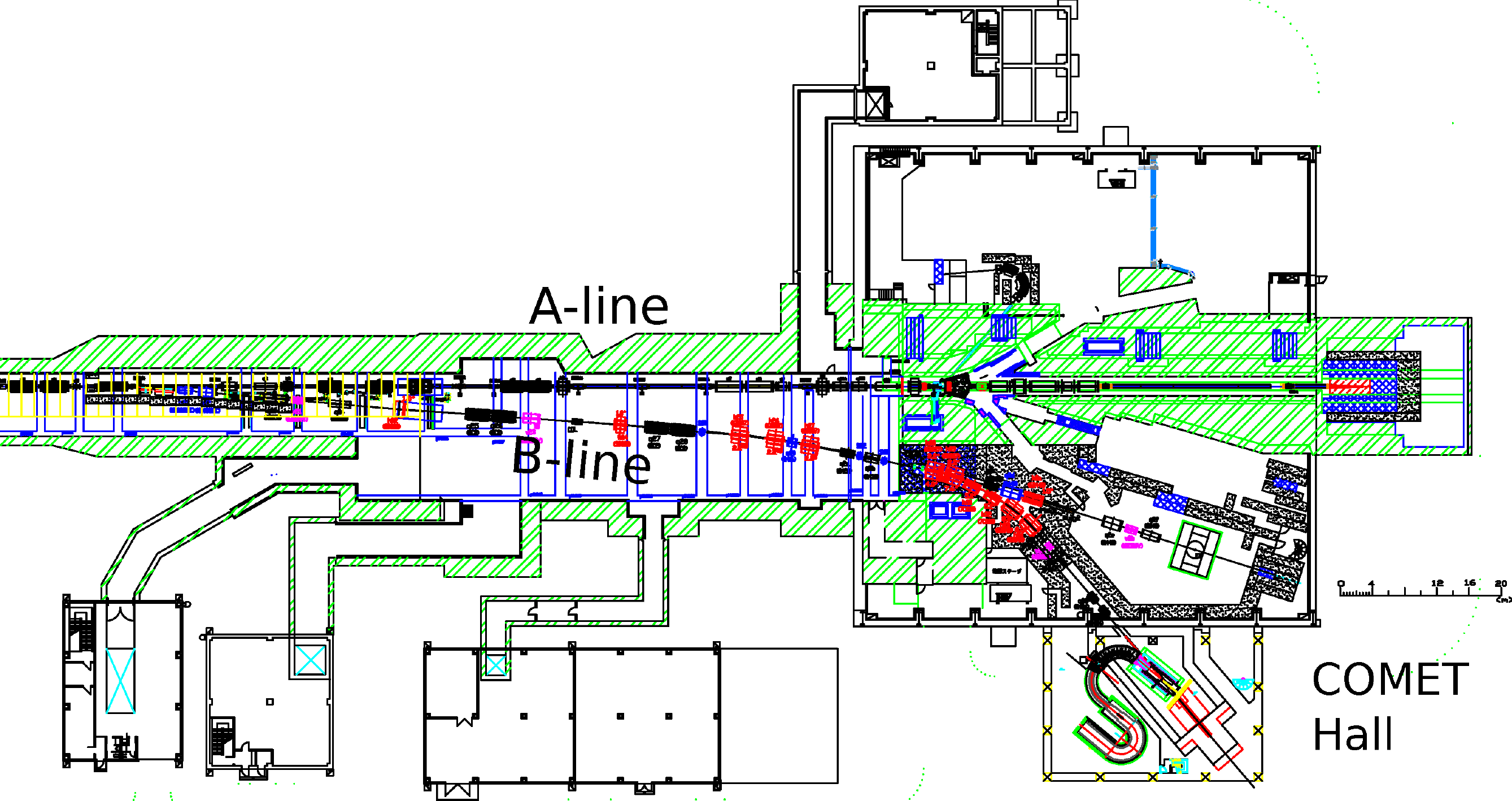}
\caption{The A and B-lines from the MR into the NP Hall. A
  schematic of the COMET experiment is shown in the bottom right.}
\label{fig:comet-beamline}
\end{figure}

The COMET experiment is being constructed in the
NP Hall. In addition to the
existing beam line (A-line) a new beam line is being built
(B-line) with two branches, one to serve  high-momentum (up to 30~GeV)
experiments and the other for COMET (8~GeV).  In the low-momentum
running for COMET the entire beam is sent to the B-line.  The
schematic of the beam lines are shown in
\cref{fig:comet-beamline}. To realise multiple operation modes, a Lambertson magnet followed by
two septum magnets are deployed to provide the A/B-line branches. The proton beam line will be common for both COMET Phase-I and Phase-II.

The proton beam dump is designed to fulfill radiation safety requirements and this
is evaluated using a PHITS~\cite{PHITS}
simulation.
The resulting size of the required
iron dump is 4~m wide and 5~m deep.

Beam profile monitors will be installed at several locations along
the beam line including: downstream of the A/B-line branch;
the boundary of the switch yard (the tunnel between the MR and
the NP Hall); and the NP Hall, as well as upstream of the COMET building
entrance. The same technology, RGIPM (Residual Gas Ionization
Profile Monitor) will be used as  for the A-line
beam monitors.
In addition to the RGIPMs, an RGICM (Residual Gas Ionization Current
Monitor) will be installed near the COMET building entrance for
beam intensity monitoring. The RGICM uses a similar technology to the RGIPM,
but precisely measures the current of ionisation electrons, which is
proportional to the beam intensity.

A diamond
detector with a fast response and high sensitivity  in a
high-radiation environment will be employed for measuring the proton beam extinction factor
and beam profile~\cite{psarin_pixel2014}.

The beam optics of the proton beam line have been optimized by a TRANSPORT
simulation.
The 3$\sigma$ beam
emittance at the extraction point used in the simulation is
1.7\,$\pi$\,mm\,mrad in the horizontal direction and 10.6\,$\pi$\,mm\,mrad in
the vertical direction, which is based on the measurement of the beam profile
in the switch yard after the beam extraction from the MR.

Beam loss due to interaction of the beam halo through the proton beam line
is evaluated to be 0.003\% using a TURTLE simulation.

\subsection{Pion Production at the Primary Target}

The proton target will be installed within the bore of the
capture solenoid and designed to maximise the capture of low energy negative
 pions produced in the backward direction.
Both the target
station and muon capture solenoid region will be designed for the Phase-II beam
power of 56~kW since once constructed and exposed to the beam, the target station
infrastructure will be activated, and cannot be modified.
However, the target itself will be replaced between the two phases, and the
target station will be designed with remote handling capability to allow for this.

While pion production is maximised with a high-$Z$ material, it is proposed
to use a graphite target for Phase-I. This will minimise the activation of the
target station and heat shield which will significantly ease the necessary
upgrades for Phase-II operation where a tungsten target will be employed.

The Phase-I beam power of 3.2~kW will deposit
a heat load of approximately 100~W in the graphite target material.
This can easily be radiated
to the solenoid shield. The target support system to accurately position the
target within the solenoid inner shield will have a low-mass design.

\begin{figure}[bth!]
 \begin{center}
  \includegraphics[trim={251px 185px 57px 44px},clip,width=0.8\textwidth]{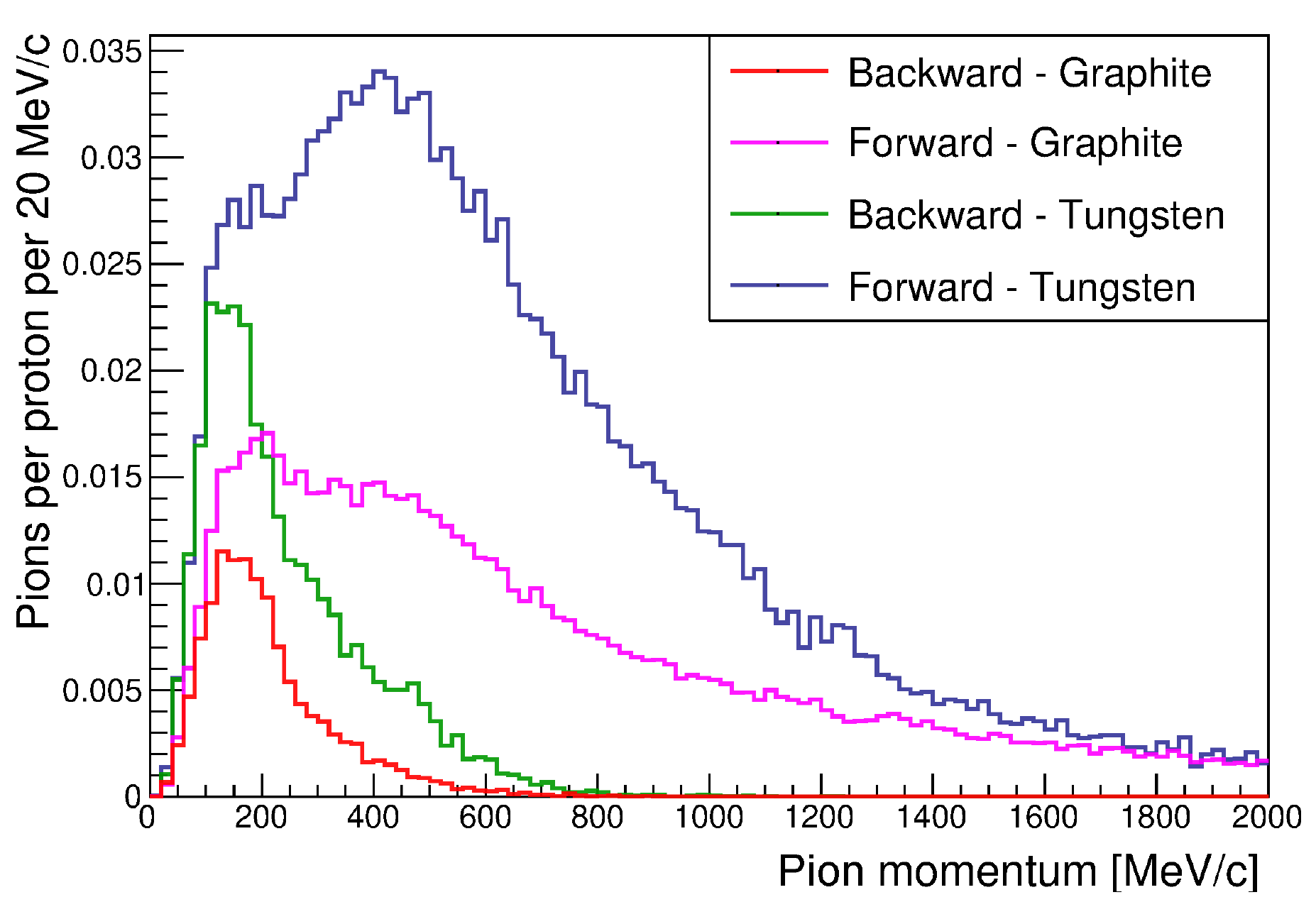}
 \end{center}
\caption{
Momentum distribution of pions exiting in the forward and
  backward regions of tungsten and graphite targets bombarded by an 8~GeV proton
  beam. The spectra are generated using Geant4 using the {\tt QGSP-BERT} hadronisation model.
  }
\label{fig:pion_bw_fw_C_W}
\end{figure}

Pion production yields from protons incident on graphite and tungsten
targets in the backward and forward regions with respect to the proton beam
direction are presented in 
\cref{fig:pion_bw_fw_C_W}.

\begin{figure}[hbt!]
\begin{center}
\includegraphics[width=0.8\textwidth]{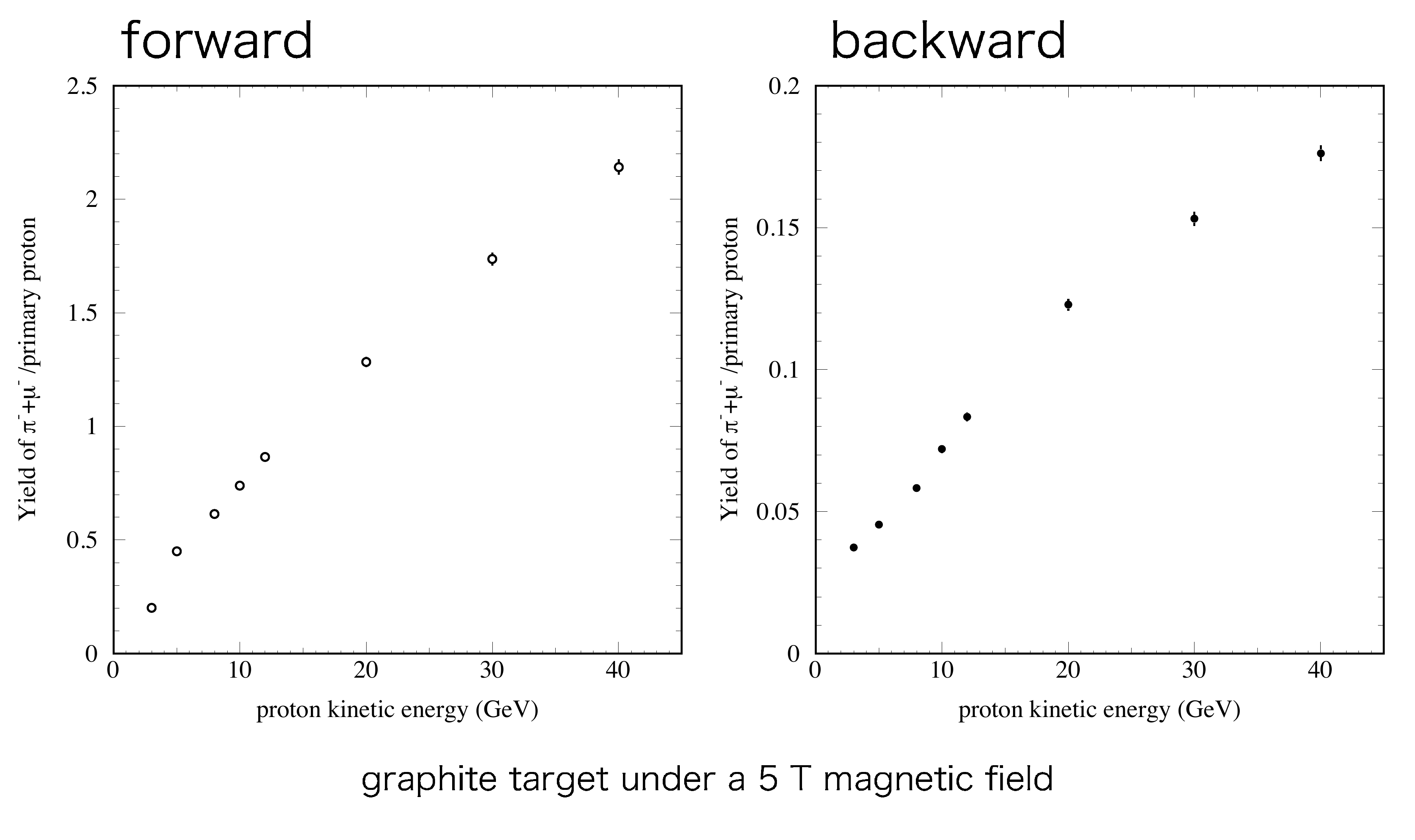}
\end{center}
\caption{Yields per proton of forward pions and muons, left, and backward pions and muons, right,
from a graphite target in a magnetic field of 5 Tesla, as a function of proton energy.}
\label{fig:pion-yield-proton-energy}
\end{figure}

\Cref{fig:pion-yield-proton-energy} shows the yields of pions and muons as a function of proton
energy, calculated using Geant4.  As seen in \cref{fig:pion-yield-proton-energy}, the pion yield
increases almost
linearly with proton energy and therefore with proton beam power.

The choice of proton energy was determined by considering the pion production yield and
backgrounds.  In particular, backgrounds from antiproton production are important. The current choice
of proton energy is 8~GeV, which is 
above the threshold energy for antiproton production, 6.56~GeV.

\subsection{Pion Capture}

\begin{figure}[htb!]
 \begin{center}
  \includegraphics[width=0.8\textwidth]{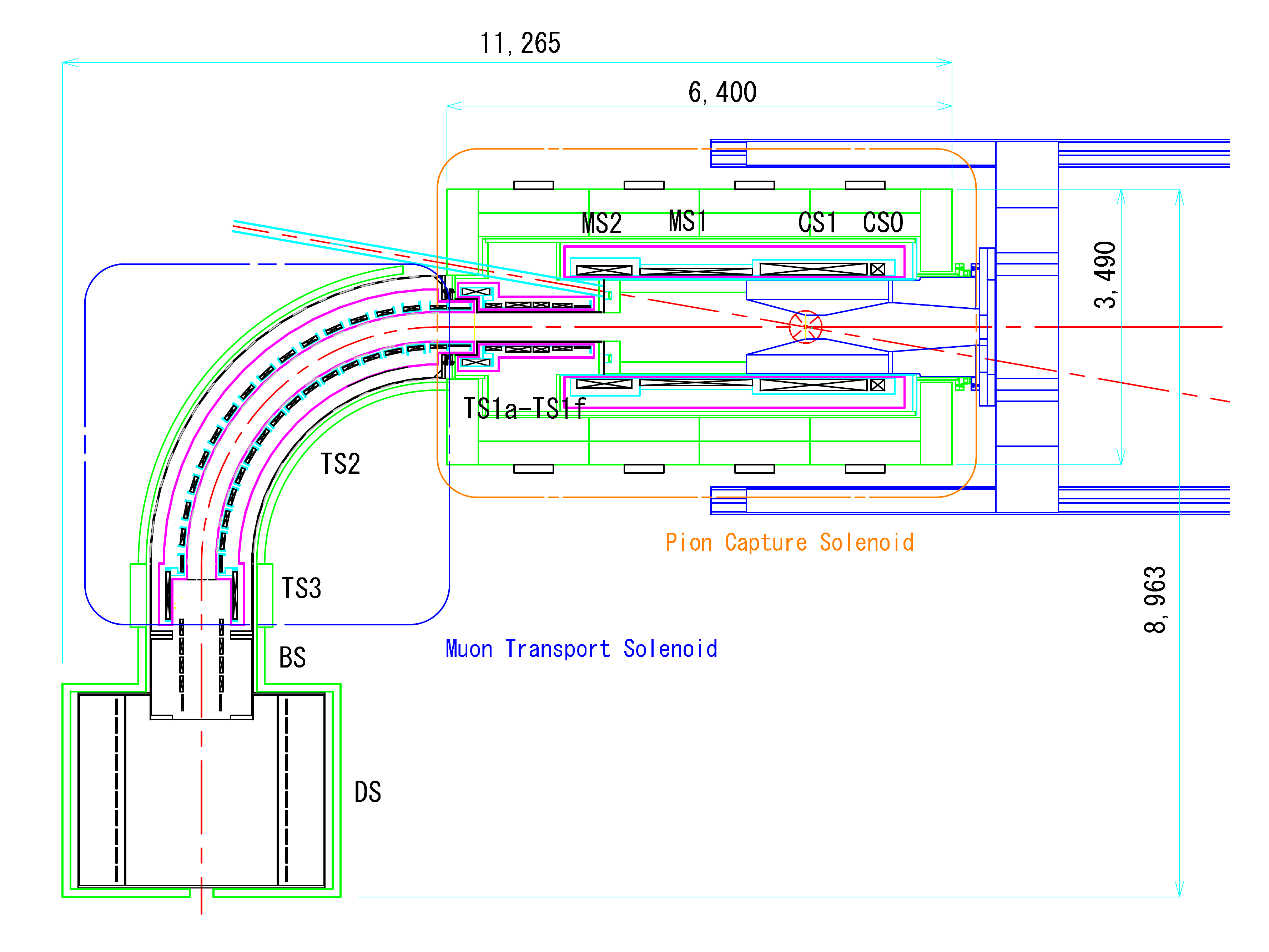} \\
  \includegraphics[width=0.8\textwidth]{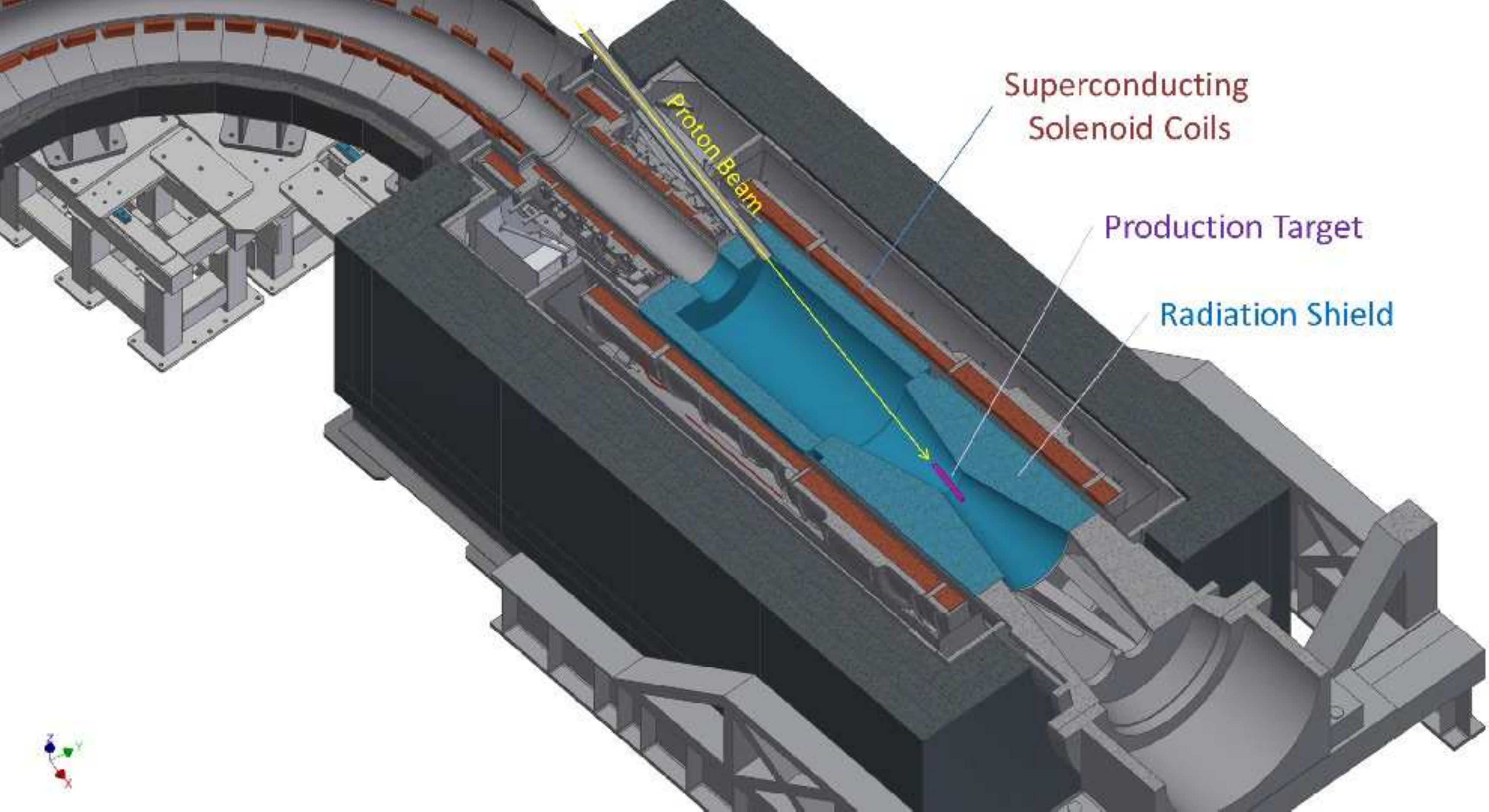}
 \end{center}
 \caption{
 (Top) Layout of the COMET Phase-I solenoid system, which consists of the pion capture solenoid (orange box), muon transport solenoid (blue box), beam dump (right part of pion capture solenoid), and detector solenoid (bottom part of muon transport solenoid). The pion  production target (proton  target) locates at the center of pion capture solenoid, marked as red circle. The slanted direction of the pion production target to the beamline is shown in red solid-dashed line. For the name of each part, CS: capture solenoid, MS: matching solenoid, TS: transport solenoid, BS: bridge solenoid, and DS: detector solenoid. 
 (Bottom) Layout of the pion capture solenoid system in 3D view. 
 }
 \label{fig:capture}
\end{figure}

The pions are captured using a high-strength solenoidal magnetic field
giving a large solid angle acceptance. \Cref{fig:capture} shows
the layout of the pion-capture system, which consists of the pion
production target, high-field solenoid magnets for pion capture, and a
radiation shield.  Pions emitted into the backward
hemisphere with a transverse momentum less than 100~MeV/$c$ are captured
by using a solenoid magnet of
5~T, and inner bore of 30~cm. This gives adequate acceptance for the parent pions of muons with momentum below 75~MeV/$c$.

The
predicted yields  three meters backwards
from the proton target from different
hadron production simulations were 
obtained by using Geant4~\cite{Agostinelli:2002hh}  and MARS15~\cite{Mokhov:2007sz} programs.
From the result  shown in \cref{tb:PionProduction}, it was found that the pion yields are different up to three times
between the hadron production  models. 
The \texttt{QGSP\_BERT} and \texttt{FTFP\_BERT} hadron
production models have the lowest yield and so the \texttt{QGSP\_BERT}  model has been used to conservatively
estimate and optimize the muon beam.
\label{sec:different_hadroncode}

\begin{table}[htb!]
\begin{center}
\begin{tabular}{lcc}\hline\hline
Models & Simulator  & N($\pi^{-}+\mu^{-}$)/$p$ at 3\,m \rule{0pt}{12pt}\\[2pt]\hline
\texttt{CEM} & MARS & $0.061\phantom{0}\pm 0.001\phantom{0}$ \cr
\texttt{CEM/LAQGSM} & MARS & $0.138\phantom{0}\pm0.001\phantom{0}$ \cr
\texttt{LAQGSM} & MARS & $0.144\phantom{0} \pm 0.001\phantom{0}$ \cr
\texttt{LAQGSM} & GEANT & $0.1322 \pm 0.0007$ \cr
\texttt{QGSP\_BERT} & GEANT & $0.0511 \pm 0.0002$ \cr
\texttt{QGSP\_BIC} & GEANT & $0.1278 \pm 0.0005$ \cr
\texttt{FTFP\_BERT} & GEANT & $0.0440 \pm 0.0002$ \cr\hline\hline
\end{tabular}
\end{center}
\caption{
\sl Comparison of the $\pi^{-}$ and $\mu^{-}$ yields three
  meters backwards from the proton target for different hadron production
  codes.  
  } \label{tb:PionProduction}
\end{table}

The captured pions have a broad directional distribution. In order to
increase the acceptance of the muon beam line it is desirable to make them more parallel to the beam
axis by decreasing the magnetic field adiabatically.
  Under a solenoidal magnetic field, the
product of the radius of curvature, $R$, and the transverse momentum, $p_T$, is an invariant:
\begin{equation}
p_T\times R \propto \frac{p_T^2}{B} = \mathrm{constant,}
\end{equation}
where $B$ is the magnitude of the magnetic field.  
Therefore, if the magnetic field decreases gradually, 
$p_{T}$ also decreases, yielding a more parallel beam.  This is the principle of the
adiabatic transition.  Quantitatively, when the magnetic field is reduced by a factor of two, $p_T$
decreases by a factor of $\sqrt{2}$. However this causes
the radius of curvature to increase by a factor of $\sqrt{2}$ and hence the inner radius of the magnet in the pion decay section has
to be $\sqrt{2}$ times that of the pion-capture solenoid.  Thus the pion beam can be made more parallel at the
cost of an increased beam size.
In COMET Phase-I solenoid system, a magnetic field of 5~T in the pion capture solenoid (CS) gradually decreases to 3~T at the  matching solenoid (MS).

 % subsection
\subsection{Muon Beam Transport}\label{sec:muon-Transport}
The muon beam transport
consists of curved and straight superconducting solenoid magnets
of 3~T and $\sim$7.6~m length.
The requirements are:
\begin{itemize}
\item the muon transport should be long enough for pions to decay to muons,
\item the muon transport should have a high transport efficiency for muons with a momentum of
  $\sim40$ MeV/$c$, and
\item the muon transport should select muons with low momentum and eliminate muons of high momentum
  ($p_{\mu} > 75$ MeV/$c$) to avoid backgrounds from muon decays in flight.
\end{itemize}
The optimal muon momentum is  $\sim 40$~MeV/$c$.  Muons with
higher momentum are less likely to be stopped and give  rise to  backgrounds in the signal region from
decays in flight.  Positive muons  are
another potential source of background. Curved solenoid
transport is used to minimise these.

 A charged particle in a solenoidal field
follows a helical trajectory and in a curved solenoid, the central axis of this trajectory drifts in the
direction perpendicular to the plane of curvature.  The magnitude of this drift, $D$, is given by
\begin{eqnarray}
D &=& \frac{1}{q  B} \left( \frac{s}{R} \right)
\frac{p_L^2 + \frac{1}{2}p_T^2}{p_L}, \\
& = &
\frac{1}{q B} \left( \frac{s}{R} \right)
\frac{p}{2}\left( \cos\theta + \frac{1}{\cos\theta}\right) \,,
\end{eqnarray}
where $q$ is the electric charge of the particle (with its sign), $B$ is the magnetic field at the
axis, and $s$ and $R$ are the path length and the radius of curvature of the curved solenoid,
respectively.  Here, $s/R$ ($=\theta_\text{bend}$) is the total bending angle of the solenoid, hence $D$
is proportional to $\theta_\text{bend}$. $p_L$ and $p_T$ are longitudinal and transverse momenta so
$\theta$ is the pitch angle of the helical trajectory.  Particles with opposite signs drift in opposite directions and this is used for charge and momentum
selection with a  collimator  placed after the curved solenoid.

To keep the centre of the helical trajectories of the 40~MeV/$c$ muons in the
bending plane, a compensating dipole field parallel to the drift direction must be applied.

The COMET Phase-I beam line uses one curved solenoid with a bending angle of
90$^{\circ}$ with a compensating dipole field
of $\sim 0.05$~T . The collimator system is designed to remove particles travelling 8.5~cm above or 10~cm below % the 8.5cm and 10cm numbers have been checked: YK 20181213
the beam height and will be realised by
installing two plates of stainless steel
at the exit of the muon-transport system.
To separate the muon stopping target region, filled with helium,
from the muon beam line in vacuum, a vacuum window of
500~\micro{}m titanium will be installed at the exit of the curved
solenoid. The muon-transport section and the Detector Solenoid (DS) are
connected by the beam Bridge Solenoid (BS), where the magnetic field changes
from 3~T to 1~T.

\subsubsection{Muon beam yields}\label{sec:muonbeamyields}

\begin{figure}[htb!]
 \begin{center}
 \includegraphics[width=0.8\textwidth]{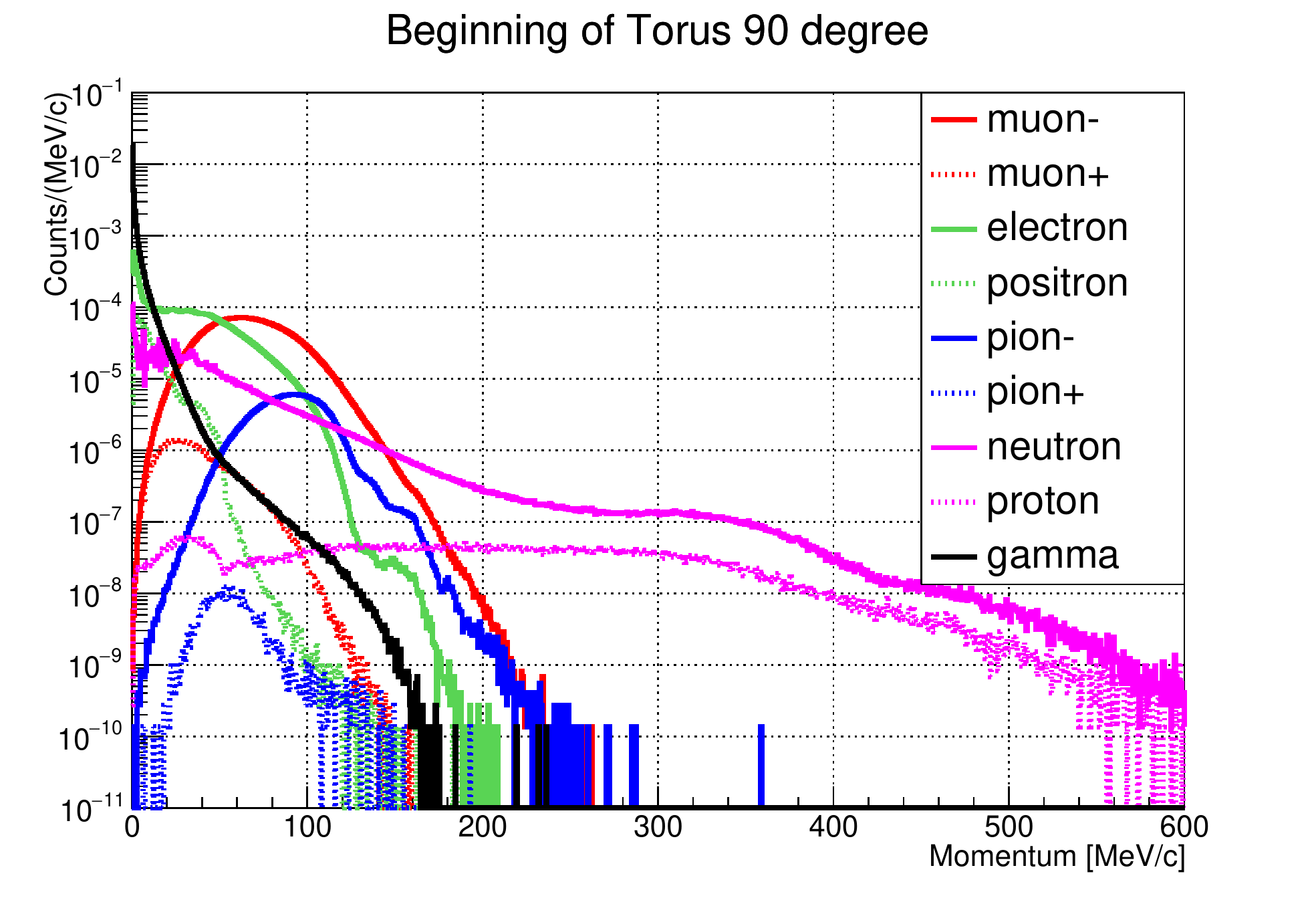}
 \end{center}
 \caption{
 Momentum distributions of various beam particles at the exit of
   the first 90$^{\circ}$ curved solenoid,
 using a graphite proton target.
 \label{fig:muonbeam-momentum-after90}}
\end{figure}

The momentum distribution of various beam particles at the exit
of the first 90$^{\circ}$ curved with the muon beam collimator is given in
\cref{fig:muonbeam-momentum-after90}.

\begin{figure}[htb!]
 \begin{center}
 \includegraphics[width=0.8\textwidth]{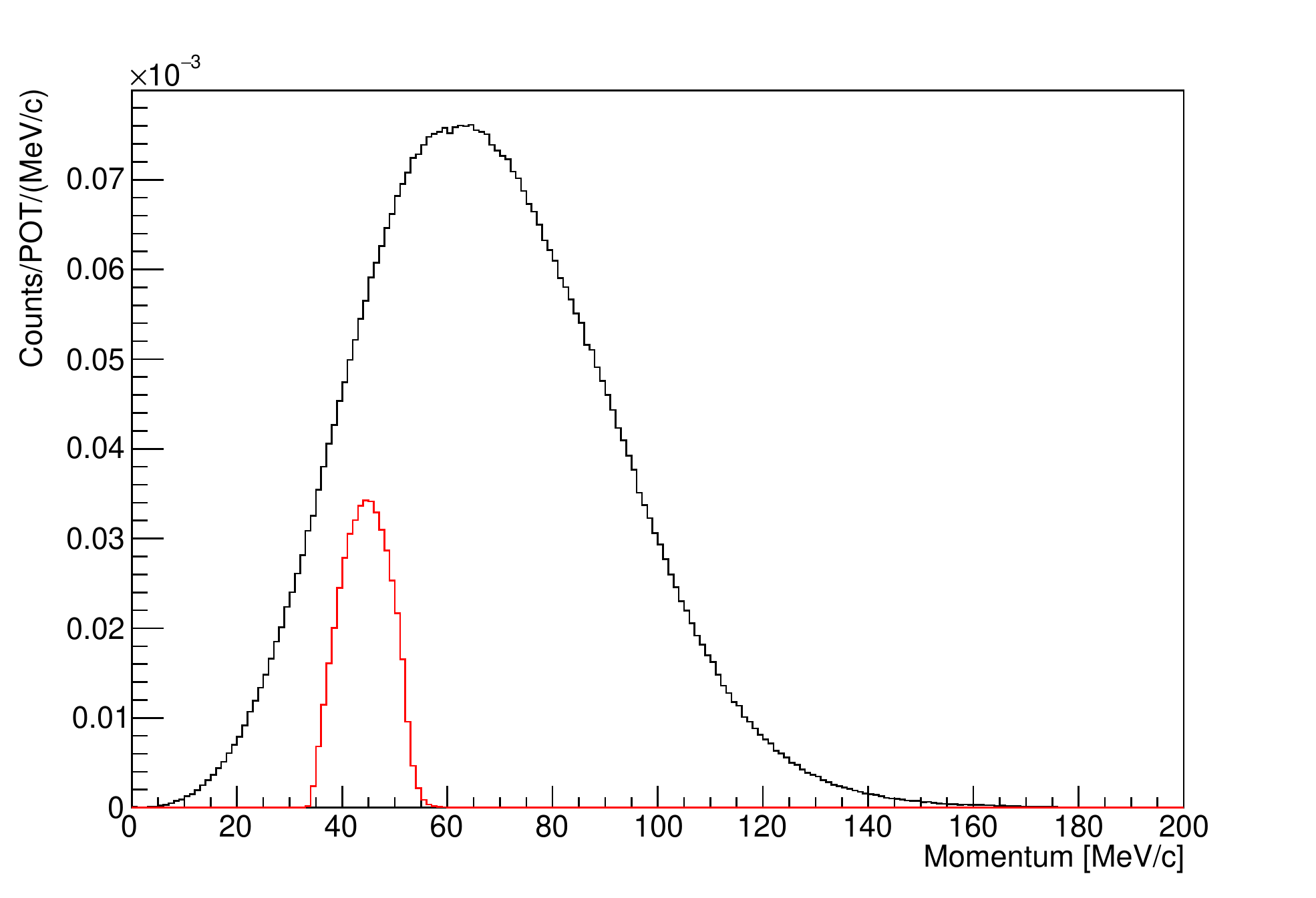}
 \end{center}
 \caption{Distributions  of momentum at the end of the muon beam transport solenoid. Black and red solid lines
   are those arrived at the end of muons transport solenoid including muon beam collimator, and stopped in the muon stopping target, respectively.}
 \label{fig:muonbeam}
\end{figure}

\Cref{fig:muonbeam} shows the distribution of the muons 
momenta
at the end of the muon-transport solenoid.
The  solid black line gives those reaching  the end of the muon-transport solenoid 
including the muon beam collimator,
and red line those stopping in the target. The low-energy cut off of the red line at about 35 MeV/c is due to absorption in the titanium safety windows.

Estimates using the {\tt QGSP\_BERT} model of the number of muons and pions per proton after the muon-transport section and on the muon stopping
target  are summarised in \cref{tb:muoncollimator}.  The number of muons
stopping in the muon stopping target is about $4.7 \times 10^{-4}$ per proton,
and so with a 0.4~$\mu$A
proton beam, the yield of stopped muons is about $1.2 \times 10^{9}$
per second.

\begin{table}[htb!]
 \begin{center}
  \caption{Muon and pion yields per proton in front of the Bridge Solenoid (BS), after the BS, and stopped on the muon stopping target.}\label{tb:muoncollimator}
 \begin{tabular}{lccc}
   \hline \hline
   Yield (per proton): & After muon-transport section & Stopped in muon target \\
   \hline
   Muons & $5.0 \times10^{-3}$ & $4.7 \times 10^{-4}$ \\
   Pions & $3.5 \times10^{-4}$ & $3.0\times10^{-6}$ \\
   \hline \hline
 \end{tabular}
 \end{center}
 \end{table}
 % subsection
\subsection{Muon Stopping Target}\label{section:muonstoppingtarget} % Hisataka Yoshida % Was A Chapter Not A Section

The muon-stopping target is placed in the centre of the DS and designed to maximise the
muon-stopping efficiency and  acceptance for the \mue conversion
electrons. The design must also minimise the energy
loss of the conversion electrons as this increases their momentum spread.

To eliminate beam-related background events arising from prompt beam
particles the measurement window will only open approximately
0.7~\micro{}s after the primary proton pulse. High-$Z$ target materials
are 
not appropriate for the stopping target since the muonic
atom lifetime decreases with increasing $Z$. Aluminium ($Z=13$)
with a muonic atom lifetime of 864~ns is the preferred target
over titanium ($Z=22$) and lead ($Z=82$) which have muonic atom lifetimes
of 330~ns and 74~ns respectively.

The configuration and dimensions of the muon-stopping target have been
optimized~\cite{COMET-doc-34} for maximum muon-stopping efficiency
and for minimal backgrounds and energy spread of the electrons. 

The current design consists of 17 aluminium disks, 100 mm in radius and 200 $\mu$m in thickness, with 50 mm spacings.

\begin{figure}[htb]
  \begin{center}
  \includegraphics[width=0.8\textwidth]{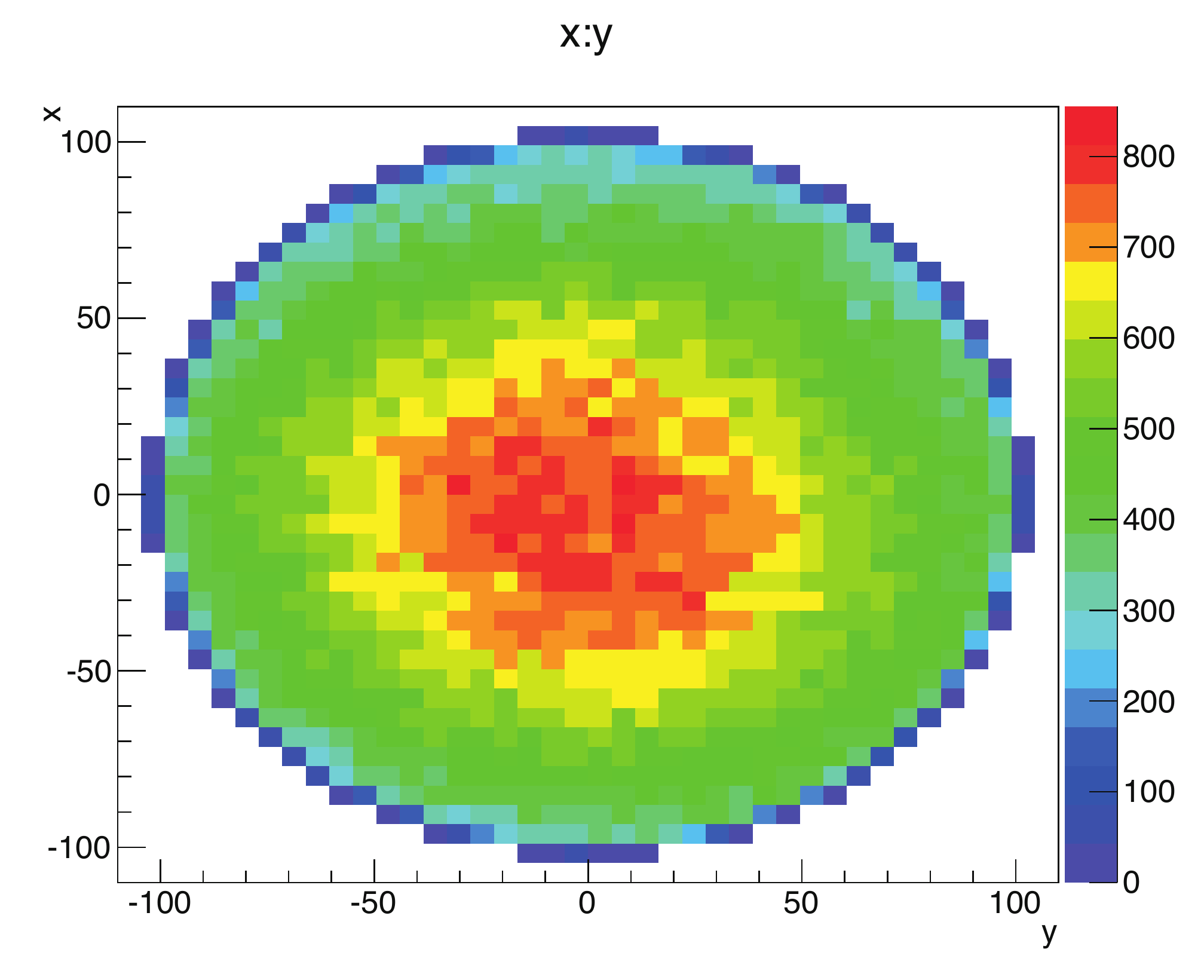}
  \end{center}
  \caption{%
  The distribution of stopped muons stopped  projected on the $x$ and $y$ axis.
  The $z$ axis (color codes) shows the number of stopped muons in arbitrary unit.
  }
\label{fig:muon-stop}
\end{figure}

\Cref{fig:muon-stop} shows the distribution of the number of muons stopped in each of 17 disks of the muon stopping target. In optimising the design, one must take into consideration that the total number of stopped muons increases with the number of target disks, but that this has a cost in the form of additional 
energy loss of the electron
in the target.
From \cref{fig:muon-stop}, the proposed target size would make reasonable radial coverage.

\begin{figure}[htb]
  \begin{center}
  \includegraphics[width=0.8\textwidth]{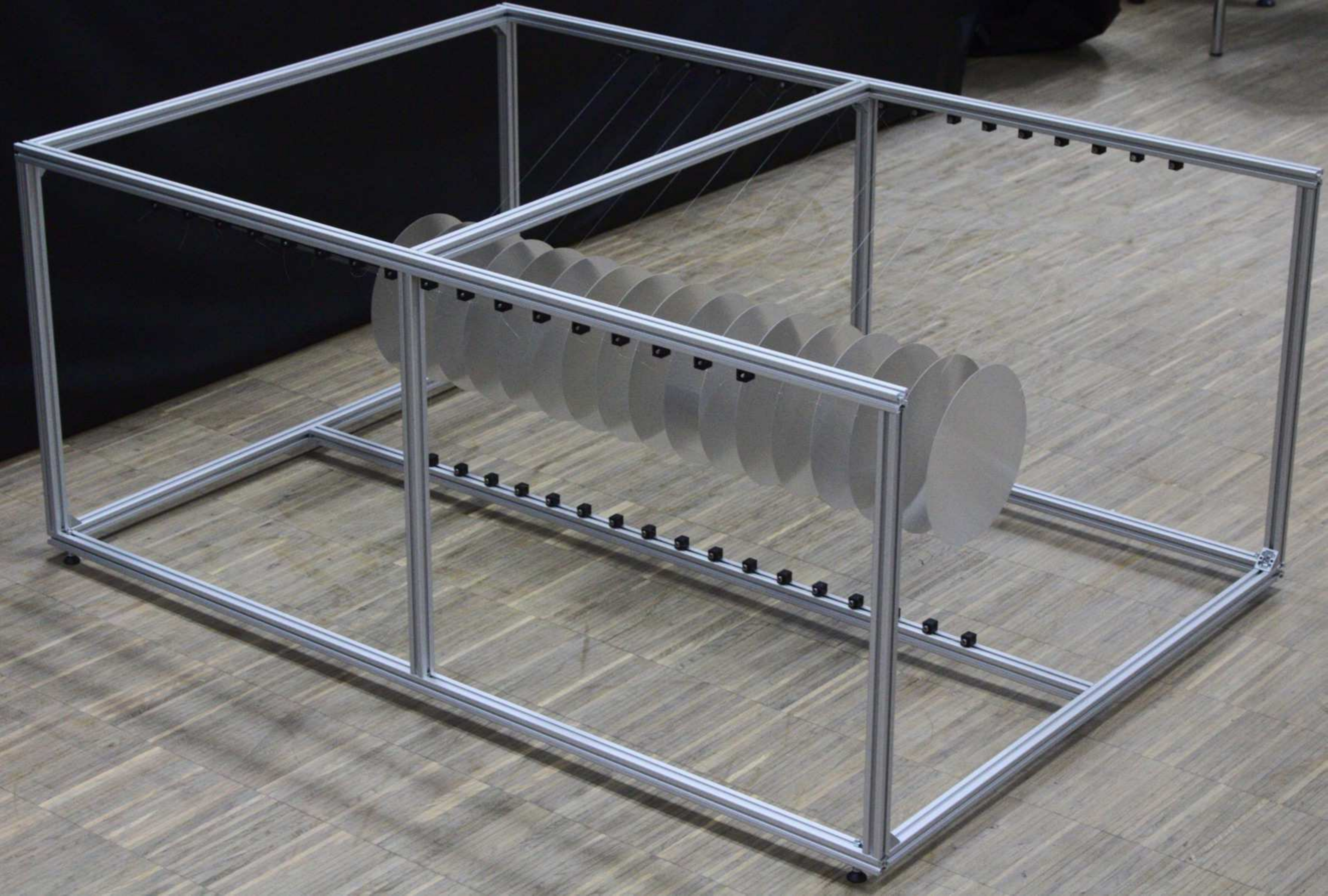}
  \end{center}
  \caption{
  A mock-up muon stopping target for test. 
    }
\label{fig:muon-target}
\end{figure}

A mock-up  muon stopping target for test is shown in 
\cref{fig:muon-target}. Each aluminium disk is supported by three
spokes. The spokes are connected to the ring structure which is placed
inside the inner wall of the cylindrical drift chamber (CyDet). The
spokes are made of aluminium or a high-Z material to avoid backgrounds
from muons stopped in the spokes.

\subsubsection{X-ray monitor}

When the muonic atoms are formed on the muon stopping target, a cascade of
X-rays are emitted as the muons drop down to the 1$s$ state. These can be used
to tag and count the formation of these muonic atoms. In turn, this can help
measure the number of muon captures that forms the denominator for the \mue
conversion rate that is the ultimate output of this experiment.

The observation of such muonic X-rays from Aluminium has been achieved in the
past in cosmic rays~\cite{doi:10.7566/JPSJ.84.034301}, and for COMET, a design
based on the principles of those of this earlier measurement is being
studied at this time. This involved a high-efficiency Germanium detector
combined with a coincidence and anti-coincidence system.

 % subsection

\section{CyDet: the Cylindrical Detector System}

\begin{figure}[htb!]
 \begin{center}
 \includegraphics[width=\textwidth]{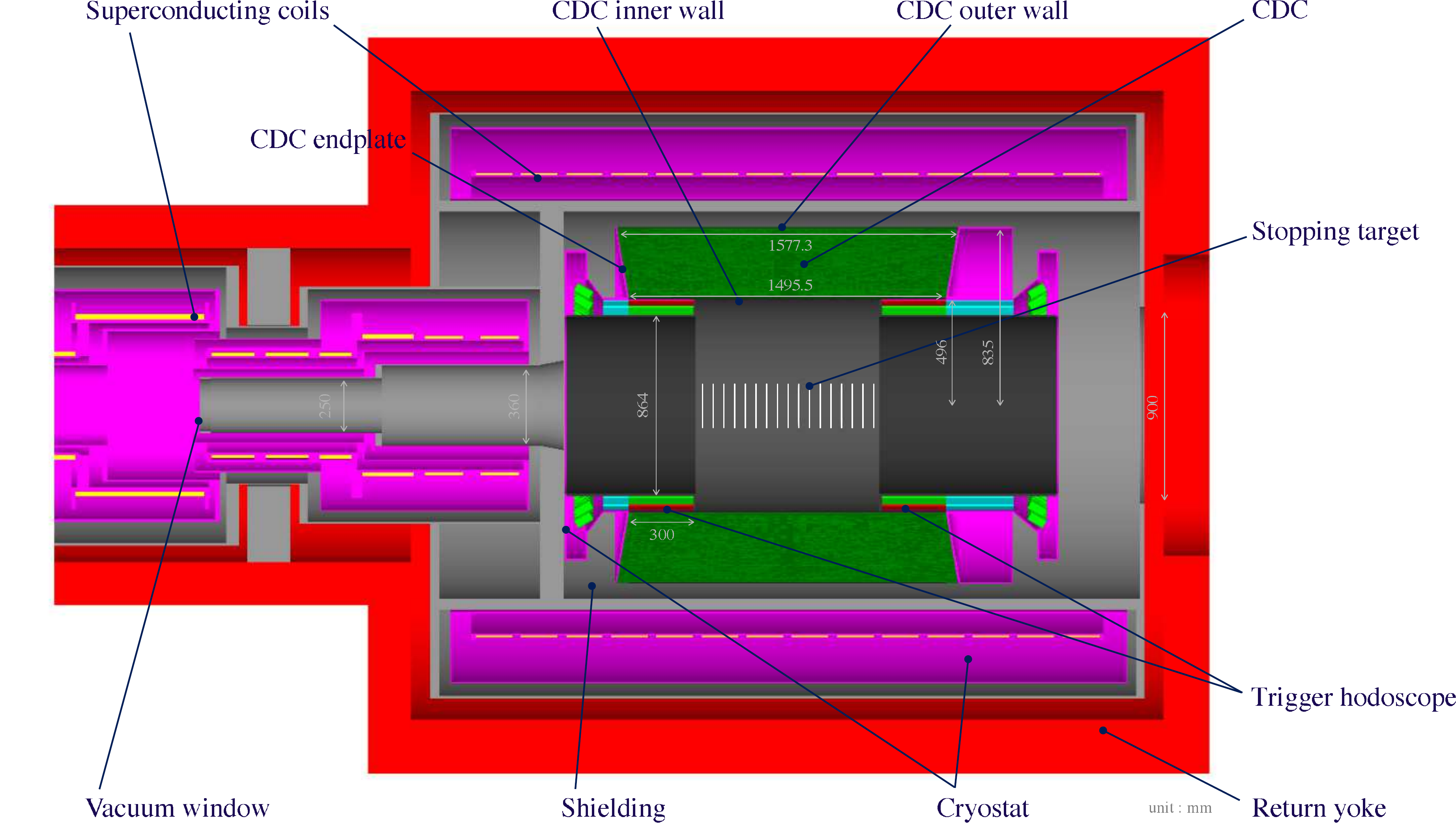}
 \end{center}
 \caption{Schematic layout of the CyDet detector}
 \label{fig:cdc2d}
\end{figure}

The cylindrical detector system (CyDet) is the main detector system for the \mue conversion search in COMET Phase-I. It consists of a cylindrical drift chamber (CDC) and a cylindrical trigger hodoscope (CTH).  \Cref{fig:cdc2d} shows a schematic layout of the CyDet.  It is located after the BS in the muon transport section, and installed inside the warm bore of a large $1~\mathrm{T}$ superconducting Detector Solenoid (DS) and around the stopping target.

This detector has been adopted for Phase-I as there is no downstream curved solenoid electron transport and so most beam particles that do not stop in the muon-stopping target will go downstream and escape from the detector region without leaving any hits in the detector system.

A key feature of COMET is to use a pulsed beam that allows for the elimination of prompt beam backgrounds by looking only at tracks that arrive several hundred nanoseconds after the prompt beam flash. Therefore, any momentum-tracking devices must be able to withstand the large flux of charged particles during the burst of ``beam flash'' particles.

\subsection{Cylindrical drift chamber}

The detector is designed to avoid high hit rates due to beam particles, DIO electrons, and low-energy protons emitted after the nuclear capture of muons. Among the small fraction of particles which eventually enter the CDC and leave hits, DIO electrons and low energy protons dominate. The protons are easily identified, because the energy deposits in the CDC cells is about 100 times larger than that of similar-momentum electrons. To achieve the required sensitivity for Phase-I, the  momentum resolution must be 
about 200~keV/$c$ for 
105~MeV electrons. 
At this energy,
the momentum 
resolution is dominated by multiple-scattering.
Consequently, the CDC must be a low-mass detector and this dictates the construction 
and the choices of  cell configuration,  wires, and the gas mixture.\label{sec:cdcconfiguration}

\subsubsection{Layer and Cell Configuration}
The CDC is arranged in 20 concentric sense layers 
with alternating positive and negative stereo angles.
Cylindrical drift chambers with only stereo layers have been constructed in the past, such as the KLOE drift chamber~\cite{ADINOLFI200251}.
The 1st and 20th sense layers have a lower HV and act as guard layers to remove the space-charge that would otherwise accumulate.

Each cell has one sense wire surrounded by an almost-square grid of
field wires. The ratio of the total number of field wires to sense wires is 3:1.  The cell size is 16.8~mm wide and 16.0~mm in height and nearly constant over the entire CDC region.  Square cells are well-suited to the low
momentum tracks which can
enter the drift cells at large angles with respect to the radial
direction. The stereo angle $\varepsilon$ is set to 64--75~mrad to achieve a longitudinal spatial resolution $\sigma_{z}$ of about 3~mm. In total there 
are
4,986 sense wires and 14,562 field wires.

The field wires are made 
of
aluminium in order to reduce multiple scattering.
Whilst it would be desirable to use
 $\phi$80~\micro{}m aluminium wires, the operation voltage for this case would then need to be below 1730~V to keep the electric field on the wire surface below 20 kV/cm, above which corona discharges and whisker growth can occur on the wires. However,
 tests with the CDC prototype have shown that 1730~V is not enough to obtain sufficient signal gain. Consequently, $\phi$126~\micro{}m wires are used, allowing operation with voltages of up to 1900~V---and simulations show
that the change in the momentum resolution does not significantly affect the physics sensitivity. A tension of 80~g is applied to the field wires to match the gravitational sag of the sense wires.

The sense wires are made of gold-plated tungsten, 25~\micro{}m in diameter, tensioned to 50~g.
The deflection due to gravity is about 
50~\micro{}m
at the centre of the CDC.
The total tension force on the end plates is 1.4~tonne.

\begin{figure}[htb!]
 \begin{center}
 \subfigure[][]{
  \includegraphics[height=0.4\textwidth]{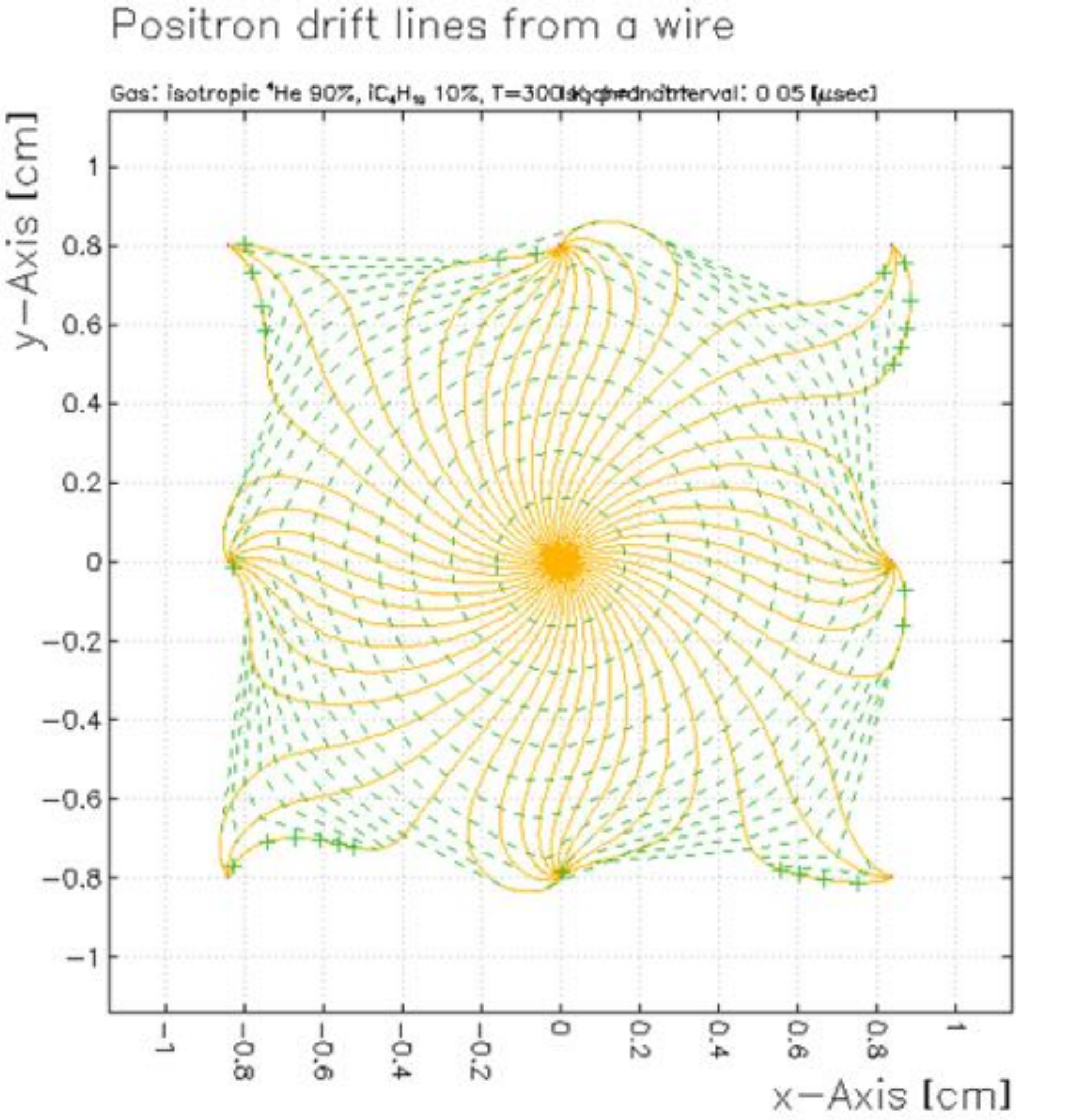}
  }
  \subfigure[][]{
  \includegraphics[height=0.4\textwidth]{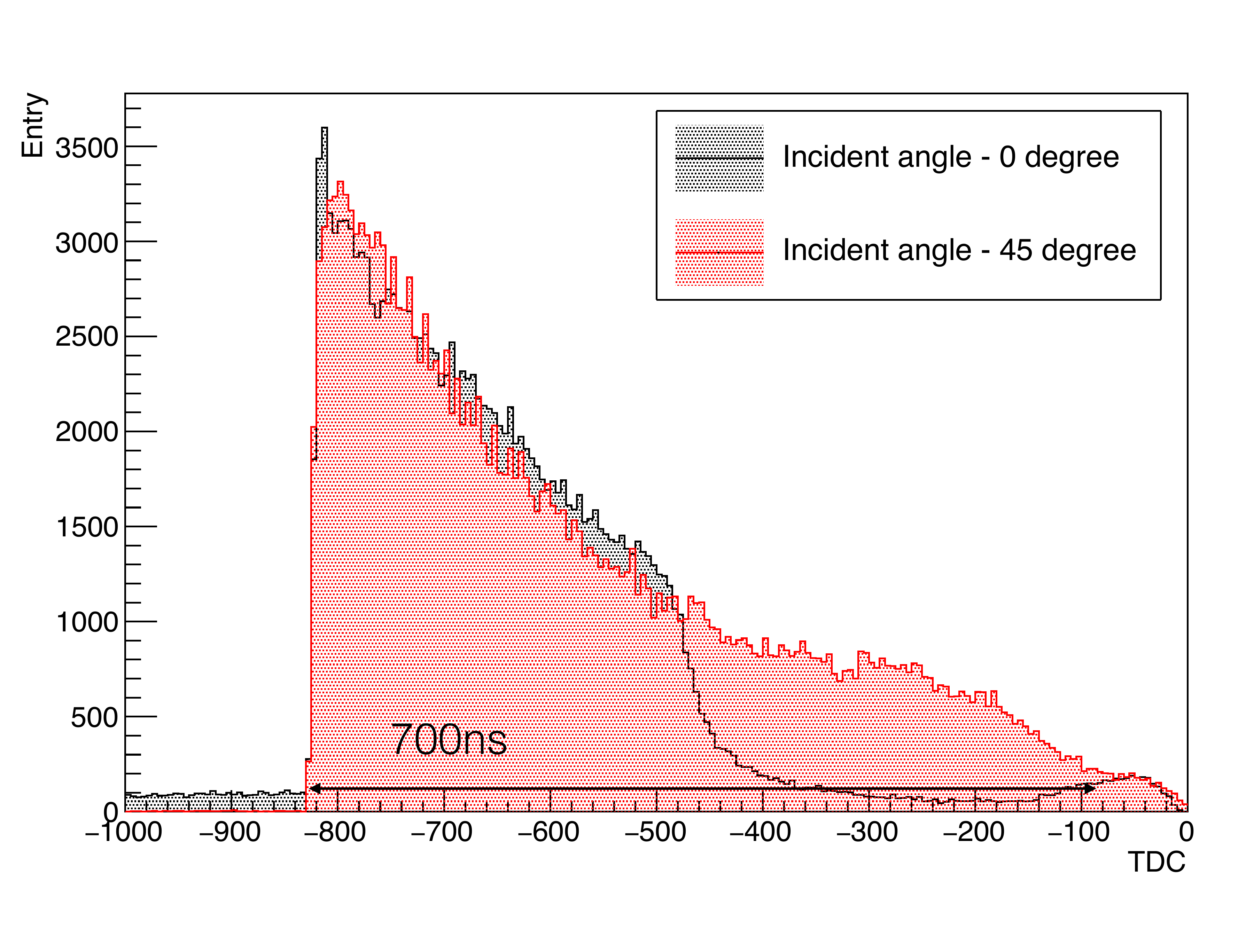}
  }
 \end{center}
 \caption{
 (a) Garfield simulation of the drift lines for a CDC cell under a 1 Tesla magnetic field, and (b) the drift time distribution for two incident angles  from the 
 fourth prototype.
 Gas composition is He:i-C$_{4}$H$_{10}$ (90:10).}
 \label{fig:CDC-DriftlineZ}
\end{figure}

The  chamber gas  is He:i-C$_{4}$H$_{10}$ (90:10), which has a radiation length of about 1300~m. The field and sense wires reduce the average value of the radiation length of the CDC tracking volume to 507~m.
A HV up to  1900 V is applied to the sense wires with the field wires at ground potential, giving an avalanche gain of approximately  $1 \sim 4 \times 10^{4}$.
Typical drift lines for a cell is shown in \cref{fig:CDC-DriftlineZ}, calculated for this 
gas
mixture.

\begin{figure}[htb!]
 \begin{center}
  \includegraphics[width=0.80\textwidth]{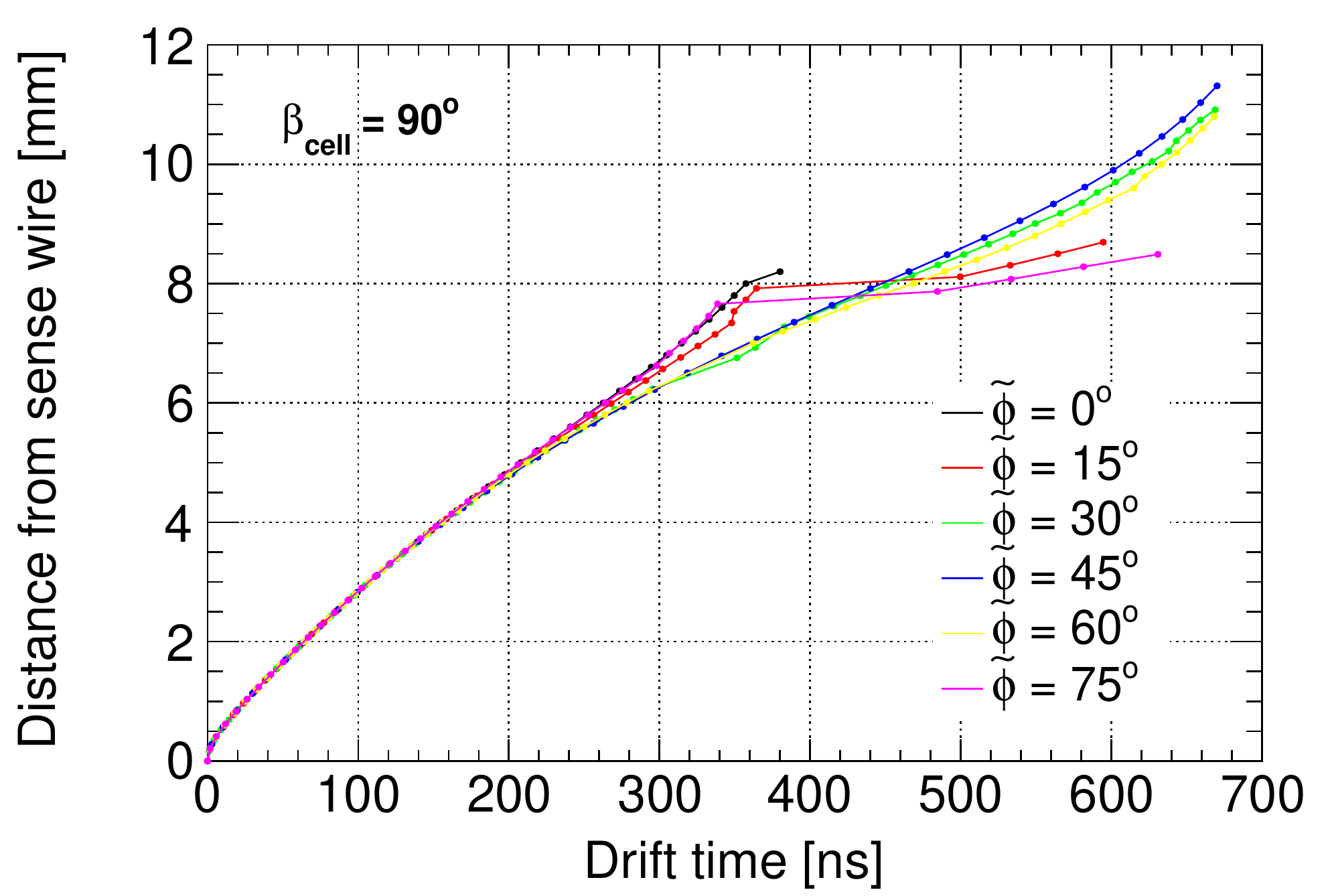} 
 \end{center}
 \caption{
Space-time correlations with different incident angles ($\tilde{\phi}$) without magnetic field, calculated with Garfield simulation.  
}
 \label{fig:CDC-garfield-xt}
\end{figure}

The Garfield program has been used to study cell properties, including
drift time isochrones, time-distance relationships, distortions and gain variations.
Typical drift time distributions are shown in \cref{fig:CDC-DriftlineZ} (b). Space-time correlations with different incident angles are shown in \cref{fig:CDC-garfield-xt}.

\subsubsection{Mechanical Design}

The main parameters of the CDC are summarised in \cref{tbl:CDC-parames}.
There are three main mechanical parts composing the CDC: 
the endplates, the
inner wall and the outer wall. 
The radii of the inner and the outer walls
are chosen to avoid DIO electrons with momentum less than 60~MeV$/c$ from hitting the CDC and to
fully cover the tracks of 105~MeV$/c$ signal electrons. The walls are made from carbon fibre reinforced
plastic (CFRP); the inner wall is 0.5~mm thick and the outer wall 5~mm. The inner and outer walls
have thin aluminium foils glued inside them to eliminate charge-up on the CFRP. 
Tapered 
aluminium endplates with 10~mm thickness 
are 
chosen for the endplates to adequately support a 1.4 ton wire tension load. % (already mentioned) The outer wall is made of CFRP with a thickness of 5~mm, and the inner wall of 0.5~mm CFRP.

\begin{table}[htb!]
  \vspace{3mm}
  \begin{center}
     \caption{Main parameters of the CDC.} \label{tb:size}
    \label{tbl:CDC-parames}
    \begin{tabular}{llc}\hline\hline
    Inner wall & Length  & 1495.5~mm \cr
                    &  Radius & 496.0--496.5~mm \cr
                    &  Thickness & 0.5~mm \cr
\hline
    Outer wall & Length & 1577.3~mm \cr
                     &Radius  &  	835.0--840.0~mm \cr
                    &  Thickness & 5.0~mm \cr
\hline
    Number of sense layers & & 20 (including two guard layers) \cr
\hline
    Sense wire & Material  & Au-plated W \cr
                       &Diameter  & 25~\micro{}m\cr
                      & Number of wires & 4986 \cr
                       & Tension & 50~g \cr
\hline
    Field wire & Material & Al \cr
                          &Diameter & 126~\micro{}m \cr
                     & Number of wires & 14562 \cr
                      & Tension & 80~g \cr
\hline
    Gas & Mixture & He:i-C$_{4}$H$_{10}$ (90:10) \cr
        & Volume  & 2084 L \cr
\hline\hline
    \end{tabular}
    \vspace{3mm}
  \end{center}
\end{table}

The mechanical properties of the design have been calculated through Finite Element Analysis using
SolidWorks.
The total wire tension load is calculated to be $F_{wire}$=12700~N/m$^2$ based on the parameters described in \cref{tbl:CDC-parames}.
The maximum deformation of the endplate is estimated to be as small as 
1.1~mm for the tapered angle of 10\Deg. 
The deformation was applied as pre-tension in advance of wire stringing, 
and then the pre-tension was released as the wires were strung to keep the constant  
deformation and sufficient wire tension.

\subsubsection{Electronics}
\label{subsec:cdcreadout}
The Belle-II CDC readout electronics board (RECBE)~\cite{ref:trigger:belleii:2}  is
used for the front-end readout of the  CDC with appropriate modifications.
 Each board has 48 input channels, 6 ASD (Amplifier Shaper Discriminator) ASIC chips~\cite{SHIMAZAKI2014193}, 
 6 ADCs and an FPGA.
Data is sent to the DAQ PC via an optical fibre cable. RJ45 connectors are used to download the firmware into the FPGA and for transmitting clock, trigger and busy signals 
between
the FCT board described in \cref{subsec:trigger}.

Eight pre-production RECBE boards underwent burn-in tests in 2015. As a result of these tests and employing the Arrhenius model to predict how time-to-fail varies with temperature, the RECBE lifetime estimate is longer 
than 2.5 years 
for one board.
The readout 
electronics
are 
located on the CDC downstream endplate and the HV cables 
are 
connected on the upstream endplate.
The production of all the readout boards (128 boards with spares) was completed by the IHEP group in China in 2015.

\begin{figure}[htb!]
	\begin{center}
		\includegraphics[width=0.8\textwidth]{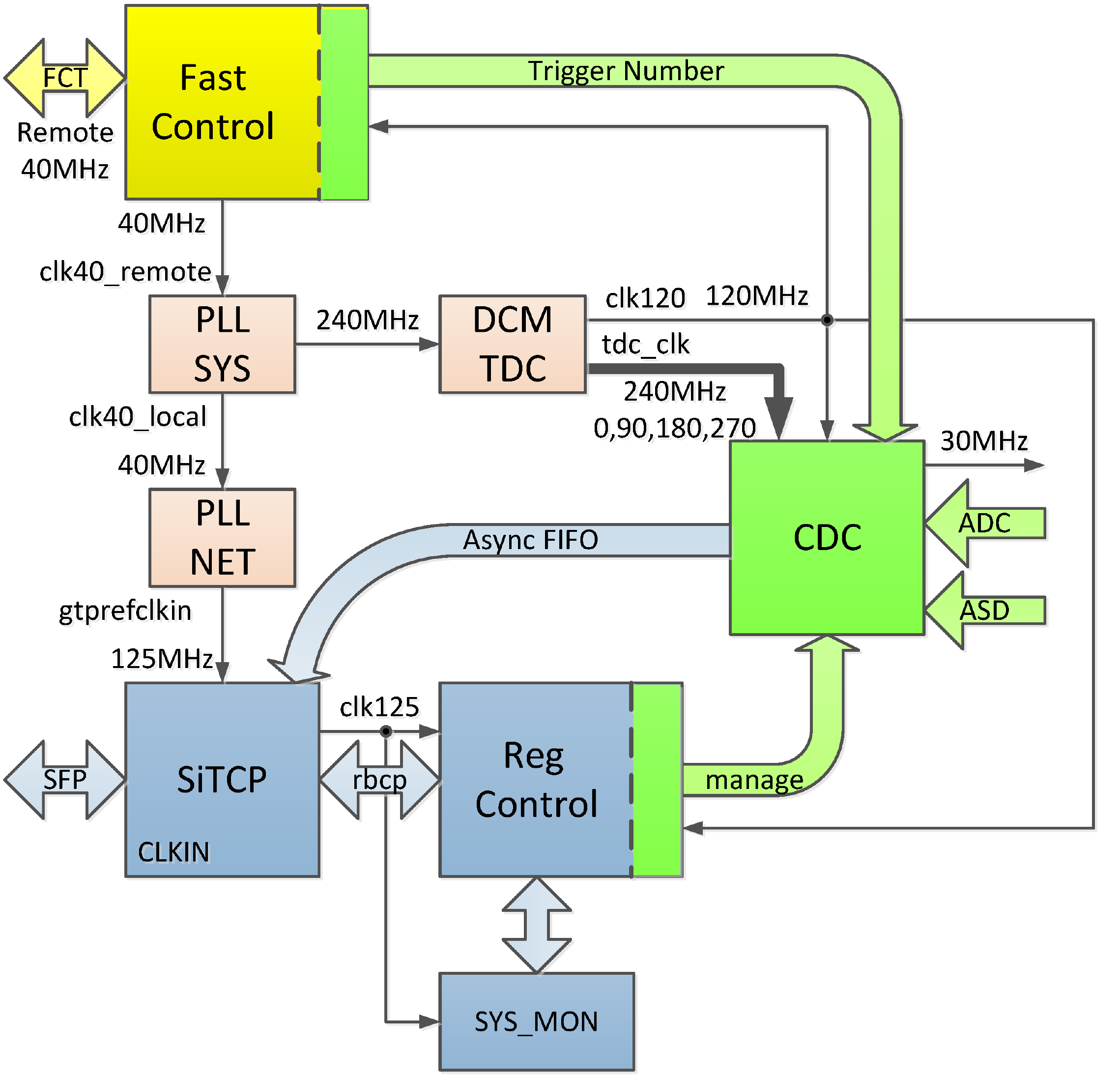}
	\end{center}
	\caption{Block diagram of the RECBE firmware. In yellow is the 40~MHz domain, which comes from the FCT board, in green is the 120/240~MHz domain, which manages for ADC and TDC and in blue  is the 125~MHz domain, which matches the data transfer rate of the Gigabit Ethernet.}
	\label{fig:cdcreadoutfirmware}
\end{figure}

\Cref{fig:cdcreadoutfirmware} shows a block diagram of the COMET CDC readout  implemented in the FPGA 
of RECBE.
The main features are:
\begin{itemize}
\item The fast control block receives the reference clock, trigger (trigger number) from the FCT board and sends a busy signal to stop receiving triggers if the buffer is full.
\item The CDC block arranges data of drift time and charge from  the TDCs and ADCs.
\item The SiTCP block is used to transmit the event data to the DAQ system via a Gigabit Ethernet fibre link.
\item The Reg Control block is responsible for configuration and status.
\item The SYS MON block is used for status monitoring, such as temperature and voltage.
\end{itemize}

The size of the ring buffer which stores the ADC and TDC data is 256 deep, corresponding to $\sim$8.533~\micro{}s.
Typically, the event window size for the CDC is 32 samples  corresponding to $\sim$1.067~\micro{}s which makes
the  trigger latency  $\sim$7~\micro{}s.
Hence typically, eight events can be stored in the buffer.

The frontend readout boards are installed near the detector region 
where the radiation level is high. The radiation effects, in particular 
from neutrons, onto FPGA hardware will be a severe problem. 
We developed 
high-reliablity
firmware with auto-recovery schemes, 
and evaluated it with a neutron beam. In the test, soft error rates 
were measured and good performances of the schemes were demonstrated~\cite{NAKAZAWA2019351}.

\subsection{CDC performance estimation and tests}

\subsubsection{CDC hit rates}
\label{sec:cdchitrates}

The CDC hit rates have been studied with {\tt Geant4} simulations. Potential sources causing noise hits are grouped into three categories:
\begin{enumerate}
\item  muons and pions at the muon stopping target and its vicinity, which create secondary (or tertiary) particles in the CDC,
\item a prompt beam flash,
\item neutrons which are either in a beam,  from the proton target or the proton dump.
\end{enumerate}

\paragraph{Hit rate contribution from stopped muons and pions}

\begin{figure}[htb!]
\begin{center}
\includegraphics[width=0.45\textwidth]{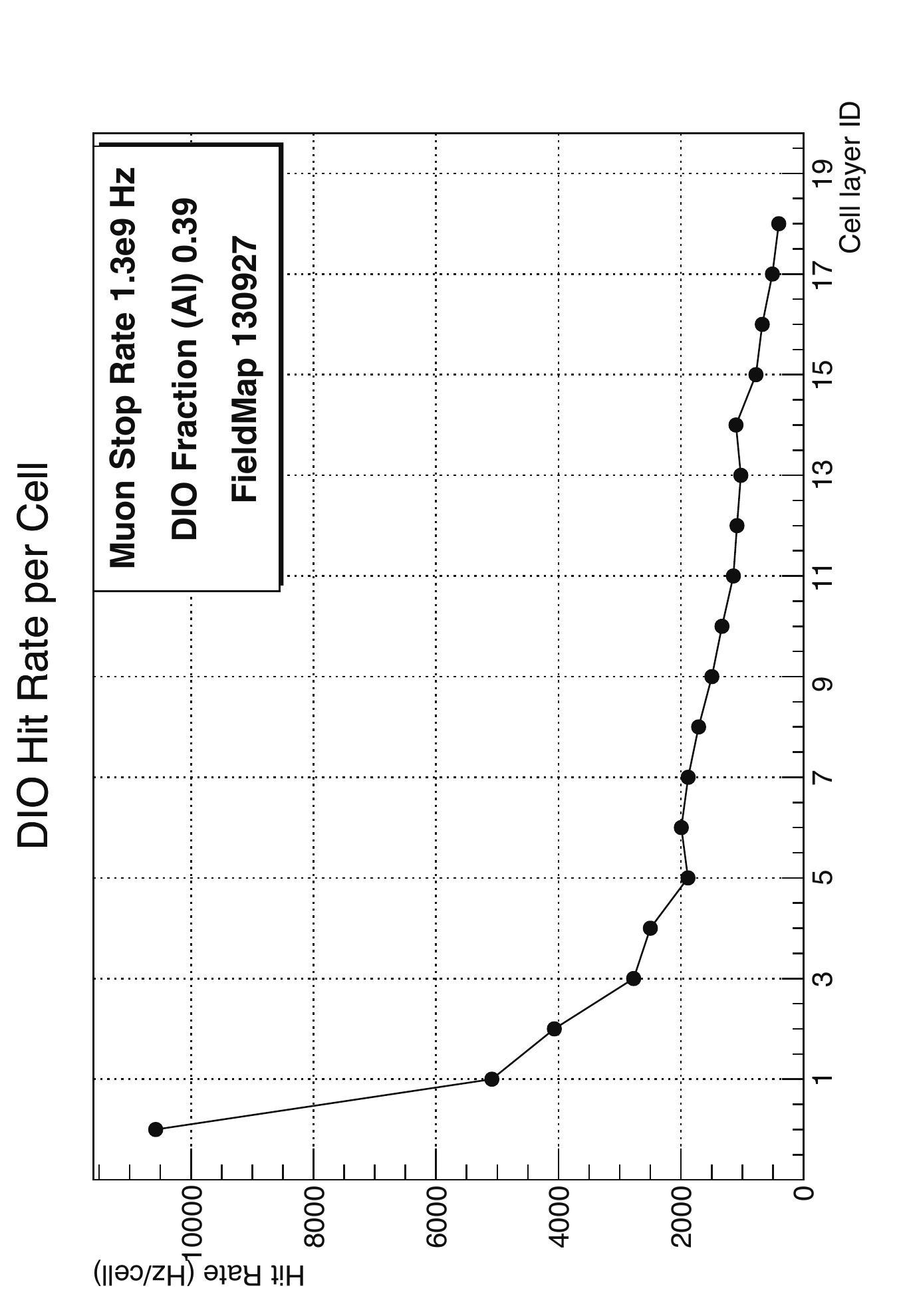}
\includegraphics[width=0.45\textwidth]{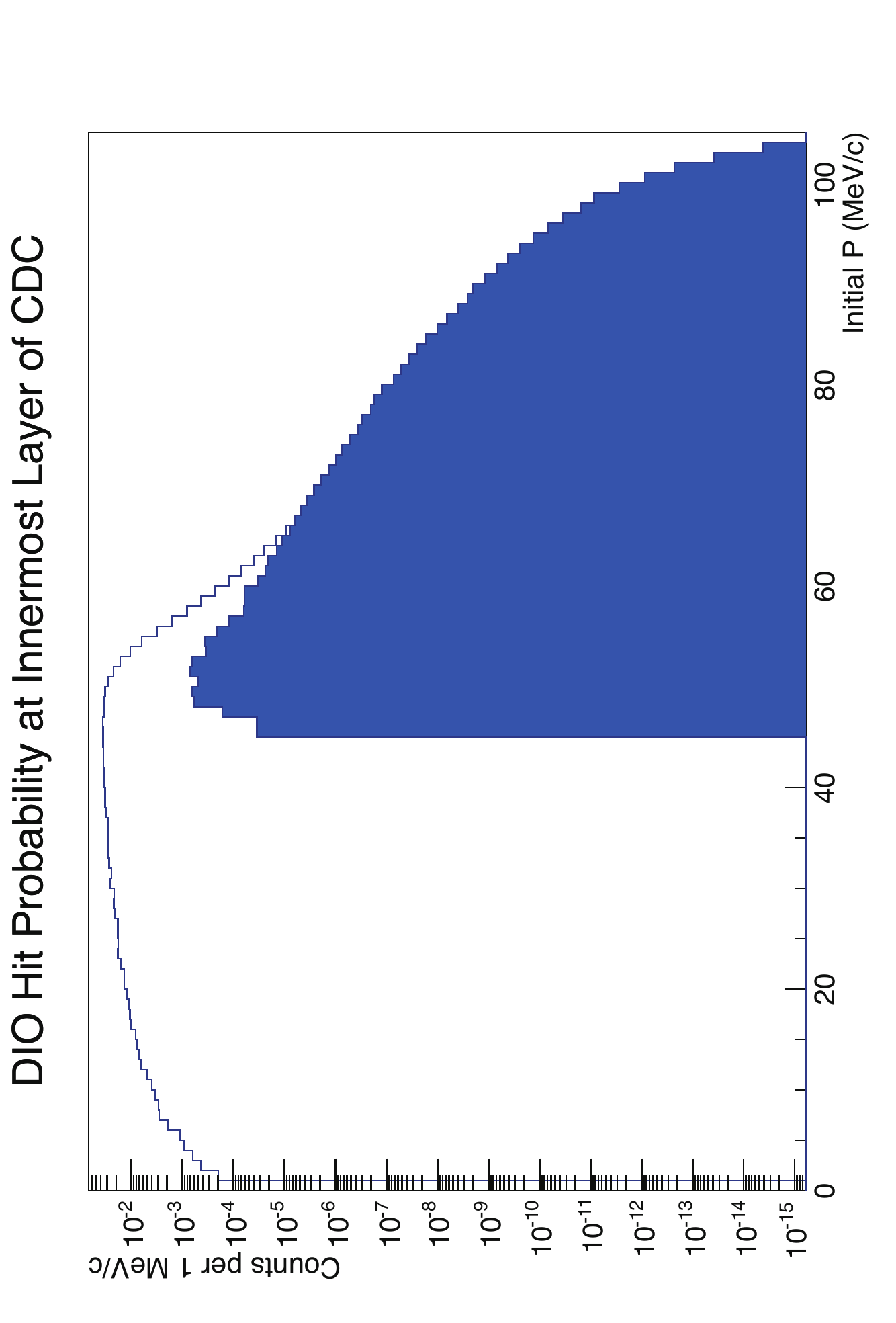}
\caption{(Left) 
DIO electron hit rate for each CDC cell layer. Note that the cell-layer ID of 0 means the guard layer.
(Right) Momentum distribution of the DIO
  electrons. The blue histogram shows those of electron hits the CDC. }
\label{fig:CDC-DIO-hitrate}
\end{center}
\end{figure}

The estimated  hit rates of each CDC cell at different layers from DIO electrons from stopped muons
are shown in \cref{fig:CDC-DIO-hitrate} 
(left).
The rate decreases quickly at deeper CDC layers, since the DIO momentum spectrum drops as a function of an electron momentum, as shown in \cref{fig:CDC-DIO-hitrate}
(right).
From \cref{fig:CDC-DIO-hitrate}
(left), 
the time-averaged rate for the innermost sense wire  is at most 
5~kHz/cell,
yielding
an instantaneous rate of about 
15.6~kHz/cell
allowing for the duty factor of the J-PARC MR proton beam cycle, which is about 3. This implies a hit occupancy for one bunch cycle of 1.17 \micro{}s of about 
1.8 \%.

\begin{figure}[htb!]
\begin{center}
\includegraphics[width=0.45\textwidth]{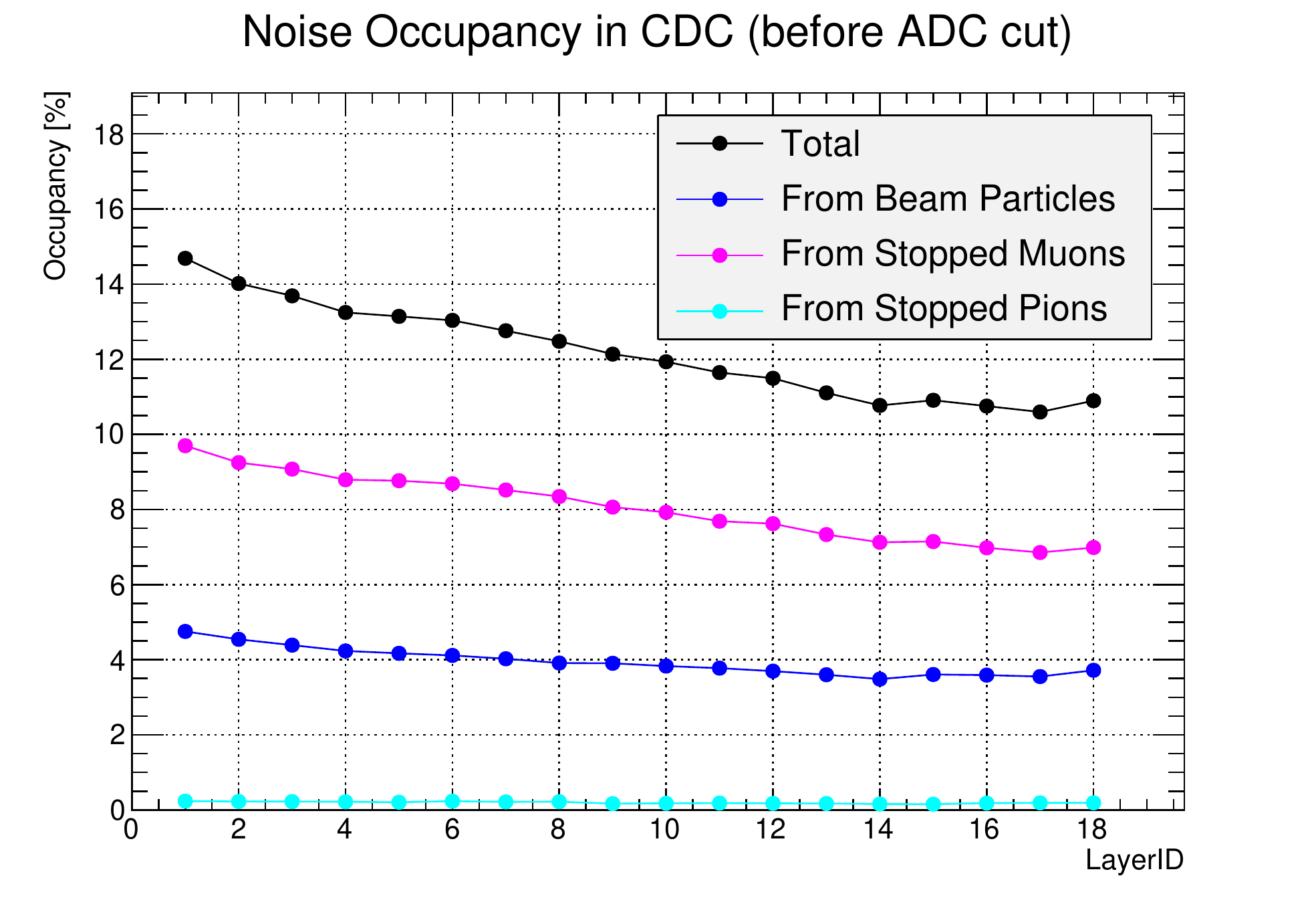}
\includegraphics[width=0.45\textwidth]{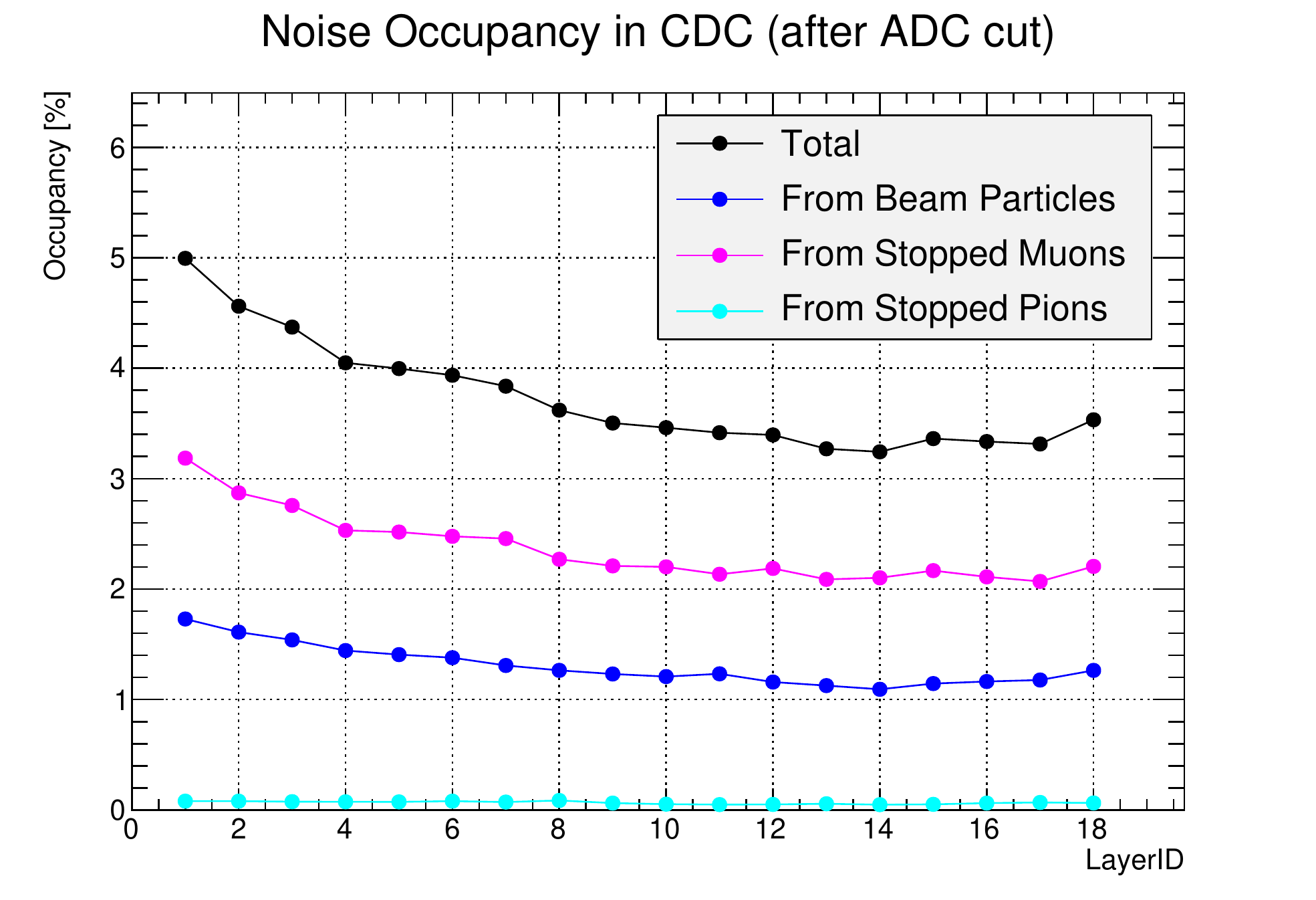}
\caption{CDC single hit occupancy as a function of the CDC layers for a gas mixture of He:i-C$_{4}$H$_{10}$ (90:10). The left and right figures are respectively those before the energy deposit cut and after the cut, which selects hits with energy deposit smaller than 5 keV. The closed circles filled with blue, magenta, cyan, and black are respectively from beam particles, stopped muons, stopped pions, and total occupancy. }
\label{fig:CDC-hitoccupancy}
\end{center}
\end{figure}

Using the results from the 
AlCap 
\footnote{The AlCap is a collaboration between COMET and Mu2e that measures the rate and spectrum of particles emitted from nuclear muon capture on aluminium~\cite{alcap:1501:04880}.  }
experiment at PSI the time-averaged hit rate on a single cell from proton emission from muon capture is  estimated to be 1.4~kHz.
Other sources of hits following nuclear muon capture, such as bremsstrahlung photons, muonic X-rays, neutrons from nuclear muon capture, $\gamma$-rays from the final state nucleus have also been considered.
The magenta points and lines in \cref{fig:CDC-hitoccupancy} summarise the CDC occupancy caused by stopped muons 
which result in a 
total occupancy of between 7\% and 10\%.

The
hit rate contribution from stopped pions  is also summarised with the cyan points and lines in \cref{fig:CDC-hitoccupancy}. Their contribution is  small compared to the other categories.

\paragraph{Hit rate contribution from beam flash}

\begin{figure}[tb!]
\begin{center}
\includegraphics[width=0.8\textwidth]{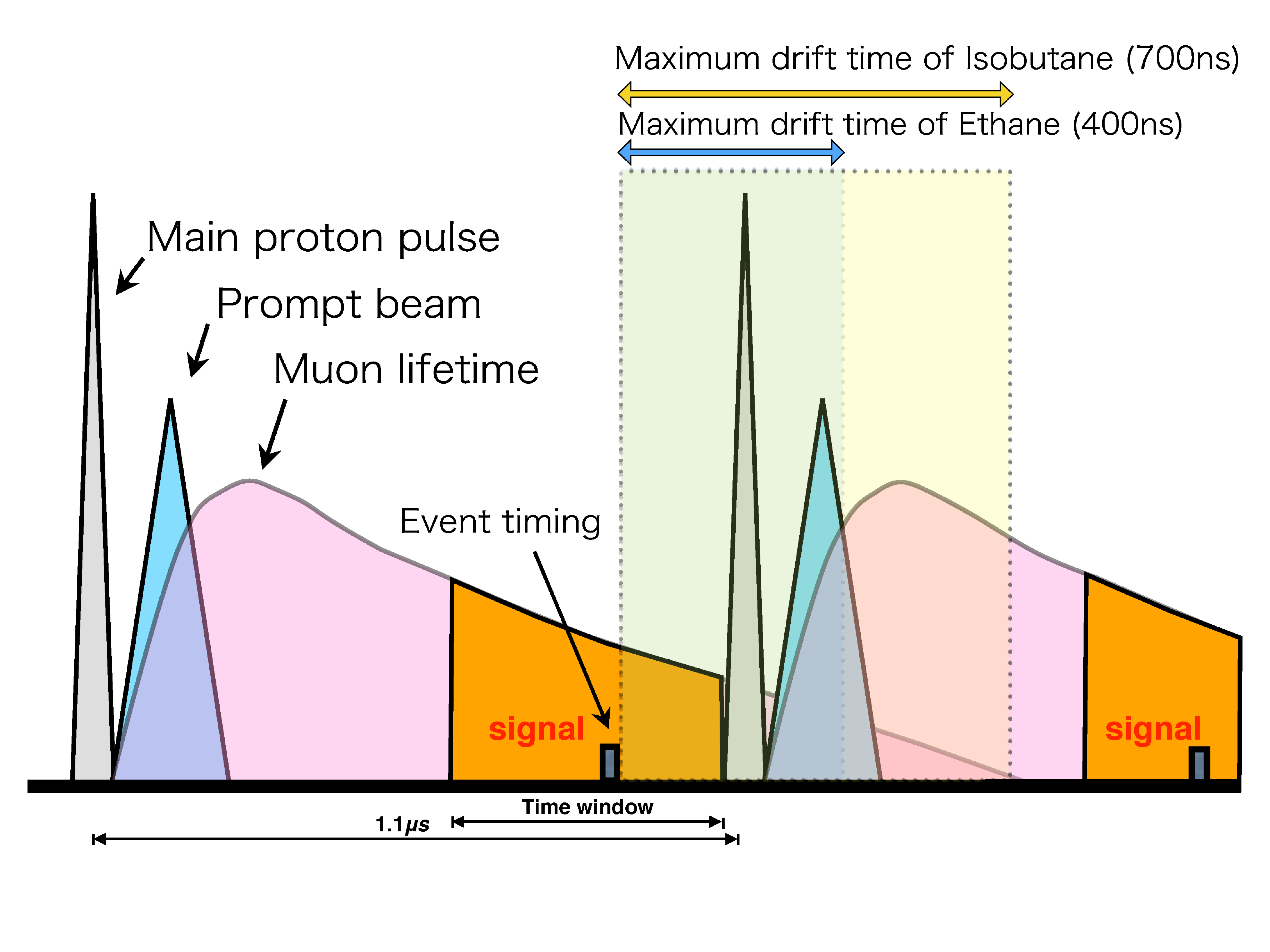}
\caption{Relation between prompt beam flash, event timing in the time window of measurement, drift time of hits in a time period of two beam cycles.}
\label{fig:CDC-DriftTimeDistribution-Prompt}
\end{center}
\end{figure}

\begin{figure}[htb!]
\begin{center}
\includegraphics[width=0.8\textwidth]{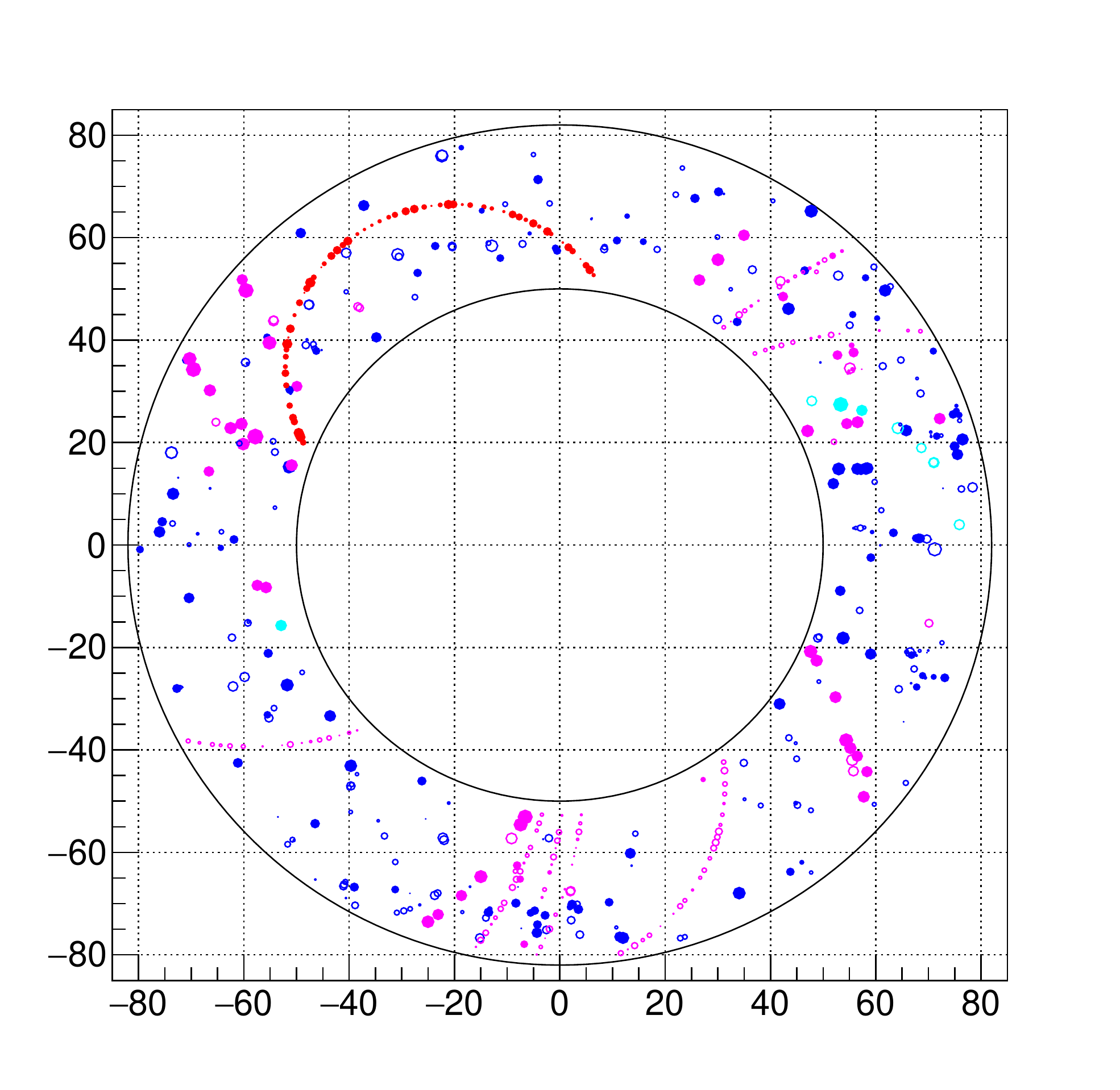}
\caption{%
Simulated CDC event display for a gas mixture of He:i-C$_{4}$H$_{10}$ (90:10). This event occurs 1090~ns after the prompt beam flash. 
Hits in red, cyan, magenta, and blue are respectively from signal tracks, tracks from pion capture, tracks from muon capture, and other noise hits.
The hits in open circles and closed circles are respectively those with energy greater and smaller than 5 keV.}
\label{fig:CDC-eventdisplay-isobutane}
\end{center}
\end{figure}

Although the beam flash is very short the associated CDC hits arrive over a period as a result of the drift time. The distribution of drift times could be different for different gas mixtures, as described in \cref{sec:cdcconfiguration}. The drift time distribution from the 
fourth prototype tests at SPring-8 are shown in \cref{fig:CDC-DriftlineZ} (b). 
The maximum drift time, coming from the cell corners, is 700~ns and \cref{fig:CDC-DriftTimeDistribution-Prompt} shows the relation between prompt beam flashes, event timing in the time window of measurement, and the drift time of the hits. As the separation of the beam pulses is 1170~ns  the following beam flash will  come within the drift time of 700~ns and therefore the CyDet detector must be able to accommodate the beam flash. The major sources creating noise hits are photon and neutron interactions which  depend strongly on the muon beam design, in particular the collimation. Radiation shielding to prevent photons hitting the CDC (in particular the endplates) is important. 
The total contribution from beam flash is estimated to lead to around 4\% occupancy as shown in \cref{fig:CDC-hitoccupancy}.  In \cref{fig:CDC-eventdisplay-isobutane} a sample event display with beam flash is shown. The event in \cref{fig:CDC-eventdisplay-isobutane} occurs  1090~ns after the prompt beam timing, and therefore the time period for open hits covers the following beam bunch.  It can be seen that noise hit rates are not too large.

\paragraph{Hit rate contribution from ambient and beam neutrons}
A fraction of the fast neutrons from the beam target will penetrate the endplates and inner and outer walls and degrade to a thermal spectrum.
These have been simulated with both {\tt PHITS}
and {\tt Geant4} which indicate that with paraffin shields installed along the beam line, the hit rate induced by ambient and beam neutrons can be made very small.

\subsubsection{Charge measurement}
\label{CDC-charge}

\begin{figure}[htb!]
\begin{center}
\includegraphics[width=0.8\textwidth]{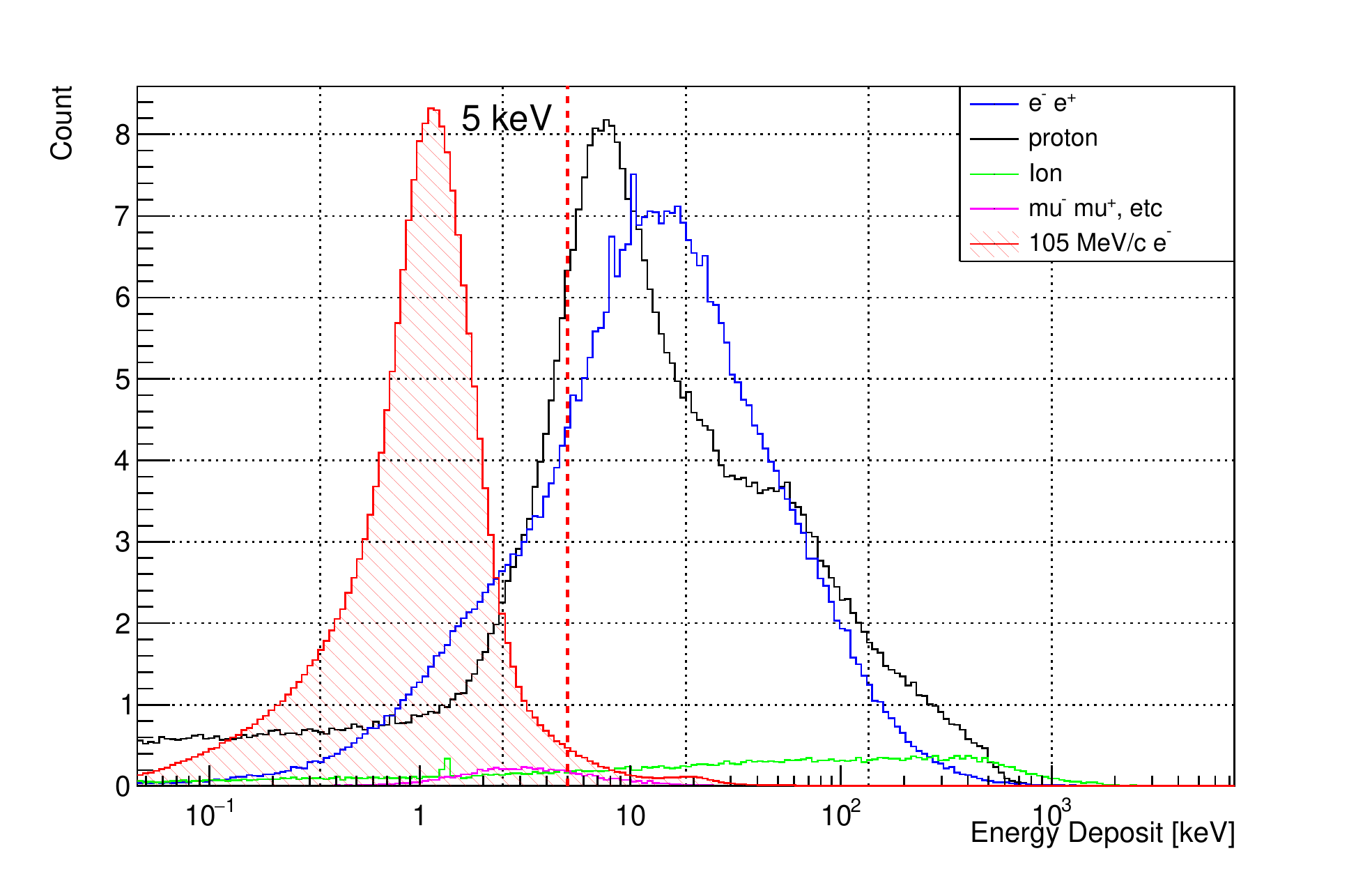}
\caption{%(Top)
  Total energy deposits per cell for signal electrons
  and noise hits.  The signal electrons are shown in the red shaded
  histogram; noise hits from various sources are shown in other
  histograms. A 5~keV threshold line (as used in the event displays)
  is also shown. 
}
\label{fig:CDC-charge}
\end{center}
\end{figure}

Most noise hits are either associated with low-energy electrons or
positrons created by photon conversion, electrons from DIO decays of
muons and protons created from nuclear muon capture. For the conversion electrons and positrons,  typical energies are a few MeV
and their helical trajectories will have a very small radius and are likely to stay in the same CDC cell for a long time.  In some
cases, when created at one of the endplates they may travel along the CDC,
and reach the other endplate.  Therefore  many of
these noise hits will have a \emph{large} charge in the hit
cell. Low momentum protons are heavily ionising and so will also
deposit a lot of energy in a CDC cell.

The CDC readout system is capable of measuring total charges with
30~MHz sampling.
\Cref{fig:CDC-charge} shows the
total energy deposit for electron signals of \mue conversion and noise
hits. Hence about 68\% of the noise hits can
be identified and removed by only retaining hits with an energy
deposit smaller than 5~keV, while 99\% of the signal hits will pass this selection. \Cref{fig:CDC-hitoccupancy}(Right) shows the hit occupancy after this selection. The original hit occupancy of 12\% would then be reduced to about 3.5\% .
This high-charge cut is carried out in the RECBE firmware as described in \cref{subsec:cdcreadout}.

\subsubsection{Prototype CDC tests}

Four small-sized prototype chambers were constructed in order to examine the performance of the  CDC, their
specifications are summarised in \cref{tb:CDC-prototype-spec}.

\begin{table}[htb!]
 \centering
 \caption{Specifications of the prototype chambers.}
 \label{tb:CDC-prototype-spec}
 \begin{tabular}{lcccc} \hline\hline
  Prototype & 1 & 2 & 3 & 4 \\ \hline
  Sense wire $\phi$ [$\mu$m]  & 30  & 25  & 30  & 25 \\
  Field wire $\phi$ [$\mu$m]    & 126 & 80 & 80  & 126 \\
  \# of sense layers               & 11  & 5   & 5    & 9 \\
  \# of readout channels         & 199 & 27  & 27 & 87 \\
  Stereo angle [mrad]              & 25  & 70  & 70  & 66 \\
  Wire length [mm]                 & 600 & 200 & 200 & 600 \\
  Angle coverage [deg]            & 30  & 8    & 8    & 15 \\ \hline\hline
 \end{tabular}
\end{table}

\begin{figure}[htb!]
 \centering
 \includegraphics[width=\textwidth]{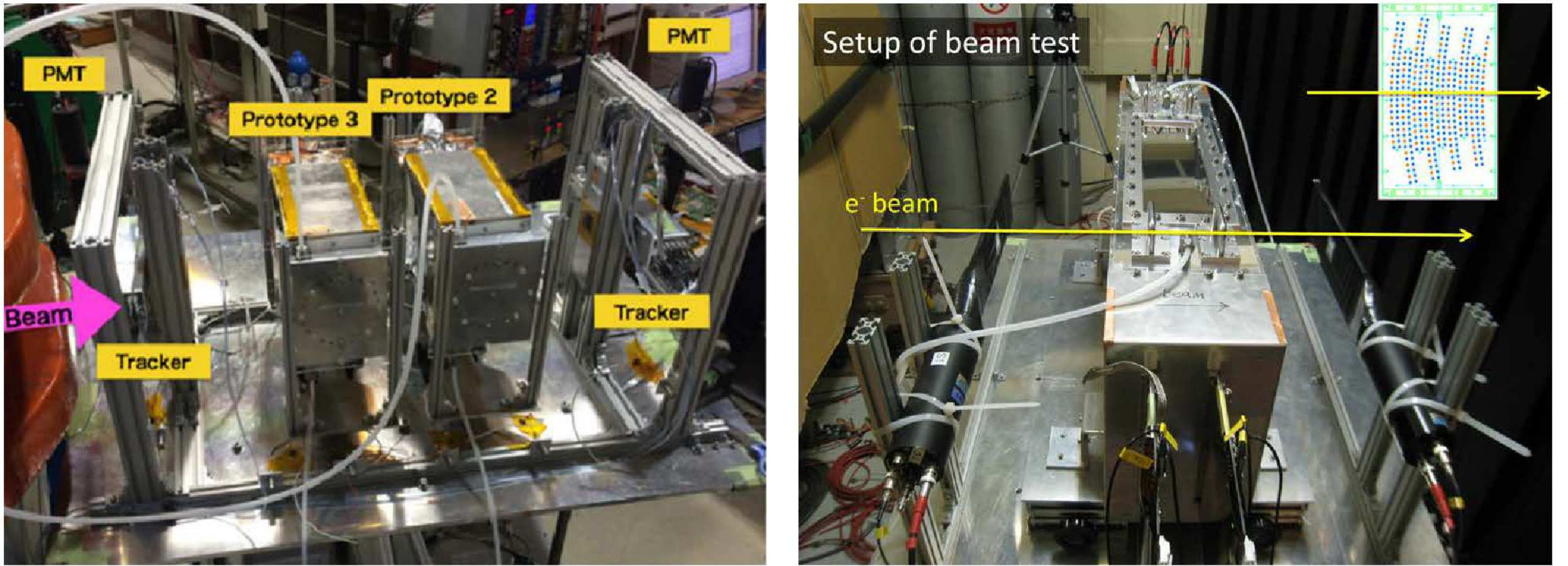}
 \caption{Setups of the beam test of the second and third prototypes at ELPH/Tohoku University (left); and the beam test of the forth prototype at LEPS/SPring-8 (right).}
 \label{fig:CDC-Prototype234setup}
\end{figure}

\begin{figure}[htb!]
 \centering
 \includegraphics[width=\textwidth]{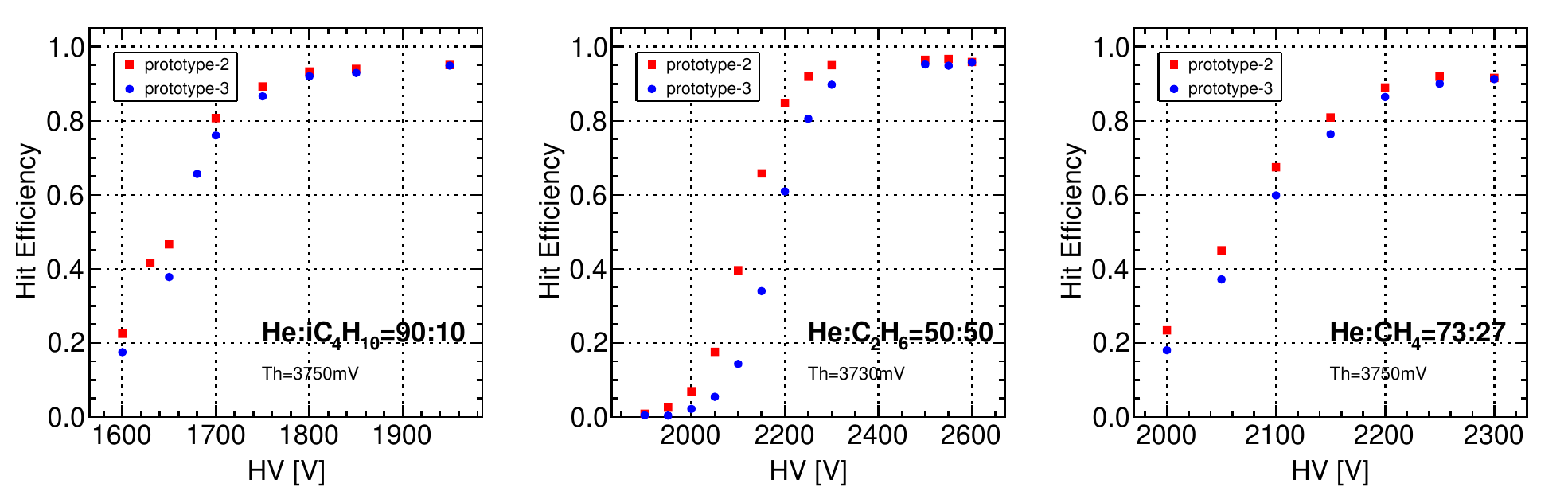}
 \caption{
 Hit efficiency of the second and third prototypes for gas mixtures of 
 He:i-C$_{4}$H$_{10}$ (90:10) (left), 
 He:C$_{2}$H$_{6}$ (50:50) (middle), 
 and He:CH$_{4}$(73:27) (right). 
 The second and third prototypes have sense wire diameters of 25 and 30~\micro{}m, respectively.
 }
 \label{fig:CDC-prototype23-eff}
\end{figure}

The first prototype chamber was constructed in order to examine the performance in a tentative design.
The second and third prototypes are simple box-type chambers with  sense wire diameters of 25 and 30~\micro{}m, respectively. They were constructed for the beam test at ELPH, Tohoku University in December 2014.
The chambers were irradiated by electron beams  for three gas mixtures as shown in \cref{fig:CDC-Prototype234setup} (left).
\Cref{fig:CDC-prototype23-eff} shows the hit efficiency as a function of the applied high voltage for each gas mixture.
This confirmed that the operation voltage should be higher than 1800 V for He:i-C$_{4}$H$_{10}$ (90:10) to achieve sufficient efficiency and led to the decision to use the  126~\micro{}m diameter field wires.
 \footnote{
 In this test, hit efficiency was saturated at nearly 95\% due to weakness of signal cables against electronic noise. This inefficiency was overcome by modifying cables after the test.
 }

\begin{figure}[htb!]
 \centering
 \includegraphics[width=0.8\textwidth]{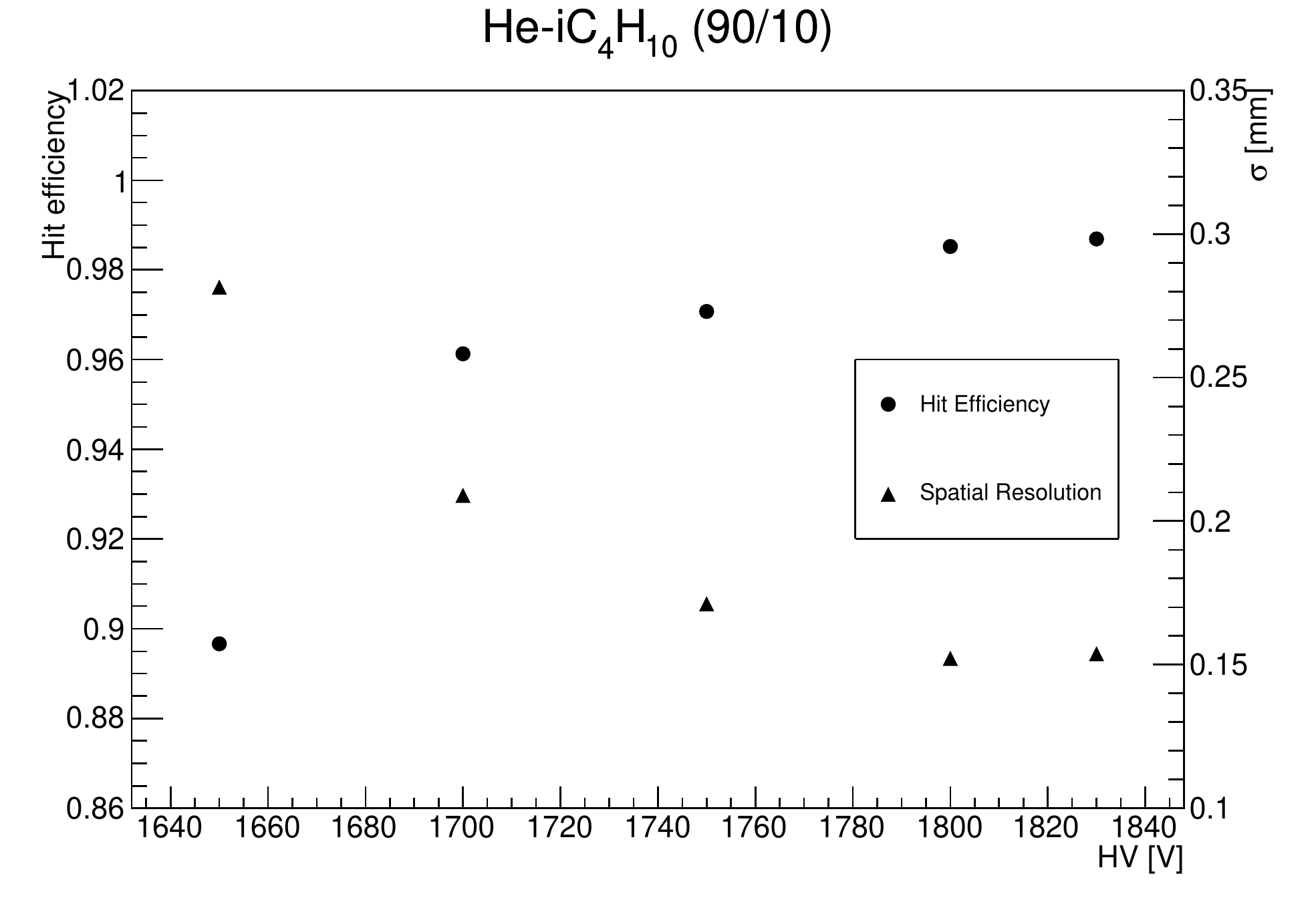}
 \caption{
 Hit efficiency (left axis, circle markers) and spatial resolution (right axis, triangle markers) for the forth prototype as a function of the applied high voltage. 
 }
 \label{fig:CDC-eff-sig-hv}
\end{figure}

The fourth (final) prototype chamber was constructed after the mechanical specification of the CDC had been determined. It  is a partial copy
of the real CDC design and hence
can be used to study the performance of the CDC under more realistic conditions.
It was tested  in an electron beam at LEPS/SPring-8 in July 2015, as shown in \cref{fig:CDC-Prototype234setup} (right), to examine the performance for three different gas mixtures.
The hit efficiency, spatial resolution, and drift velocity for different applied high voltage and different threshold values were investigated.
\Cref{fig:CDC-eff-sig-hv}  shows the hit efficiency and  the 
spatial resolution
as a function of applied high voltage.
The spatial resolution was extracted from the standard deviation of residual distributions by fitting tracks excluding one layer. Therefore, the resolution here includes a tracking uncertainty.

The conclusion from the prototype studies is that the He:i-C$_{4}$H$_{10}$ (90:10) gas mixture satisfies the requirements for efficiency and spatial resolution. These results, together with the Garfield predictions are now used in the current Geant4 simulations.

\subsubsection{Ageing tests}

A preliminary ageing test was performed in July 2014 at Osaka University.
A test chamber  was produced with Au-W sense wires of $\phi 25~\mu m$ and Al field wires of $\phi 80~\mu m$ and a He:i-C$_{4}$H$_{10}$ (90:10) gas mixture.
A central sense wire is the wire to be tested for the ageing effect.
There are two holes on each of the sides to irradiate X-rays with a $\mathrm{^{55}Fe}$ source.
One, ``Side A'', is used to give charges with two $\mathrm{^{90}Sr}$ sources,
and the other, ``Side B'', is used for reference as a not-aged sample wire.

To accelerate the ageing (charge accumulation), the applied HV value of the test wire was set to 2600~V when the wire is exposed to $\mathrm{^{90}Sr}$ sources,
and a HV of 1500~V  applied to the other 6 wires.
In this condition, we obtained an electric current of $\sim 6.5~\mu A$ on the test wire.
After every 30 minutes of exposure to two $\mathrm{^{90}Sr}$ sources,
the $\mathrm{^{90}Sr}$ sources were replaced by a $\mathrm{^{55}Fe}$ source for the gain measurement with 5.9~keV X-rays.

\begin{figure}[htb!]
\centering
\includegraphics[width=0.8\textwidth]{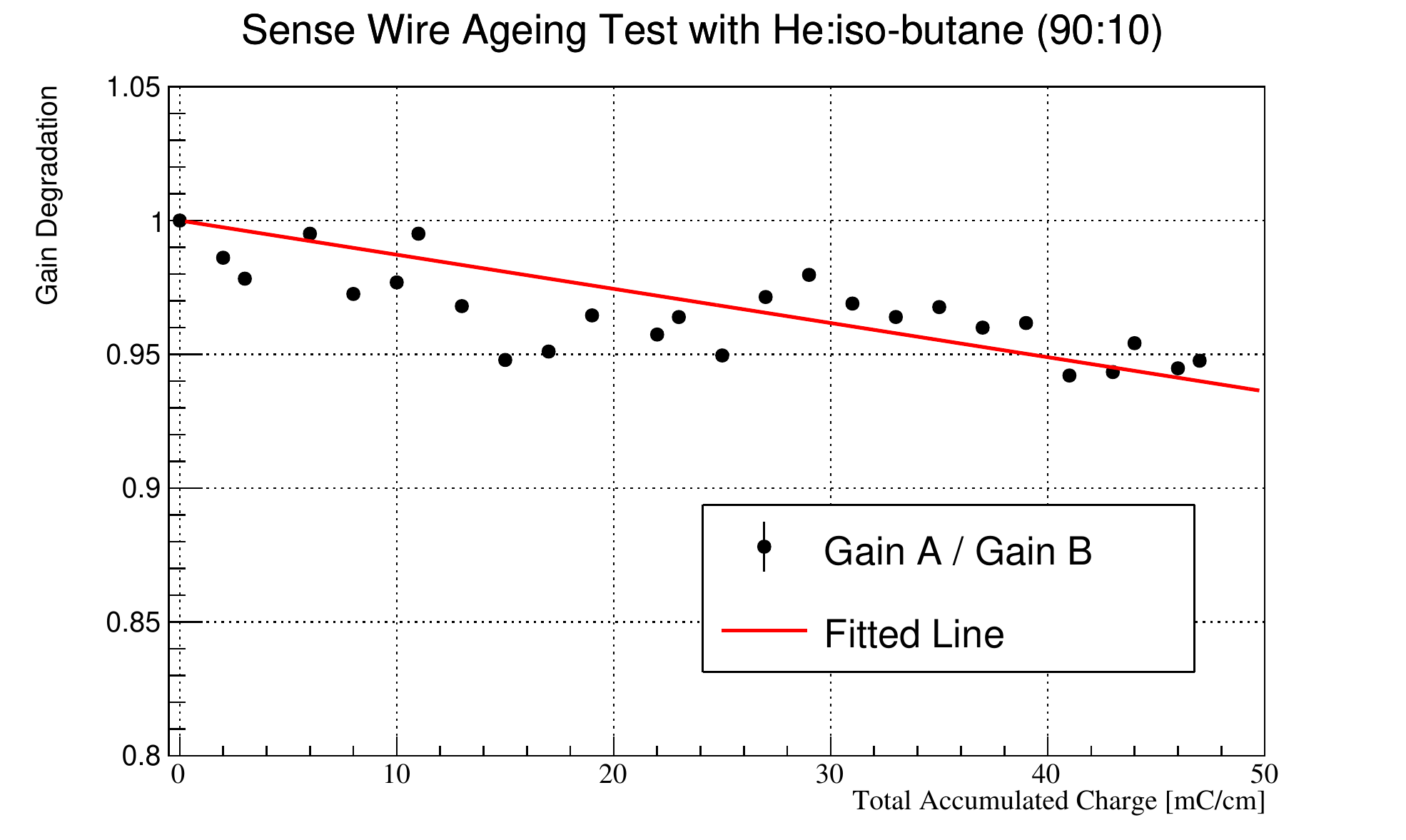}
\caption{The first result of ageing test with He:i-C$_{4}$H$_{10}$ (90:10). The black points indicate the gain drop as ratio of the measured gain against the gain not-aged area (Side B). The red line shows the fitted result with the data. The best fit value of the gradient is $-0.00132$, corresponding to the gain drop of 0.13\%/mC/cm.}
\label{fig:CDC-aging-plot1}
\end{figure}

After about 21~hours, the accumulated charge reached about 47~mC/cm and the
 ratio of the ADC values of ``Side A'' against ``Side B'' is shown in \cref{fig:CDC-aging-plot1}.
The black points indicate the results of the gain measurement with only statistical uncertainties,
and the red line is a fitted function with the points and a fixed offset ($ = 1.0$).
The systematic uncertainty is not considered here.
The best fit value of the gradient is $-0.0013$, corresponding to a gain drop of $0.13$~\%/mC/cm.
The temperature, humidity and atmospheric pressure were recorded in the experimental room during the measurement
and no significant changes observed.

By considering our running period in Phase-I (O$(100)$ days), the accumulated charge should be less than 2~mC/cm/wire with an
 estimated gain drop at the end of COMET Phase-I of only 0.3\%, which is sufficiently small.

\subsubsection{Cosmic-Ray Tests}
\begin{figure}[htb!]
\begin{center}
\includegraphics[clip,width = 0.9\textwidth]{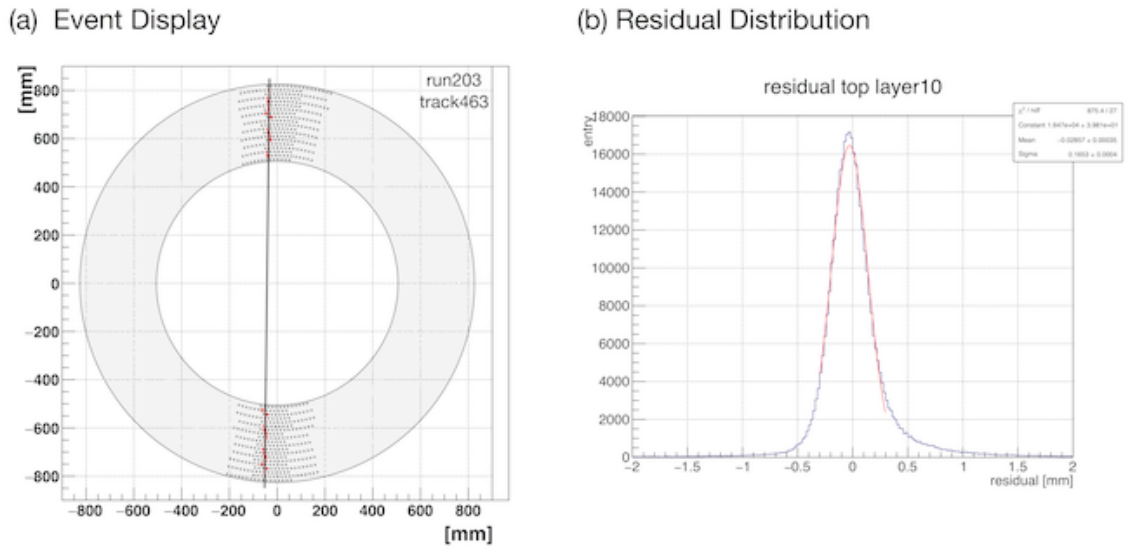}
\caption{
(a) Typical event display in cosmic-ray tests. (b)  Residual distribution for the layer-10 at 1825 V. The distribution is fitted with a Gaussian.
}
\label{fig:cdc-crt-results}
\end{center}
\end{figure}

The CDC 
was 
constructed in 2016, and its 
performance evaluation test using cosmic rays started in summer 2016~\cite{moritsu:ichep2018}. 
Stable operation of the CDC was achieved with the He:i-C$_{4}$H$_{10}$ (90:10) gas mixture and an applied high voltage up to 1850 V. 
\Cref{fig:cdc-crt-results} (a)  shows a typical 
event display where a clear cosmic-ray track can be drawn. 
From the deviation of drift distance from the distance of closest approach between a hit wire and a reconstructed track, a residual distribution was obtained as shown in \cref{fig:cdc-crt-results}(b), 
giving a position resolution of 170~\micro{}m including a tracking uncertainty.

\subsection{CyDet Trigger Hodoscope} 
\label{sec:cth}

\begin{figure}[htb!]
 \begin{center}
 \includegraphics[clip,width=0.8\textwidth]{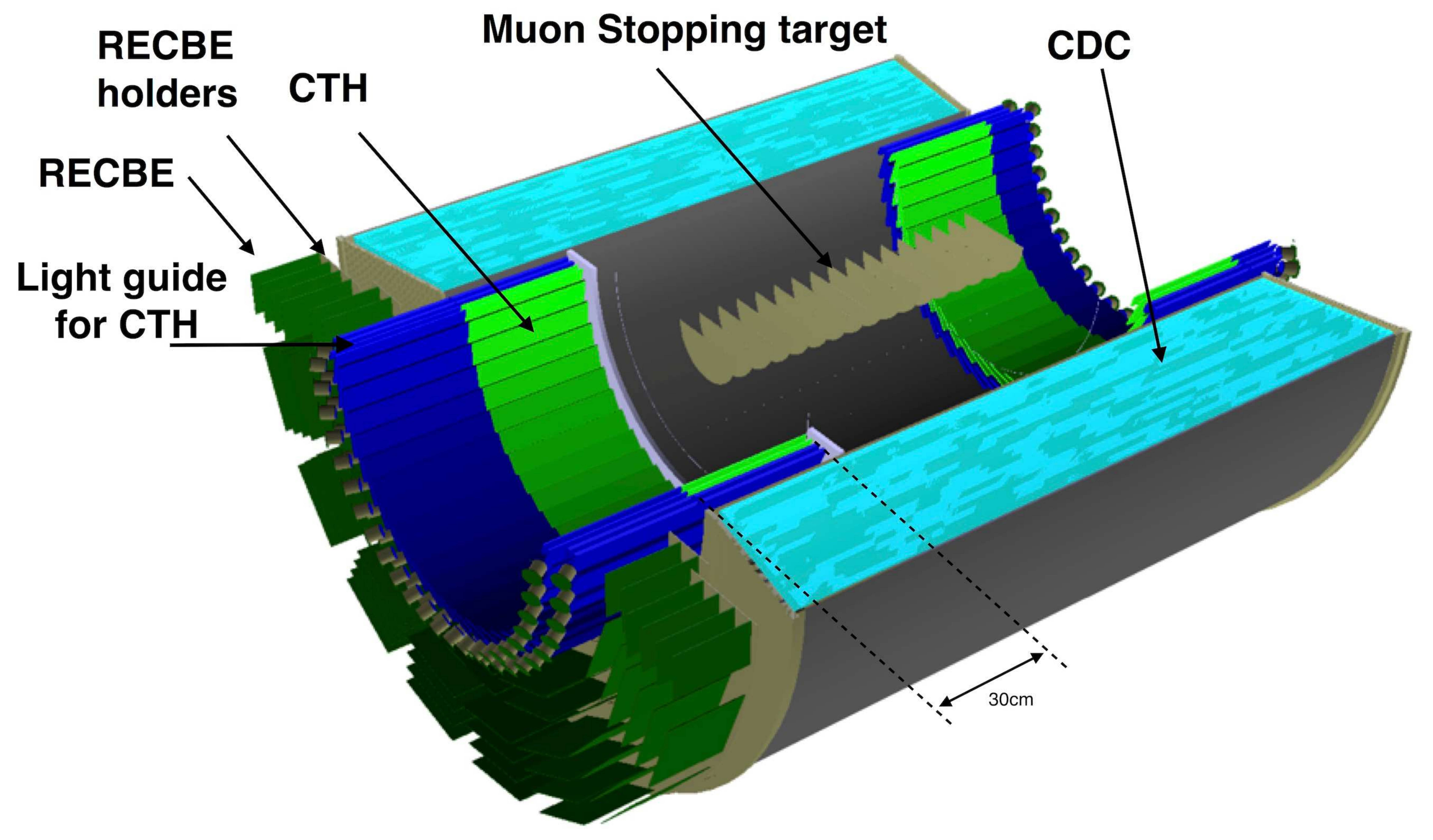}
 \end{center}
 \caption{
 A cross sectional view of CyDet
 showing the layout of the CTH. 
   }
 \label{fig:CDC-CTH-layout}
\end{figure}

\begin{figure}[tbh!]
\centering
	%\begin{minipage}[t]{0.60\linewidth}
 	\includegraphics[width=0.8\textwidth]{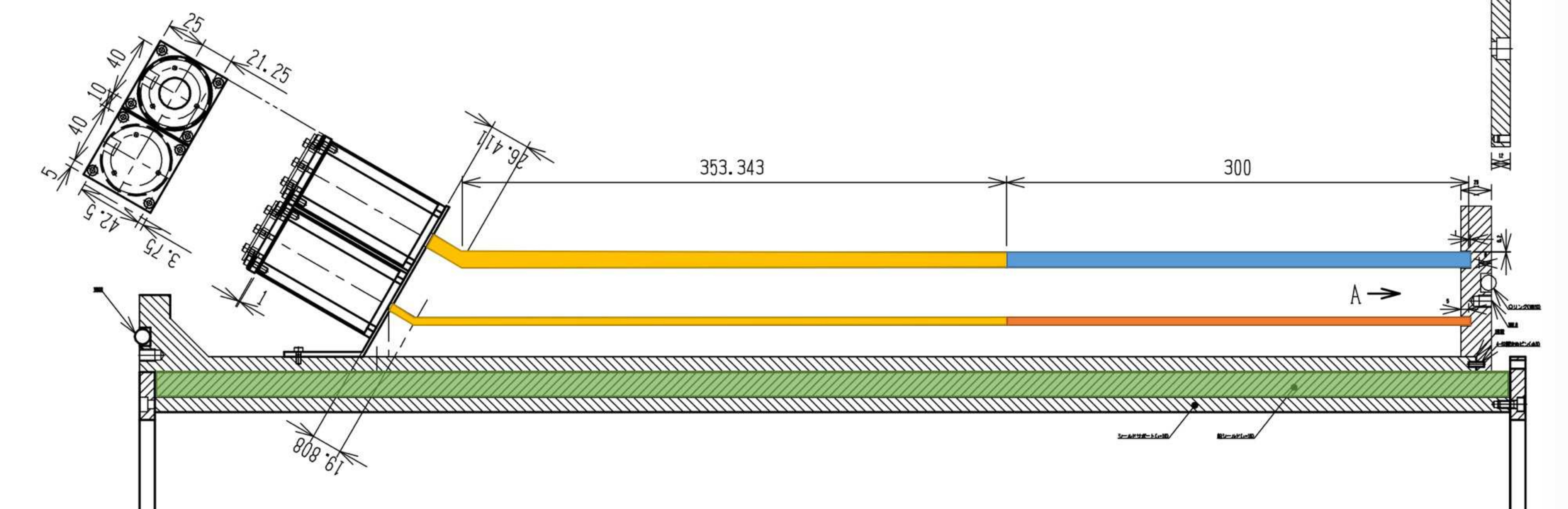}
	\caption{ 
  Drawing of downstream trigger hodoscope module. The yellow, blue, and orange parts correspond to the light guide, 
  Cherenkov radiator, and plastic scintillator, respectively. For the upstream part, the design is the same but the length of light guide is shorter.
  }
	\label{fig:CTH_design}
	%\end{minipage}
\end{figure}

\begin{figure}[tbh!]
\centering
	%\begin{minipage}[t]{0.60\linewidth}
    \includegraphics[width=0.8\textwidth]{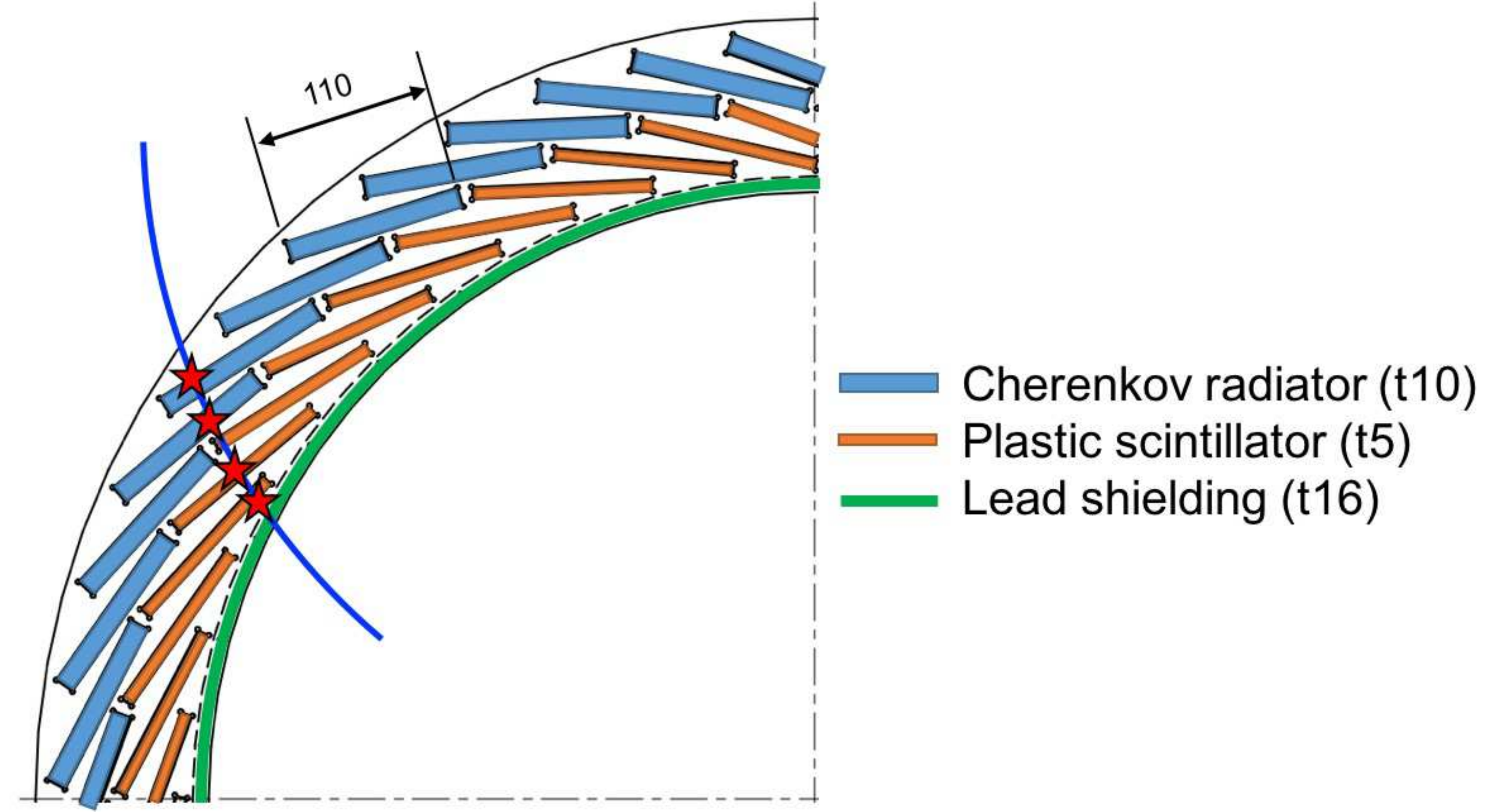}
    \caption{
  A quarter of hodoscope ring, also showing the example 4-fold coincidence by a signal electron. 
    Counters are tilted and located shifting half width so that four-fold
  coincidence with the neighbouring counters can be required, in order
  to reduce accidental coincidence.}
  	\label{fig:CDC-CTH-angle}
  	%\label{fig:cdcwiretensionfield}
  	%\end{minipage}
\end{figure}

The CyDet trigger hodoscopes (CTH) are placed at the upstream and
downstream ends of the CDC to 
generate
the first level trigger. \Cref{fig:CDC-CTH-layout} 
shows the location of CTH in the cross sectional view of CyDet.  
Their position defines the fiducial region  which is for tracks entering the CDC between the CTH counters and then triggering after the first or subsequent turns.
Each hodoscope consists of 
48
modules, each module  comprising 
a plastic scintillators and a Lucite
Cherenkov counter, 
separated by a few cm,
as shown in \cref{fig:CTH_design}.
The Cherenkov counters, together with the scintillation counters, identify electrons
from the protons from nuclear muon capture and cosmic-ray muons.
The Cherenkov and scintillation counters are tilted by specific angles to the tangent of the concentric circles so that a
four-fold coincidence (two-fold in both Cherenkov and scintillator
rings) can be made with a high acceptance for the signal electrons and a reduction in the fake triggers caused by $\gamma$-rays
as shown 
in \cref{fig:CDC-CTH-angle}.
A simple two-fold coincidence would be insufficient to reduce the fake trigger rates
from energetic $\gamma$-ray conversions.

	 %\hspace{0.01\linewidth}

\subsubsection{Design of CTH}

 The trigger hodoscopes must be
operated in a 1~T solenoidal magnetic field and  a high
neutron-fluence of about $10^{11}$ (1~MeV-equivalent) neutrons per
cm$^2$.
The signal-noise ratio $S/N$ is required to be larger
than 20, and the time resolution  less than
1~ns.

Despite the high magnetic field, the photosensor that best
meets these requirements is a fine-mesh photo multiplier tube (PMT).   MPPCs would not survive the neutron irradiation
and APDs would not provide a sufficiently good $S/N$.

Each module has two layers: $300 \times 110 \times 5\,\mathrm{mm}^3$
ultra-fast PVT-based scintillator (ELJEN
EJ-230)~\cite{cth:elgen_datasheet} and $300 \times 110 \times
10\,\mathrm{mm}^3$ UV-transparent acrylic plastic as a Cherenkov
radiator to identify the electrons.  The acrylic plastic and the
plastic scintillator are separately wrapped and 
connected via a light guide 
to a Hamamatsu
H8409-70~\cite{cth:pmt_datasheet} PMT.
This PMT has a small transit time spread of 0.35~ns and a high gain of
$\sim10^7$ and can operate  in 1~T magnetic field,
although the gain is somewhat reduced.  To compensate for this, the signals are subsequently amplified.

The fiducial region is defined by the trajectories of signal electrons which enter the CDC on their first or multiple turns, and subsequently hit the CTH to generate trigger.
The length of the CTH counters has been optimized 
and chosen to 30~cm, in order 
to maximize the acceptance
of the signal electrons.

The support structure of the trigger hodoscopes must both support the modules and  also isolate them
from the helium gas that surrounds the muon stopping target as helium
 causes degradation of PMTs due to increased after-pulsing.

 \subsubsection{Prototype tests}

\begin{figure}[htb!]
\centering
\includegraphics[width =  0.8\textwidth]{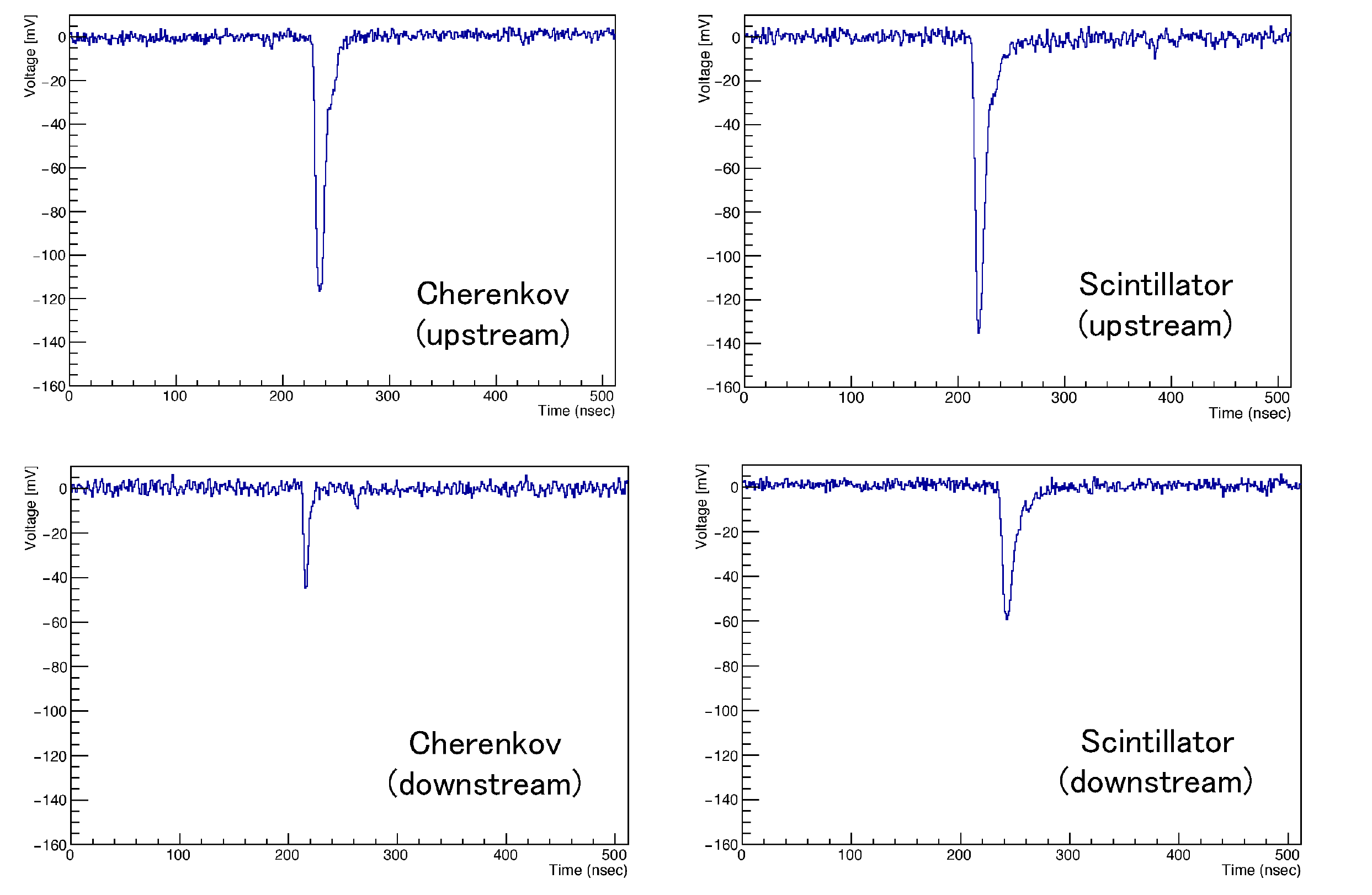}
\caption{Waveforms from the Cherenkov detector (left) and the
  scintillator (right).}
\label{fig:CTH_WF}
\end{figure}

A beam test using 155~MeV/$c$ electrons was carried out to evaluate
detector performance without a magnetic field.   \Cref{fig:CTH_WF} shows typical
waveforms measured in the beam test.  The
scintillator modules produce more light than the Cherenkov modules, and higher
light yields are  recorded for the upstream modules, which
have shorter light guides. For all four modules types, the readout meets
the required $S/N$ ratio.

\begin{figure}[htb!]
\centering
\includegraphics[width=0.9\textwidth]{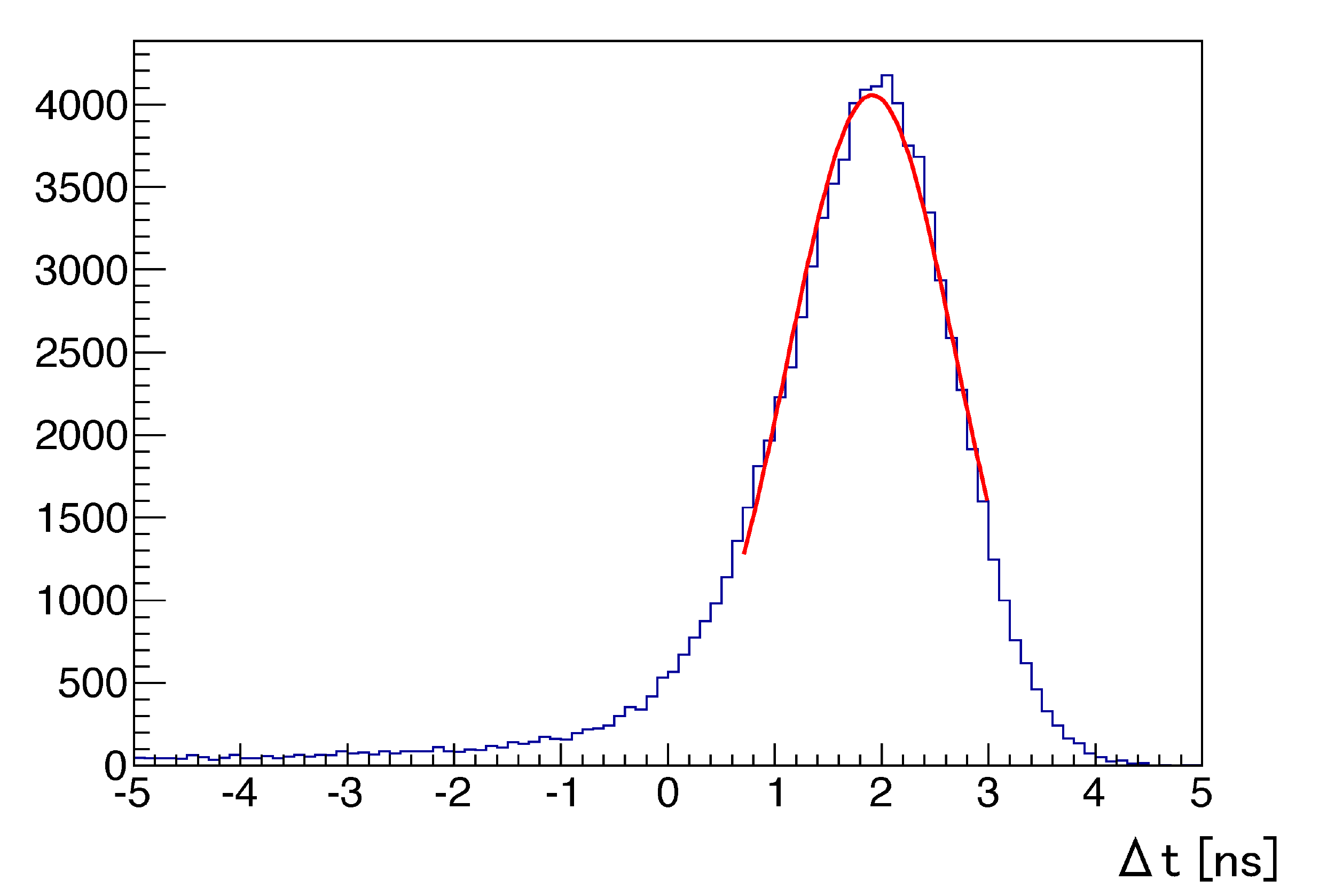}
\caption{Detection time difference between the Cherenkov detector
  and the scintillator.}
\label{fig:CTH_Time}
\end{figure}

\Cref{fig:CTH_Time} shows the distribution of the difference in
detection time between the Cherenkov detector and the scintillator
(Both were arranged closely so that the electron beam
hit both counters).
From fitting the distribution, the combined $\Delta T$ resolution is
measured to be 0.8~ns which meets the requirement of resolution better than
1~ns.  

\paragraph{CTH trigger rates}\label{sec:triggerrates}

A CyDet trigger is made by a four-fold coincidence of two adjacent CTH pairs of a scintillation  and a Cherenkov counter. An example event is shown in \cref{fig:CDC-CTH-angle}. The trigger rate was estimated with a trigger coincidence window  set at 10~ns and the time window of measurement  either from 500~ns to 1170~ns or 700~ns to 1170~ns, as described in \cref{sec:signaltimewindow}. The major background sources for fake trigger signals come from photon conversion in or near the CTH, with  most photons 
coming from 
bremsstrahlung from Michel electrons produced in muon decay at rest in the  stopping target. To reduce the fake trigger signals additional lead (Pb) shielding, about 16~mm thick, is required beneath the CTH. With this  shielding,  trigger rates of 26 kHz and 19 kHz are estimated for the time window of measurement from 500~ns and 700~ns respectively. These rates are the sum of the separate upstream and downstream CTH rates. 
As these trigger rates  result in a rather high data rate, an online trigger selection using the CDC hit information will be implemented; this is discussed further in
\cref{subsec:trigger}.

\paragraph{CTH hit rates}\label{sec:CTHhitrate}

\begin{figure}[htb!]
\begin{center}
\includegraphics[width=0.9\textwidth]{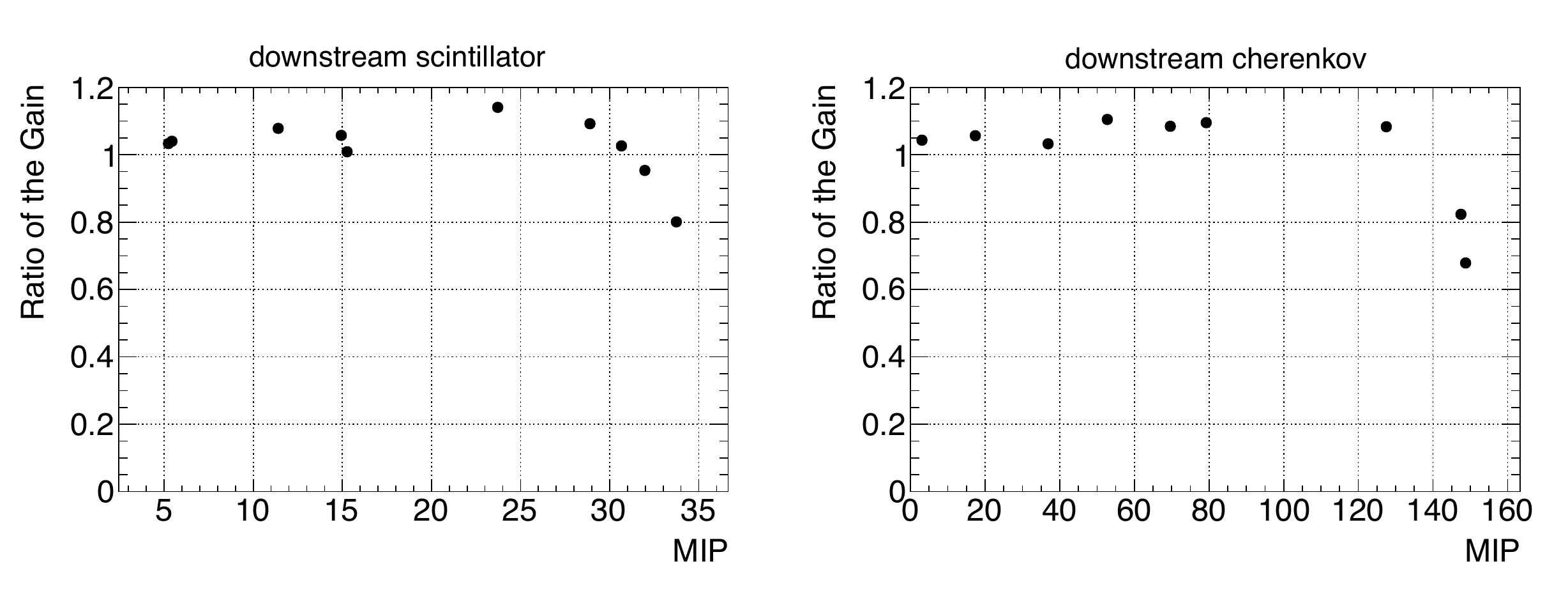}
\end{center}
\caption{Gain drop of the CTH scintillator counter (left) and CTH Cherenkov counter (right), measured by the second LED pulses, as a function of the pulse height of the first LED to simulate different amplitude of a beam flash. The gain drops of the CTH plastic scintillator and CTH Cherenkov counter start from 25 MIPs and 120 MIPs equivalent respectively.}
\label{fig:cthprompttest}
\end{figure}

The effects of this beam flash in the CTH have been examined experimentally. It was found that
the  gain of the CTH scintillator counter and of the CTH Cherenkov counter begin to degrade if the beam flash is greater than 25 MIPs for the scintillator and 120 MIPs for the Cherenkov respectively. The test results are summarised in \cref{fig:cthprompttest}. 
From the simulations, the average beam flashes are less than these limits  and it is concluded that the effect of beam flash will not cause any loss.

The instantaneous rates after the prompt beam flash are estimated for each counter of the upstream CTH and downstream CTH separately as shown in \cref{tb:cthafterhitrate}.
It should be noted that they are average instantaneous rates during the time period.

\begin{table}[thb!]
\caption{Average instantaneous hit rates during the time period after the prompt beam flash, from 200 ns to 1170 ns.}
\begin{center}
\begin{tabular}{lcccc}\hline\hline
& upstream & upstream & downstream & downstream \\
& scintillator & Cherenkov & scintillator & Cherenkov \\ \hline
Average rate (MHz) & 3.5 & 1.5 $-$ 2 & 4 & 3 \\ \hline\hline
\end{tabular}
\end{center}
\label{tb:cthafterhitrate}
\end{table}
 % subsection
\subsection{CDC Tracking}

The CDC track reconstruction consists of track finding process
followed by track fitting process. The former  selects  good
hits in order to identify track-like structures and eliminate
background noise, whereas the latter does track fitting with Kalman
filtering to determine which hits are most 
probable
part of the track,
whether a single continuous track is a good interpretation of the hit
pattern, and (assuming it is)  find the best estimate of the
momentum of the charged particle that made the track.  In the fitting
process it is necessary to consider both single-turn and
multiple-turn tracks.

\let\vaccent=\v % rename builtin command \v{} to \vaccent{}
\renewcommand{\v}[1]{\ensuremath{\mathbf{#1}}} % for vectors

\paragraph{Track finding and reconstruction}
\label{sec:trackfinding}

The CyDet offline track finding algorithm outlined in this section
filters out background hits
using three main stages~\cite{EwenTrackFindingWithGBDT, EwensThesis}.
First,  a Gradient Boosted Decision Tree (GBDT) to classify the
hit as signal or background based on the properties of the hit itself (local features),
and the properties of neighbouring hits (neighbour features).  Second, it  performs a
circular Hough transform on the output of this GBDT, reweights the result, and inverts the
transform to recover information about which hits form a circular path with signal-like hits.
Third, a new GBDT uses this information, the local features, and the neighbour features to
classify hits as signal or background. The output at this hit-filtering level suppresses nearly
98\% of background hits while keeping 99\% of signal hits\footnote{This analysis makes use of the \texttt{scikit-learn}~\cite{scikit-learn} package in Python 2.7.}.

\begin{figure}[tbh!]
\centering
    \includegraphics[width=0.8\textwidth]{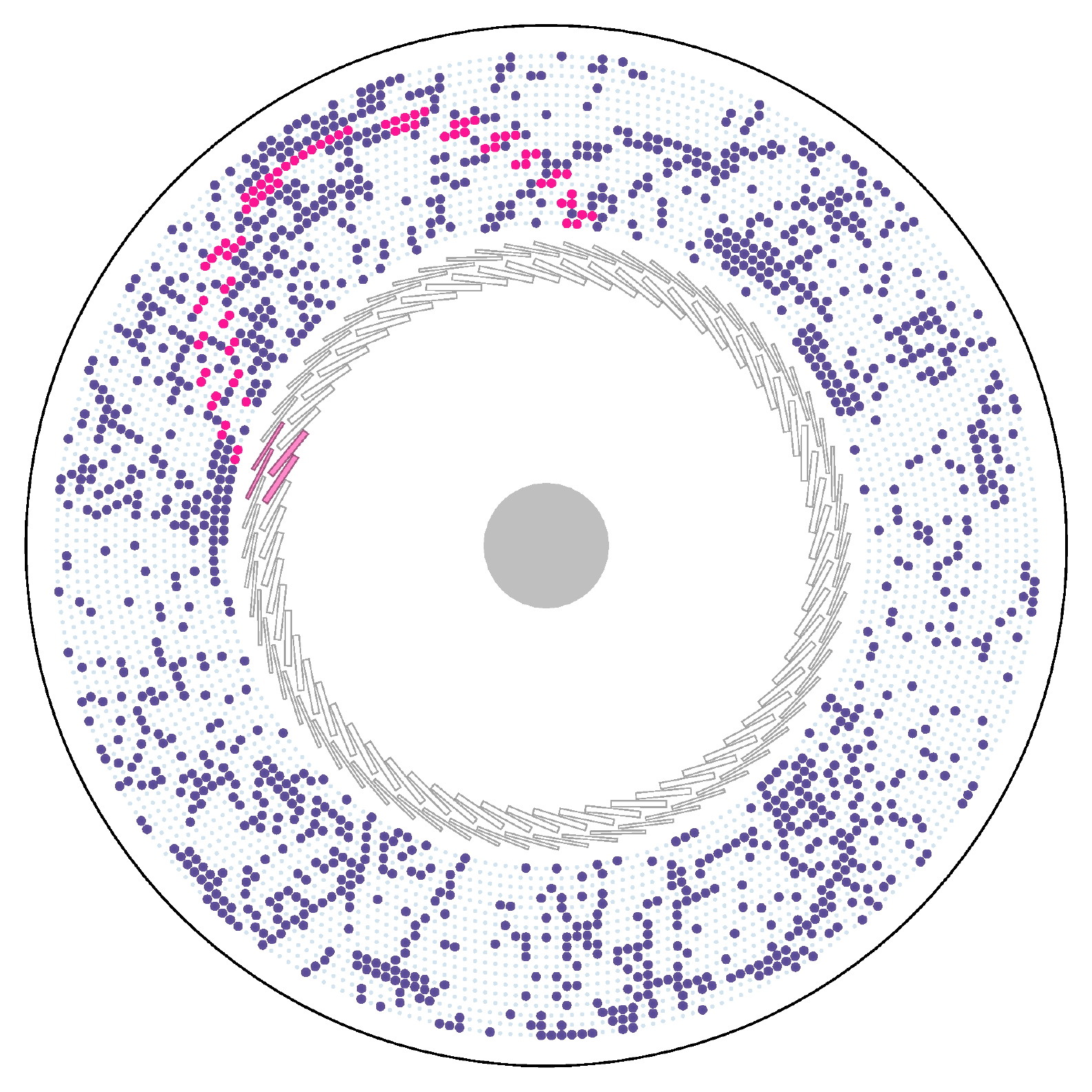}
    \caption{A 15.4\% occupancy event in the CyDet.  This is a projected view from the
    central plane of the detector, looking in the direction of the beam line. The dark blue points are hits
    caused from background processes, while the magenta hits correspond to the signal
    electron. 
    The magenta-filled boxes represent the CTH hits by signal electron.
    }\label{fig:tracking:algo_0}
\end{figure}

The algorithm has been tested on simulation data generated using
Geant4 and ICEDUST. \Cref{fig:tracking:algo_0} gives an example of a truth-labelled input event. 
The magenta
points are the signal hits, while the dark blue points are the background hits.  The event shown has
an uncommonly high occupancy of  15.4\%, to
illustrate the performance of the algorithm.

\begin{figure}[tbh!]
\centering
    \includegraphics[width=0.8\textwidth]{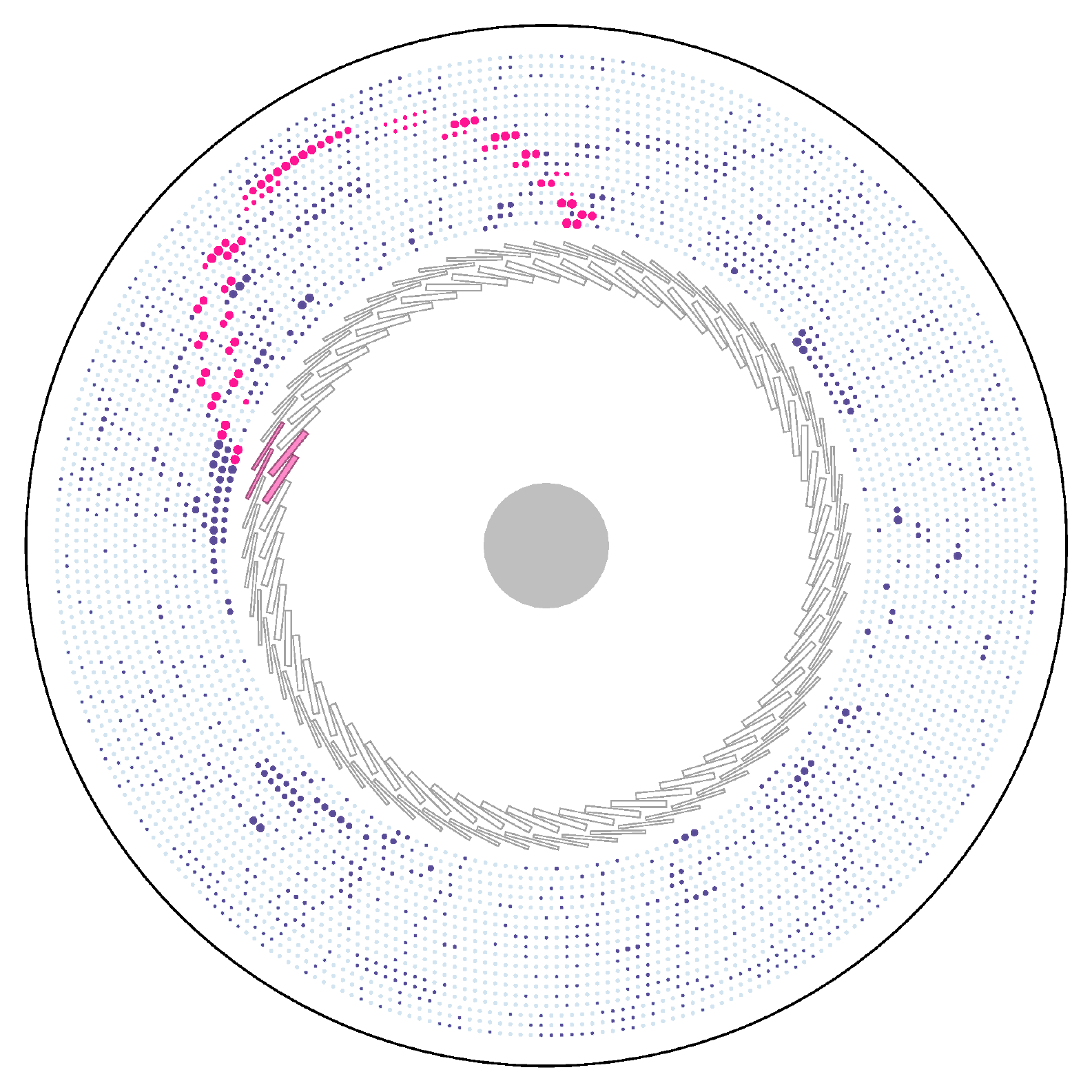}
    \caption{A visual representation of the neighbour-level GBDT applied to the event
    shown in \cref{fig:tracking:algo_0}.  The locations of the hits are shown by the
    outlines of the hits.  The fill is scaled with the output of the GBDT, where a full circle
    corresponds to a signal-like response.}\label{fig:tracking:algo_1}
\end{figure}

\paragraph{Neighbour-level GBDT}\label{sec:tracking:neigh_level}

The algorithm begins by considering three ``local-level'' classification features of each hit
wire.  The first of these is the energy deposition of the hit.    A cut on this feature alone can reduce the background
hits by 68\% while retaining around 99\% of signal hits.   The second feature is the timing of the wire hit
relative to the timing of the hit in the CTH trigger system.  Signal hits tend to occur soon
after these trigger hits, while background hits occur randomly with respect to the trigger
timing.  The third feature is the hit's radial distance from the centre.  The magnetic field
and geometry are tuned so that signal tracks curve through the fiducial volume, rarely reaching
the outer layers, yet always passing through the inner ones.  The background hits are
distributed more evenly throughout the layers, peaking slightly at the inner and outer
layers. 

The separation power of these features are further exploited by defining features that
describe the neighbouring wires of a hit, i.e.\ the ``neighbour-level'' features.  Due to the
alternating stereo angles, the features on the neighbouring wires in the same layer are more
powerful than adjacent layers, referred to as the left and right neighbours of a
hit.  Along with the local features, the left-right timing and energy deposit features are also
used. This defines seven input features for the GBDT,  referred to as the neighbour-level
GBDT\@.  Its output is visualised in \cref{fig:tracking:algo_1}.

\begin{figure}[h!]
  \centering
  \includegraphics[width=0.8\textwidth]{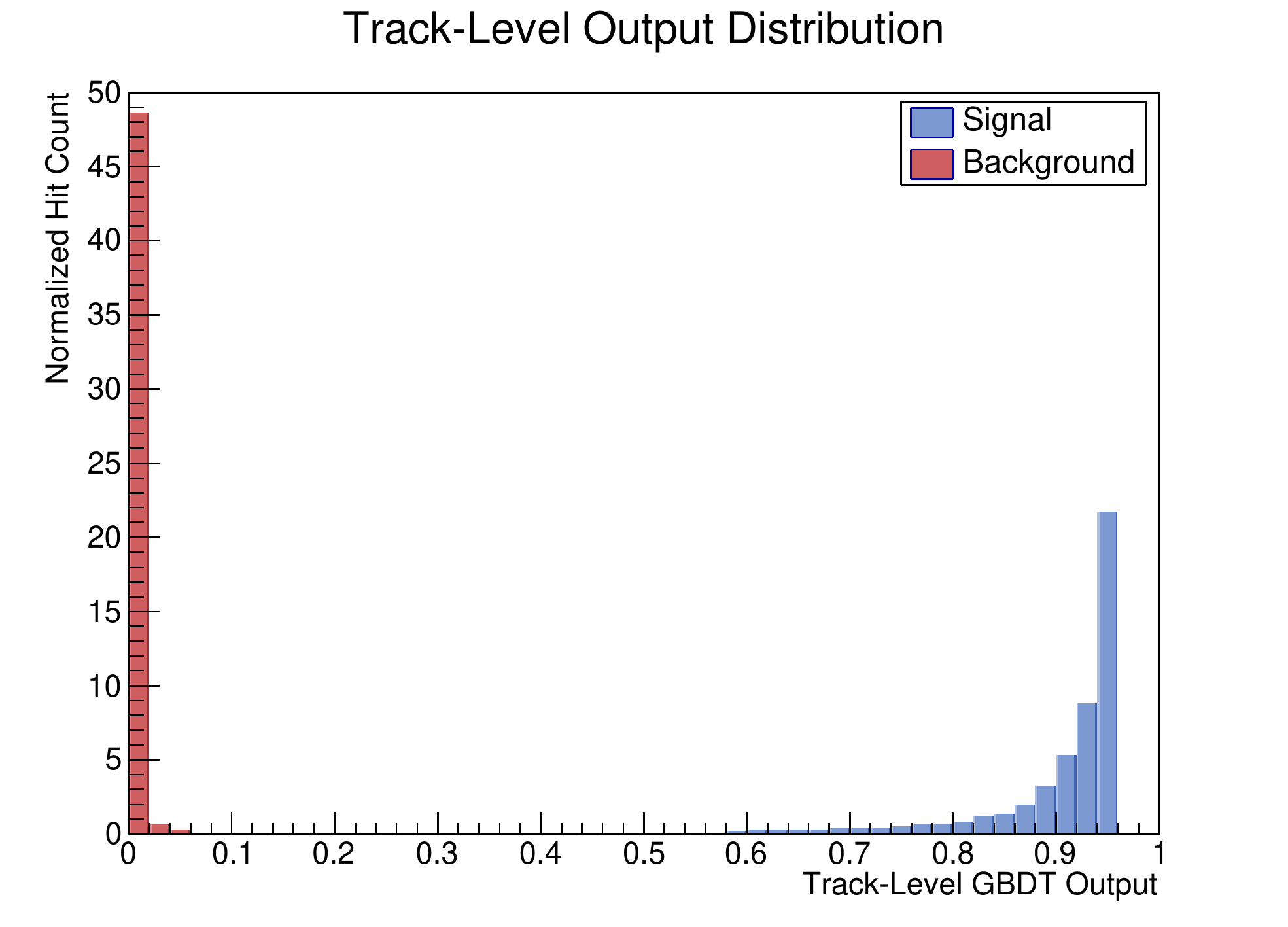}
  \caption{Distribution of the output of the track level GBDT, comparing response from signal
  hits to the response from background hits.}\label{fig:tracking:track_output_dist}
\end{figure}

While local and neighbour features alone yield promising results, there are still some
isolated clusters of misclassified background hits, as well as a diminished response for
isolated signal hits.  To correct this, a circular Hough transform is used on the output of
the GBDT to determine which hits lie in a circular pattern with other signal-like hits. 
Weights can then be defined for each wire hit based on the Hough transform and the GBDT output.
Finally, a second GBDT, the track level GBDT, is used to discriminate signal from background hits. 
The output from this GBDT, shown in \cref{fig:tracking:track_output_dist} on the Monte Carlo sample used, demonstrates excellent separation between signal and background

\begin{figure}[h!]
  \centering
  \subfigure[][ROC curves at full scale.]{
  \includegraphics[width=0.45\textwidth]{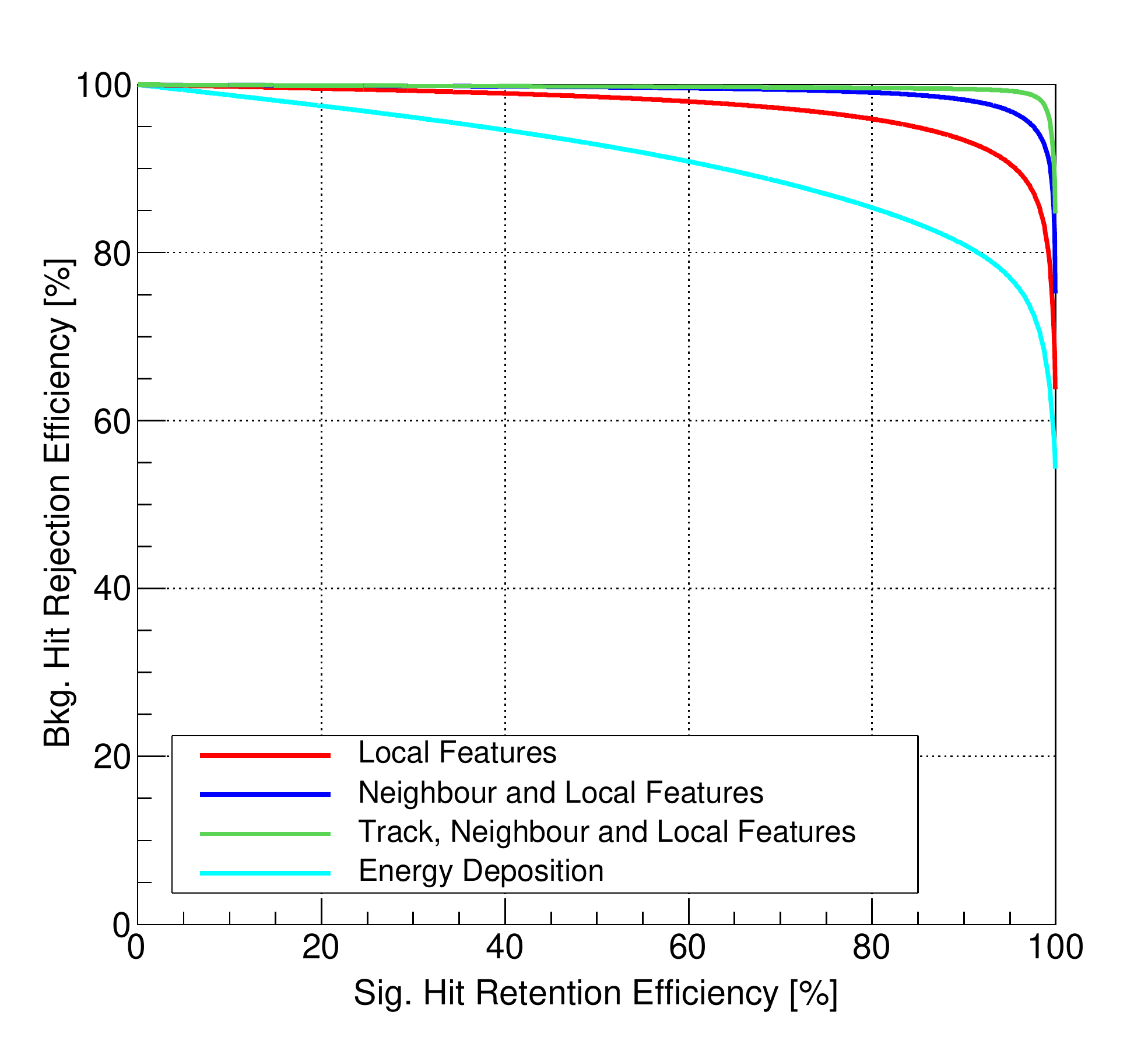}
  \label{fig:tracking:roc}
  }~
  \subfigure[][ROC curves with zoomed scale.]{
  \includegraphics[width=0.45\textwidth]{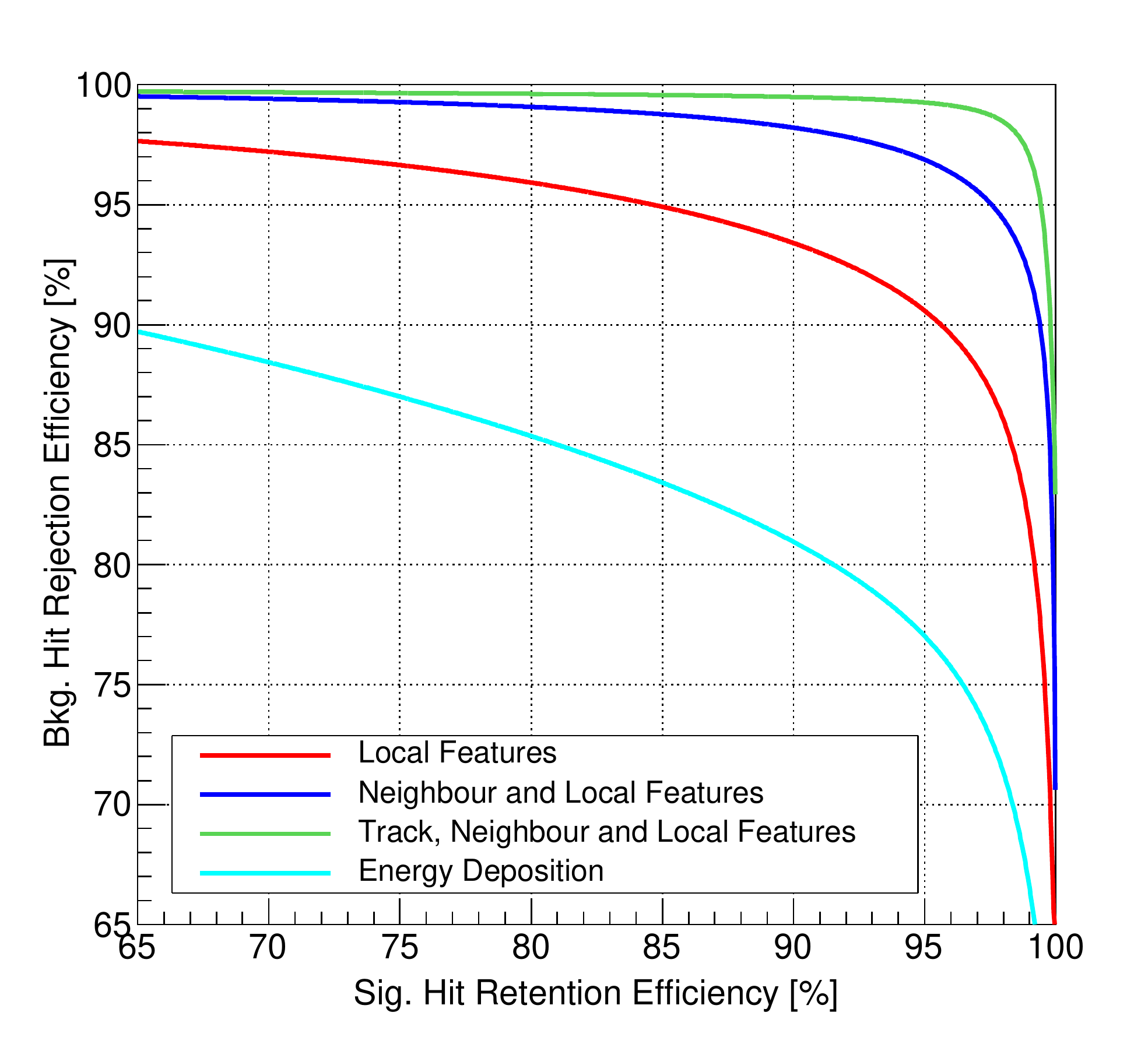}
  \label{fig:tracking:roc_zoom}
  }~
  \caption{ROC curves for four independent classifiers.  The red curve is from a GBDT trained
  on energy deposition alone, the blue curve is from the neighbour-level GBDT, and the green
  curve is from the track-level GBDT. The cyan curve represents the cut-based case using energy deposition alone. }\label{fig:tracking:both_rocs}
\end{figure}

ROC curves are plotted in
\cref{fig:tracking:both_rocs} to demonstrate the  ability  of the algorithm to reject
background as a function of its ability to retain signal.    
The plots compare the performance of the neighbour-level
GBDT,  the track-level GBDT and the baseline separation power of solely the energy deposition of a
hit.  They show that the background rejection rate for a signal efficiency of 99\% is 98\% with the track-level GBDT\@ but just 68\% using  energy deposition
alone.

 % subsection
\section{StrECAL: the Straw-Tracker ECAL Detector System}

After the physics data taking with CyDet during Phase-I running, 
another detector system, StrECAL,  will be installed instead of
the CyDet. The StrECAL is a combination of a straw-tracker (a
low-mass detector consisting of planes of gas-filled straws acting as drift
chambers) and a crystal electromagnetic calorimeter. 
The primary purpose of
this detector in Phase-I is to make direct measurements of the composition of
the muon beam, but the detectors are very similar to those which will be
employed in Phase-II, and act as a prototype for the Phase-II detectors.

\subsection{Straw Tracker} \label{ch:straw}
The Straw Tracker to be developed for Phase-I % is a  prototype for
will make direct measurements of the
particles in the muon beam line, and the rate of particle production (in particular
anti-protons), as a function of the beam energy, and other backgrounds.
It will be placed inside the vacuum vessel and the Detector Solenoid (DS)
which has a field strength of 0.8$-$1.1~T.  The detector will provide a precise measurement of a particle's momentum
and its identity, through $\dv{E}{x}$, $E/p$ and the time of flight information
in combination with the calorimeter.  For Phase-I, as shown in
\cref{fig:muonbeam-momentum-after90},
many kinds of particles will reach and
enter the DS.  For both phases, the
volume inside the magnet will be evacuated to enable good-quality measurements of the beam particles in Phase-I and to minimise the amount of material in Phase-II.

\subsubsection{Overall structure}
\label{sec:design}
\begin{figure}[htb!]
  \begin{center}
    \includegraphics[width=\textwidth]{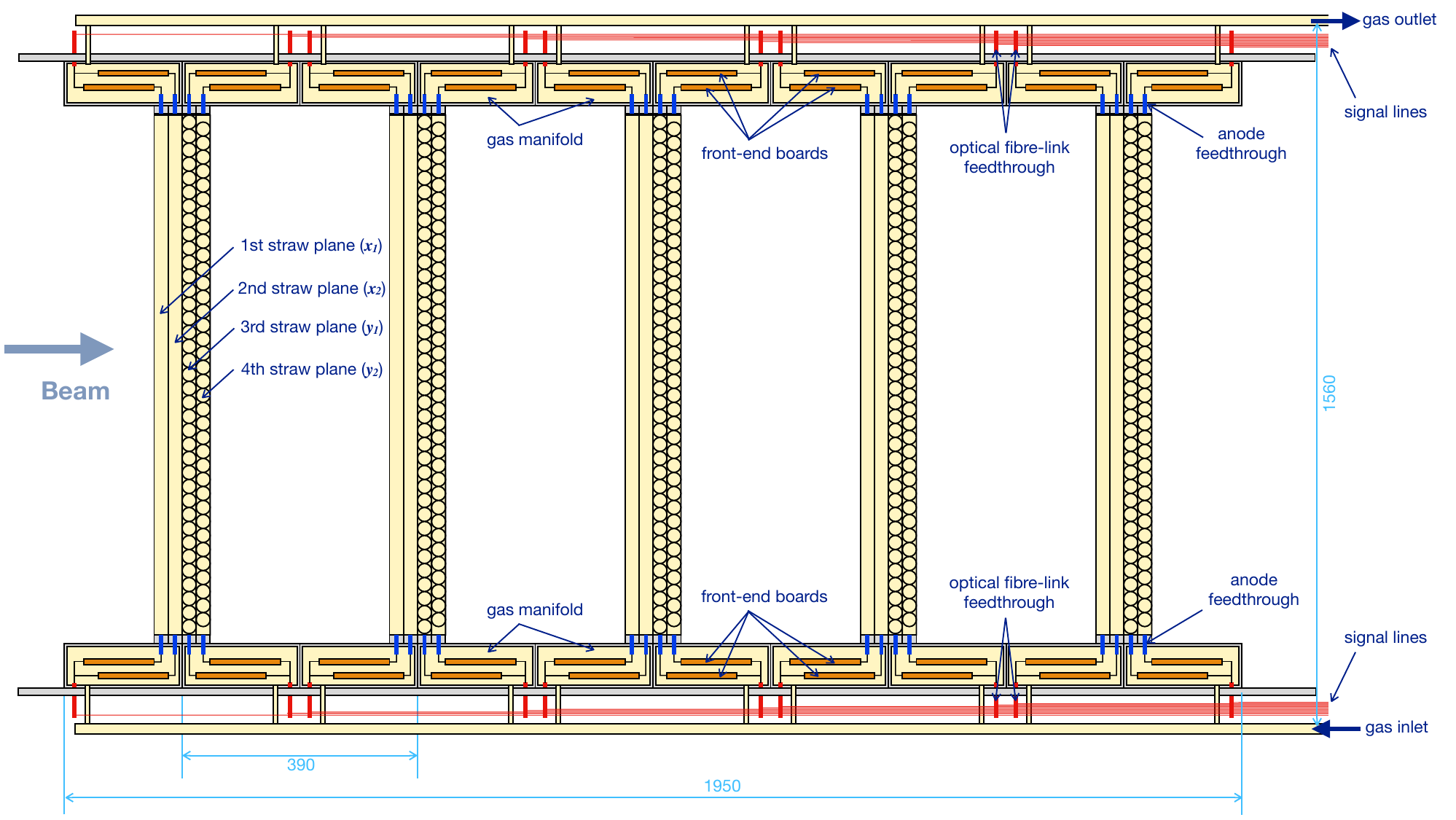} \\
    \includegraphics[width=0.8\textwidth]{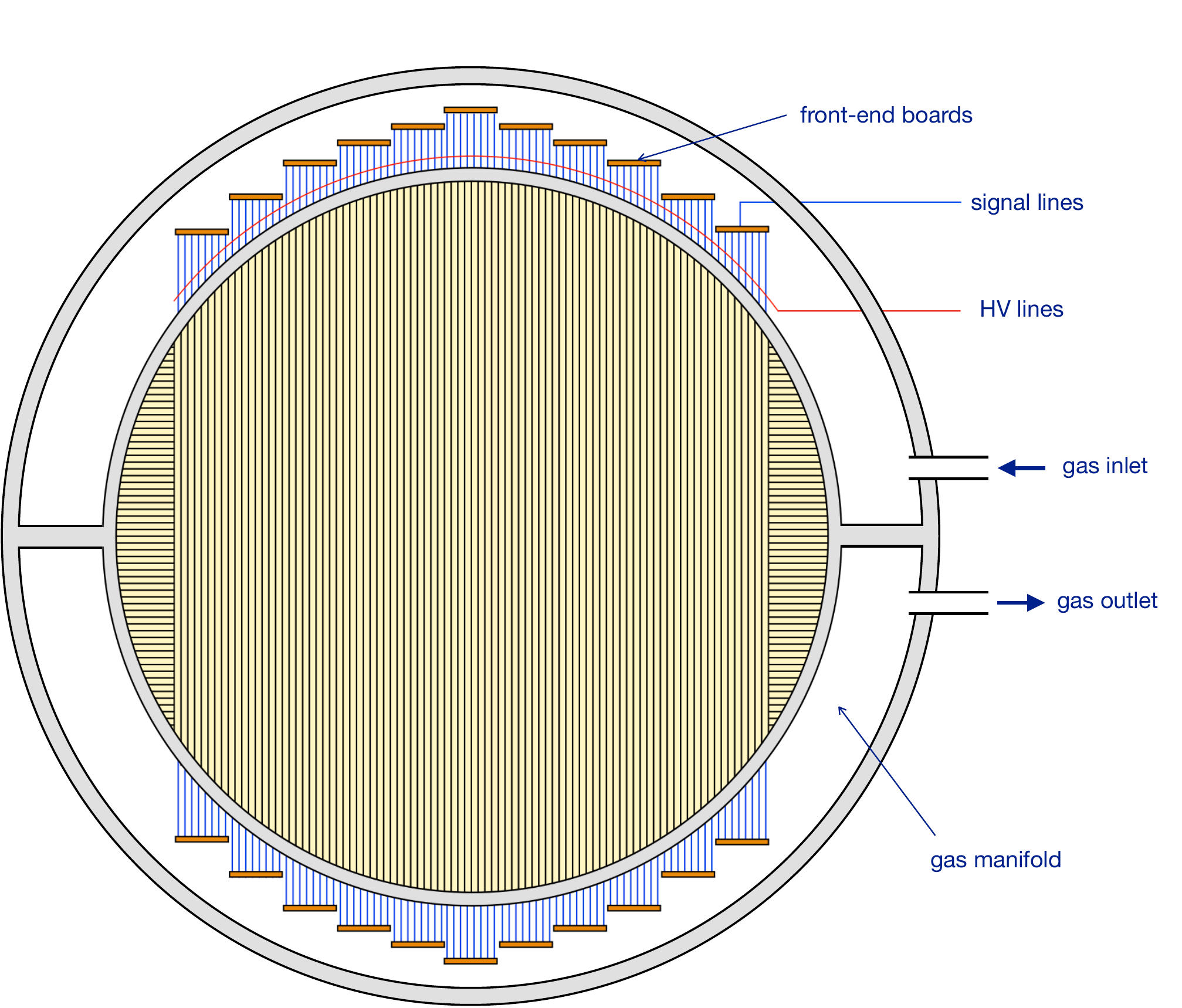}
    \caption{Schematic view of the Straw Tracker;
             (Top) Side view.
             The straw dimensions is scaled by a factor of three for clarity.
             (Bottom) Cross-sectional view of a plane.}
    \label{fig:strawtracker}
  \end{center}
\end{figure}

The overall structure of the Straw Tracker is schematically shown in
\cref{fig:strawtracker}.  Each of the five tracker super-layers,
or ``stations'', consists
of four planes; two to measure the $x$ coordinate and two to measure
the $y$ coordinate.  Each pair of planes is staggered by half a straw
diameter in order to resolve any left-right ambiguities.  Each layer is
constructed as a stand-alone unit and mounted on the detector frame
which is inserted and removed from the DS on rails
and linear bearings.  A spare
layer will also be built.
Anode wires, made of gold-coated tungsten, are extracted via a
feedthrough into the gas manifold as shown in
\cref{fig:strawtracker}.  The anode wires are held at high voltage
and the straw wall is grounded, to act as the cathode.  A gas mixture of
50\%-Ar and 50\%-\ethane is provided from this gas manifold to the
straw tube.
The straws have a diameter of 9.75~mm, range in length from 692 to 1300~mm, and
are mounted on aluminium ring supports.

\subsubsection{Mechanical construction}
The straw walls conduct electricity, and are
made of a metalised Polyethylene Terephthalate (PET) film of
20 $\micro$m thickness\footnote{R\&D is currently ongoing on reducing the wall thickness~\cref{sec:straw_tube}.}.
The support rings for the straws have inner
and outer radii of 65 and 78~cm, respectively.
Gas manifolds and electrical connections are also attached to the supports.
The gap between them provides space to mount the front-end electronics,
the power distributors
and HV circuit. Each of the five stations are equally spaced
and rigidly attached to each other. 

\begin{figure}[htb]
  \begin{center}
    \includegraphics[width=0.45\textwidth]{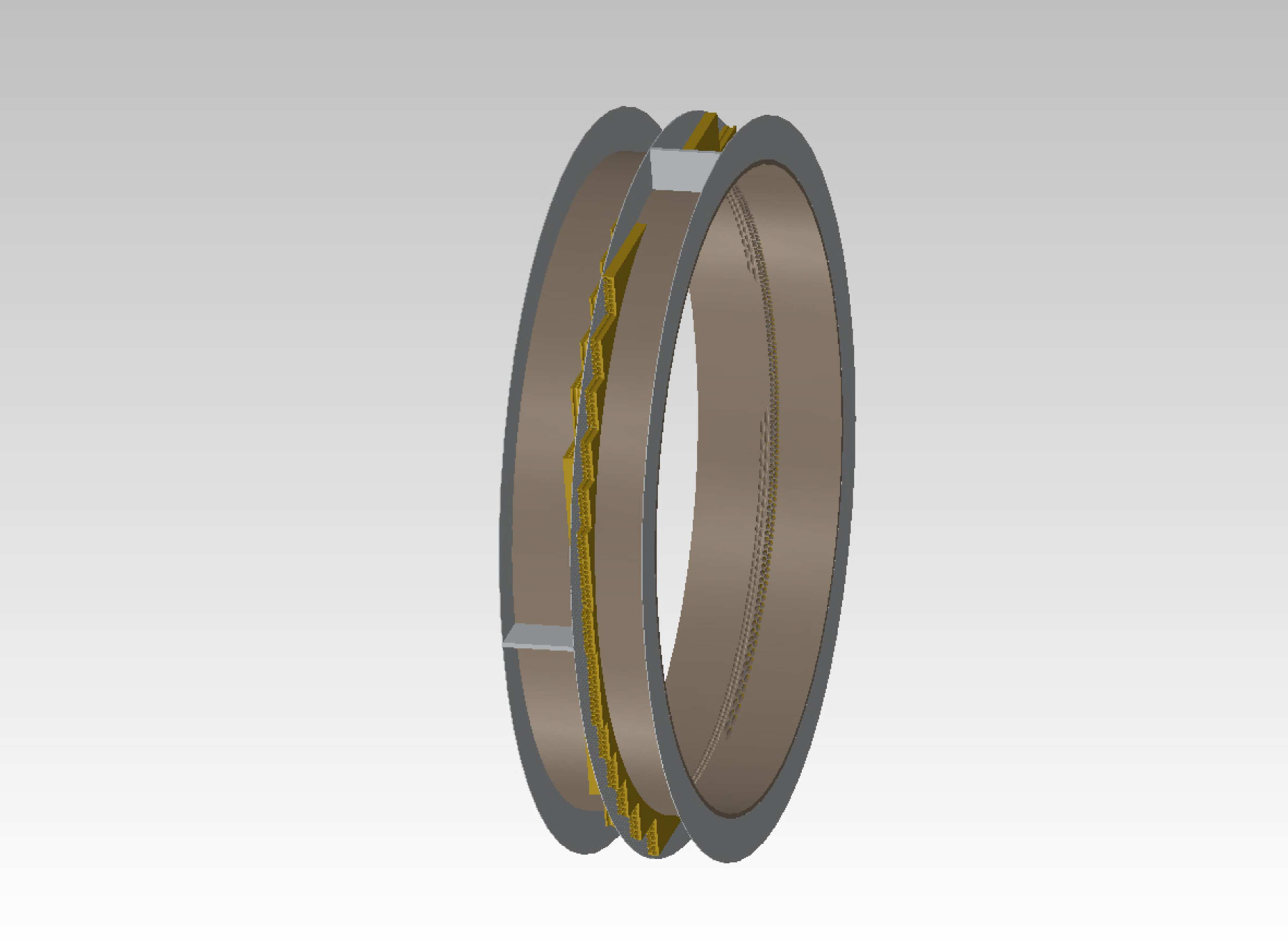}
    \includegraphics[width=0.45\textwidth]{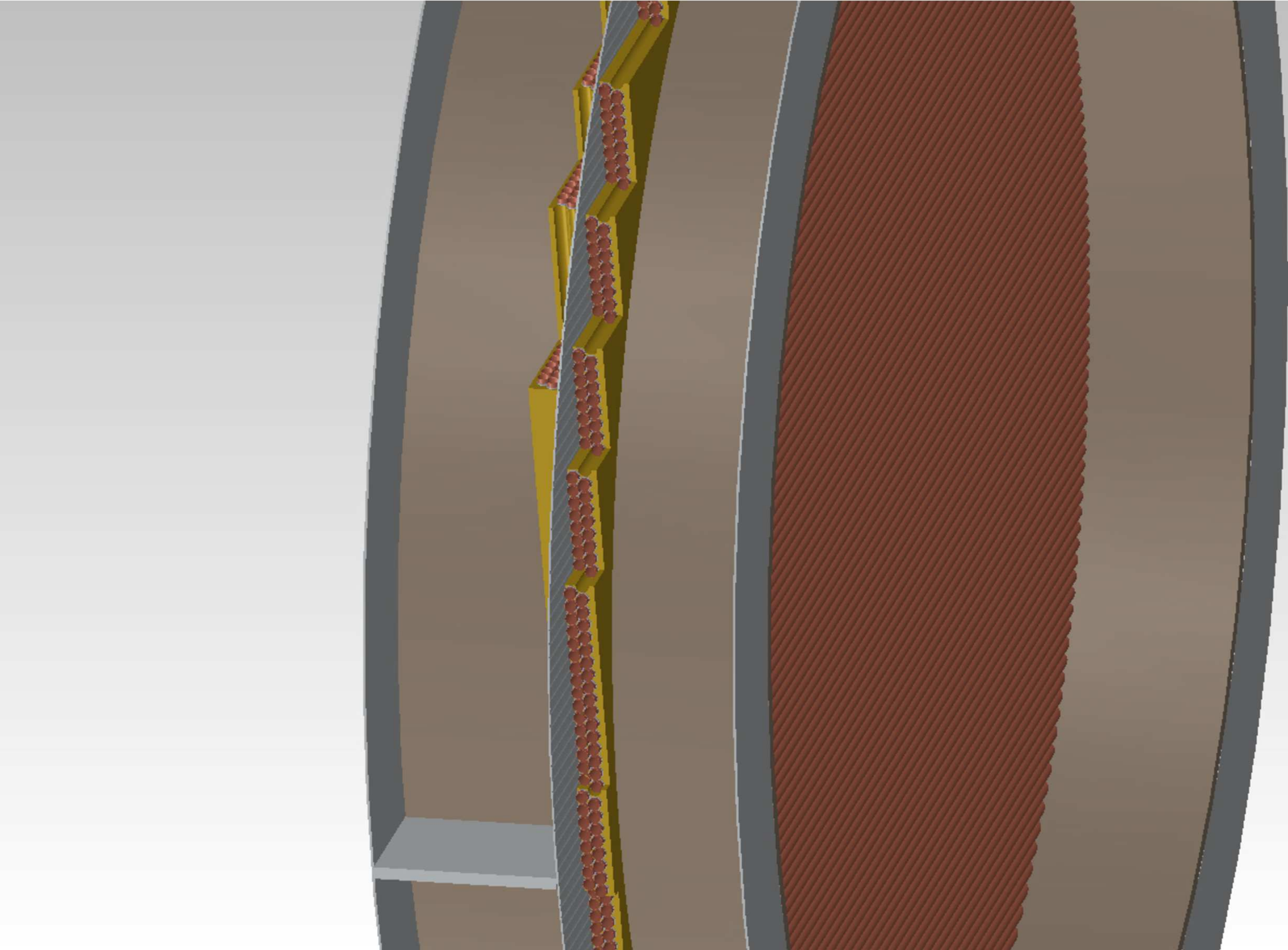}
    \caption{Design of the support structure for one tracker station.
             (Left: Overview of one station without straws,
              Right: Close-up view with straws)}
    \label{fig:straw_support_structure}
  \end{center}
\end{figure}

Finite Element Analyses found an excess tension of 1.7~\kgf on each
straw compared to the original expectation, but also indicated that this results in acceptably small deformations.

\subsubsection{Straw tube}
\label{sec:straw_tube}

\begin{figure}[tbh!]
\centering
    \includegraphics[width=0.8\textwidth]{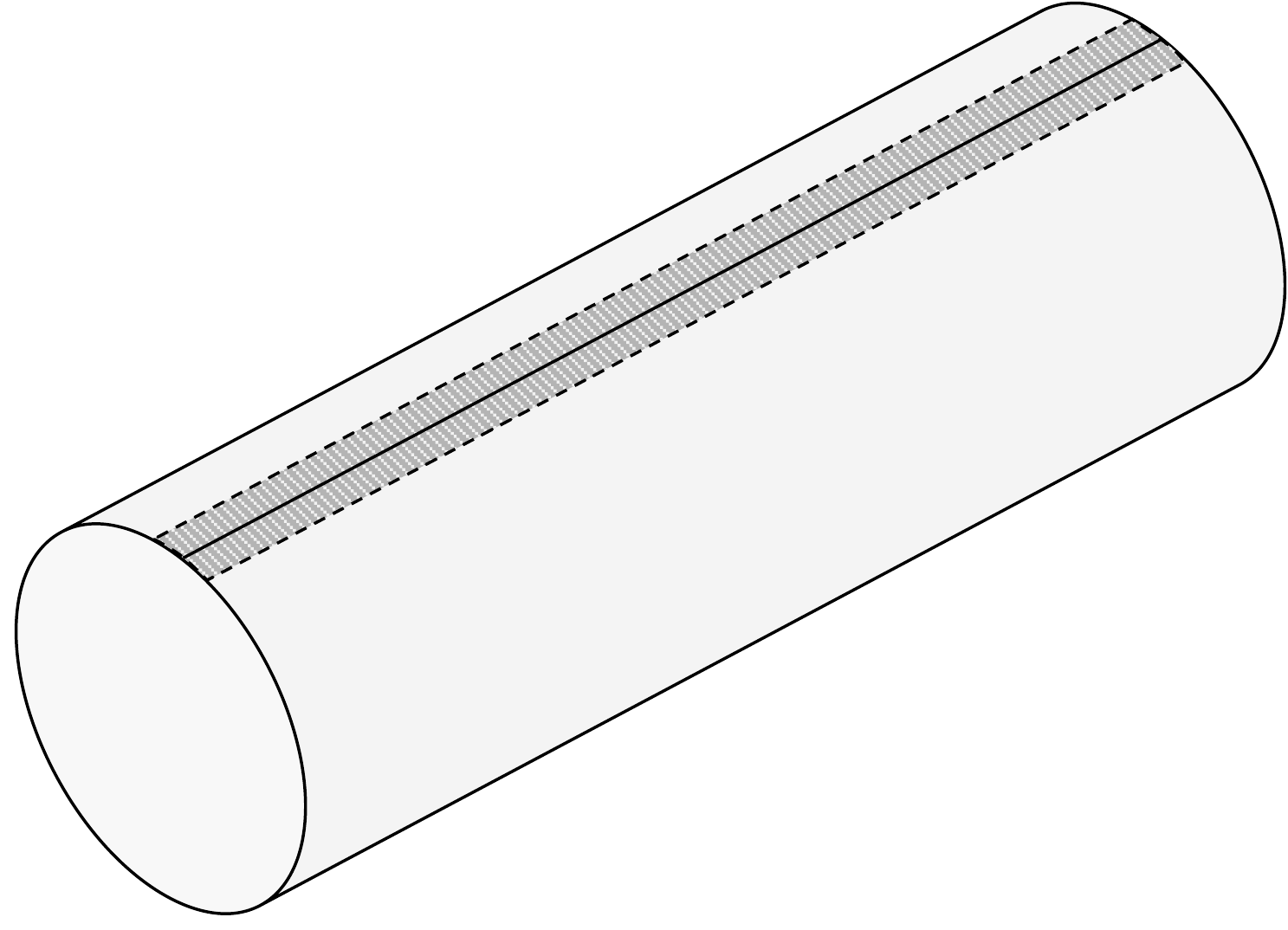}
    \caption{The new straight-adhesion style for straw construction
    \label{fig:straw_dimension}}
\end{figure}
\begin{figure}[tbh!]
\centering
  \includegraphics[width=0.8\textwidth]{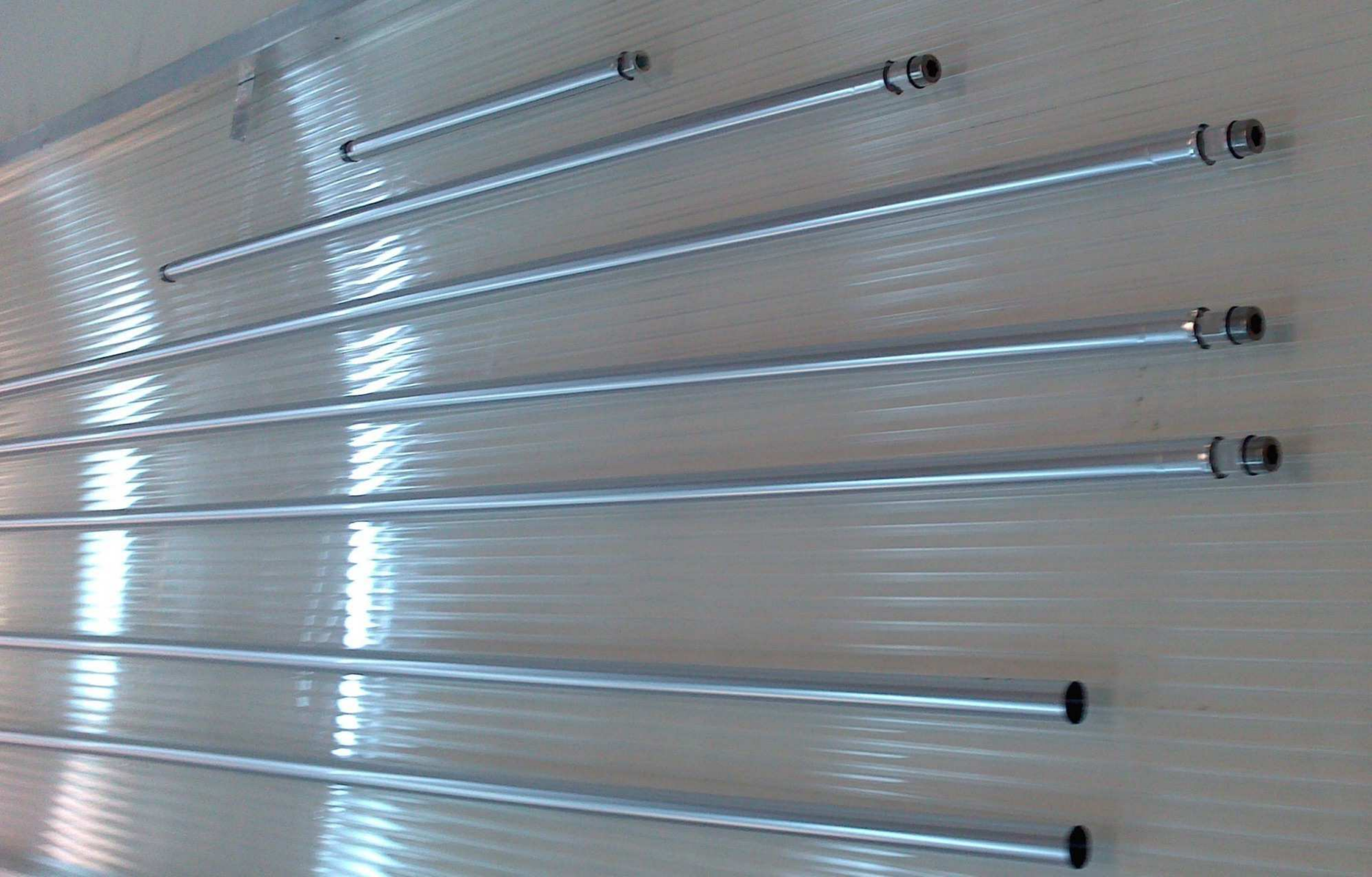}
  \caption{Sample tubes with 20 $\micro$m-thick walls with 70~nm-thick aluminium layers.\label{fig:straw_20um}}
\end{figure}

A method of straw production which does not require multiple
over-woven layers has been developed by the JINR group for the NA62
experiment at CERN~\cite{na62straw}.  In this method, a single layer
is rolled and attached to itself in a straight line using ultrasonic
welding as schematically shown in
\cref{fig:straw_dimension}.
JINR-COMET group have tried to develop the new welding station
in order to produce the COMET-design tube which is thinner than NA62 tube,
and finally succeeded to produce 20$\mu$m-thick 
wall
straw tube,
after several R\&D with JINR-NA62 group.
The provided straws, shown in \cref{fig:straw_20um}, were mechanically tested
and confirmed to be robust enough.
The mass-production of straw tube for Phase-I, 2,900 tubes including 500 of spare tubes,
have been already completed by JINR-COMET group.

\begin{figure}[tbh!]
  \begin{center}
  \includegraphics[width=0.9\textwidth]{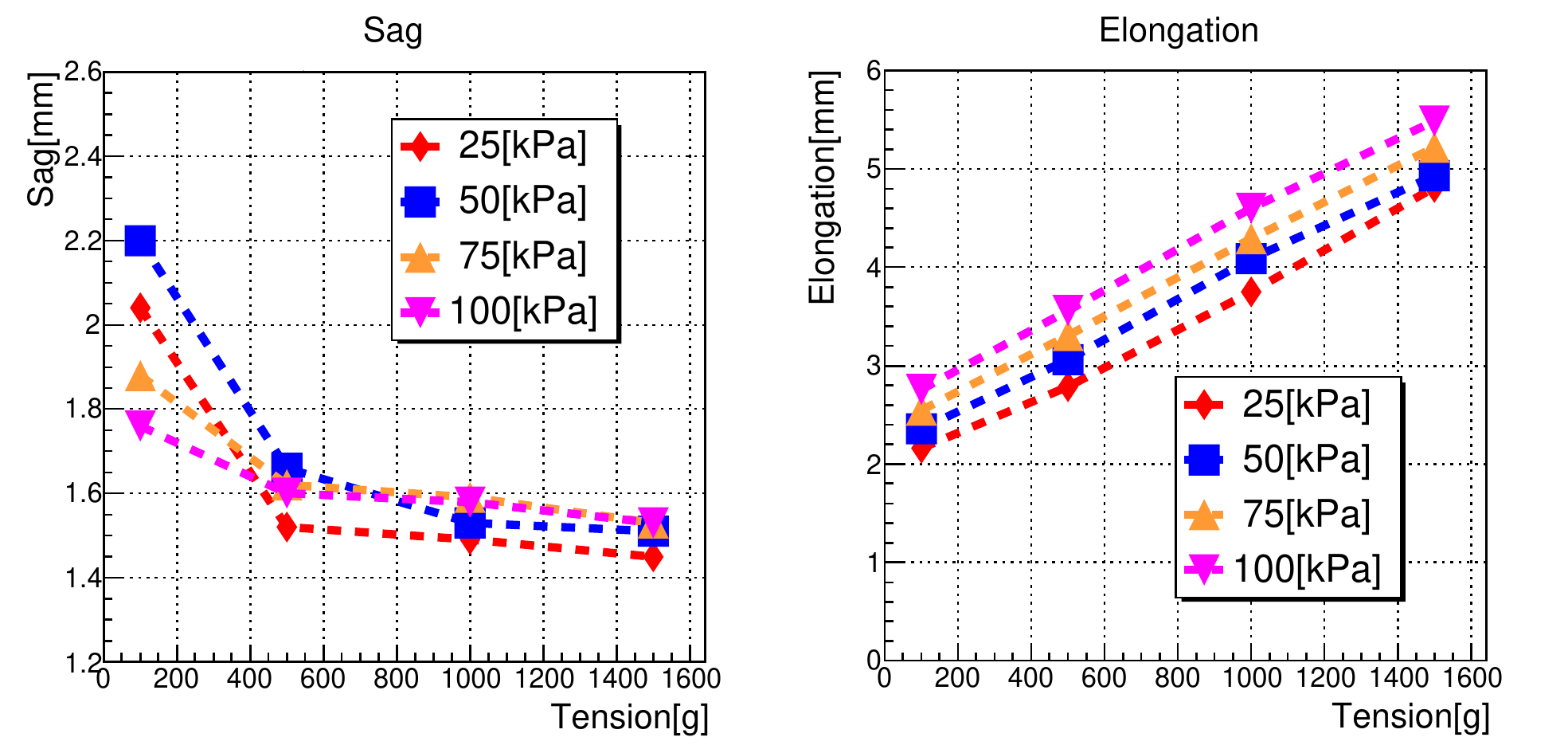}
    \caption{Straw-pre-tensioning study results;
     (Left) the sag that is found for various pressure differences across the straw wall
     as a function of the applied tension,
     (Right) elongations of the straws. 
     The sag results include 1.5~mm measurement offset, which should be subtracted from the data points in order to get the actual sag values. The elongation results also include 2~mm measurement offset. 
    \label{fig:straw_tension_results}}
  \end{center}
\end{figure}

Possible deformations of the straw as a function of the  pre-tensioning value
were investigated
since the pre-tensioning must be quite high to avoid
deformations when it is operated inside the vacuum\footnote{
As a point of reference, the NA62 tracker
is pre-tensioned at 1.5-\kgf on each straw tube.}.
The measurement results of
sag (defined by the deformation made by gravity from the normal position without gravity, also including 1.5~mm measurement offset) and elongation (including 2~mm measurement offset)
for 1~m straw 
are shown in \cref{fig:straw_tension_results}.
They show that tensions  higher than 1~\kgf prevents sagging, and this results in elongations of 1.7--2.0~mm.
Therefore, straw deformations can be avoided by stretching the straws by 2~mm
during assembly. 

The sense wires are chosen to be gold-plated tungsten containing 3\%
rhenium.  Additional supports for the anode wires are not required; wire stability can be estimated from
the electrostatic force on a
off-centre anode wire~\cite{XuAndHungerford2006}:
\begin{equation}
      L_{c} = \pi R (C V) [2 \pi \epsilon_{0} T]^{1/2},
\end{equation}
where $T$ is the tension on the wire, $V$ is the applied voltage, $C$ is
the capacitance per unit length, $L_{c}$ is the critical wire length
for a given tension, and $R$ is the straw radius.  Assuming a straw
radius of 4.9~mm, an anode wire radius of 12.5 $\micro$m, a
capacitance/length of 10.5~pF/m, a maximum voltage of 2.2~kV, and a
critical length of 2~m, the required tension on the wire is
found to be approximately 70~g.

\subsubsection{Simulation studies for the Straw Tracker}
\label{sec:garfield}
The processes occurring
in the straw chamber are simulated using three simulation tools,
HEED~\cite{heed}, MAGBOLTZ~\cite{magboltz}, and GARFIELD~\cite{garfield}.

\begin{figure}[htb!]
  \begin{center}
    \subfigure[][]{\includegraphics[trim={0 0 14pt 0},clip,width=0.45\textwidth]{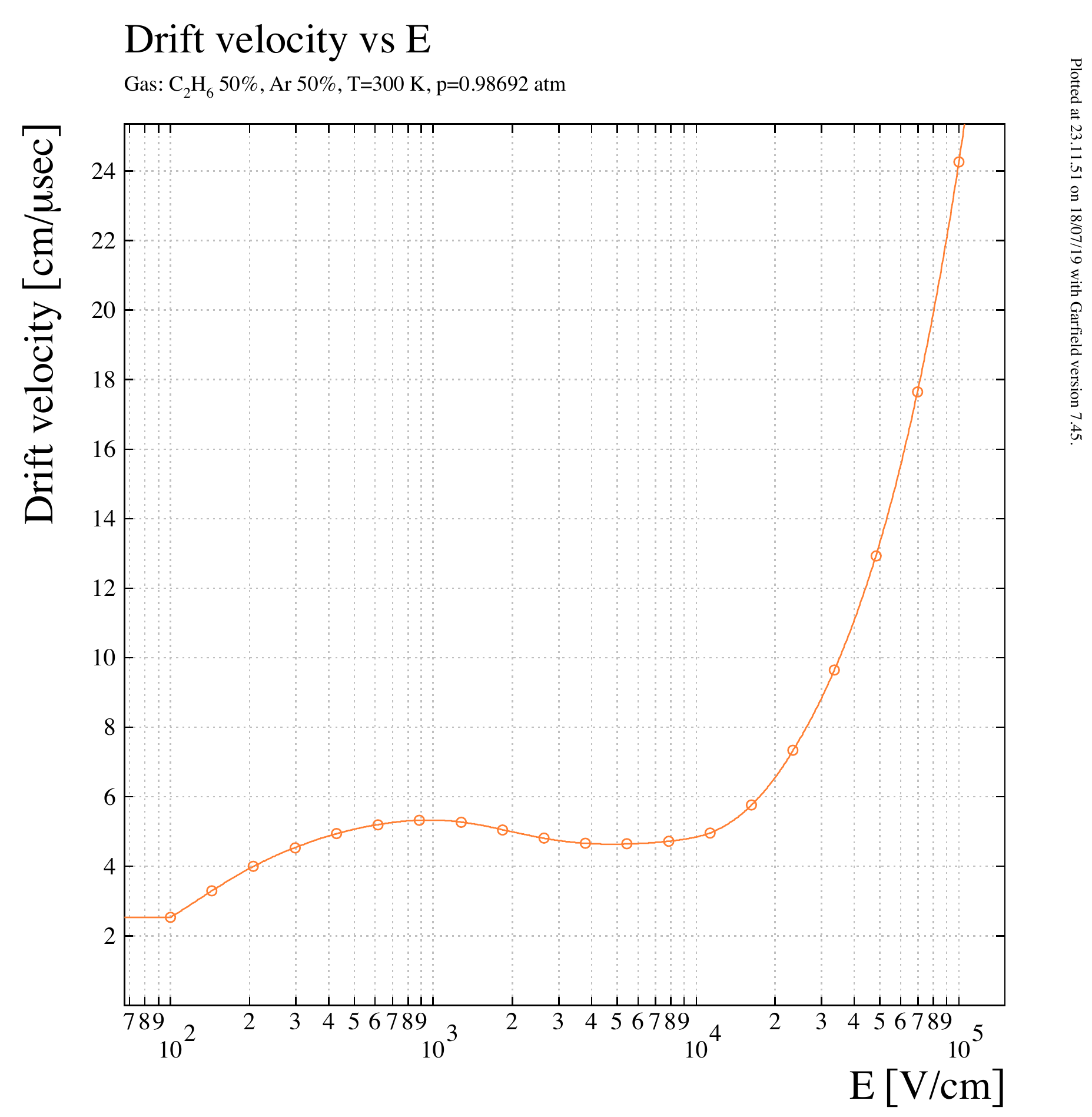}}
    \subfigure[][]{\includegraphics[trim={0 0 14pt 0},clip,width=0.45\textwidth]{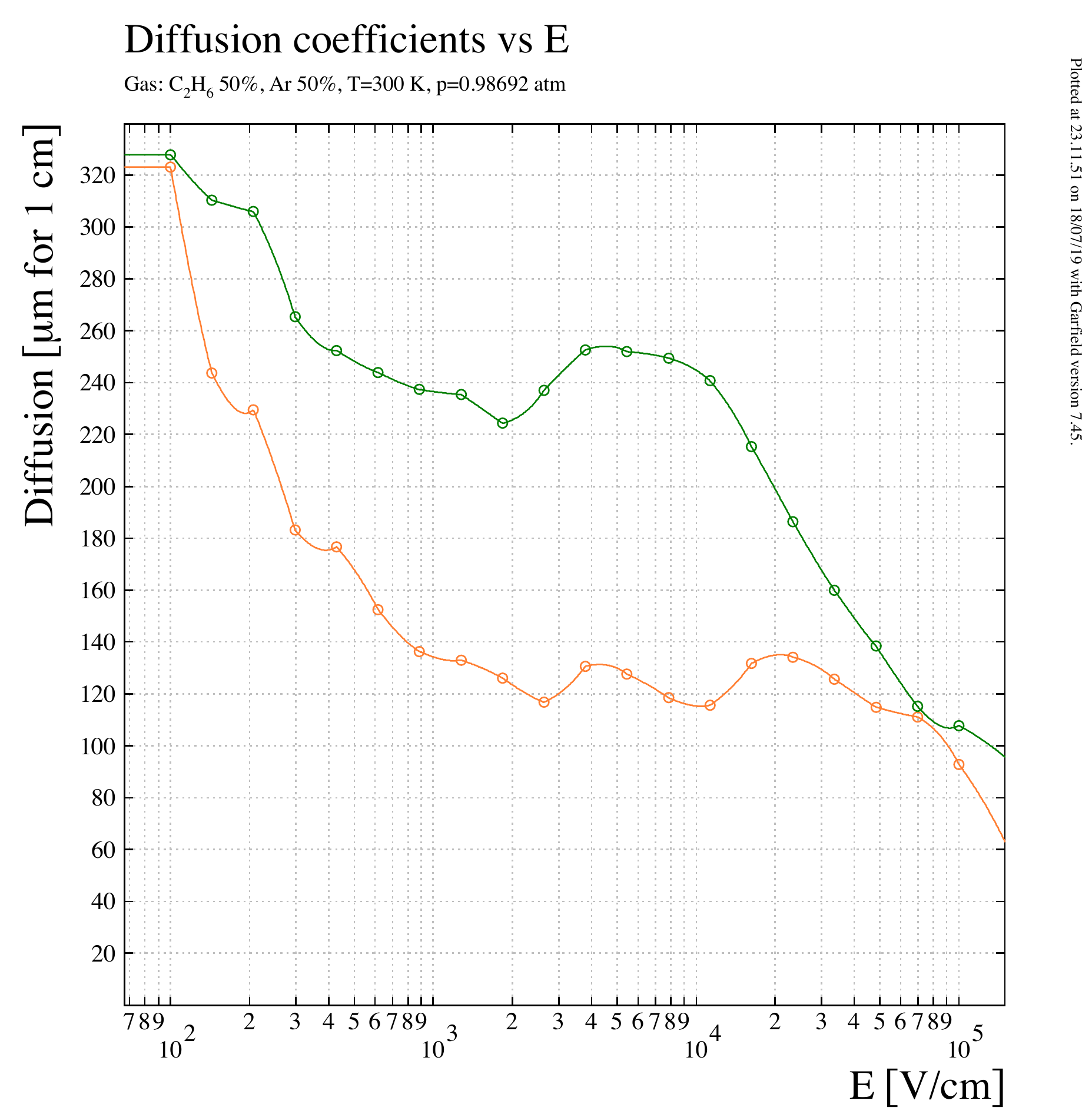}} \\
    \subfigure[][]{\includegraphics[width=0.45\textwidth]{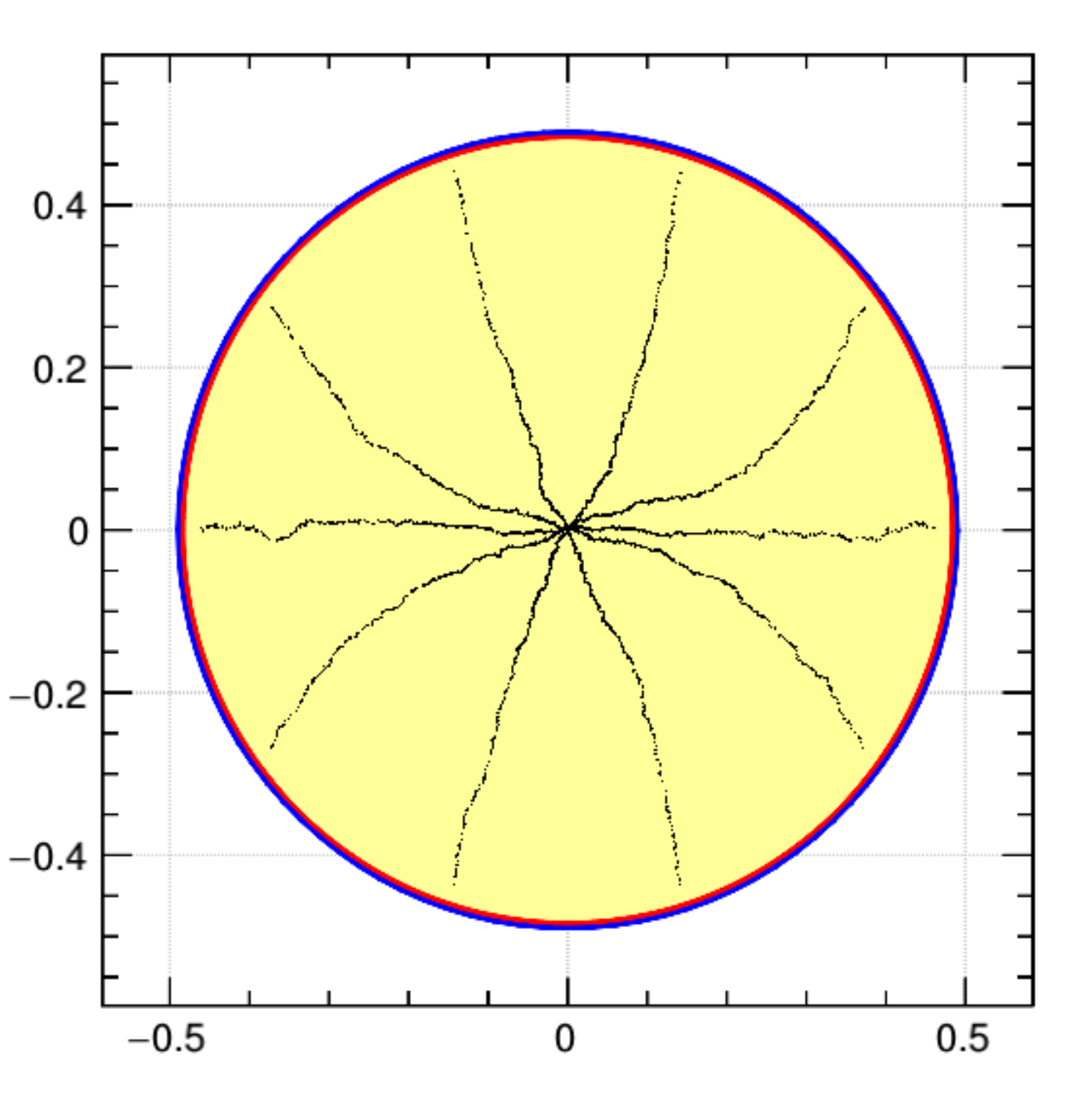}}
    \caption{%{\bfseries FIXME: Coordinate system wrong} 
    Gaseous detector simulations for the COMET Straw Tracker with GARFIELD,
    for an Ar-\ethane 50:50 mixture in 300~K and 0.98692~atm condition. 
              (a) Drift velocity and
              (b) Diffusion coefficients as functions of the applied electric
              field and (c) Drift lines and isochrones.
      For the diffusion coefficients, the orange and green lines correspond to the transverse and longitudinal coefficient, respectively.
      \label{fig:Straw_garfield}}
  \end{center}
\end{figure}

\Cref{fig:Straw_garfield} shows the results of these simulations.
\Cref{fig:Straw_garfield}(a) shows the calculated drift velocity
for an Ar-\ethane 50:50 mixture as a function of the applied electric
field, \cref{fig:Straw_garfield}(b) shows the diffusion
coefficients, and \cref{fig:Straw_garfield}(c) shows the drift lines,
where the strength of the magnetic field is 1~T.  According to this 
study, the drift velocity is expected to be saturated at approximately
5~cm/$\micro$s for an electric field higher than
$10^3$~V/cm. Relatively low diffusion coefficients, roughly
100--300\,$\micro$m/cm, are expected for an electric field of order of
$10^3$~V/cm; hence good spatial resolution can be expected.

\subsubsection{Spatial resolution estimation}
\label{sec:garfield2}
\begin{figure}[h]
  \begin{center}
    \includegraphics[width=0.8\textwidth] {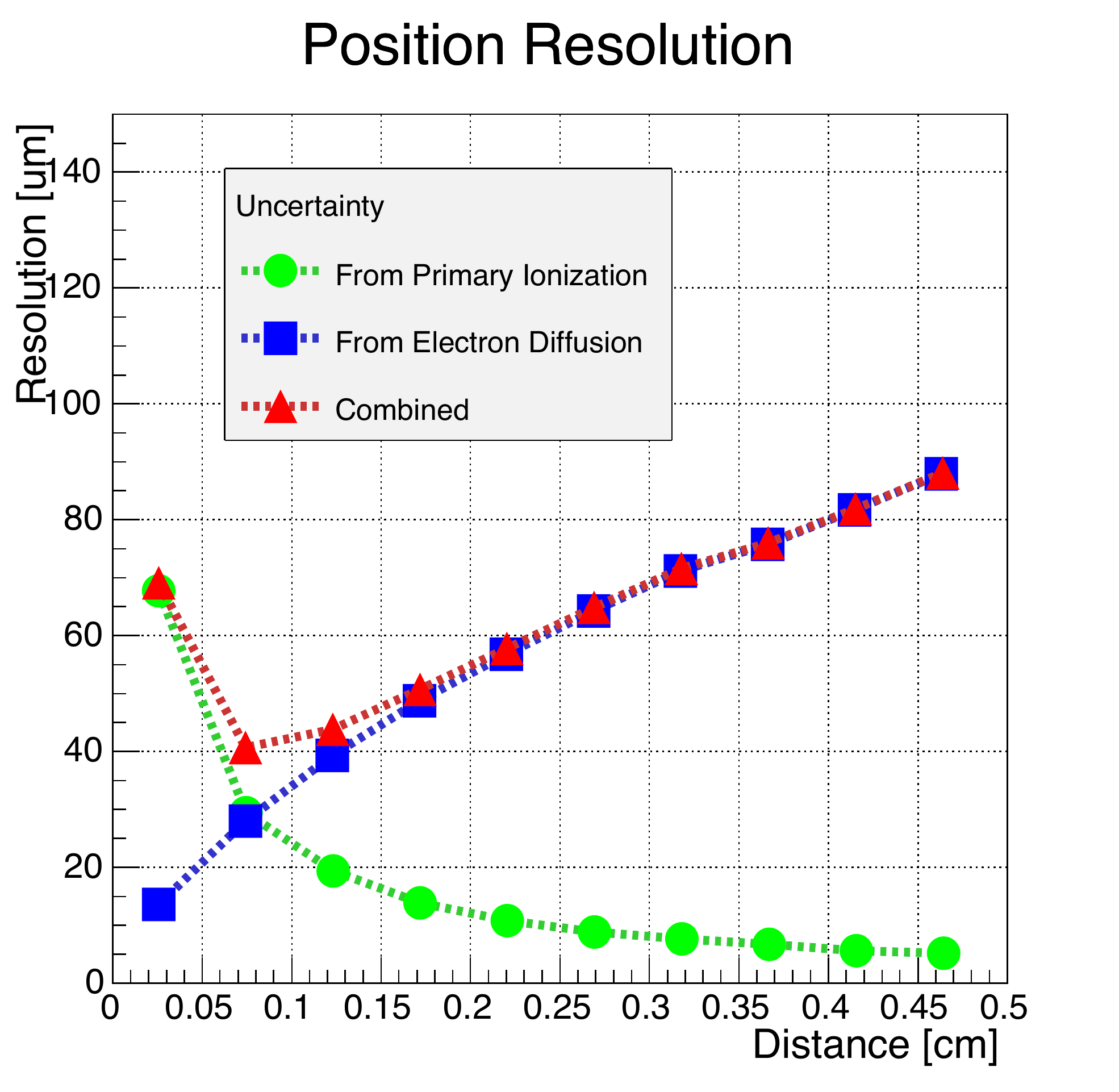}
    \caption{Expected spatial resolution as a function of the distance from wire, simulated by Garfield++ (Ar:C$_{2}$H$_{6}$ (50:50), 1900 V)
    \label{fig:garfpp_reso2}}
  \end{center}
\end{figure}

To estimate the  intrinsic spatial resolution,
GARFIELD++~\cite{garfield++}
simulations, validated where possible through comparisons with real data, are employed.
The expected spatial resolution as a function of the distance from the wire for the gas mixture of
Ar:C$_{2}$H$_{6}$ (50:50) and a HV of 1900~V, where the incident particle is assumed to be an 100 MeV/$c$ electron, is shown in \cref{fig:garfpp_reso2}.
Electrical noise is not taken into account
as it is not easy to predict the actual noise level.
The results indicate that 
the Straw Tracker will satisfy the required performance for the beam
background measurements.

\subsection{Readout Electronics}
\label{sec:ROESTI}

The readout electronics boards will
be installed in the gas manifold, so the front-end electronics
must operate in the vacuum inside the DS to measure the
analogue signal from the anode wires.
All signals are digitised at the front-end boards, and
stored in digital pipelines to allow for trigger latency.
Once a trigger is issued, only those channels with signals
above a set threshold are read, stored in buffers,
and then serially transferred to the data acquisition system.
The events are then rebuilt, analysed, filtered, and finally
committed to permanent storage.

To achieve a momentum resolution better than 200~keV/c,
a spatial resolution of $\sim$100~\micro{}m is desired; this requires a timing resolution of better than 1~ns in the readout board.
\begin{figure}[h]
  \begin{center}
    \includegraphics[width=0.6\textwidth] {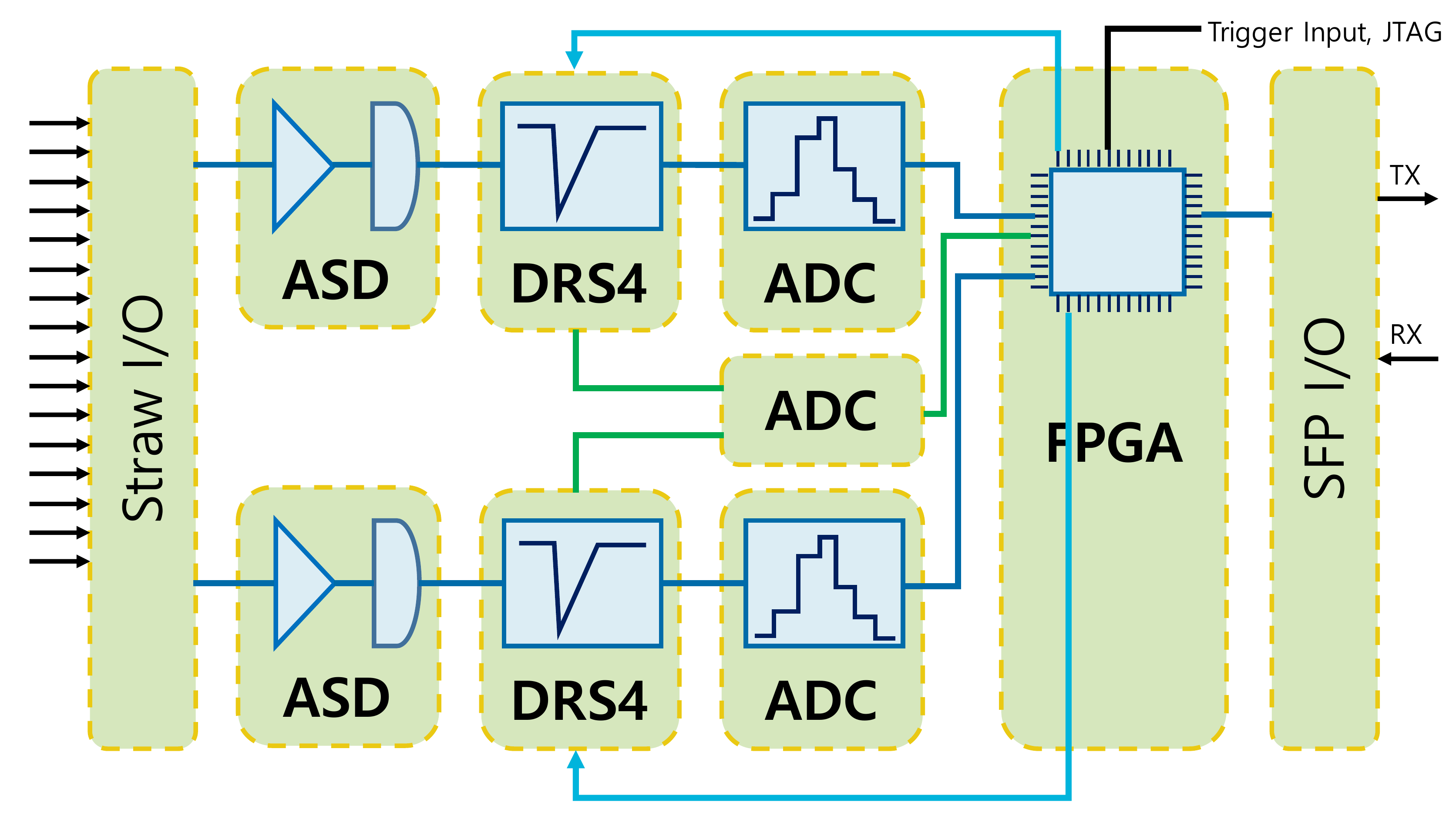}
    \caption{The ROESTI
             front-end board for the straw Tracker.}
    \label{fig:Straw_ROESTI_diagram}
  \end{center}
\end{figure}

The front-end boards, termed ROESTI (Read Out
Electronics for Straw Tube Instrument),
contain all the front-end processing; pre-amplification and pulse
shaping, discrimination, and digitisation,
controlled by an FPGA-based readout controller, as
shown in \cref{fig:Straw_ROESTI_diagram}. Pre-amplification,  pulse shaping and
signal discrimination are performed by the ASD
(Amplifier-Shaper-Discriminator) chip\footnote{
  This ASD chip has been developed for ATLAS MDT/TGC
  front-end electronics originally, and recently modified for Belle-II
  CDC electronics; this version is adapted from Belle for the COMET
  straw front end.}, and the amplified signal is then digitised by a
DRS4 chip~\cite{drs4}.
The digitised
waveform data, correction data, and relevant metadata are then sent out via an optical fibre.
The FPGA also has other input/output lines for triggering and JTAG connections.
Development of the ROESTI board is
supported by the KEK Electronics Group and the Open Source Consortium of
Instrumentation (OpenIt).

\paragraph{FPGA firmware design}
A modern FPGA design, the Artix-7 (XC7A200T-2FBG676C, Xilinx) is used on the ROESTI board.
The firmware is composed of five blocks; Network Interface, Monitor, Module Control, Trigger Interface, and Data Interface.
In the Network Interface block, the input/output signal can be transmitted/received between a PC and
several boards.
A UDP connection for parameter control between the board and PC is also found in this block.
In the Monitor block, the temperature and voltage in the FPGA are monitored and detection and correction of
SEU (Single Event Upsets) and URE (UnRecoverable Errors) are also handled here.
In the Module Control block, all chip parameters are controlled, including the ASD threshold for the DAC, the offset voltage, sampling speed for the
DRS4 module and, following a trigger
signal, the start signal for sending  information from the DRS4 to the ADC is issued.
In the Trigger Interface block the trigger signal is handled and the information  sent to the Module Control and Data Interface
blocks.
In the Data Interface block, ADC and monitor data are received and converted to packet data, which is then sent to the Network Interface block.

\subsubsection{Prototypes}

\paragraph{The single-straw prototype}
\label{sec:single-straw-prototype}
A small prototype (the
``single-straw prototype'') was built
to investigate the
gas tightness, operation in vacuum and noise shielding.

A new feedthrough system was developed which
 provides not only  electrical connections but also the straw
tensioning scheme.
The straws are stretched by
rotating the bushing part of the feedthrough to pull the straw tube by about
2~mm,  equivalent to a pre-tension of 1 \kgf.
\begin{figure}[h]
  \begin{center}
    \includegraphics[width=\textwidth] {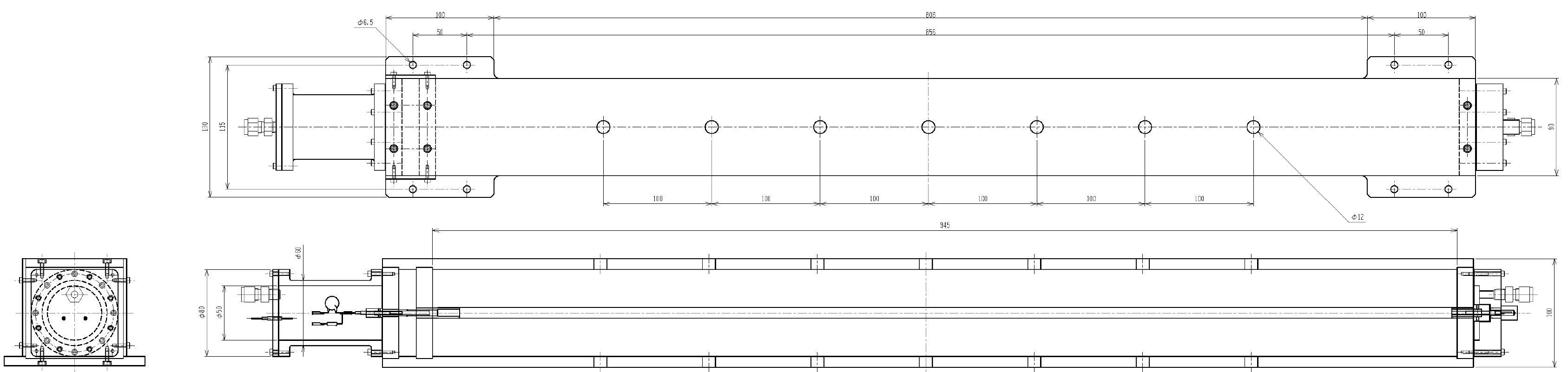}\\
    \vspace{0.01\textheight}
    \includegraphics[width=0.8\textwidth]{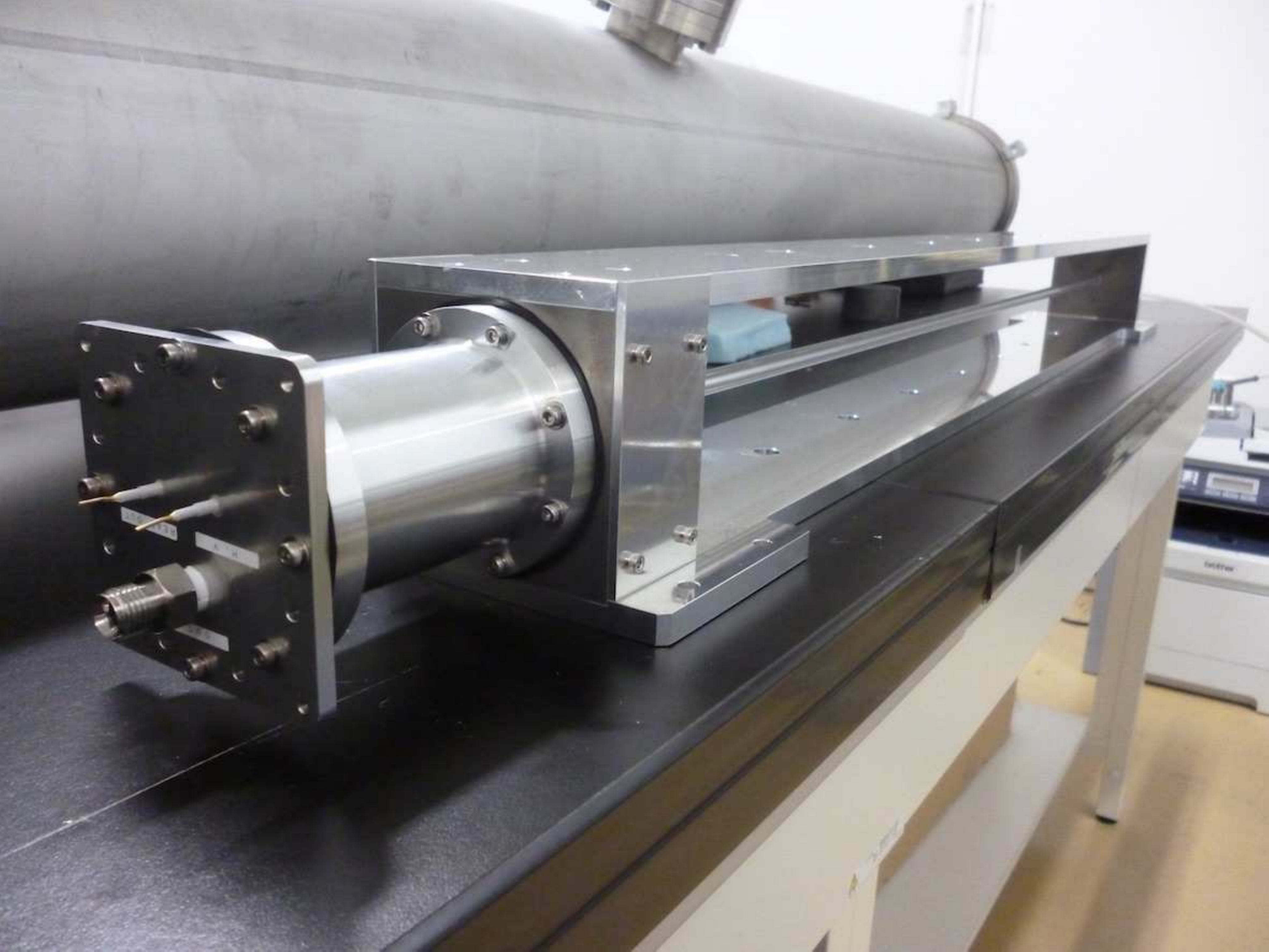} %comment out???
    \caption{The Single-Straw Prototype.
            (Top) Drawing,
            (Bottom) Photo of the whole view,
          \label{fig:straw_smallprototype}}
  \end{center}
\end{figure}
\Cref{fig:straw_smallprototype} shows a drawing and photograph of the single-straw
prototype and the installation
of the single-straw prototype into the vacuum vessel.

\begin{figure}[h]
  \begin{center}
    \includegraphics[height=0.45\textwidth]{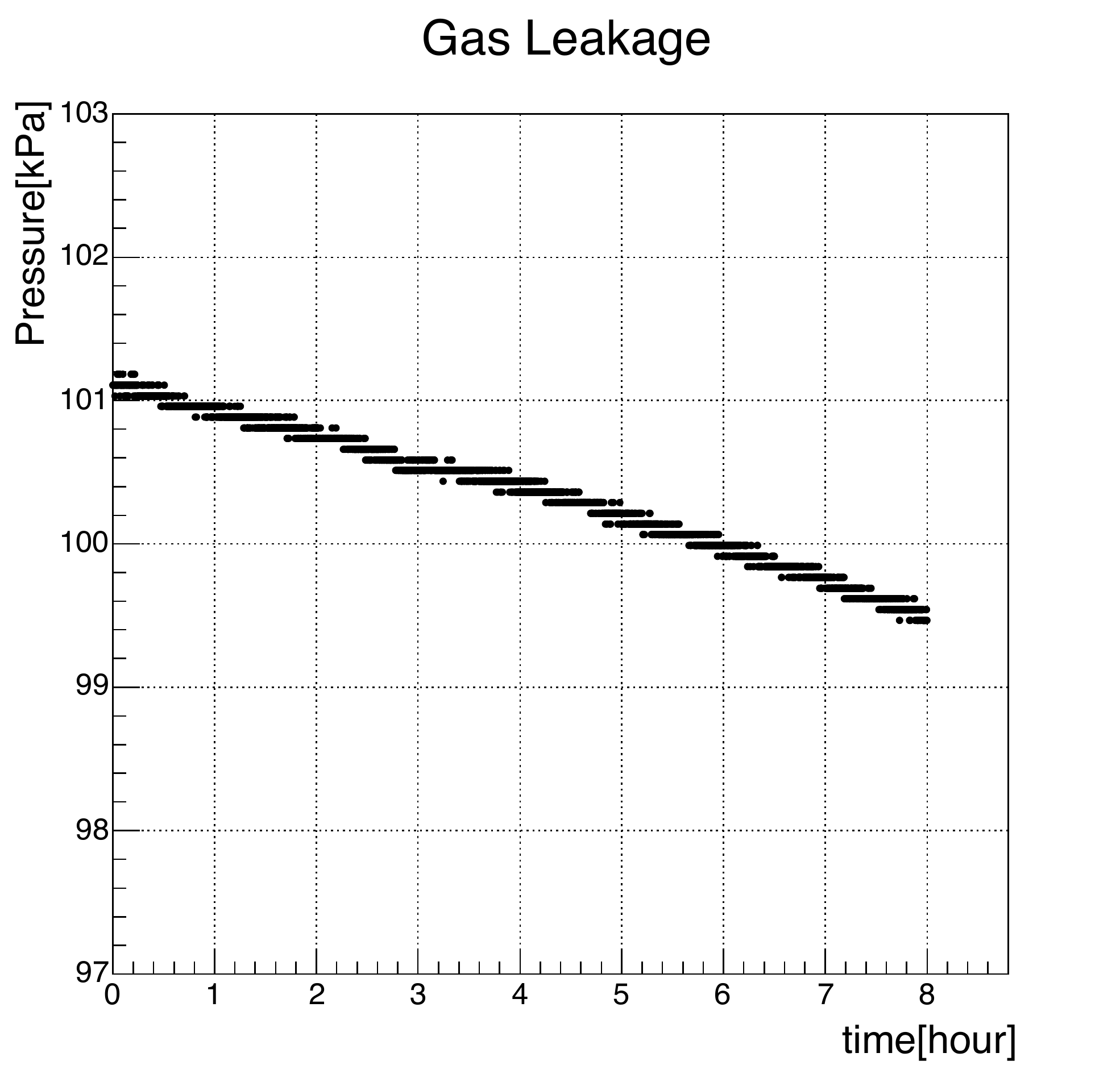}
    \includegraphics[height=0.45\textwidth]{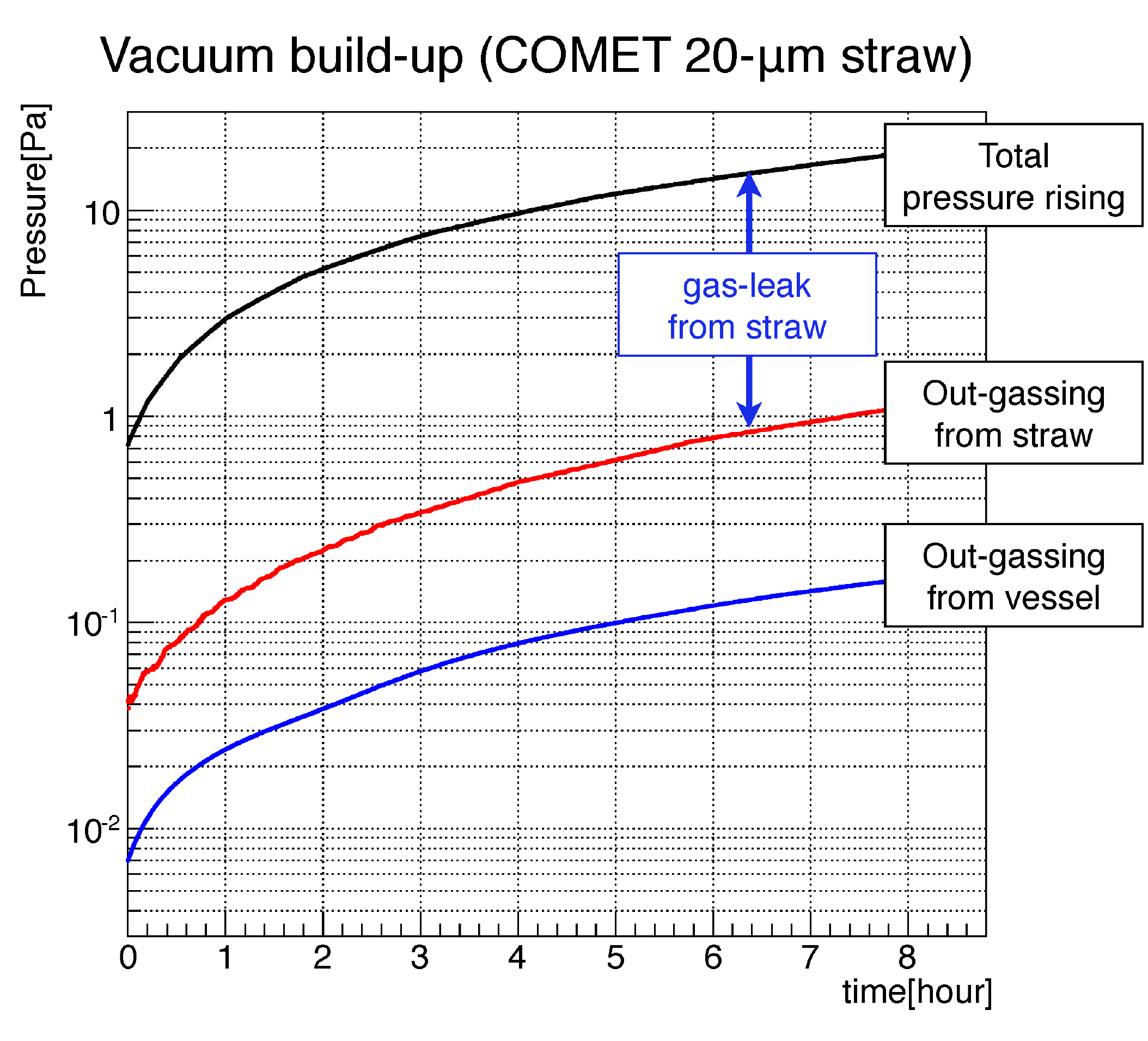}
        \caption{Measured gas leakage:
        (Left) Pressure drop inside the straw tube as a function of time
            after it is over-pressurised to 2 bar,
        (Right) Pressure build-up as a function of the time after pump close.
          \label{fig:Straw_vacuum_operation}}
  \end{center}
\end{figure}
Measurements of  the gas tightness
revealed
a leak rate of 0.0035 cm$^3$/min/m, which, % [assuming this is what 'unit length' means]
when scaled to the full spectrometer, is
well within what is needed to keep pumping rates at modest levels,
as shown in \cref{fig:Straw_vacuum_operation}. The leakage test using a full scale prototype is also conducted, which resulted smaller leakage than expected from the single straw leakage test.  
Electrical shielding properties have also been tested using this prototype.
Several gas mixtures were tested using $^{55}$Fe as an X-ray source.
By changing the applied HV, the
gas gain was measured and the good gas amplification performance confirmed.
These results from the single-straw prototyping validate the
use of the newly-developed straws with 20~$\mu$m walls for COMET Phase-I.

\paragraph{Full-scale prototype}
\label{sec:kek_prototype}

\begin{figure}[ht]
  \begin{center}
    \includegraphics[width=0.8\textwidth]{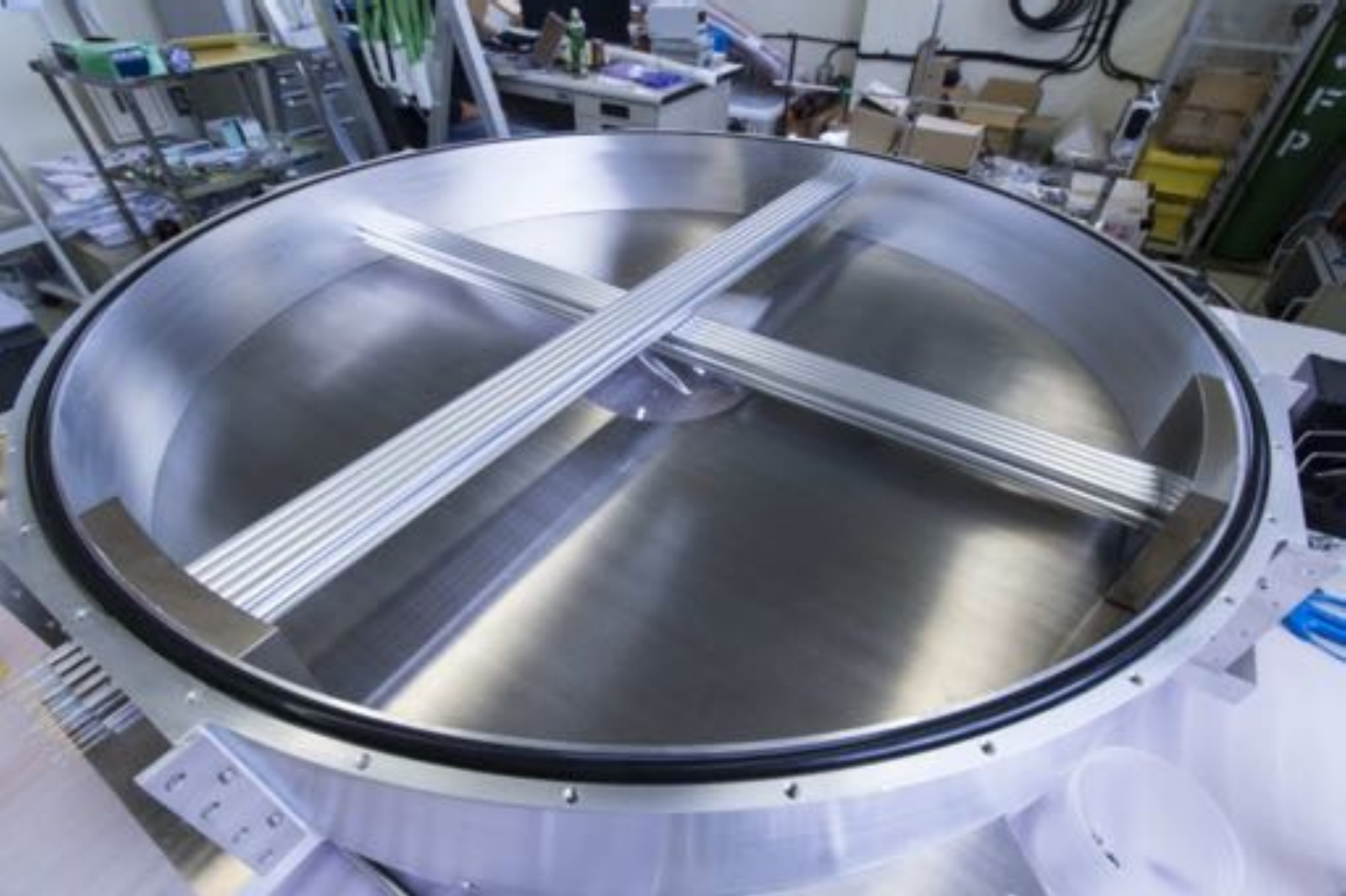}
    \caption{The Full-scale prototype,
              partially completed without the vacuum wall.
    \label{fig:straw_fullproto_photo}}
  \end{center}
\end{figure}

A second prototype,
the ``full-scale prototype'', has similar dimensions to a final tracker station but with fewer straws.
It has six straw-tube planes, three for  the $x$-coordinate and three for  the $y$-coordinate, with  each coordinate  measured by 16
straw tubes. \Cref{fig:straw_fullproto_photo} shows a partially completed prototype.
The 20~$\mu$m-wall straws are mounted
using a newly-developed  feedthrough system and the entirety of the exterior
is covered with a vacuum wall so that it
can be evacuated, allowing the behaviour  in vacuum to
be investigated.
The  prototype is constructed of aluminium so that it will not be
affected by magnetic fields.

It has been operated in a 50--300~MeV/$c$ electron test beam  at the
Research Centre for Electron Photon Science (ELPH),
Tohoku University.

\begin{figure}[tbh!]
\centering
    \includegraphics[width=0.8\textwidth]{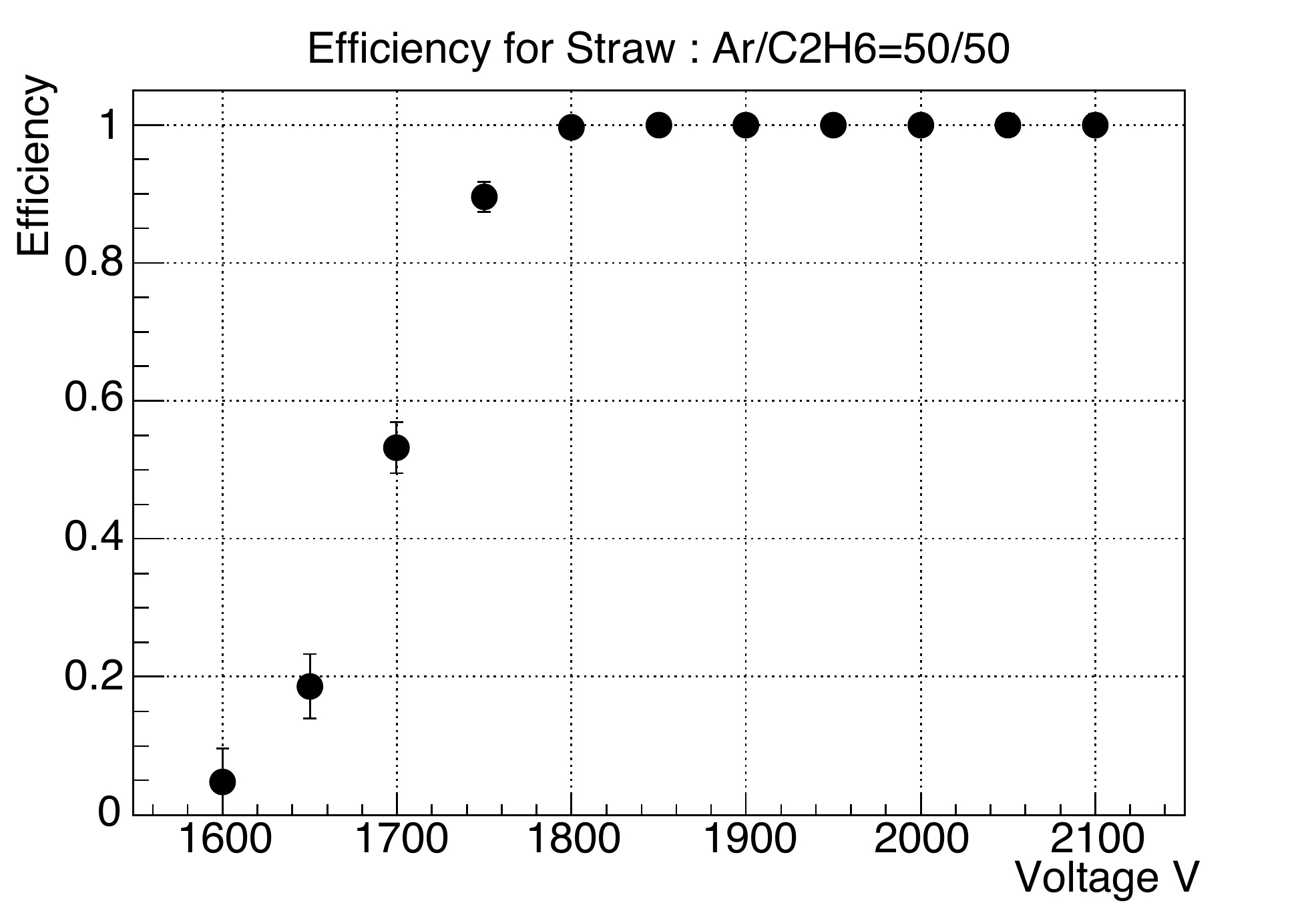}
    \caption{Measured single straw detection efficiency.
    \label{fig:straw_beamtest_efficiency}}
\end{figure}
\begin{figure}[tbh!]
\centering
  \includegraphics[trim={279px 0 0 0},clip,width=0.8\textwidth]{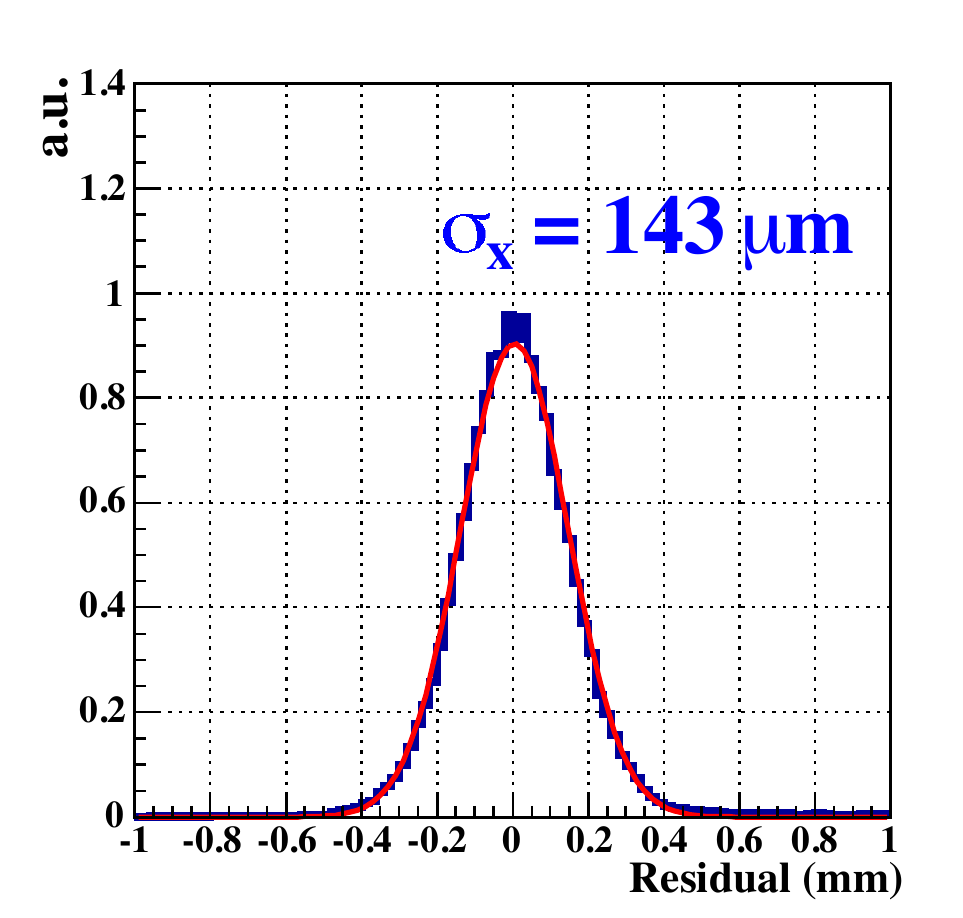}
  \caption{Residual distribution for
    Ar:C$_{2}$H$_{6}$(50:50) gas mixture at HV = 1900 V.
  \label{fig:straw_beamtest_xt}}
\end{figure}

\Cref{fig:straw_beamtest_efficiency} shows the measured single straw
detection efficiency for the Ar:C$_{2}$H$_{6}$(50:50) gas mixture
as a function of applied HV.
\Cref{fig:straw_beamtest_efficiency} shows that a voltage
 higher than 1800~V results in full efficiency for a single straw although gaps between straw tubes\footnote{
The full-scale prototype has a small gap of 0.5 mm between each straw tubes.
} can lead to a small overall efficiency loss.
\Cref{fig:straw_beamtest_xt} shows the residual distributions for tracks.
A spatial resolution of 143.2~\micro{}m is obtained
for a HV of 1900~V.
This value includes the uncertainties
arising from the precision of track reconstruction, 
and if this is taken into account the true spatial resolution is estimated
to be 119.3~\micro{}m.
\begin{figure}[ht]
  \begin{center}
    \includegraphics[width=0.45\textwidth] {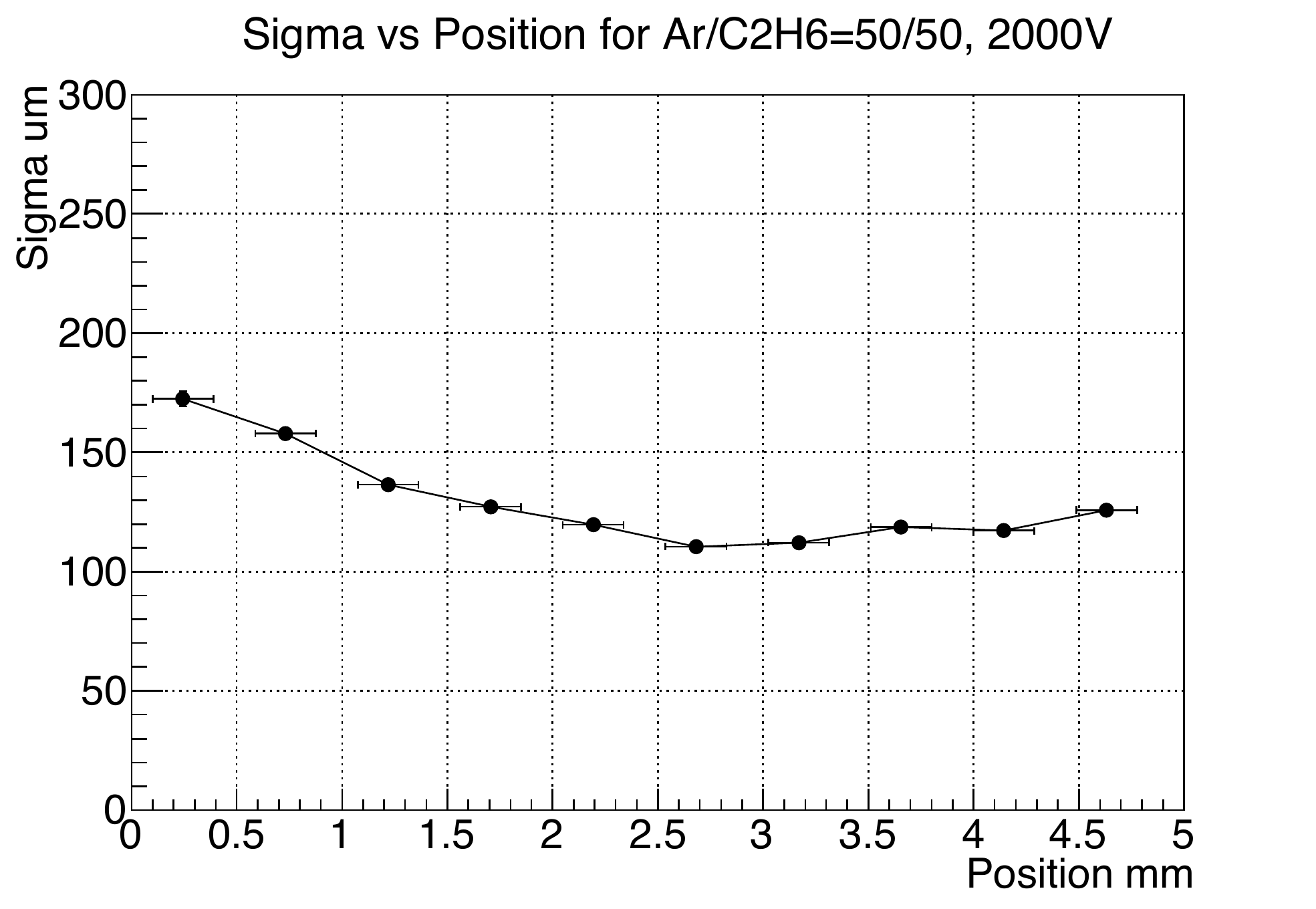}
    \includegraphics[width=0.45\textwidth] {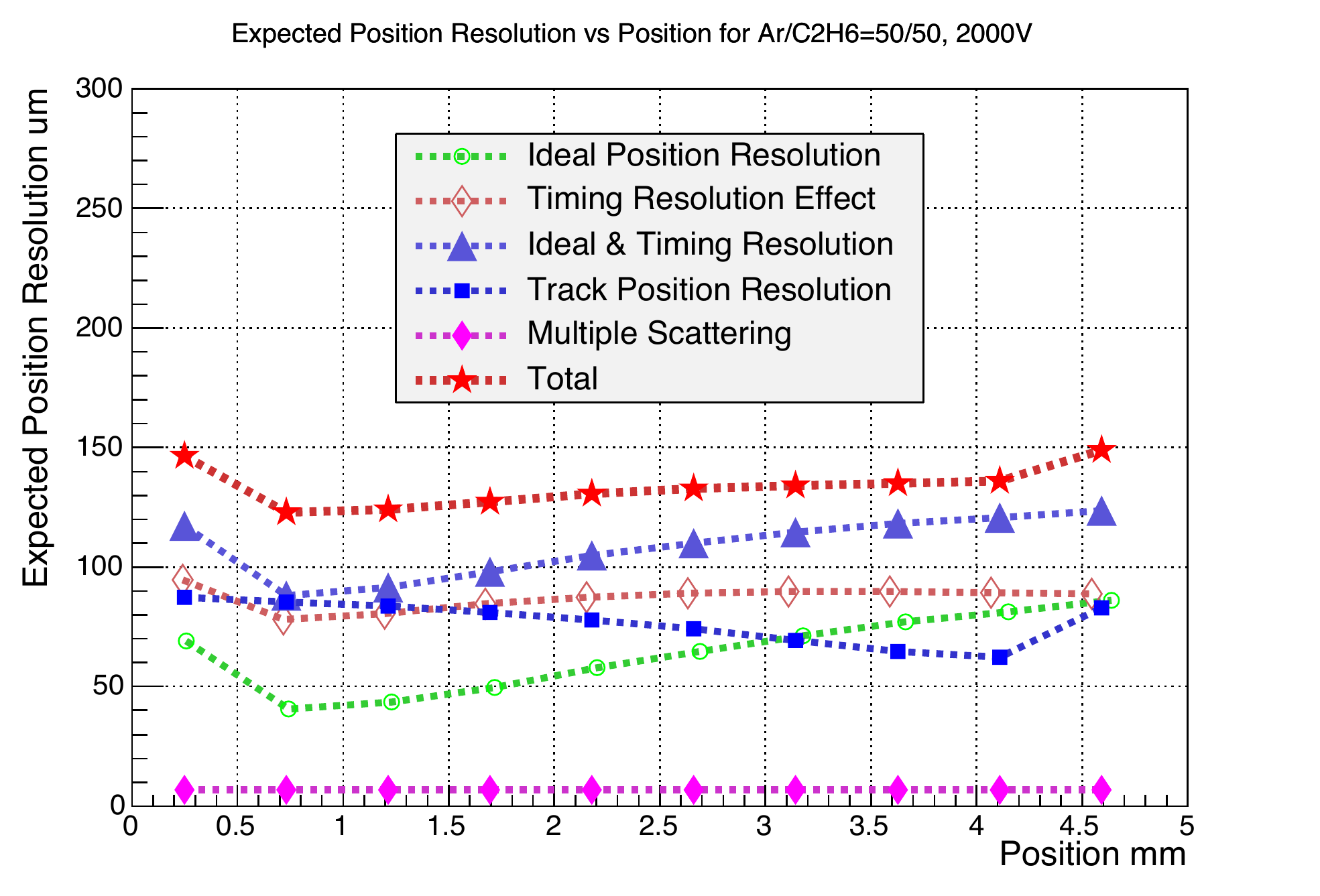}
    \caption{Incident position dependence of
    the obtained spatial resolution,
    gas mixture = Ar:C$_{2}$H$_{6}$(50:50), HV = 2000 V.
    (Left) Data, (Right) Garfield++ simulation
    \label{fig:straw_beamtest_reso_ArEth}}
  \end{center}
\end{figure}
The left plot of 
\cref{fig:straw_beamtest_reso_ArEth}  shows the dependence of the  spatial resolution on the
incident position
for  Ar:C$_{2}$H$_{6}$(50:50) and a HV of 2000 V, and
the right plot 
shows the
expected spatial resolution simulated with
GARFIELD++.
Here the green (open circle) plot shows the ideal spatial resolution.
The left plot in \cref{fig:straw_beamtest_reso_ArEth} shows the measured incident-position dependence, which is
well-reproduced in the simulation as shown in the right plot of \cref{fig:straw_beamtest_reso_ArEth}.

In conclusion, the detection efficiency
and intrinsic spatial resolution are confirmed to meet requirements.
The ability to maintain a vacuum inside the DS
is also confirmed and expected to be better than what is required.

 % subsection
\subsection{Electron Calorimeter (ECAL)}\label{ch:ecal}

The electron calorimeter (ECAL) system consists of segmented scintillating crystals.
It is placed downstream of the Straw Tracker to measure the energy of electrons with good resolution and hence add redundancy to the electron momentum measurement. It will also provide an additional hit position on the electron track trajectory and provide the trigger signals.

\begin{figure}[htt!]
\centering
\includegraphics[width=0.8\textwidth]{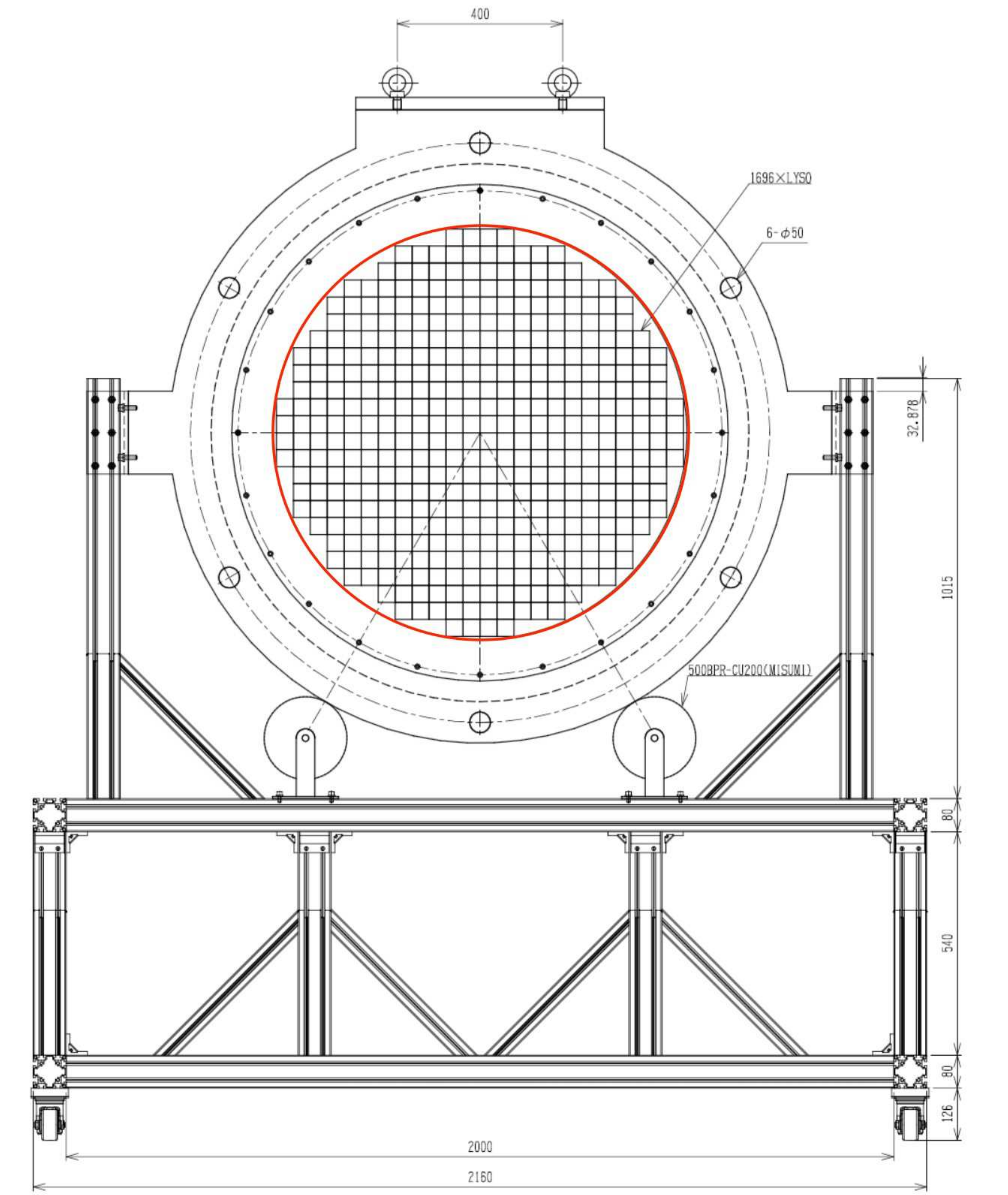}
\caption{A schematic layout of the electron calorimeter system. The matrix structures inside the red circle represent the LYSO crystal array.  }
\label{fig:ECAL-Layout}
\end{figure}

The specifications for the ECAL are determined by its requirements for Phase-II running, which are an energy resolution of better than 5\% at 105 MeV and a cluster position resolution  that is better than 1~cm.
The latter will enable the shower topology to  be used also to discriminate electrons from neutrons and 
low-energy photons.
The crystals need to have a good light yield, and fast response and decay times in order to reduce pileup.
A schematic layout of the ECAL system is shown in \cref{fig:ECAL-Layout}.

\paragraph{Scintillating crystals}

The properties of candidate crystal types are summarised in \cref{tb:crystal}. Taking into account both performance and cost, LYSO 
(Lutetium-yttrium oxyorthosilicate, $\rm{Lu}_{2(1-x)}\rm{Y}_{2x}\rm{SiO}_5$)
has been chosen for the ECAL.
\begin{table}[htb!]
\caption{
Characteristics of inorganic scintillator crystals. The superscript of $f$ and $s$ represent the fast component and the slow component, respectively.}
\label{tb:crystal}
 \begin{center}
  \begin{tabular}{lcccc}
   \hline\hline
                                  &         GSO(Ce) & LYSO & PWO                 & CsI(pure)  \\ \hline
Density (g/cm$^{3}$)              &            6.71 & 7.40 & 8.3                 & 4.51       \\
Radiation length (cm)             &            1.38 & 1.14 & 0.89                & 1.86       \\
Moliere radius (cm)               &            2.23 & 2.07 & 2.00                & 3.57       \\
Decay constant (ns)               & 600$^s$,~56$^f$ & 40   & 30$^s$,~10$^f$      & 35$^s$,6$^f$ \\
Wave length (nm)                  &             430 & 420  & 425$^s$,~420$^f$    & 420$^s$,~310$^f$ \\
Refractive index at peak emission &            1.85 & 1.82 & 2.20                & 1.95       \\
Light yield (NaI(Tl)=100)         &   3$^s$,~30$^f$ & 83   & 0.083$^s$,~0.29$^f$ & 3.6$^s$,1.1$^f$ \\
References                        & \cite{gso:4337375, gso:682439, gso:NAKAYAMA199834, gso:TANAKA1998283} & \cite{csi:lyso:saintgobain} & \cite{pwo:ANNENKOV200230} &  \cite{csi:lyso:saintgobain} \\
\hline\hline
\end{tabular}
\end{center}
\end{table}
High segmentation is required both to reduce pileup and provide good position information.
The ECAL will consist of crystal modules which have a 2$\times$2~cm$^{2}$ cross-section and whose length is 12~cm corresponding to 10.5 radiation length.
The ECAL covers the cross-section of the 50-cm radius detector region and 1,920 crystals are needed.

\paragraph{Photon detector}

The photon detectors for the ECAL must be able to operate in the  1~T magnetic field, have a high quantum efficiency around the wavelength range of LYSO scintillation and  excellent linearity. The Hamamatsu S8664-1010  avalanche photodiode (APD) with an active area of 10$\times$10 mm$^{2}$ satisfies these requirements; its characteristics are summarised in \cref{tb:apd}.
\begin{table}[htb!]
\caption{
The characteristics of APD, Hamamatsu S8864-1010~
\cite{ecal:apd_datasheet}.
}
\label{tb:apd}
 \begin{center}
  \begin{tabular}{lccc}
   \hline\hline
Type & S8664-1010\\
   \hline
Active area (mm$^{2}$)      &	10 $\times$ 10\\
Package size (mm$^{2}$)    & 14.5 $\times$ 13.7 \\
Spectral response range (nm)  & 320--1000\\
Peak sensitivity wavelength (nm)  & 600\\
Quantum efficiency at 420\,~nm (\%)      & 70\\
Breakdown voltage (V)  & 400\\
Nominal gain  & 50\\
Typical dark current (nA)  & 10\\
Maximum dark current (nA)  & 100\\
Terminal capacitance (pF)  & 270\\
\hline\hline
\end{tabular}
\end{center}
\end{table}
Laboratory tests have been made to check the noise performance with a suitable preamplifier  which confirm that the requirements are met.

\subsubsection{Readout electronics}
\label{sec:readout-electronics}

\begin{figure}[htt!]
\centering
\includegraphics[width=0.8\textwidth]{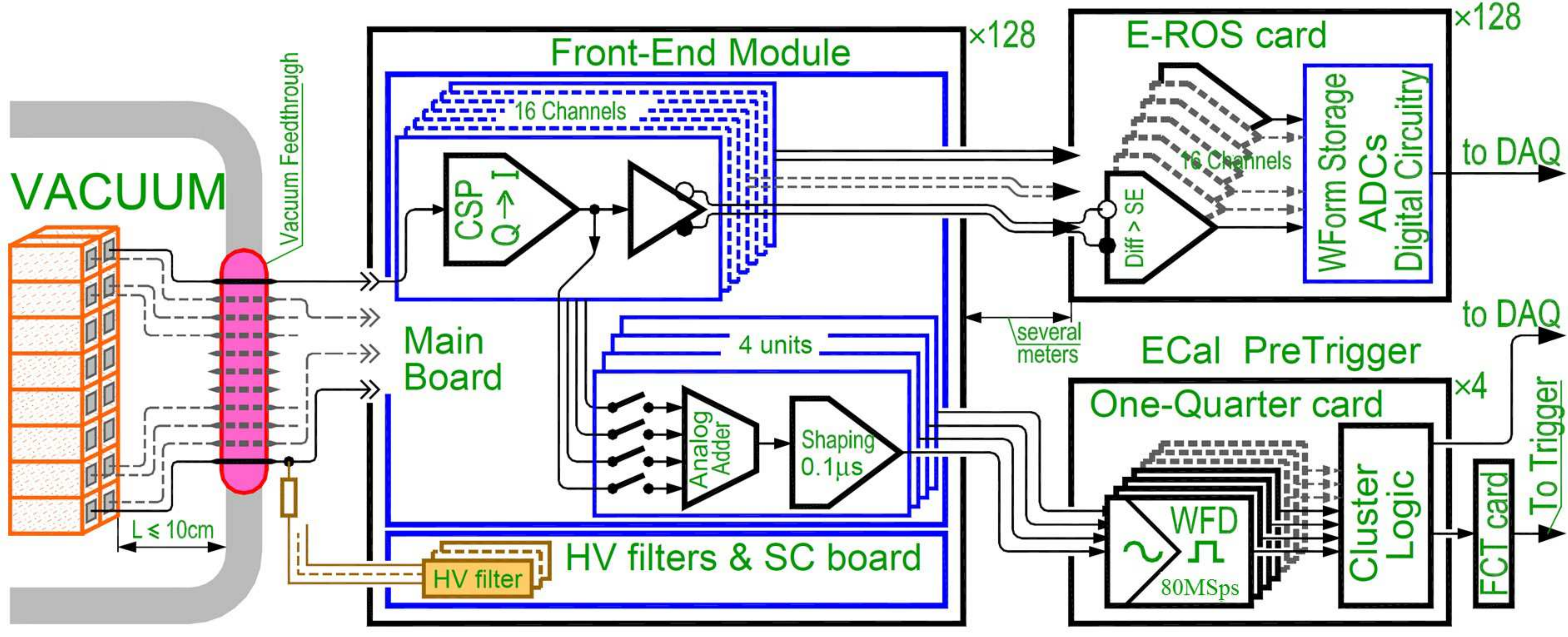}
\caption{A schematic of the ECAL readout electronics.}
\label{fig:ECAL-Readout}
\end{figure}

A schematic diagram of the readout electronics for the ECAL system is shown in \cref{fig:ECAL-Readout}.
The crystals and the APDs are located inside a vacuum vessel.
The Front-End card houses 16 full-bandwidth amplification channels which produce signals for energy measurements and four 4-input analogue adders which derive signals for triggering. In the full-bandwidth channel the input signal is amplified by a Charge-Sensitive Amplifier (which integrates the signal), and then the signal shape is restored so that it becomes close to the shape of the input signal. The peaking time of the output signal is about 15~ns, and the falling slope is exponential, with the decay time constant equal to that of the LYSO. In this way, noise level are kept low and the pileup of signal is minimized. These full-bandwidth differential signals are transported to EROS boards for sampling. The EROS board is similar to the ROESTI board (described in \cref{sec:ROESTI}), but a differential-to-single ended signal converters are connected to the inputs. 
For the trigger, analogue signals from each block of $2\times 2$ crystals are summed up by an analogue adder (in the Front-End cards), and the summed signal is then shaped. All summed signals are then fed to the Pre-Trigger. The number of trigger cells in the full ECAL will be too large to be processed in one module; therefore, four identical Pre-Trigger modules will be used, each of which will process the signals from one quarter of the crystal matrix. 
The latest prototype of the electronics is designed 
so that
it has the same structure as above and improved noise performance, and has the appropriate form-factor for the ECAL mechanical design. To confirm the expected performance, 
a
beam test was carried out and the results are described in \cref{sec:rd-status}.

\subsubsection{Module, readout and mechanical structure}
\label{sec:structure}

The basic unit of the ECAL is a 2$\times$2 crystal matrix module with 480  modules to cover the full cross-section of the detector region.

\begin{figure}[htt!]
  \begin{center}
    \includegraphics[width=0.45\textwidth]{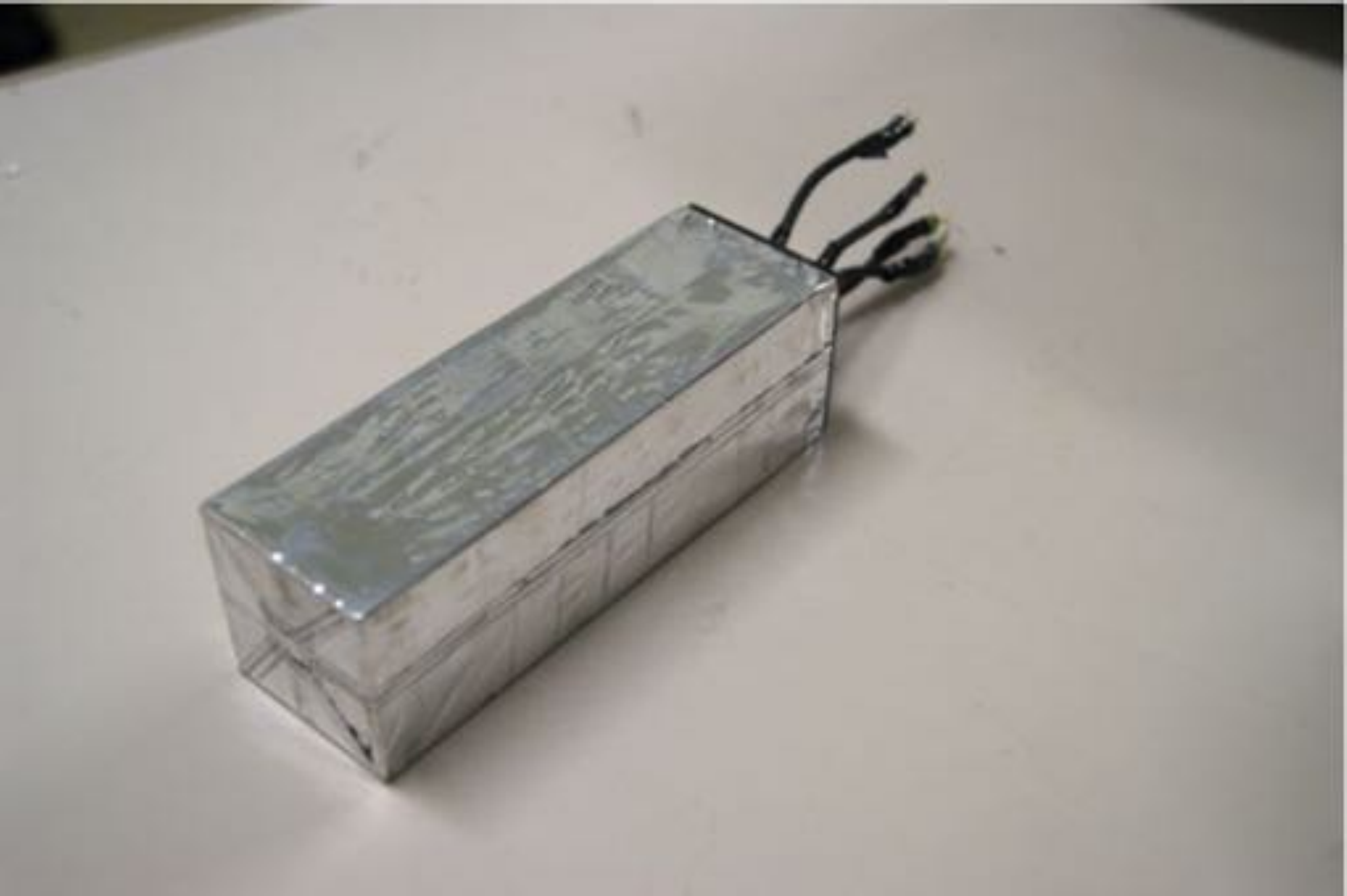}
    \includegraphics[width=0.45\textwidth]{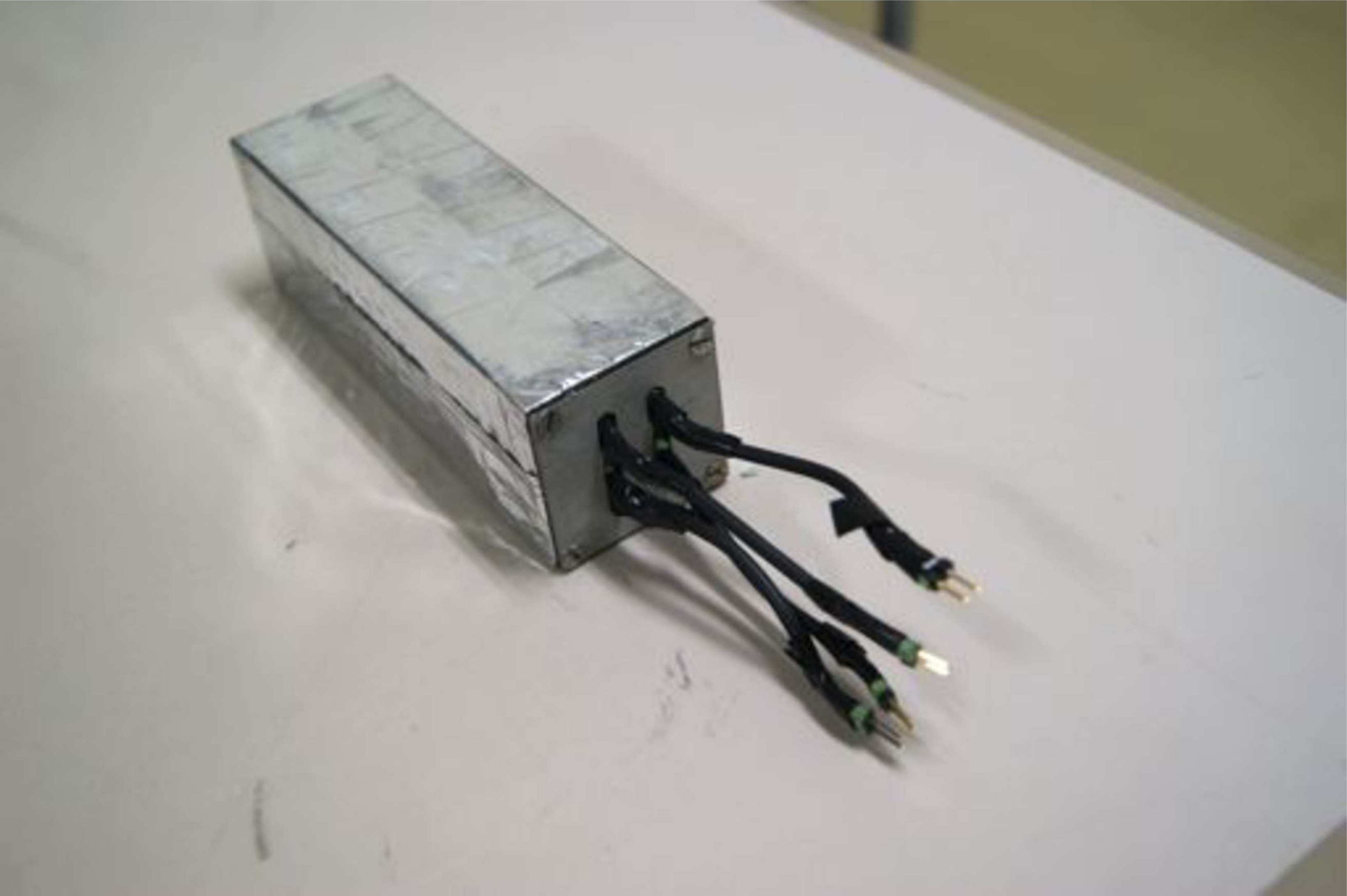}
    \caption{A prototype of the 2$\times$2 crystal matrix module (without the preamplifier board).}
    \label{fig:ECAL-Module1}
  \end{center}
\end{figure}

\begin{figure}[htt!]
  \begin{center}
    \includegraphics[width=0.8\textwidth]{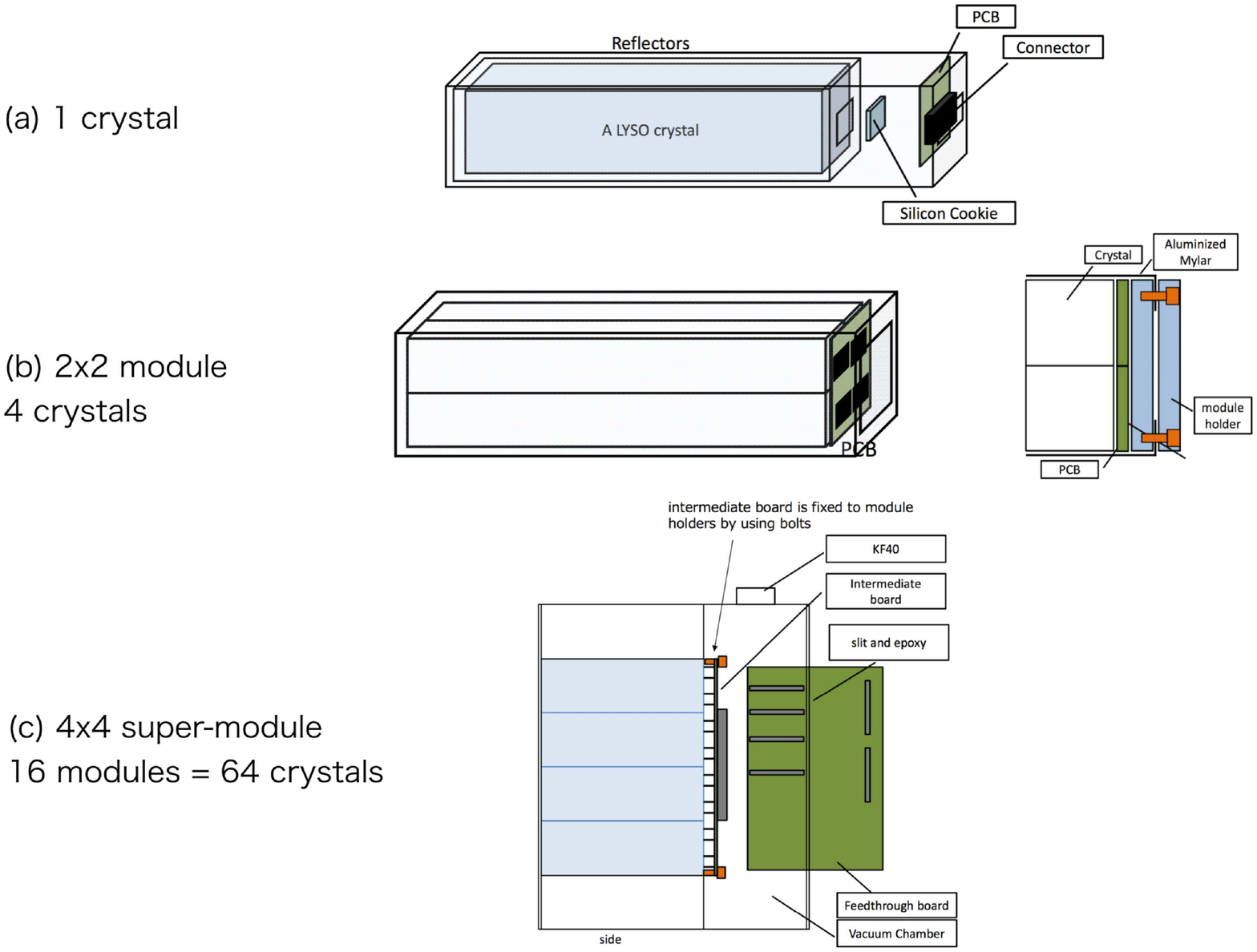}
    \caption{The module structure in the ECAL. (a) 1 crystal + 1 APD on PCB, (b) 2$\times$2 crystal matrix module, (c) super-module consisting of 4$\times$4 modules (= 64 crystals) and feedthrough}
    \label{fig:ECAL-Module3}
  \end{center}
\end{figure}

\begin{figure}[htt!]
  \begin{center}
    \includegraphics[width=\textwidth]{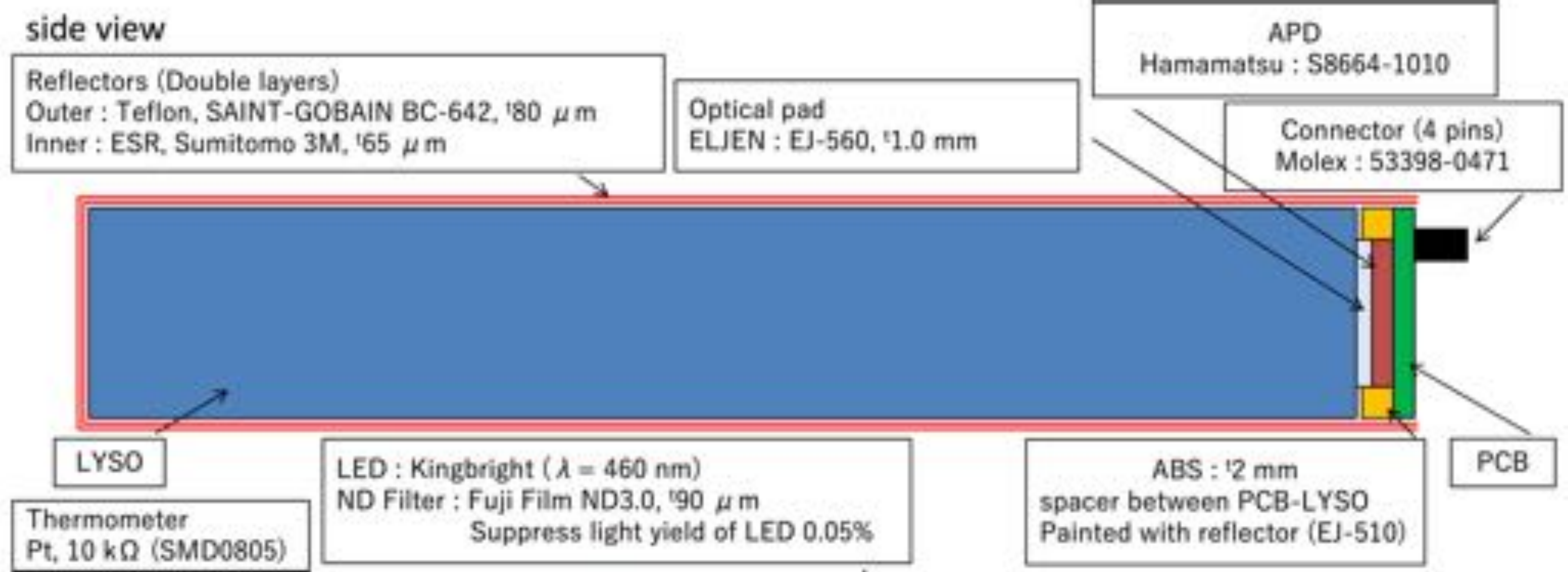}
    \caption{Detail of the one crystal structure.}
    \label{fig:ECAL-Module4}
  \end{center}
\end{figure}

A prototype module is shown in \cref{fig:ECAL-Module1}.
The current design of the module structure for the ECAL is shown in \cref{fig:ECAL-Module3} and the
detail of  one crystal structure is shown in \cref{fig:ECAL-Module4}.
A
polished crystal is first wrapped by 
a reflector film (3M ESR)
together with 
a silicone rubber optical interface
(ELJEN Technology, EJ-560) and 
a
PCB on which the APD (Hamamatsu S8664-55 APD, similar but smaller than the S8664-1010) is attached.
An LED
with a wavelength similar to that of the LYSO scintillation photon (420 nm), is also placed on the PCB and is used to flash light for monitoring purpose.
This one crystal structure is then wrapped by 
a layer of 
Teflon tape from Saint-Gobain.
Four wrapped crystals are then used to construct the 2$\times$2 matrix module, which is wrapped by 
an aluminized Mylar film.
The modules are further arranged to form a super-module (\cref{fig:ECAL-Module3} (c)).

\subsubsection{Prototype studies}
\label{sec:rd-status}
A first  ECAL prototype  was tested in a 65--145~MeV/$c$ electron beam at Tohoku University. It consisted of 7 $\times$ 7 crystals with 7 preamplifier boards and the prototype electronics but with the  Hamamatsu S8664-55 APD with an active area of 5$\times$5 mm$^{2}$ rather than the currently preferred S8664-1010. The resolution was obtained by converting the signal from each of the 49 crystals to an energy deposit  and then the energy deposit for the prototype ECAL  obtained with a simple clustering algorithm. Tests were conducted with both GSO and LYSO crystals.

\begin{figure}[htt!]
\centering
\includegraphics[width=0.8\textwidth]{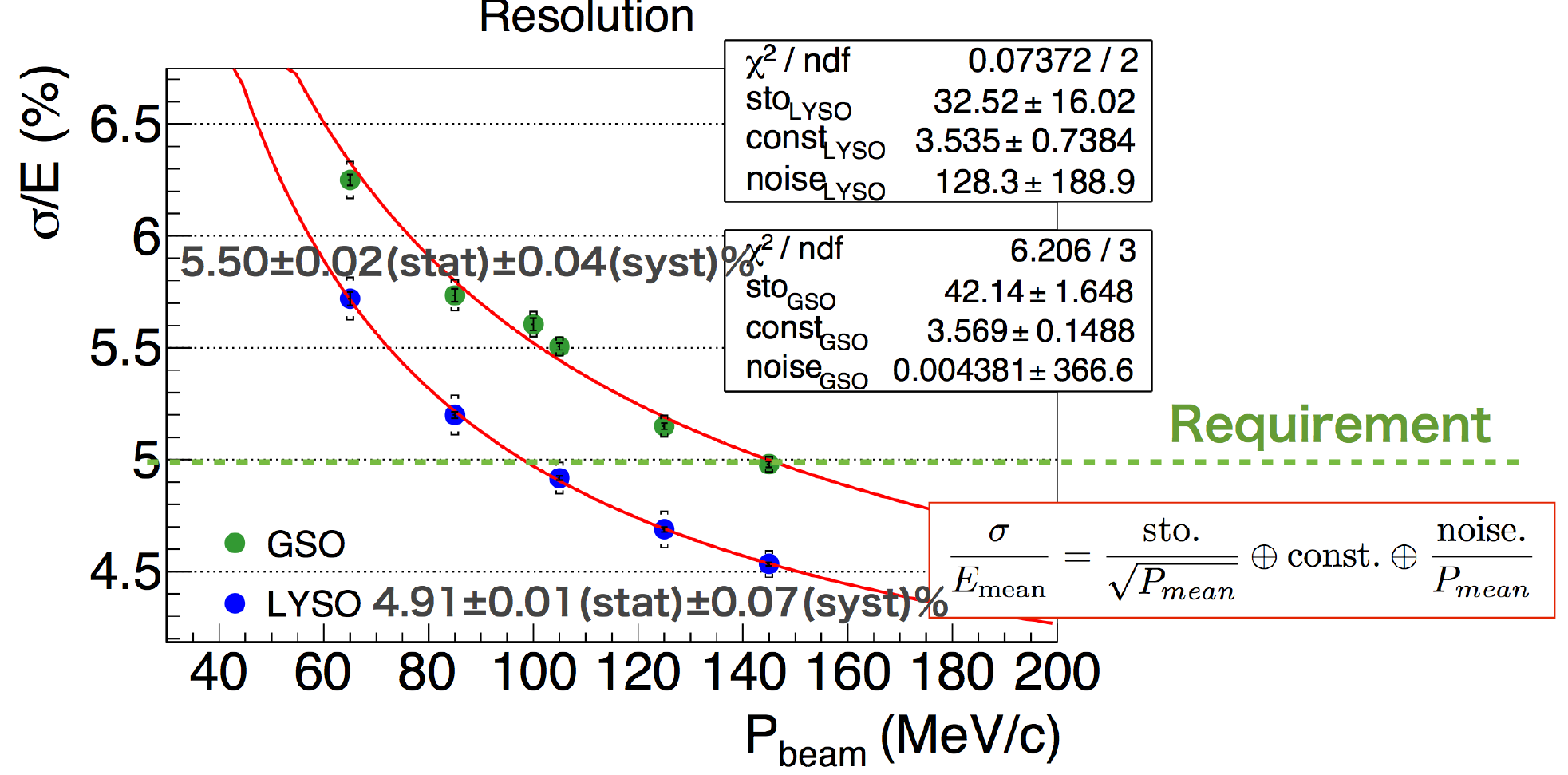}
\caption{The measured energy resolution as a function of beam momentum.}
\label{fig:ECAL-EnergyResolution1}
\end{figure}

\Cref{fig:ECAL-EnergyResolution1} shows the energy resolution as a function of beam 
energy.
The resolution at 105~MeV
was 5.50 $\pm$ 0.02 (stat) $\pm$ 0.04 (syst)\,\% for GSO, and 4.91 $\pm$ 0.01 (stat) $\pm$ 0.07 (syst)\,\% for LYSO. The LYSO crystals are found to
meet the required energy resolution of better than 5~\% at 105~MeV. These preliminary tests also confirmed that both GSO and LYSO could meet the position resolution requirement of less than 
1~cm.

Based on the prototype ECAL results in the test experiment, the cost-performance evaluation on GSO and LYSO has been made and our decision of the crystal choice for the ECAL is LYSO.

Achieving the energy resolution and position resolution requirements by using the LYSO crystal in the first ECAL prototype system, the further performance improvements have been studied towards the determination of the final design.
A second LYSO prototype  has been constructed and tested at the GeV-$\gamma$ Experimental Hall in 
ELPH
of Tohoku University. For this
 the APD used was the Hamamatsu S8664-1010 which has a larger active area of 10$\times$10 mm$^{2}$, and hence the capability to collect more scintillation photons.

\begin{figure}[htt!]
  \begin{center}
    \includegraphics[width=0.8\textwidth]{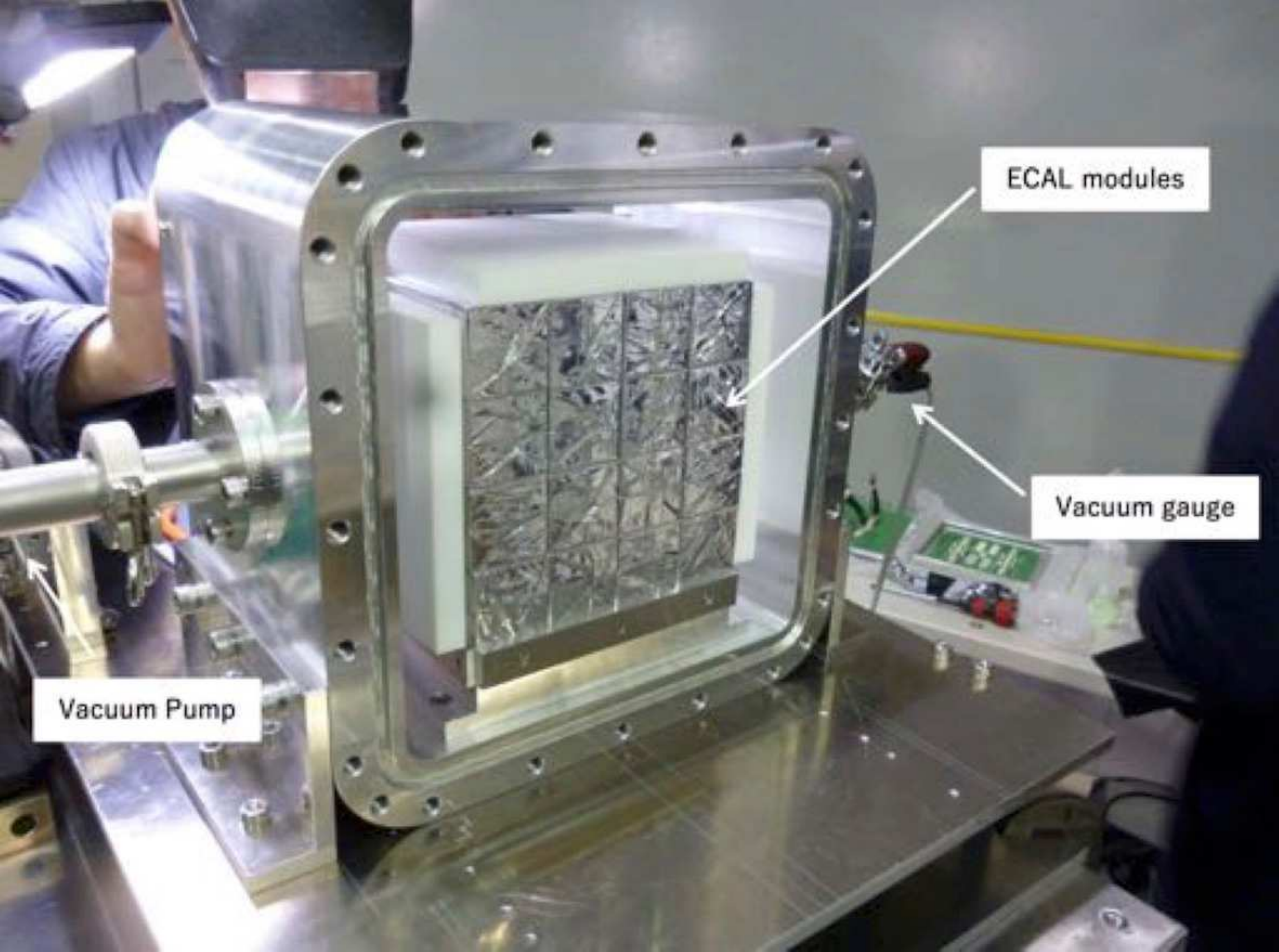}
    \caption{A vacuum chamber for the prototype ECAL system.}
    \label{fig:ECAL-Chamber}
  \end{center}
\end{figure}

A vacuum chamber was constructed to evaluate the prototype performance in a realistic environment.
The prototype modules were installed inside the vacuum chamber together with the intermediate board and the feedthrough board.
\Cref{fig:ECAL-Chamber} shows the vacuum chamber and the modules 
installed
in the chamber.

The front-end preamplifier board was re-designed  to match the form factor to the ECAL prototype and the
 noise performance  optimised for the larger area APD.

\begin{figure}[htt!]
\centering
\includegraphics[width=0.9\textwidth]{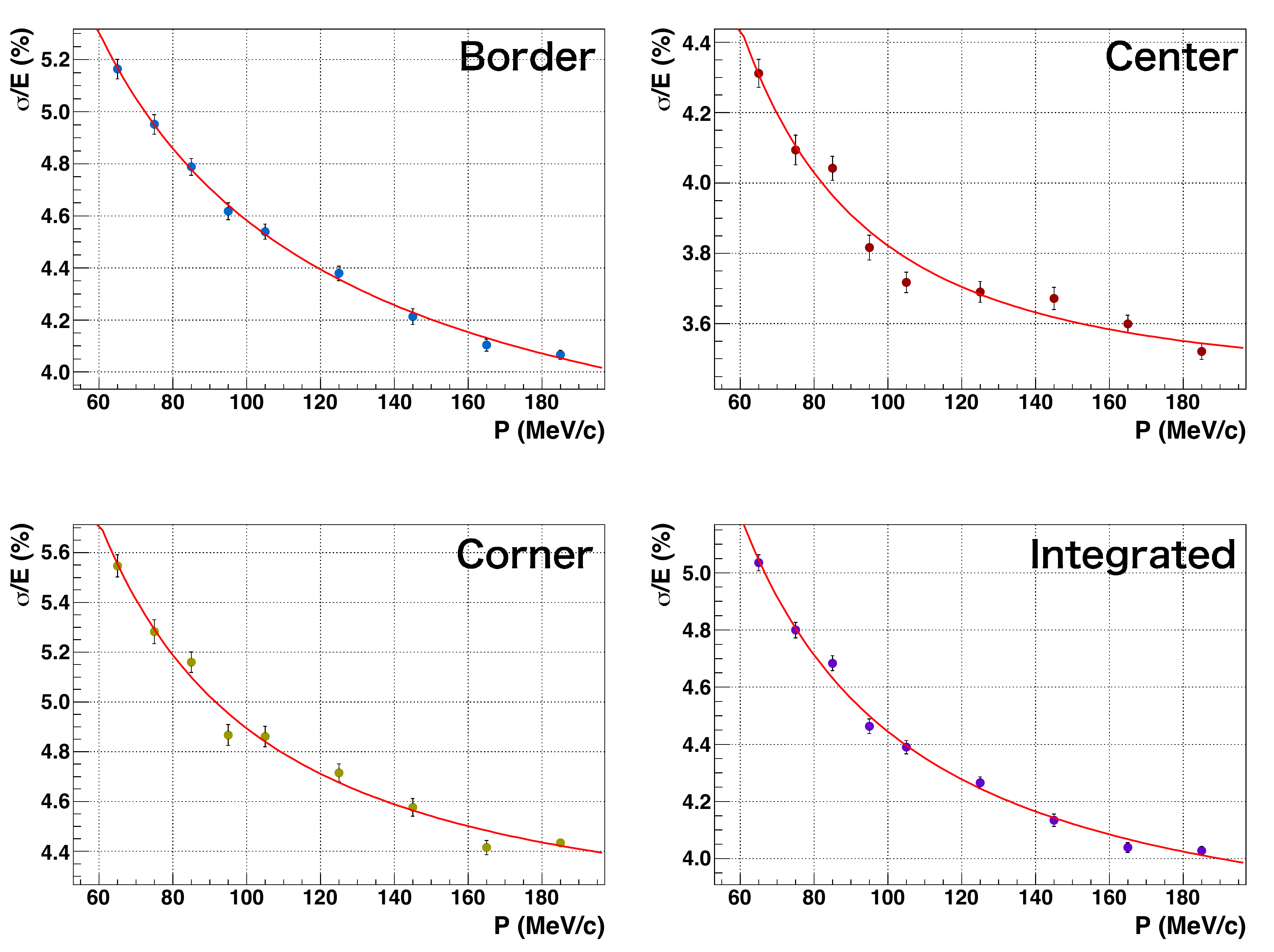}
\caption{A preliminary result of the measured energy resolution of ECAL as a function of beam momentum. The terms ``Border'', ``Corner'' and ``Centre'' relate to the position of impact---the boundary between two and four crystals, and the centre of a single crystal, respectively. Bottom-right: energy resolution with the position dependence integrated.}
\label{fig:ECAL-EnergyResolution3}
\end{figure}

\begin{figure}[htt!]
\centering
\includegraphics[width=0.8\textwidth]{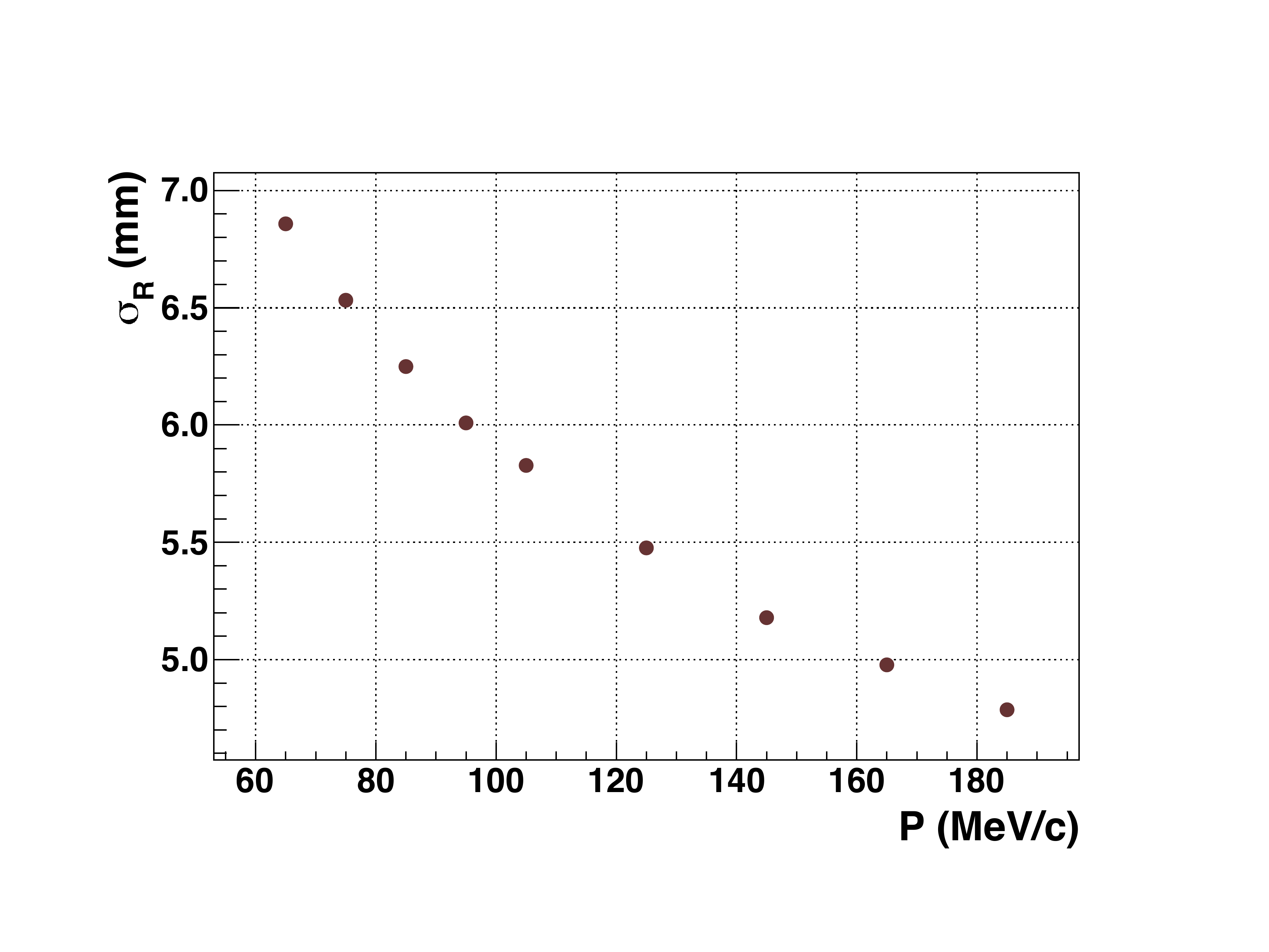}
\caption{A preliminary result of the measured position resolution of ECAL as a function of beam momentum. Position dependence was integrated in this result.}
\label{fig:ECAL-PositionResolution3}
\end{figure}

The energy resolution and position resolution measurements are shown in \cref{fig:ECAL-EnergyResolution3} and \cref{fig:ECAL-PositionResolution3}, respectively.
At 105~MeV/$c$, the resulting overall energy resolution is 4.4~\%, varying from 3.8~\% to 4.8~\% depending on where the electron impinges on the ECAL. The overall position resolution is found to be 5.8~mm.

 % subsection
\section{Cosmic-Ray Veto}

\subsection{Physics Requirements and General Layout}

Cosmic Ray  muons (CRM) can decay in flight or interact with the materials around the area of the muon-stopping target and produce signal-like electrons in the detector region.
In order to have control over this background, a Cosmic Ray Veto (CRV) system  is required for COMET (see \cref{sec:cosmicraybackground} for cosmic-ray induced background).
The CRV has to identify cosmic ray muons  with an average inefficiency that is lower than $10^{-4}$.

For COMET Phase-I, two types of cosmic-ray shielding will be used: passive and active. The passive shielding consists of concrete, polyethylene and lead, as well as the iron yoke of the DS. The flux of low-angle cosmic particles is also attenuated  by the surrounding sand as the detector is located underground.

The active shielding is provided by a CRM detection system covering the CyDet area.
Detailed studies of CR-induced backgrounds in \cref{sec:cosmicraybackground} indicate that the Bridge Solenoid (BS) area must also be covered by a CRV, because  interactions of CRM in the BS could produce electrons that scatter off the BS and enter the CDC, hit the cylindrical trigger hodoscope (CTH) and mimic signal events.
A suppression factor of $10^4$ is needed for this CRM background and it is obtained by using---in the offline analysis---the  signature left in the CRV by the CRM.
The active veto system covering the CyDet is made of scintillator-based detectors, whereas Glass Resistive Plate Chambers (GRPC) are envisaged in the BS area.
\\

\subsection{Scintillator-Based Cosmic-Ray Veto Design }
The CyDet CRV has four layers of active material. Its basic element  is a strip made of a polystyrene-based organic scintillator . This detector is named the Scintillator-based Cosmic Ray Veto (SCRV).

\begin{figure}[thb!]
 \begin{center}
  \includegraphics[width=0.8\textwidth]{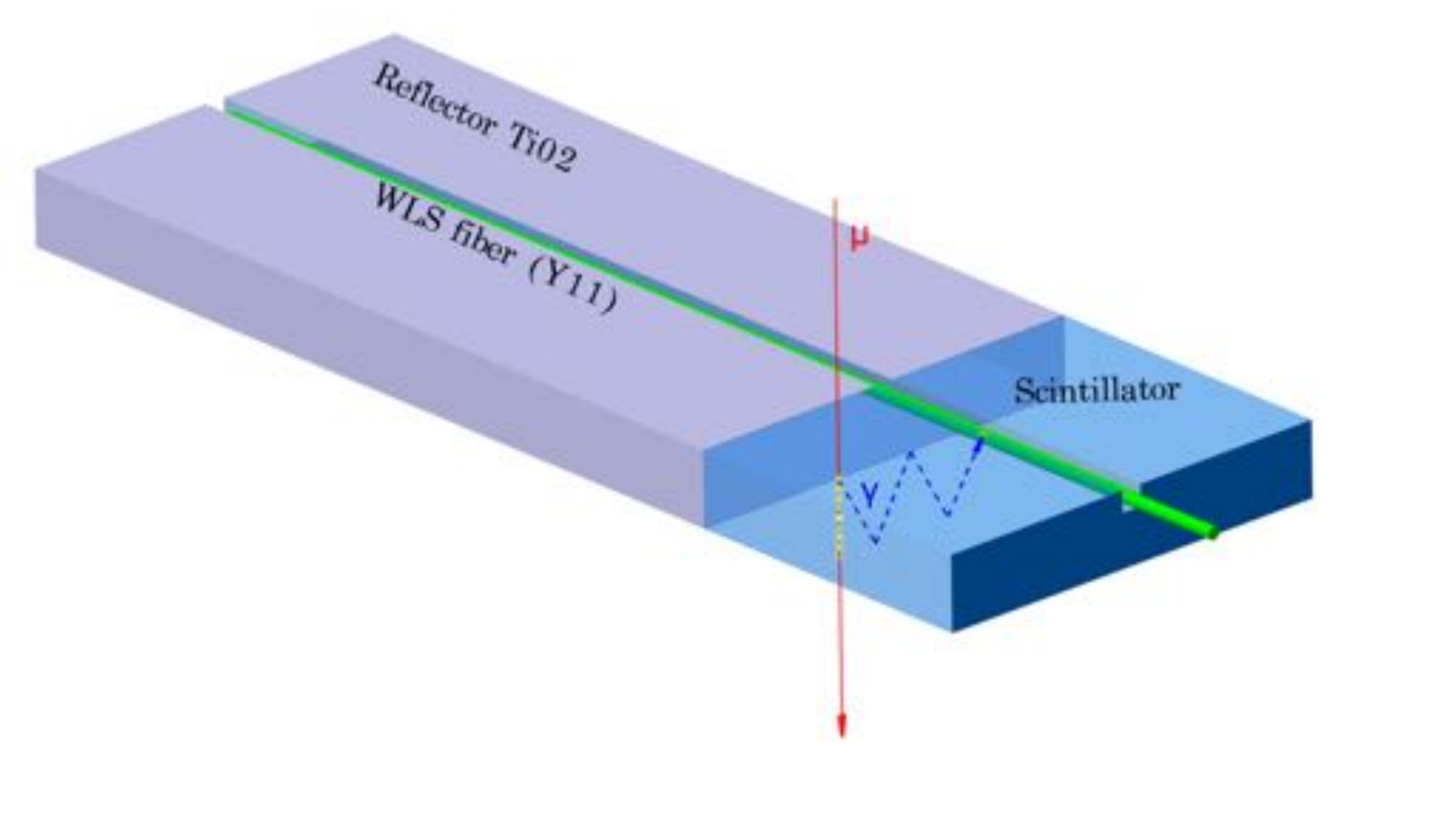}
 \end{center}
 \caption{A sketch of the design for a single channel and the principles of particle detection.}
 \label{fig:CRV_Scintillator}
\end{figure}

\paragraph{Scintillator and light transport}
The principle for particle detection and the general design of a single SCRV channel are shown in \cref{fig:CRV_Scintillator}.
The single scintillator strip has a cross-sectional area of $ 0.7\times 4 ~\mbox{cm}^{2}$ and a length up to 360~cm.
It is made of polystyrene (Styron 143E) acting as ionization and photon carrier medium with 2\% scintillating fluors (p-terphenyl) and 0.05\% POPOP.

\begin{figure}[tbh!]
 \begin{center}
  \includegraphics[width=0.481\textwidth]{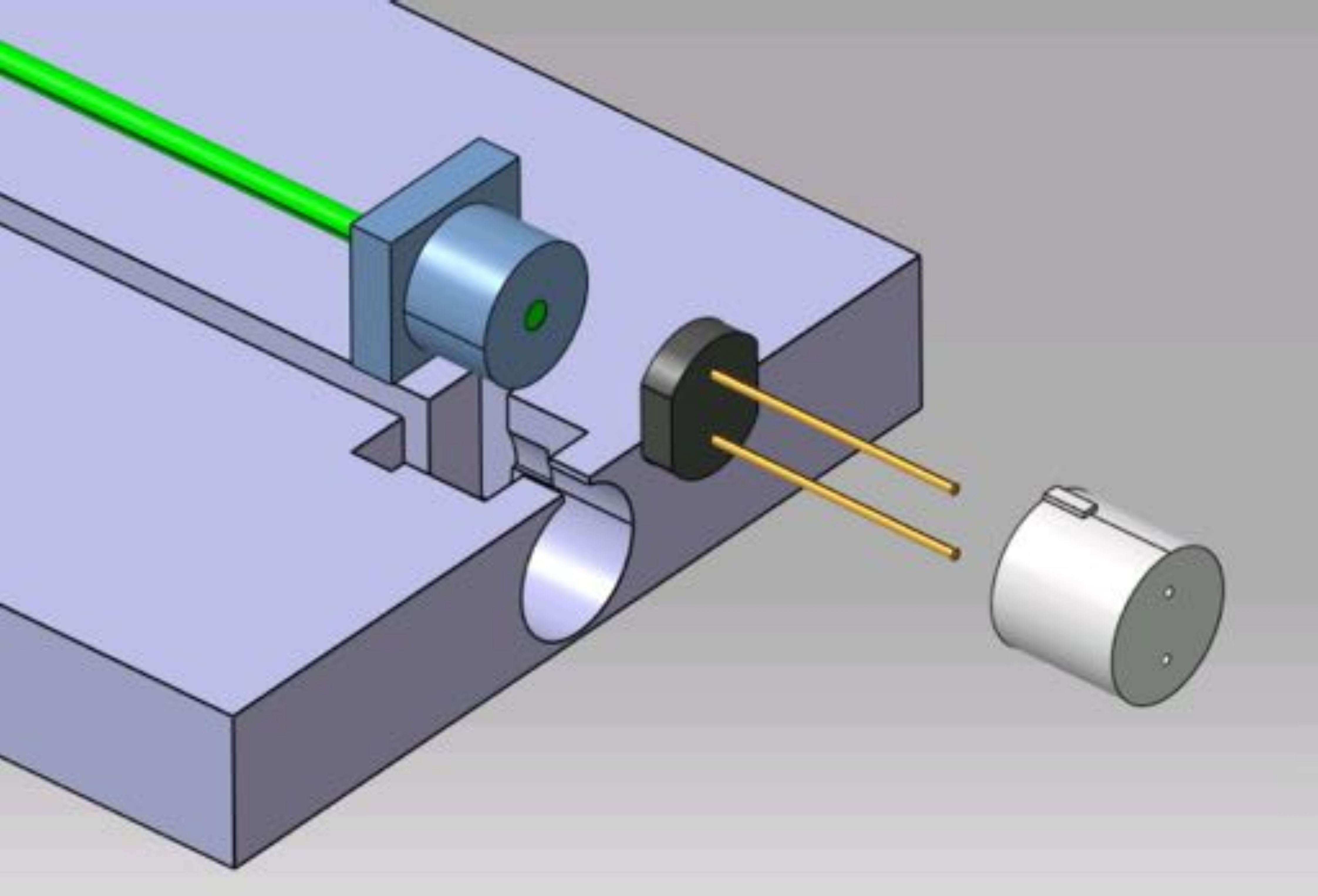}
  \includegraphics[width=0.5\textwidth]{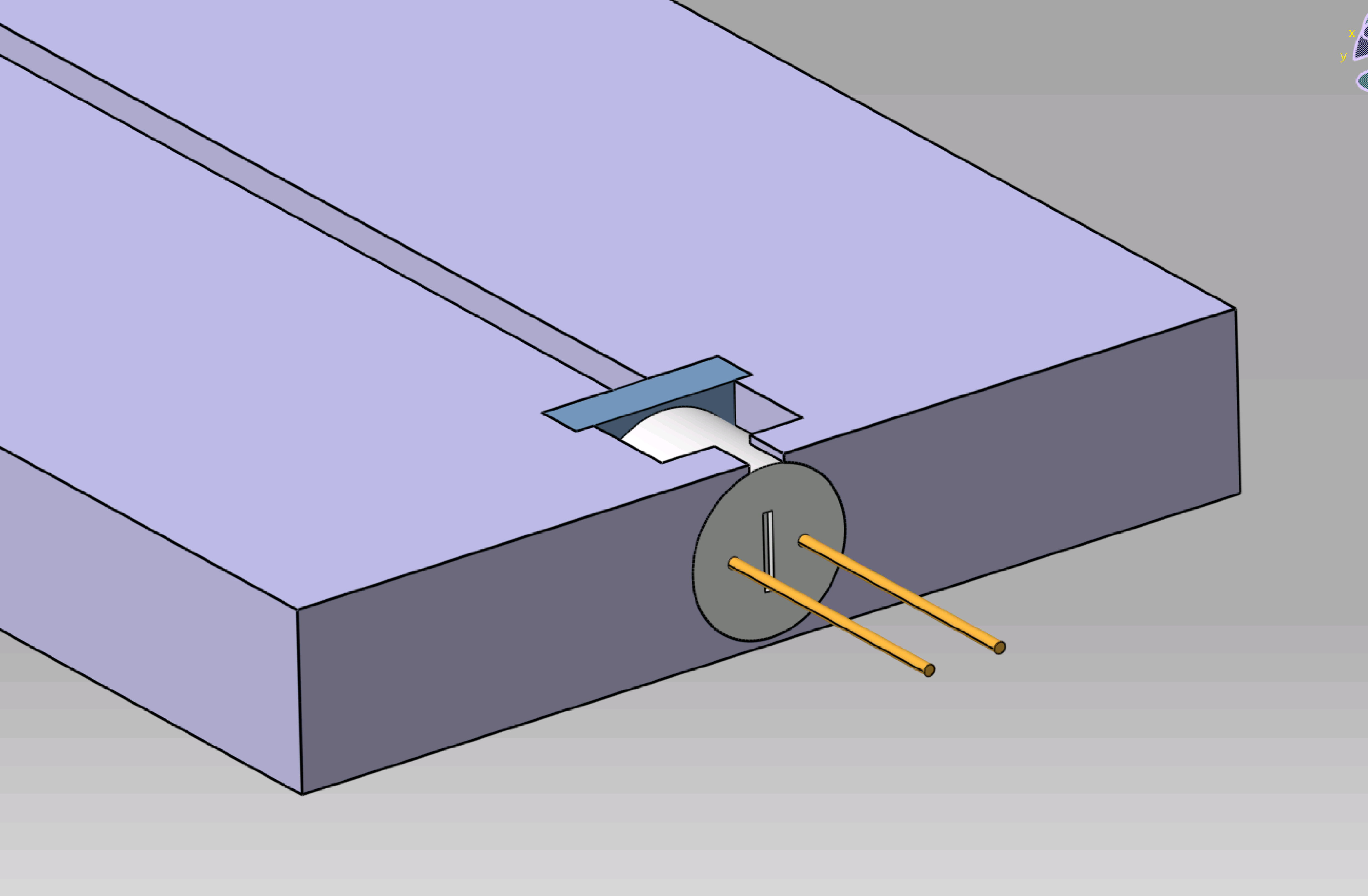}
 \end{center}
 \caption{One of the baseline designs for the coupling mechanism of SiPM to WLS fibre.}
 \label{fig:CRV_OpticalConnector}
\end{figure}

SCRV strips are read out by wavelength-shifting (WLS) fibres which transport light to the photodetectors. The use of WLS fibres is necessary in order to compensate for the short attenuation length of the scintillators and to optically connect the scintillators to the photo detectors. The WLS fibre is placed along the strip length in a surface groove  (see \cref{fig:CRV_OpticalConnector}) of a rectangular shape. Several different groove dimensions have been studied and the optimal one was determined to be $1.5 \times 3.5 ~{mm}^{2}$.
A good optical coupling between the scintillator strip and the WLS fibre is ensured by the use 
of
a highly transparent optical glue, BC600 (Bicron optical cement).

Several different fibre types from Bicron and Kuraray have been investigated and the preliminary choice is the 1.2~mm diameter, multi-cladding fibre from Kuraray (Y11), which delivers a high photon yield to the photo-detector.
The fibres shift the blue scintillation light to wavelengths between 470 and 570~nm.
This not only improves the attenuation length significantly, but also brings the light signal into the green range where the quantum efficiency of modern photo-detectors is much larger with respect to blue light.

The WLS fibres are read out by Silicon Photo-multiplier (SiPM) detectors at both ends. 
The double-ended read-out design allows one to determine the muon impact point 
along the strip with an accuracy of 
a few cm, 
by measuring the time difference between the SiPM signals,
or by measuring the difference of light yield of both ends.
Consequently, the required spatial accuracy of a few cm is achieved without introducing longitudinal segmentation.
As CDC will be able to provide much better tracking of cosmic muon, this spatial resolution of CRV  is enough. 
Special plastic connectors  housed in carefully designed receptacles at the ends  of the strips (See \cref{fig:CRV_OpticalConnector}) couple the SiPM to the WLS fibre.

Results from R\&D measurements which were made for material selection are summarised in \cref{table:LYSummary}.
In the baseline design, BC-600 optical cement will be used to glue Kuraray Y11 fibres in the grooves, 
with one fibre for each strip which will have a TiO$_2$ reflective coating.
During the design selection, 
two strip designs have been considered: the current one described above and a wider strip which is read out by several parallel WLS fibres. The narrow strip design with a single fibre was chosen since it has the following advantages:
\begin{itemize}
\item  Light from a MIP is not shared between different SiPMs resulting in a  very high
efficiency even with a high signal threshold.
\item  The efficiency of each strip can be measured using coincident signals recorded in other strips.
\item In case of problems with one channel only a small part of the detector is affected.
\item A time resolution of about 1~ns can be achieved.
\end{itemize}

\begin{table}[thb!]
  \caption{Comparison of the light yields measured for the different samples considered in the R\&D studies.}
  \label{table:LYSummary}
  \centering
  \begin{tabular}{ l  l }
  \hline
  \hline
    Sample design & light yield (au) \\
    \hline
    1.5$\times$1.6 mm$^2$ groove    & 194.5 \\
    1.5$\times$3.6 mm$^2$ groove    & 220.1 (+13\%) \\
    \hline
    1 WLS   & 194.5 \\
    2 WLS   & 243.3 (+25\%) \\
    3 WLS   & 267.3 (+10\%) \\
    \hline 
    BC408   & 194.5 \\
    UniPlast    & 124.6 (-46\%) \\
    \hline
    TiO2    & 111.6\% \\
    AluMilar & 124.4 (+11\%) \\
    \hline
    \hline

  \end{tabular}
\end{table}

\begin{figure}[tbh!]
 \begin{center}
  \includegraphics[width=0.8\textwidth]{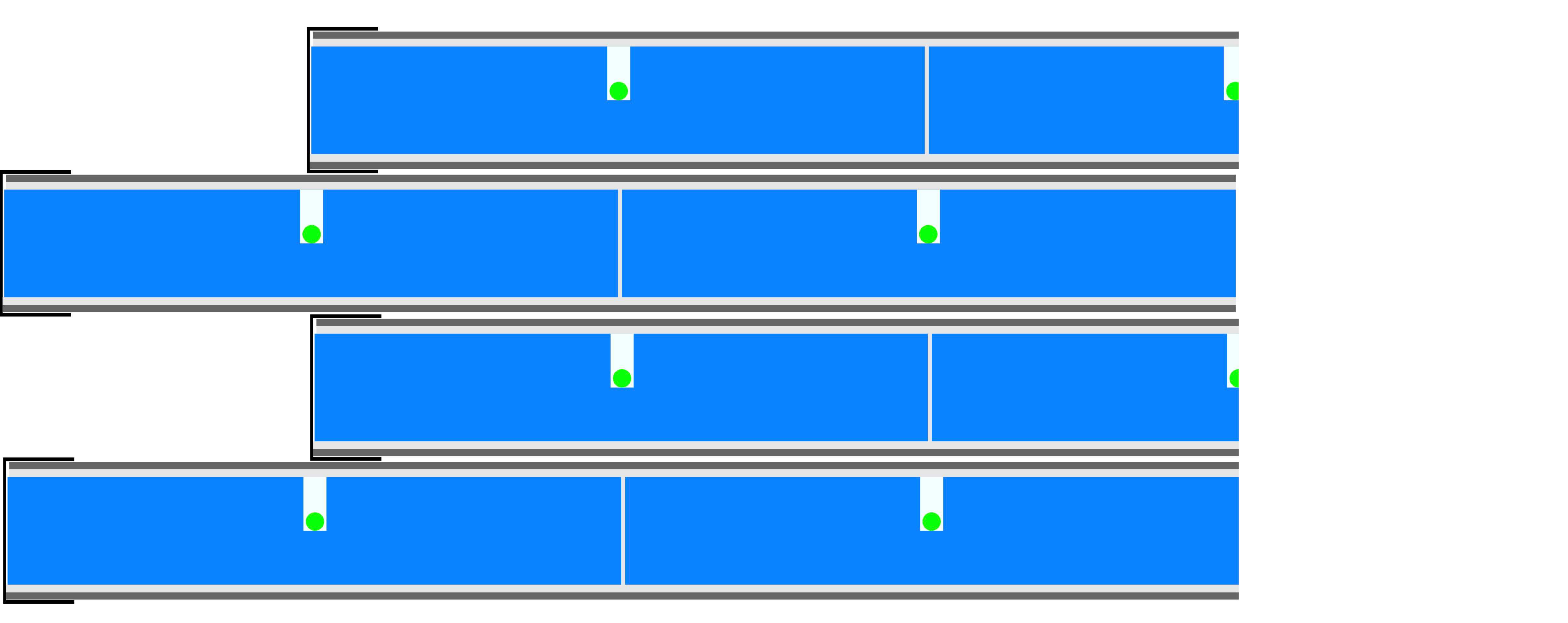}
 \end{center}
 \caption{CRV strip layout.}
 \label{fig:CRVzones}
\end{figure}

\begin{figure}[th!]
 \begin{center}
  \includegraphics[width=0.8\textwidth]{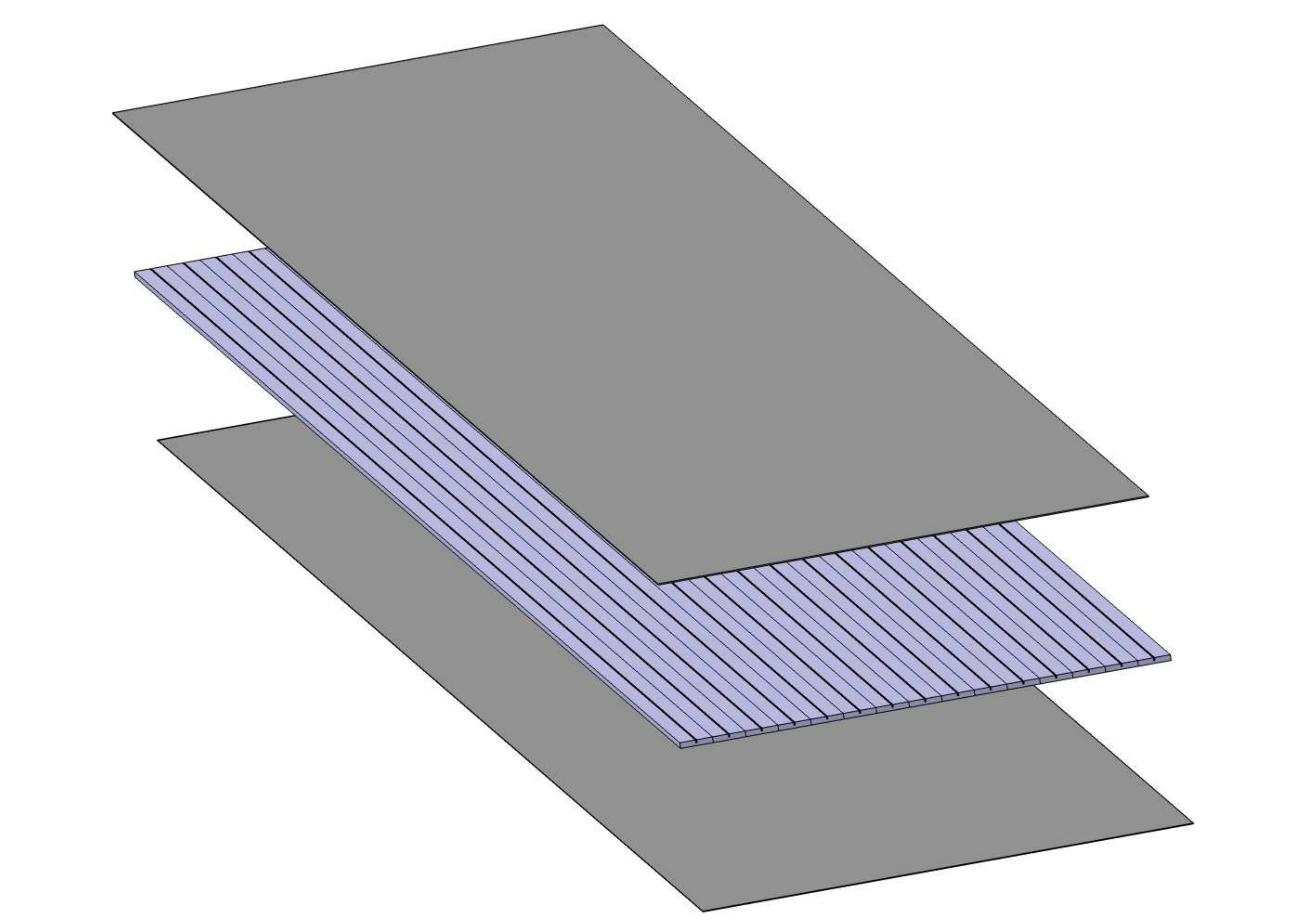}
    \includegraphics[width=0.8\textwidth]{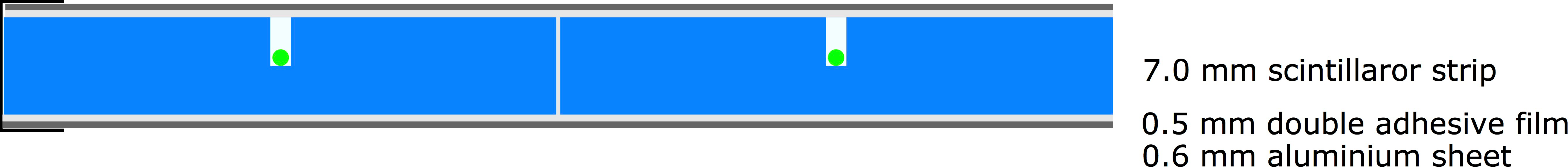}
 \end{center}
 \caption{Design of the SCRV module (top). SCRV module cross section (bottom)}
 \label{fig:CRV_ModuleDesign}
\end{figure}

\paragraph{SCRV modules and layers}
Fifteen strips form an SCRV module of dimension  $0.7\times 60\times 300 (360)$~cm$^{3}$. The relatively low weight  of the SCRV module of about 10~kg give it good handling properties. Strips are accurately placed on a 0.6~mm thick aluminium sheet, which is covered with double-sided adhesive tape on the strip side. After placing the strips next to each other, they are tightly glued onto the aluminium sheet. The  mechanical encapsulation of the module is obtained by using another sheet  as a cover (See \cref{fig:CRV_ModuleDesign}). The mechanical strength of the module is given  both by the strips being glued together  and by the aluminium sheets enveloping it.

The short sides which run along the module are physically protected by a thin, U-shaped, stainless-steel layer which is glued to the aluminium sheets on both sides. The steel mechanical envelope also shields the strips from external light sources.

\begin{figure}
\centering
    \includegraphics[width=0.7\textwidth]{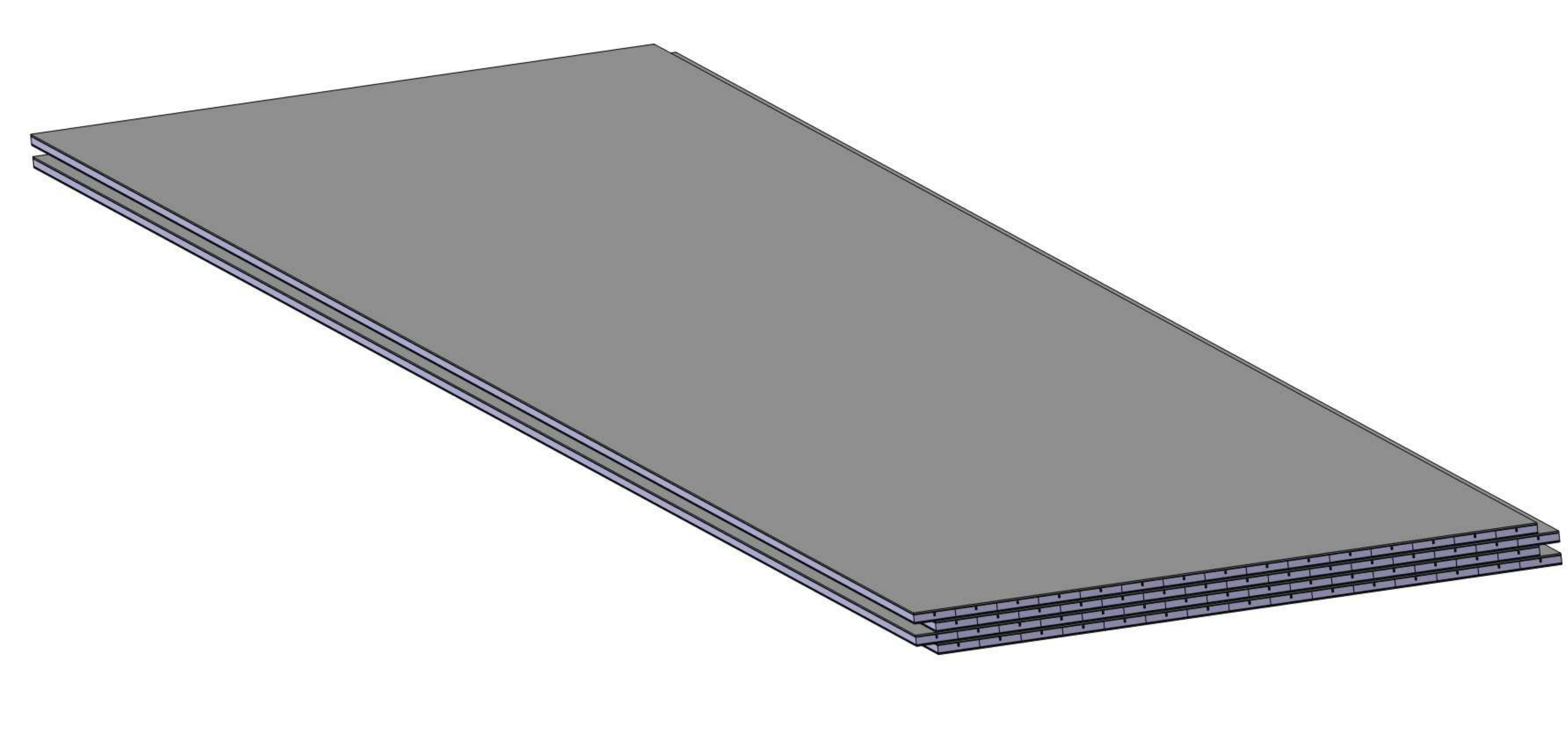}
    \includegraphics[width=0.7\textwidth]{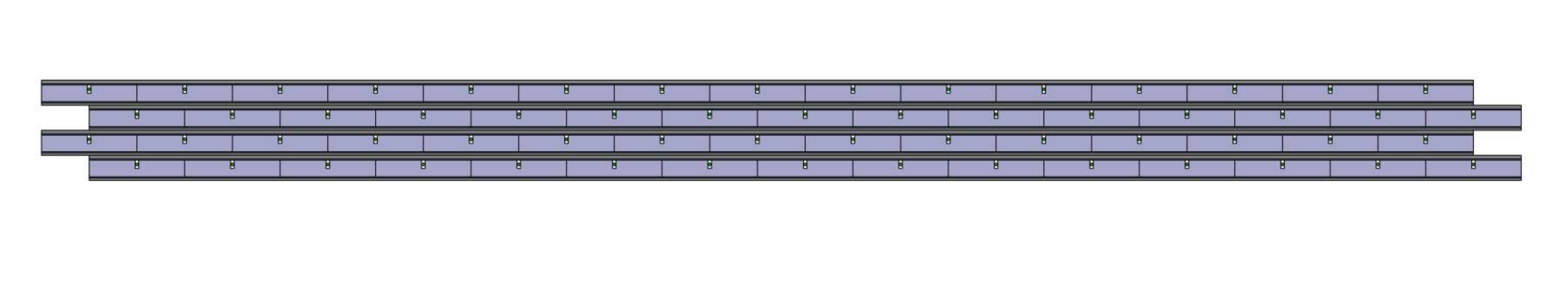}
  \caption{SCRV Layer (top) and its cross section (bottom).}
  \label{fig:CRV_Layer}
\end{figure}

Modules are placed side-by-side in order to form a SCRV layer. The cosmic ray rejection power of the SCRV is ensured by deploying four successive detection layers. The modules are shifted by 
2~cm 
from layer to layer in order to avoid the vertical alignment of gaps between strips, as well as  between modules  (see \cref{fig:CRV_Layer}).
Four thousand six hundred  strips are needed  to cover the required space.

\paragraph{Neutron shielding}

The neutron flux in the experimental hall can cause problems for the cosmic veto system by inducing noise  and causing radiation  damage to the SiPMs. The neutrons originate from the pion production
target (above 1~MeV) and the beam dump (below 1~MeV).

The dark current of the SiPM increases for
 irradiation above $10^8$ neutrons/$\textrm{cm}^2$
\cite{Angelone2010921}. However even after $7 \times 10^{11}$
neutrons/$\textrm{cm}^2$ irradiation~\cite{Angelone2010921},
the decrease in the SiPM gain does not  exceed 50\%  and the overall detector efficiency can be retained by adjusting the threshold.  It has been shown that operating with a threshold level above seven pixels  maintains the fraction of dead time at the few-percent level. At the same time,  the muon detection efficiency  is still 99.99\%  when operating with  an 11-pixel threshold.  The noise and the neutron detection efficiency are both lower when operating at high thresholds  resulting in a smaller dead time.

\begin{figure}
\centering
\includegraphics[width=0.8\textwidth]{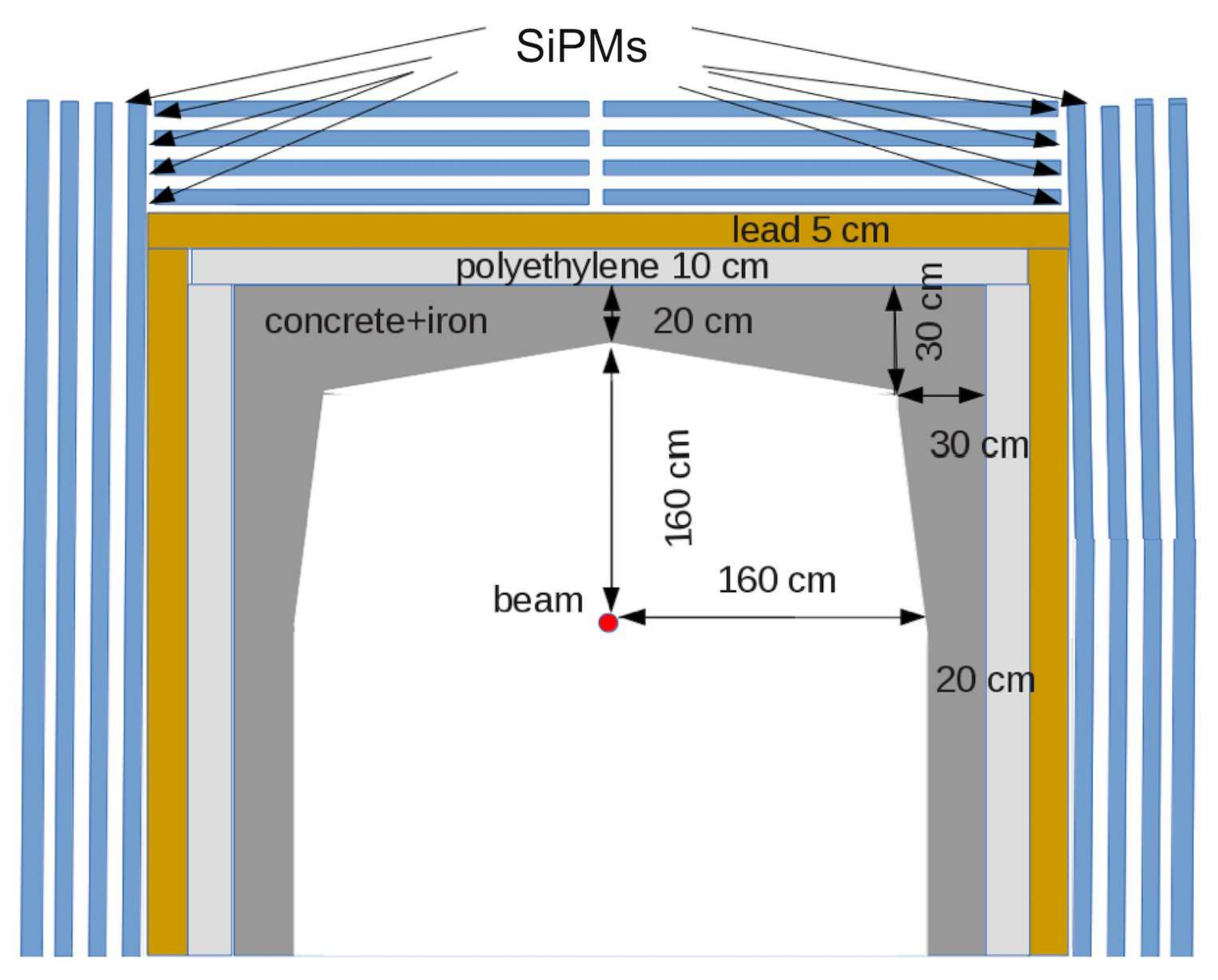}
\caption{CR veto inner shield with an arched shape (not to scale) showing the read out (in this case SiPMs) locations.}
\label{fig:inner_shield}
\end{figure}

To reduce the neutron flux in the scintillator and  the damage they induce on the SiPMs,  an inner shield will be employed using layers of iron,
polyethylene and lead.  The baseline configuration of the shield is an arched shape as shown in \cref{fig:inner_shield}, and
consists of 20 to 30~cm of an iron-concrete mix, 10~cm of
polyethylene and 5~cm of lead and will reduce the flux of fast
and more energetic neutrons at the photo-detectors by 2 orders of
magnitude.

The fraction of fast neutrons ($E\geq$ 1~\mbox{MeV}) in the spectrum is larger for the back and front sides, at 
44.8\% and 24.6\% respectively. Optimisation of the shielding is ongoing  to reduce the fluxes and the fraction of  fast neutrons
to avoid worsening photo-detector efficiencies over the full period of data taking.

\subsection{BS-Area CRV}

The region around  the BS  that requires active shielding has  a surface of $3\times 1900\times 600$~mm$^2$.
Simulations indicate that this area suffers from a larger neutron contamination compared with that affecting the CyDet-CRV.

\begin{figure}
\centering
\includegraphics[width=0.8\textwidth]{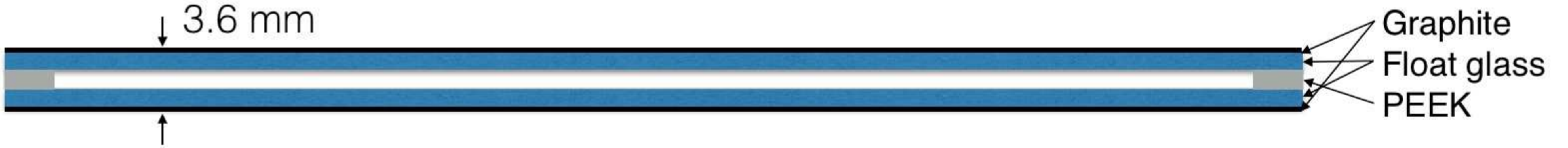}
\caption{Single-gap GRPC made of two layers of float glass, glued together using 1.2~mm 
PEEK
spacers. On the outer part of the glass sheets, a resistive painting (graphite-based) is used to apply an electric field of about 6.5~kV/mm across the gap. A mixture of  a gas with high fluor content (forane) and a quencher ($SF_6$) at a ratio of around 98:2 is flushed permanently through the chamber at a low rate  of about 15~cc/min.}
\label{fig:crv:grpc}
\end{figure}

GRPCs are a natural candidate  for operating in such high neutron flux areas; they can be built to the required size and provide an uniform tracker, without dead areas between adjacent active volumes. Moreover, their segmentation can be easily modulated to fit the required tracking performance on muons.  The baseline design is based on single gap (1.2~mm thick) chambers (\cref{fig:crv:grpc}), operated in avalanche mode. These are thin detectors of less than 3.6~mm, with nanosecond time resolution, operated at average efficiencies of 95\%  and with an intrinsic position resolution of a few mm. The design envisaged for COMET is based on R\&D performed for the detectors for the International Linear Collider~\cite{crv:SDHCAL} and used since 2012 for muography studies~\cite{crv:TOMUVOL}.

\begin{figure}
\centering
\includegraphics[width=0.8\textwidth]{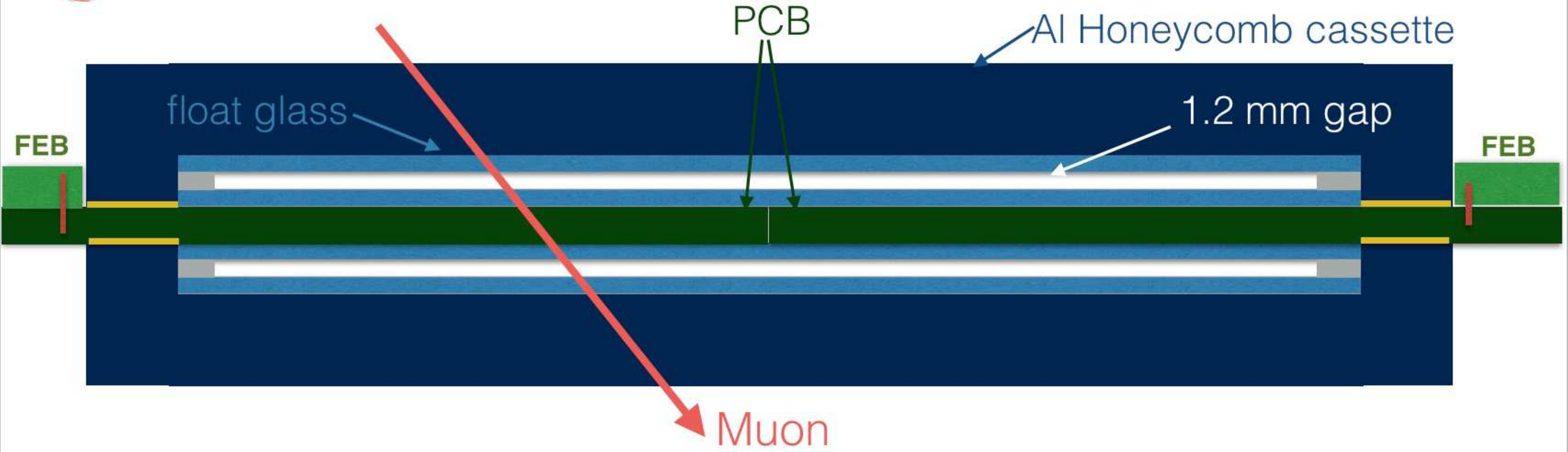}
\caption{A GRPC module made of two single-gap GRPCs housed in a unique aluminium honeycomb cassette and sharing a $XY$ readout layer. To make possible its outsourcing, the readout layer is made of two individual 60$\times$95~cm$^2$ PCBs; these are double layered and they have orthogonal readout strips on the outer faces. Each GRPC-resistive layer closest to the readout layer is grounded and the second one used to apply the polarisation high voltage. The strips signals are read on one side using mezzanine boards labelled FEB in the figure. The figure is not to scale horizontally.}
\label{fig:crv:grpc-module}
\end{figure}

The  BS-CRV is based on three trackers to be deployed on the top and the sides of the BS respectively.  Each tracker is made of  six GRPC-modules as represented in \cref{fig:crv:grpc-module}.  Two single-gap GRPCs housed in an aluminium honeycomb structure share  a centrally-placed  readout layer. The readout layer is made of two adjacent PCBs, double-layered,  with $X$ strips on the upper layer and $Y$ strips on the lower layer. This leads to 60~cm-long strips in one direction and $\sim$95~cm in the other, with single-end readouts.  This design of two single-gap GRPCs is commonly used in high energy physics experiments and increases the efficiency of each module from 95\% to 98\%.

\begin{figure}
\centering
\includegraphics[width=0.8\textwidth]{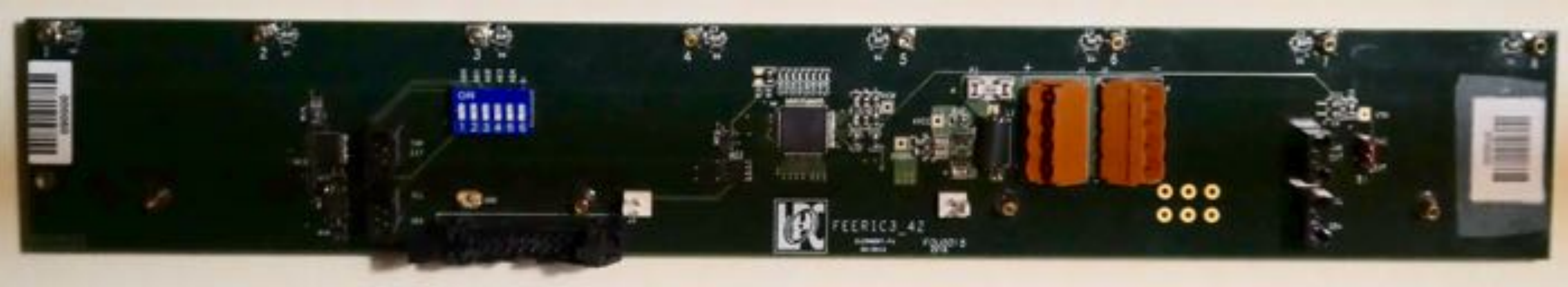}
\caption{A Front End board, developed for ALICE, housing the FEERIC ASIC in the centre of the board. At the top, the eight pins connecting the readout strips (designed for 4.5~cm readout-strips in this case) are visible. The four pins on the bottom of the picture are for grounding. The board shown is 32.5~cm long and 5~cm wide, but the final dimensions will be determined by the optimal BS-CRV segmentation.}
\label{fig:crv:grpc-feb}
\end{figure}

The readout chip in the baseline design is the FEERIC, developed for the ALICE experiment at the LHC~\cite{CERN-LHCC-2015-001}. It is an eight-channel, double-polarity chip with a LVDS output that can handle charges varying from 20~fC to a few pC. The chip is housed on a Front-End board (FEB) developed for ALICE and mounted as a mezzanine on the readout PCBs (\cref{fig:crv:grpc-module}). One such FEB, designed for 4.5~cm readout strips, is shown in \cref{fig:crv:grpc-feb}. It allows the remote setting of the data-acquisition thresholds for the ASIC through an I2C connection, and testing and calibration of the electronics using a charge injection mechanism. The output signals, in LVDS format, are 
fed
into a local DAQ board that communicates with the COMET clock and trigger system. It also implements a local trigger, based on coincident signals recorded in several GRPC modules.

Depending on the optimal segmentation for the BS-CRV, which is still under study, the total number of FEERIC chips per GRPC-module will be between 78 (4.97~mm readout strips) and 8 ( 48.44~mm readout strips).

As in the case of the scintillator CRV,  an inner shield will be deployed to reduce the beam-induced radiation on GRPCs.

\section{Trigger and DAQ }
\label{chapter:daq-trigger}

Phase-I will have two distinct running modes. One with the StrECAL as main detector to measure backgrounds and characterise the beam and the other with the CyDet as main detector to search for \muec. There will be distinct but similar DAQ and trigger systems for the two modes.
Detectors
such as a beam monitor and an X-ray monitor (to determine the muon
beam profile and number of muons captured in the target, respectively)
will be employed for both modes. Similarly, the CRV will provide a veto whilst running with beam (which can be applied offline), but can also provide a calibration trigger.

Each system consists of six main parts:
\begin{itemize} 
  \item The fast control system, which distributes a common 40~MHz clock and all the
    time-critical signals, such as triggers, to the detectors in each system.
  \item The trigger system, which determines when to read out the detectors.
    This is distributed with a fixed latency.
  \item The readout system, which transfers the event data from the detectors
    to disk when a trigger occurs.
  \item The configuration system, which transfers data to the detectors to
    set parameters to control their performance.
  \item The online software, to operate the system.
  \item The slow control and monitor system.
\end{itemize}

\subsection{Trigger Systems}
\label{subsec:trigger}

\begin{figure}[htb!]
\begin{center}
\includegraphics[width=0.44\textwidth]{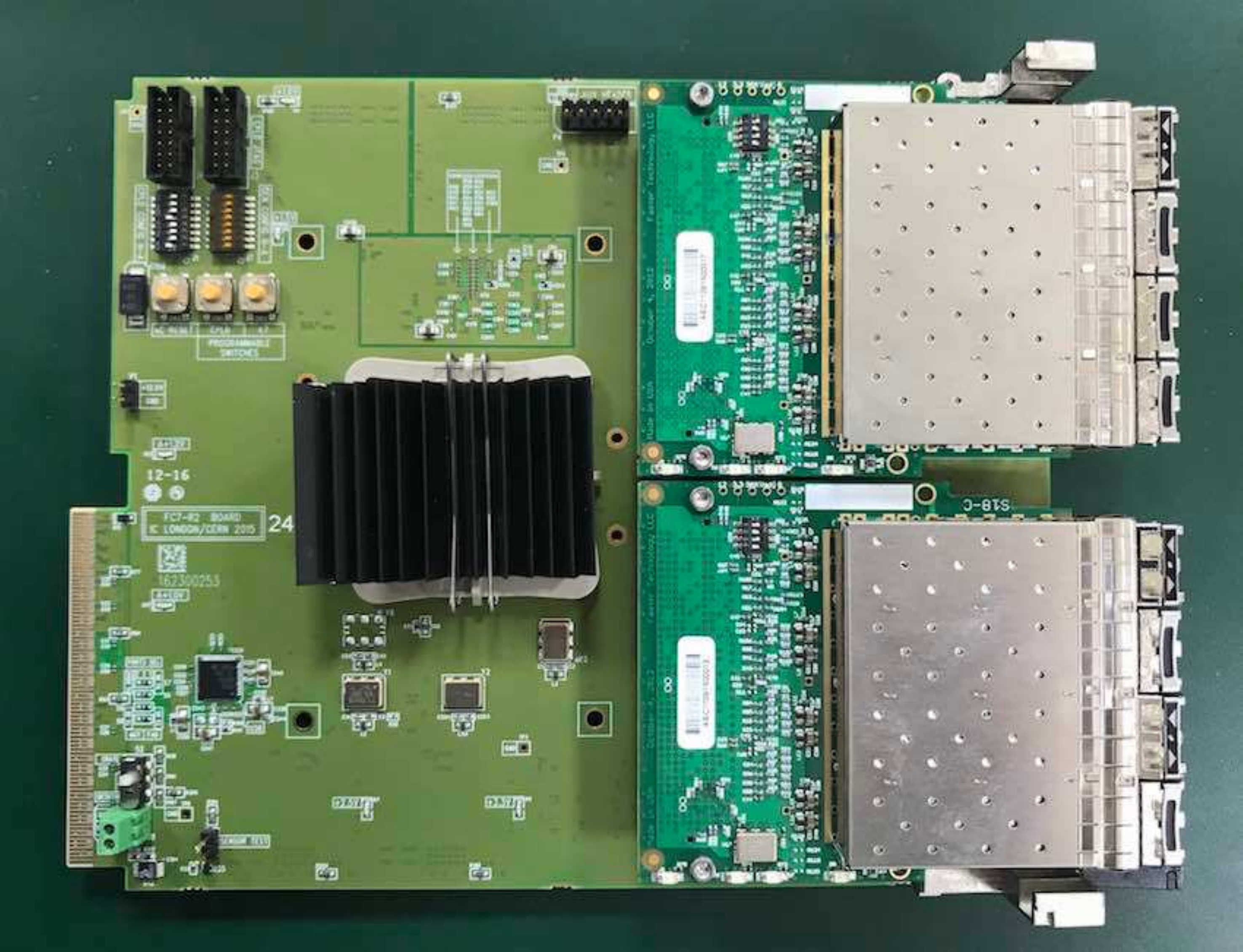}
\includegraphics[width=0.45\textwidth]{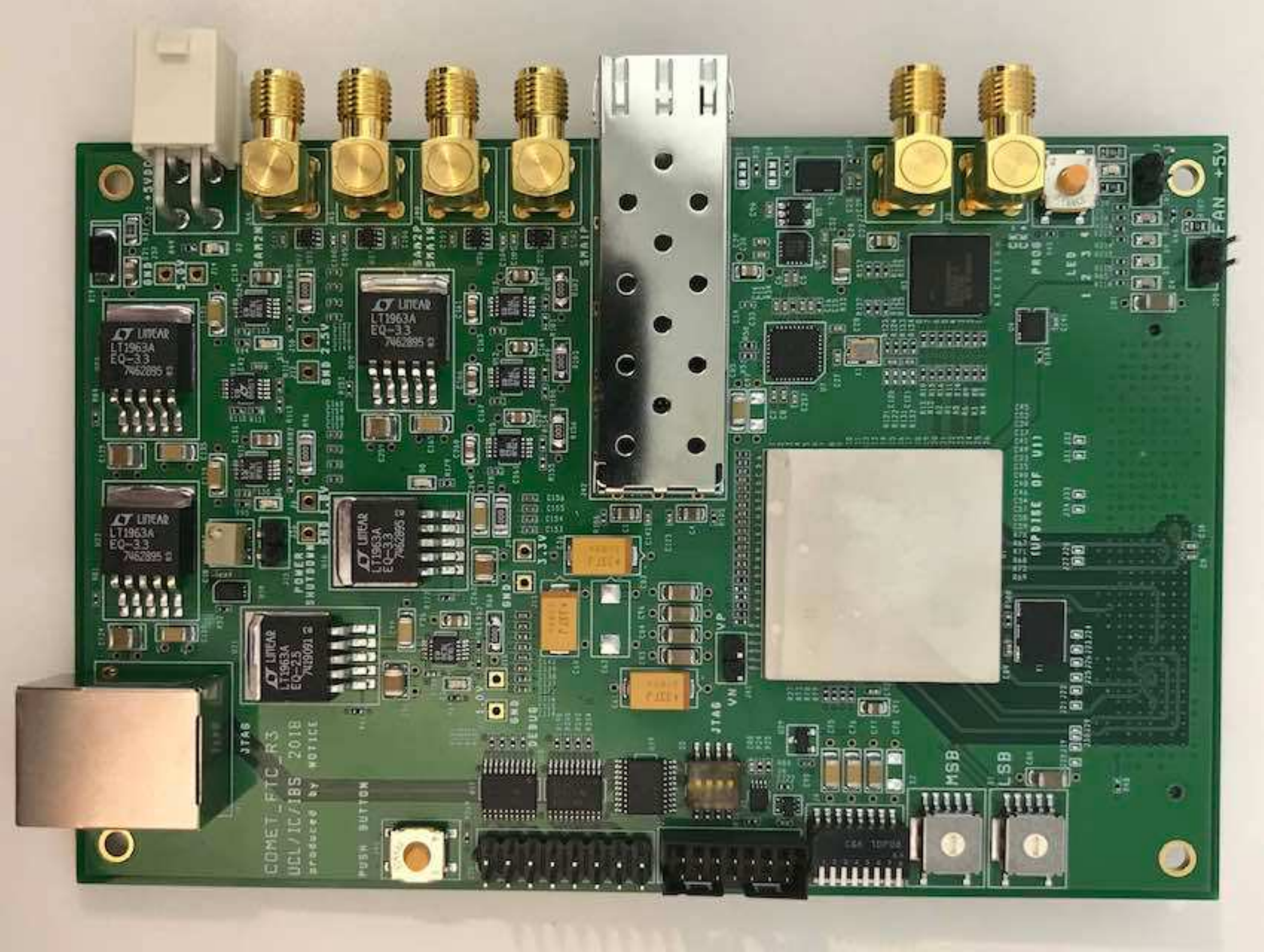}
\caption{
(Left) FC7 board developed for CMS experiment and adopted as the central trigger processor board 
in COMET trigger system. A FPGA for trigger processing and board control locates at the center,
and two commercial mezzanine cards for connecting subdetector trigger system are mounted on the right side of the board.
(Right) Custom-designed FCT board for interfacing with FC7 board.
}
\label{fig:fastcontrol:triggerboards}
\end{center}
\end{figure}

The fast control and trigger systems are composed of: 
the FC7 board 
(left figure of \cref{fig:fastcontrol:triggerboards}) 
developed for the CMS experiment 
at the LHC~\cite{Pesaresi_2015} as a central trigger processor; 
the FCT (Fast Control and Trigger) custom-designed board 
(right figure of \cref{fig:fastcontrol:triggerboards}) 
to interface between the central systems and the subdetector 
trigger or readout components; 
and a custom Multi-Gigabit Transceiver (MGT) protocol connecting these systems. 
While the subdetector trigger system for CyDet and StrECAL are different, 
the core parts shares the same system mastered by FC7 board.

\subsubsection {The CyDet Trigger}

\begin{figure}[htb!]
\begin{center}
\includegraphics[width=0.80\textwidth]{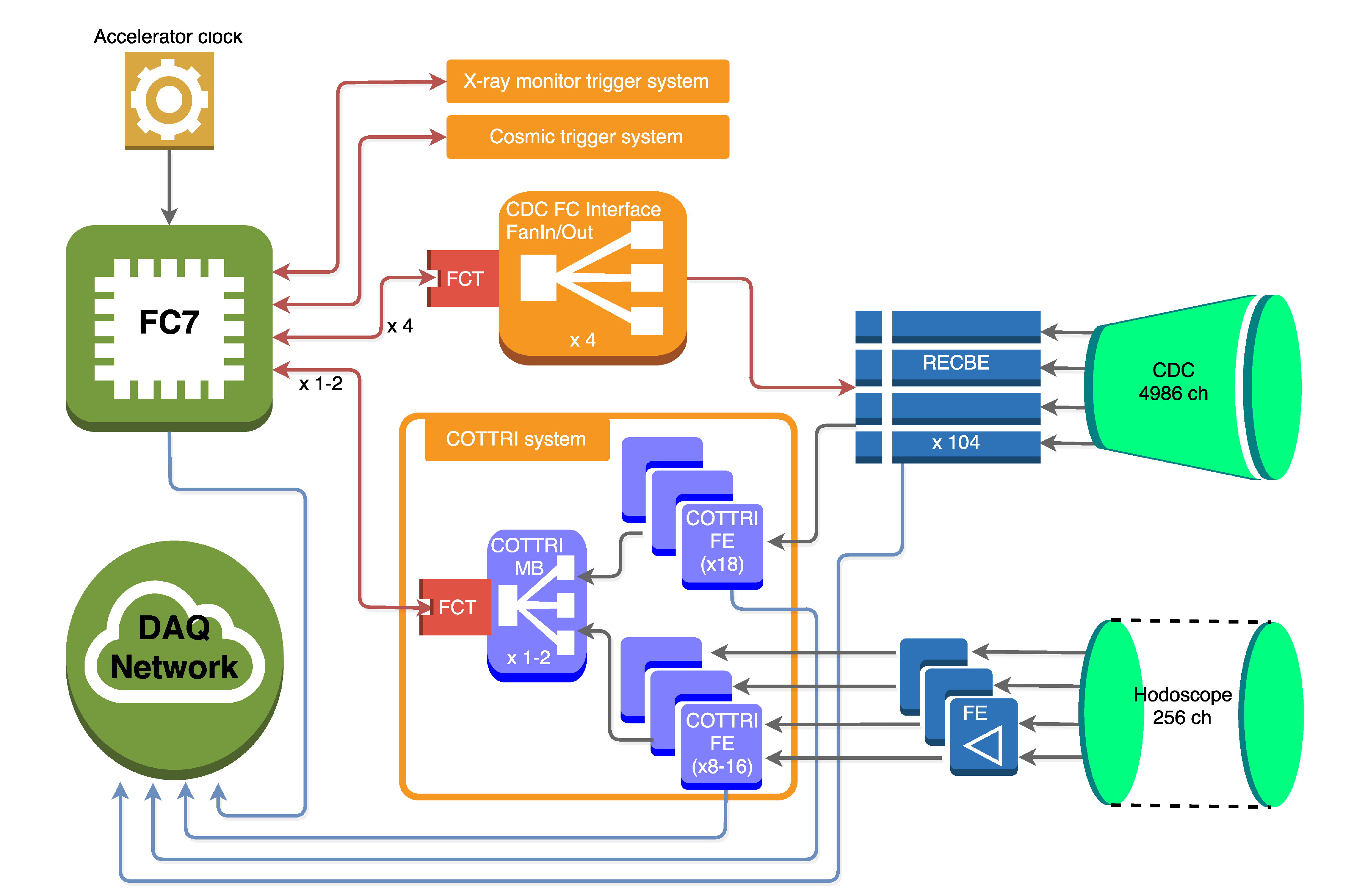}
\caption{Block diagram of the CyDet fast control and trigger systems.}
\label{fig:fastcontrol:CyDet}
\end{center}
\end{figure}

A schematic of the CyDet fast control and trigger systems is shown in \cref{fig:fastcontrol:CyDet}.
The main trigger when operating in CyDet mode is provided by requiring 4-fold coincidence on neighbouring counters from the CTH detector. This is supplemented by using the track patterns from the CDC hits as these are quite
different for high momentum electrons (signal or DIO)  than the low-momentum particle noise hits. 
For the CyDet component a simple combination of
hit pattern and energy deposition can yield a sufficiently fast trigger
with high efficiency and background rejection power, resulting in an overall
 trigger rate of a few kHz.

\begin{figure}[htb!]
\begin{center}
\includegraphics[width=0.8\textwidth]{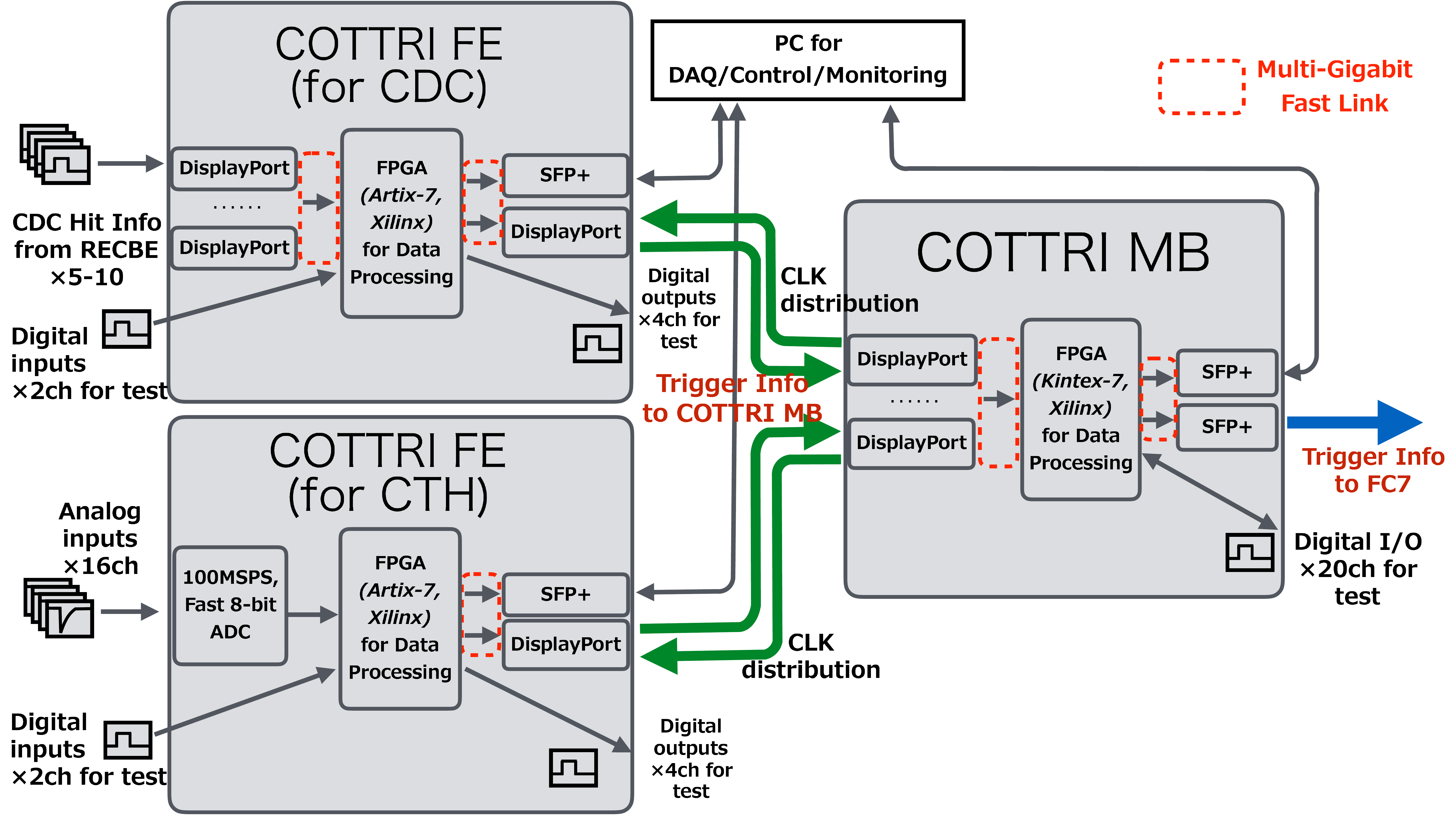}
\caption{Conceptual drawing of COTTRI system}
\label{fig:trigger:cottrifemb}
\end{center}
\end{figure}

\paragraph{The COTTRI system}
The COTTRI system provides the main logic for the CyDet trigger. It is divided into  front-end boards (COTTRI FE) and
a mother board (COTTRI MB), as  shown in
\cref{fig:trigger:cottrifemb}.
The COTTRI FE boards perform the initial processing of the analogue or digital inputs and then the COTTRI MB combines the signals to
generate the higher level trigger which is sent to the central
trigger system and distributed to the detector readout system.
The logic for the combination algorithm  resides in
the FPGA on the COTTRI MB.

12 COTTRI FE boards will
be required to process the CTH signals. The analogue signals are first amplified and digitised and then the FPGA  discriminates them and sends a digital trigger signal to the MB. They
 will be located inside the CTH
support structure and, therefore, a radiation hard design is necessary.

For the CDC application 18 FE boards will be
required to process the 104 RECBE board trigger signals where  the digitised hit information will be multiplexed and
sent to the MB. These boards will be
located inside the CDC readout box.

\label{sec:trigger:algorithm}
The proposed trigger algorithm using COTTRI will use the CTH-provided
trigger to search for CDC wire hits near the CTH hit
and count the number of CDC hits in that region.  A simple
track reconstruction can also be performed using these CDC hits.  Based on
these features, the COTTRI system makes the trigger decision and sends
it to the central trigger system with the relevant CTH and CDC hit
information.

\subsubsection{StrECAL Trigger}

\begin{figure}[htb!]
\begin{center}
\includegraphics[width=0.80\textwidth]{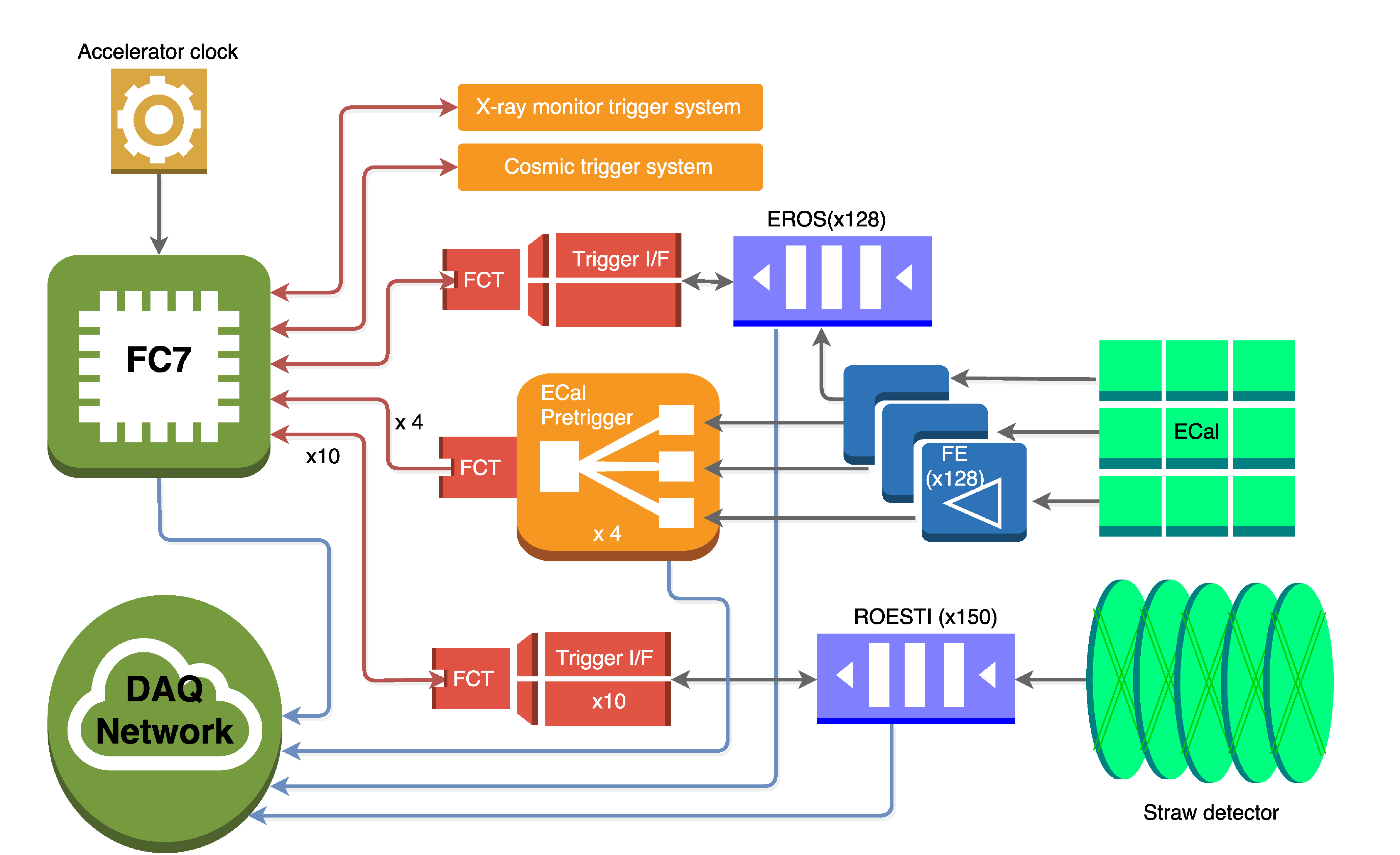}
\caption{
Block diagram of the StrECAL fast control and trigger systems. Note that, in Phase-I experiment, the number of channels of ECAL will be smaller than Phase-II, therefore, the number of Pretrigger boards is also smaller than this diagram. }
\label{fig:fastcontrol:StrECAL}
\end{center}
\end{figure}

\begin{figure}[htb!]
\begin{center}
 \includegraphics[width=0.8\textwidth]{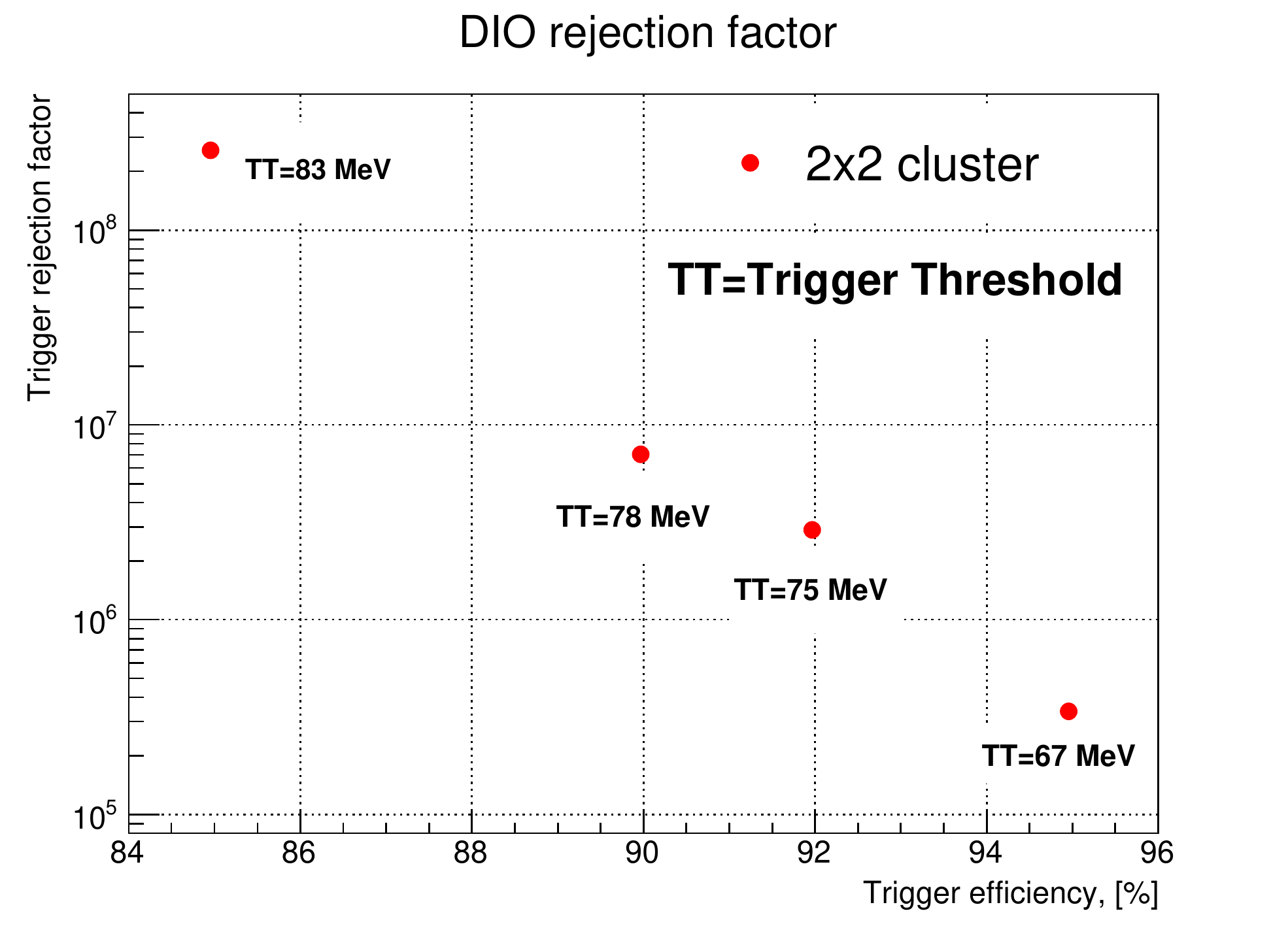}
 \caption{DIO rejection versus CE trigger efficiency on various energy threshold or ECAL pretrigger energy summation. }
 \label{fig:trigger:ecalpretriggereff}
\end{center}
\end{figure}

In the StrECAL mode (right figure of \cref{fig:fastcontrol:StrECAL}), the trigger is provided by the ECAL.
The
energy deposition from a single track can  be divided among several crystals and so a summation is necessary to reconstruct the full
energy. 
The summed energy over crystals which form a $4 \times 4$ square can effectively include almost all the energy deposited by  electrons with energies of about 100~MeV.
The basic trigger unit
(cell) will therefore be a group of $2\times 2$ crystals ( one ECAL
crystal module), and the total energy determined by
using the sum of an array of $2\times 2$ trigger cells  referred to as a trigger group.    The effectiveness from simulation is shown in
\cref{fig:trigger:ecalpretriggereff} with at least a $10^6$ DIO
rejection for around a 90 \% 
conversion electron detection 
efficiency.

The  structure of the ECAL electronics system is shown in
\cref{fig:ECAL-Readout}.  % This figure is in the ECAL section
The signal from each trigger cell is formed by analogue-summing the preamplifier outputs and hence 16 signals going to the digitising readout  and 4
signals   to a dedicated pretrigger board.

\paragraph{ECAL pretrigger board}

\begin{figure}[htb!]
\begin{center}
\includegraphics[width=0.8\textwidth]{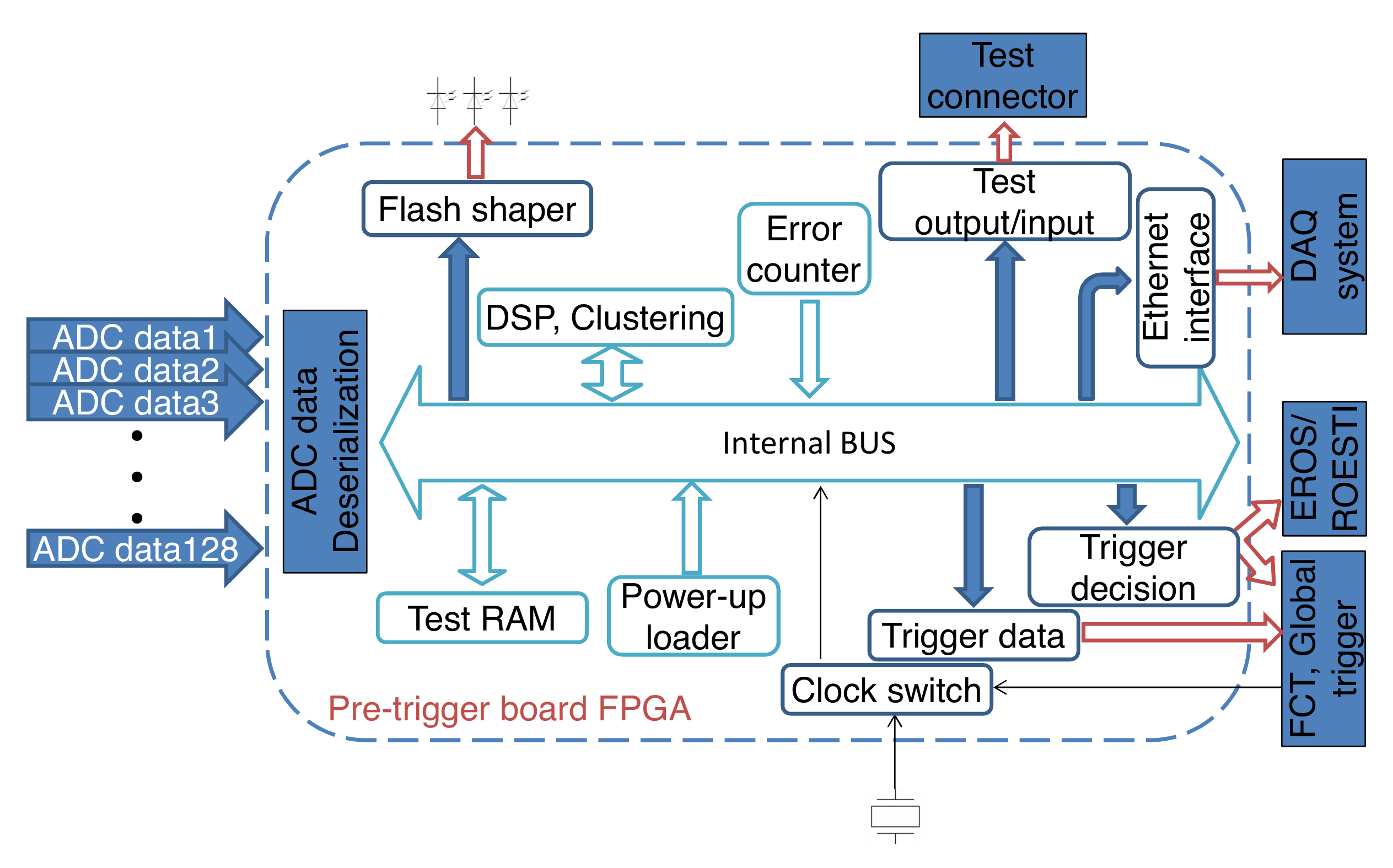}
\caption{The conceptual drawing of FPGA functionalities of ECAL pretrigger board.}
\label{fig:trigger:ecalpretriggerconcept}
\end{center}
\end{figure}

The ECAL pretrigger boards  digitise the analogue signals from the trigger cells
and pass the resulting waveforms through  filters in an FPGA. The conceptual design of the  FPGA logic is shown in
\cref{fig:trigger:ecalpretriggerconcept}.

Four trigger cells are summed in all possible combinations of such trigger groups.
And from these possible combination of pre-triggers, the group with largest energy is found, and the size of this signal is sent, together with the group number, to the FC7 for a final decision to be made.

The energy resolution of the ECAL pretrigger system is measured to
4.5 MeV for 105~MeV electrons,  which is
sufficient for trigger performance.

To avoid any inefficiency at
the edges of crystal coverage (a quarter of ECAL), digitised data of 12 trigger cell
locating at one edge of quarter of ECAL
are transmitted to the next board.
The pretrigger board also contains Ethernet connection to control PC,
so that the board can be controlled and monitored independently,
along with control and monitor through fast control and timing system.

The ECAL pretrigger boards provide a fast trigger signal but the final decision to record or not a particular event is made by the central trigger system.
The fast trigger decision is based on energy windows, which can be individually prescaled, e.g. for the signal region there is no prescaling whereas for the side bands it depends upon the expected background. These are set in the FPGA. If necessary the number of energy windows can be enlarged.

\paragraph{Other trigger system in StrECAL}
A StrECAL  cosmic trigger is also required for tests and
calibrations when not running with beam.
It will be based on the cosmic veto system with
simple coincidences of hits in different layers of bars close to
each.

\subsubsection{Trigger Performances}

During the prototyping of trigger boards, the maximal trigger rate capacity and the trigger latency are estimated. 
The final productions of those boards and integrated tests with detectors  are ongoing. 

\paragraph{Trigger rate}

Given a bunch separation of 1.17\micro{}s, the maximum rate of interesting events is around 850kHz. However, the serial data width of the trigger information is 50 bits, and this is returned over the 40~MHz MGT, so the trigger system is essentially dead for 1.25\micro{}s when a trigger is taken.

For the CyDet trigger the deadtime introduced by the RECBE board is less than $1$~\micro{}s and hence the actual maximum trigger rate in CyDet mode is $440$~kHz, whereas for the StrECAL trigger the ROESTI and EROS introduce a 36.7 \micro{}s deadtime leading to a maximum trigger rate of $26$~kHz. The effective trigger rate is, however, dictated by the DAQ system which is not greater than 20~kHz.

\paragraph{Trigger latency}

When operating in CyDet mode the required trigger latency is around 5~\micro{}s, due to the
buffer size of RECBE (8~\micro{}s). Currently the processing time in the COTTRI MB is not known, however the latency from the other components is estimated to be 1.1\,$\mu$s and so the requirement should be comfortably met.
In StrECAL mode the limit is set by the EROS board and a conservative target would be 700~ns. 
However  ~1~\micro{}s latency was measured in the current design, where the main bottleneck was signal encoding and decoding for MGT. 
It was possible to decrease the latency down to 700~ns by employing separate faster trigger line from the ECAL pretrigger to EROS which avoids MGT encoding and decoding bottleneck.

\subsection{DAQ System}\label{subsec:daq}

The DAQ system covers the data transfer from the frontend readout electronics to the data storage, through the event builder. 
The system mostly relies on the off-the-shelf equipments, so the selection of
proper equipment meeting the requirement of the COMET DAQ described here is on progress. The DAQ software
will be based on MIDAS framework. 

\begin{figure}[htb!]
\begin{center}
\includegraphics[width=0.8\textwidth]{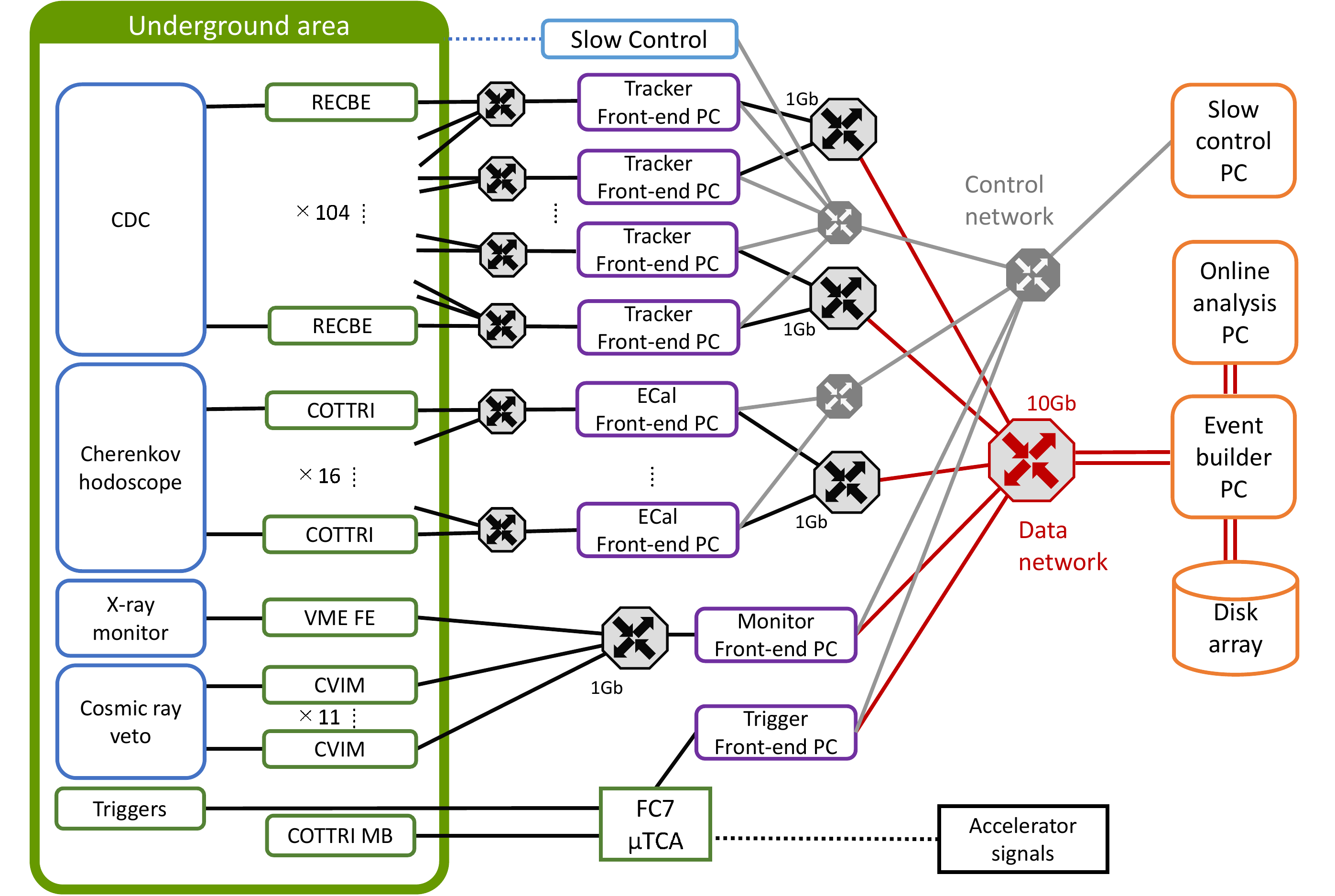}
\caption{Block diagram of the CyDet readout and configuration system.}
\label{fig:readout:CyDet}
\end{center}
\end{figure}

\begin{figure}[htb!]
\begin{center}
\includegraphics[trim={107pt 0 74pt 0},clip,width=0.8\textwidth]{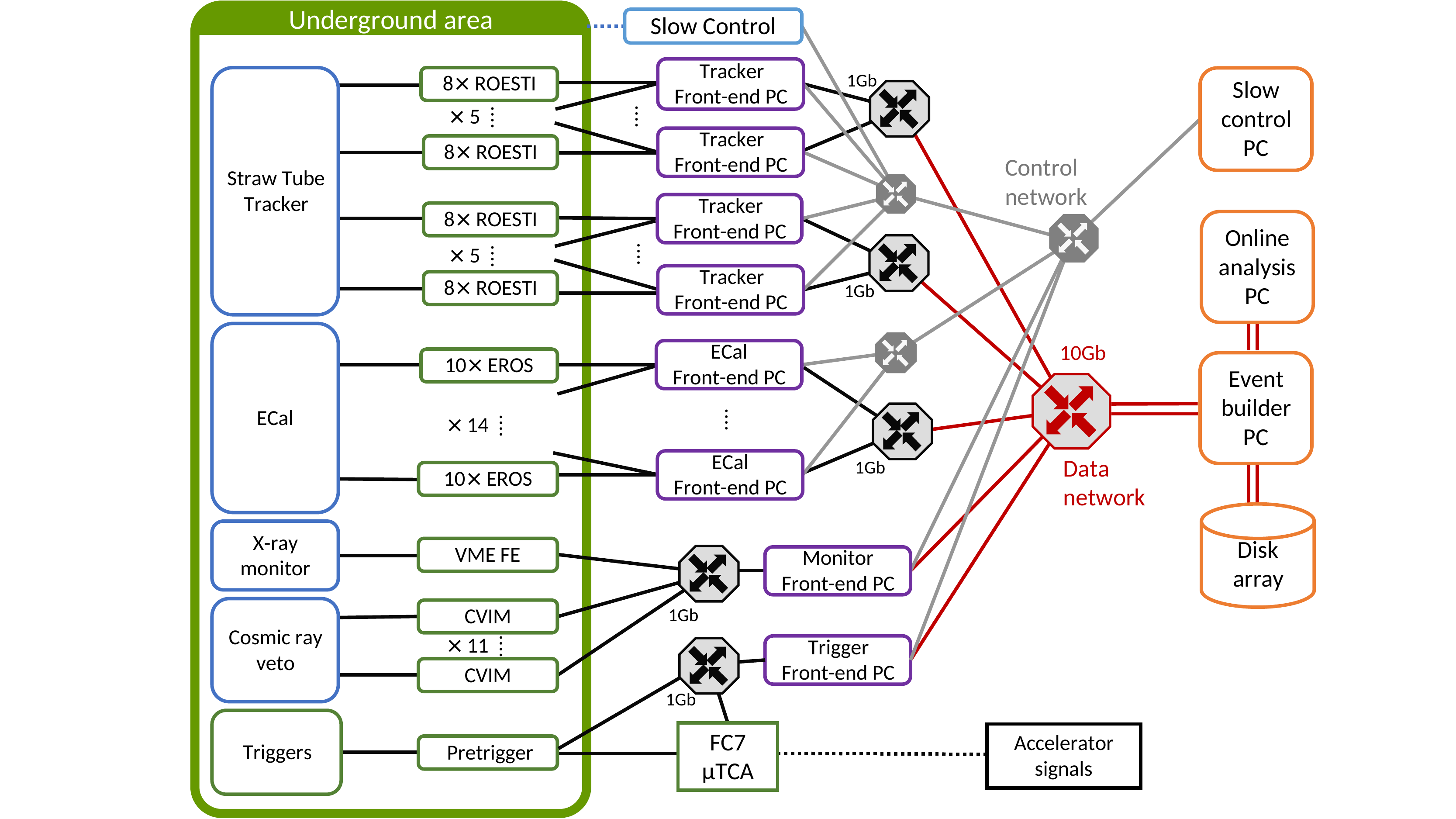}
\caption{Block diagram of the StrECAL readout and configuration system.}
\label{fig:readout:StrECAL}
\end{center}
\end{figure}

The readout systems for both CyDet and StrECAL modes are similar employing
standard Ethernet networking with commercially
available 
components.  The readout and
control networks for the CyDet are shown in
\cref{fig:readout:CyDet} and for the StrECAL in
\cref{fig:readout:StrECAL}.

An event builder PC acts as the run controller and
sends commands to start and stop runs, etc.
using  MIDAS~\cite{MIDAS1999} DAQ software control protocols.
These commands are distributed via network switches to a set of PCs  dedicated to the
readout of particular parts of the detectors.  Data are transferred on the network using standard
protocols (Ethernet, UDP, TCP/IP).

When a trigger occurs, the event data are stored in buffers in the
front-end electronics and  when the buffer has a whole
event it will be sent as a packet (or packets) of data to a PC.   Event packing will be conducted in
such a way that minimal translation is required.   As an illustration, a PC will assert (i.e. send a message to the
front-end electronics) 
``ready for a packet'', meaning it has
enough resources available to receive the largest possible packet (and
one more).  While receiving the packet it can assert/send a ``not
ready for packet'' signal/message and the front-end will continue and
finish the current packet, but then wait for an update from the
PC. Once the data is
collected on a PC all further transmissions are over standard computer
networking.

\subsubsection{MIDAS Front-Ends and Back-End}
The basic unit of
readout for the MIDAS DAQ is  a single ``equipment'' which wraps all the
activity of a subset of the readout electronics and communicates it
back to MIDAS in  standard MIDAS format.  It also acts as the
receiver for MIDAS commands.
For the main
detectors (CDC, ECAL and Straw tracker) this is accomplished using intermediate
PCs designated \textit{front-end PCs}.

Both the ROESTI and the RECBE will communicate with the front-end PCs using SiTCP,
an FPGA-based implementation of TCP~\footnote{In addition the COTTRI-FE
boards will also use a firmware-based TCP implementation}. The data
content of the TCP packets has no special restrictions, and is
assembled into an appropriate format by the SiTCP firmware.  The
front-end PCs can handle multiple boards, receiving data packets from the electronics and
decoding them.

\begin{figure}
\centering
\includegraphics[trim={0 194pt 0 0}, clip, width=0.8\textwidth]{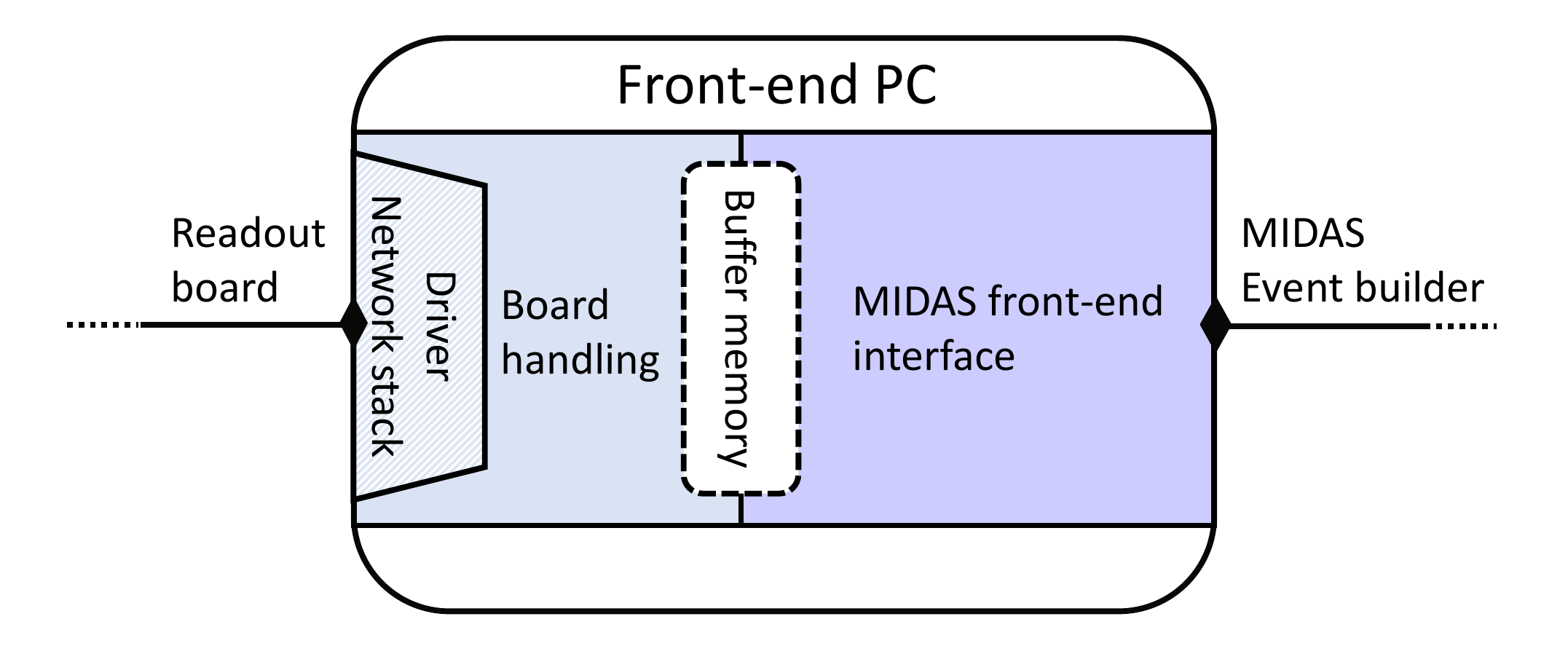}
\caption{Schematic of the processes running on a front-end PC}
\label{fig:readout:front-end}
\end{figure}

The  front-end PCs also communicate with MIDAS, responding
 to commands sent by the
DAQ and
transmitting readout data to the central Event Builder PC.
These two roles are executed concurrently as threads of the same process on the PC.
The non-synchronous devices, such as the X-ray monitor,
also can be read the same structure front-end process,
because of the supporting multiple type trigger by MIDAS.
A schematic of a basic front-end process running on
a front-end PC is shown in \cref{fig:readout:front-end}

For the main detectors, the front-end PC also performs a partial event building to reduce the load on the back-end event building
and hence increase  the performance of the  DAQ.
This front-end partial event building is a generalization of the basic front end using several threads.

\begin{itemize}
\item Reader. The reader handles the  front-end electronics,  reads the data and writes it to an internal buffer.
\item Partial event builder.
This  manages the event fragments from the  reader threads with the same trigger number and writes them to a back-side buffer

\item Sender.
This sends the built event fragments to a MIDAS server on the back-end PC using the  MIDAS remote protocol.
\end{itemize}

\begin{figure}
\centering
\includegraphics[width=0.8\textwidth]{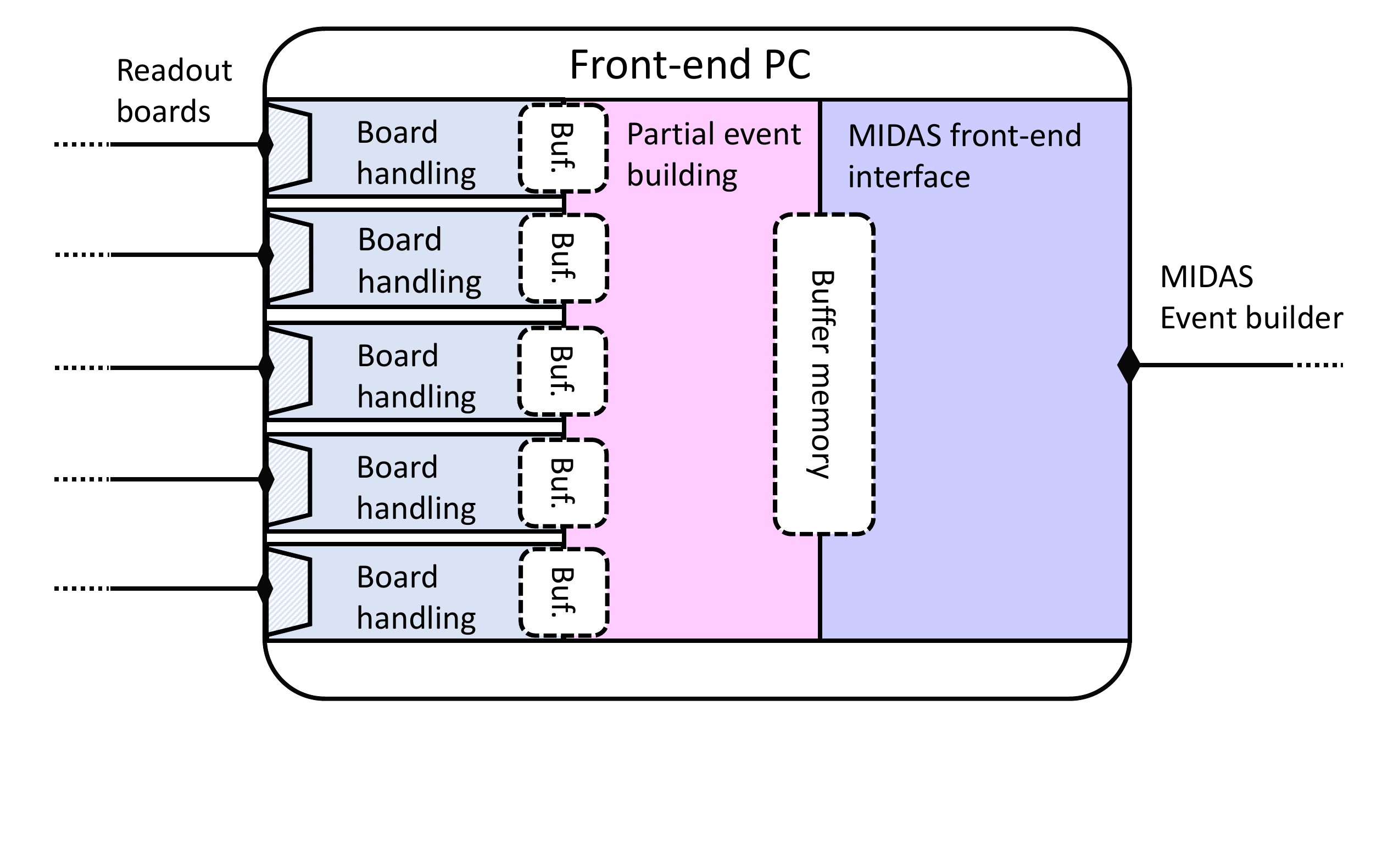}
\caption{Schematic of the event building front-end processes running on a front-end PC}
\label{fig:readout:peb}
\end{figure}

A schematic of the partial event building front-end process is shown in \cref{fig:readout:peb}

The backbone of the data network will be 10 Gigabit Ethernet as it
needs to channel data from all of the front-end PCs.
The central element of the MIDAS back-end is a PC that runs the
event builder process. This requests and collects ``fragments'' directly
from all the front-end systems to a memory buffer, and when all
expected fragment are received, it sends the completed event to a logger
that is responsible for writing the event to disk.
The PCs for the online analysis  copy the event data from the event building PC via a direct connection and distribute them to several types of analytical processes.
There will be 100~TB class storage on the local site and then the experimental data will be transferred  to
the primary data archive provided by KEK computing in Tsukuba
using 10~Gbps Tsukuba-Tokai network.

\subsubsection{Data Rates}
Both detector systems need to operate assuming the accelerator is
operating continuously for an indefinite period of time. However as the
beam is not continuous,  peak rates are higher than the time averaged
rates.  While the electronics
needs to operate at the peak rate, buffering allows the higher levels of the DAQ  to operate at a lower
average rate.
The operating parameters assumed in the following
section are shown in \cref{tab:daq:machine}. At present
 the 4-bucket mode is used for data rate estimates.

\begin{table}[tbh!]
\centering
\caption{Experiment operating parameters assumed in this section.  In
  most cases these are not fixed and will be optimised. 
In tests of 8~GeV SX, the accelerator cycle time is 2.38~s, during which each spill lasts around 0.5s.}
\label{tab:daq:machine}
\centering
\begin{tabular}{lcc}
\hline \hline
Accelerator cycle time  & \multicolumn{2}{c}{$\mathcal{O}(1)\,\mathrm{s}$} \\
Slow extraction duty factor &\multicolumn{2}{c}{0.5}\\
\hline
 & 4 bunch & 3 bunch\\
During slow extraction: \\
\qquad Normal bunch spacing & 1170~ns & 1755~ns \\
\qquad Signal window & \phantom{1}800~ns & 1300~ns\\
\qquad Signal window active fraction & 0.68 & 0.74\\
\qquad Number of bunches extracted &\multicolumn{2}{c}{$\mathcal{O}(10^6)$}\\
\hline \hline
\end{tabular}
\end{table}

\paragraph{For the CyDet} %The basic trigger rate  derives  from the CTH which,
The basic trigger rate for the CTH is about 26~kHz.
Even with the shielding the rate is still mostly due to low-energy particles.  Using
the CDC hit information,
this can be reduced to a more manageable 1.3~kHz.

Signal-like triggers (from the high energy tail of DIO) should pass
the high level trigger with probability close to 1, and as such
represent a minimum floor for the online readout rate.  Older simulations that don't incorporate the latest
software developments suggest a rate of a few hundred per second.
 The CDC
 imposes a minimum threshold momentum for electron tracks to
be observed of at least $70$~MeV$/c$, and the rate of DIO
electrons above $70$~MeV$/c$ is around $600$~Hz.  Summing both
triggers from background and signal-like electrons  an
overall trigger rate of around 2~kHz is estimated.

The event sizes for the CyDet are calculated assuming the existing
\emph{Suppress}[ed] readout of the RECBE boards (including a high-charge cut-off) is used.  This is based
on a Belle-II mode which essentially corresponds to reading out an
integrated ADC value and time stamp for channels above some
zero-suppression threshold.

The data volume per trigger assumes an
occupancy of 20\% out of around 5000 sense wires, corresponding to
$\sim1000$ hits in a signal window.  However the majority of these
hits will be eliminated by the high-charge cut leading to an estimate of around 3.5\% channels to be read out and consequently about 200 hits in the signal window.  Because the
drift time in the CDC is projected to be around 0.4~\micro{}s, it
is necessary to consider a coincidence timing about as large as
the signal window, so there can be no significant reduction
from applying a trigger window.  Overall, in the Suppressed mode, it is estimated that  the CDC gives 1.7~kiB per
trigger.

For the
trigger hodoscope readout the readout window does not need to be as wide
as for the CDC, as there is no drift delay,  however in initial
running it will be useful to retain the CTH information from the
entire readout period, which is approximately 1~\micro{}s. In this
case, the most efficient readout format is a fixed-ordering.  The data
for each counter would consist of a one byte `header' indicating how
many hits were recorded followed by the same number of TDC-ADC
pairs.  In this
scheme, the total data read out would be approximately 2.5~kiB
($=256\times 1~\textrm{header bytes}+800\times 3~\textrm{TDC-ADC
  bytes}$), which is  a substantial contribution to
the overall event size.
Once experience is gained  the CTH readout window could probably
be reduced to a few tens of nanoseconds around the trigger time, which
would reduce its contribution below 1~kiB.

\paragraph{For the StrECAL} The even sizes are larger but the trigger rate is lower, and it 
can be more easily
tuned in the run configuration by adjusting the energy threshold of the trigger logic.

For both detector configurations, a variable trigger prescale will also be implemented,
allowing some fine tuning of the rates and  a busy mechanism will
provide the ultimate guarantee of a maximum data rate.

\subsubsection{Online Software}
MIDAS~\cite{MIDAS1999} will be used for the DAQ software, which will be made 
of several distinct parts.
Front-end PCs will run detector-specific programs to read directly from the hardware and asynchronously send packed data to the back-end PC using MIDAS RPCs (Remote Procedure Calls) to store the data in the form of MIDAS banks.

The back-end PCs will run the following:
\begin{itemize}
\item An event builder  to combine the MIDAS banks from several front-ends into a single MIDAS file.
\item An ``online converter and monitor''  to convert the data into ROOT format and produce histograms for live monitoring and offline analysis.
\item A ``logger'' to write the MIDAS data to disk.
\item  A ``run control'' program to start, stop and monitor runs.
\end{itemize}
Most of these programs will be implemented within COMET's ICEDUST software
framework which has the
benefit of being based on the T2K-ND280 framework, which
also uses MIDAS, and as such already has libraries to perform these
transformations.

Data monitoring will take place on a separate dedicated PC which %(the ``Global DQM'' PC in
will run the online conversion and monitor program to read the data on
disk (in MIDAS format), convert to ROOT format and produce standardised plots for detector
status monitoring. This incorporates libraries which allow access to the information in the files directly from ROOT. The offline event display (albeit with preliminary alignment and
calibration constants) will also be used for online monitoring through this mechanism.

\subsubsection{The Radiation Environment}
The RECBE board for CDC readout will be located inside the Detector Solenoid and get affected by the radiation
specially the neutron radiation. 
Therefore,  understanding on the radiation environment around the RECBE board and preparing for the possible
situation of DAQ that may arise due to the radiation effect is very important. 

The effect of neutron radiation to the FPGA and its firmware, and mitigation efforts in COMET Phase-I experiment 
are reported in \cite{NAKAZAWA2019351}. 
Most of single event upsets due to the radiation will affect the transient data 
which can be isolated in the offline process, while it may lead to a FPGA firmware malfunctioning sometimes. 
In this FPGA malfunctioning case, the DAQ and frontend hardwares can be restarted, which will result in 
a long deadtime when happens with high rate. 

\paragraph{Gaps in the data stream}
The RECBE FPGA firmware  and DAQ is designed so that it can be reloaded 
in case that the  FPGA is not recovered automatically from the
effect of radiation. 
Again, this reloading is triggered automatically by DAQ, however,
a few seconds of gap in the data stream is not avoidable. 
This will  be handled at the central event builder process, 
so that the event building will be completed without missing fragment of data stream when
the fragment buffer remains empty for a specific timeout period. 
When the RECBE board will be rebooted and resuming to send the event data, 
the event builder will 
realign this buffer with the remainder of the detector using  the trigger number.

\paragraph{Loading of runtime parameters}
It is likely that the RECBE will need to reload various configurable
parameters after a  firmware upset. To reinitialise run, the parameters will be cached on  the front-end PC 
at run start, and the relevant parameters will be  provided when requested by the RECBE.

\section{Offline Software and Physics Analysis}
\label{section:software-offline-analysis}

The main software suite that is in use by COMET is called ICEDUST (Integrated Comet Experimental Data User Software Toolkit).
Below we describe ICEDUST and its use in analysis work in preparation for Phase-I.

\subsection{ICEDUST Framework}\label{sec:software:icedust}
For the calibration, reconstruction and analysis of data from COMET Phase-I, it
is essential to have an offline software framework that treats this data in the
same way as for simulated data.  Since the development and testing of a new
framework is a significant undertaking, it was decided that the COMET framework
be based upon an existing framework that has been well tested in a real
data-taking scenario.

A number of existing frameworks were compared against the
requirements of COMET.
And the framework used for the near detector for the T2K
experiment~\cite{Abe:2011ks} at J-PARC, known as ND280, would be the best choice upon which to develop the ICEDUST,
which had already seen several years of data-taking and debugging and use in published physics analyses.
It also includes a novel interface with the MIDAS online data acquisition system.

\begin{figure}[!ht]
\begin{center}
\includegraphics[width=0.8\textwidth]{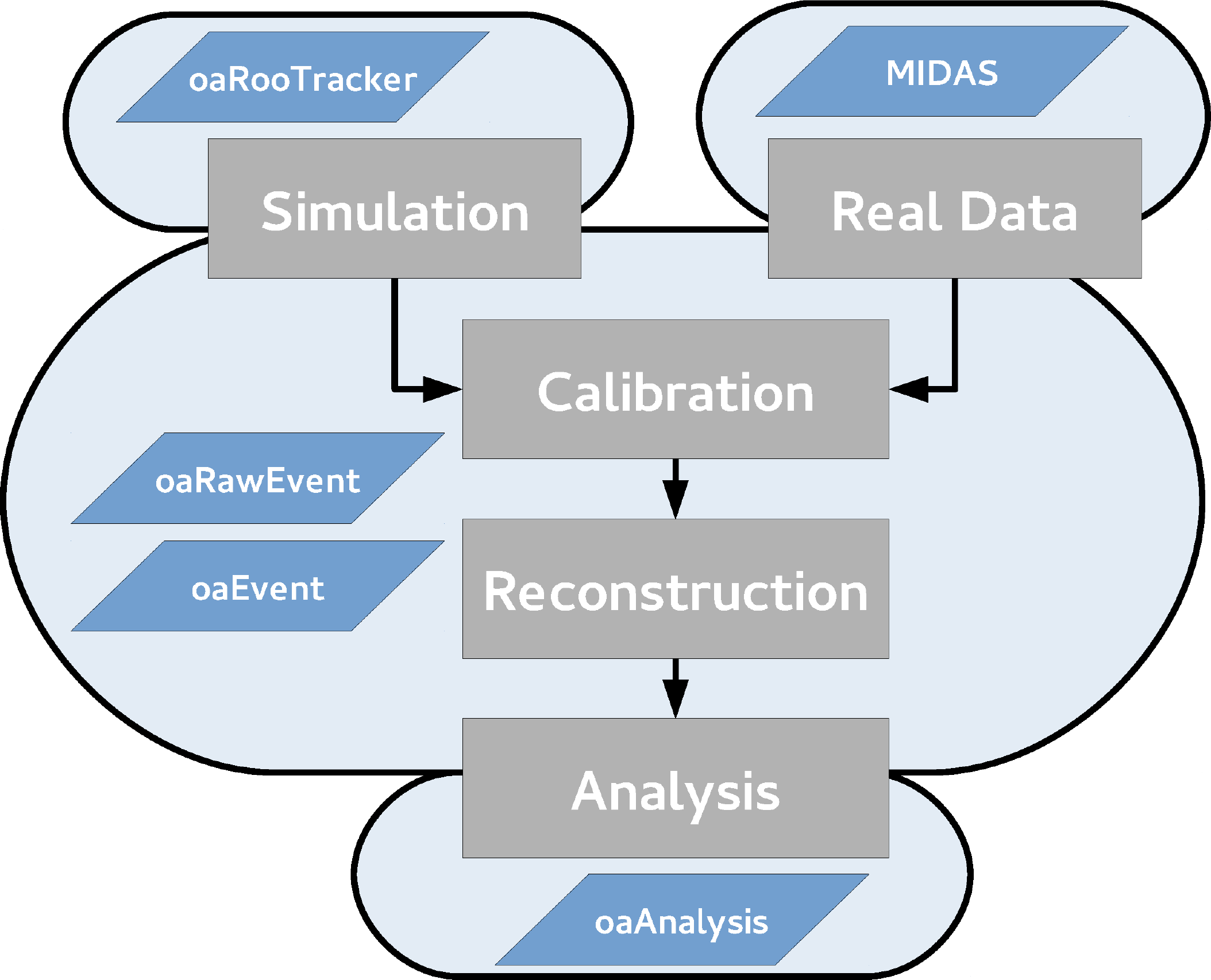}
\caption{ 
An outline of the ICEDUST framework. The larger blue regions represent parts of the framework 
that share a common data format, which is specified in the parallelogram.
}
\label{fig:ICEDUST}
\end{center}
\end{figure}

The structure and the data flow of ICEDUST are illustrated
in \cref{fig:ICEDUST}. 
Notably, data from the experiment and the simulation are processed in the same way, following the same steps through the chain.
The important components of ICEDUST are as follows:
\begin{itemize}
\item Data formating: The {\tt oaEvent} package provides the event data format both for simulated and real data,
as well as the geometry information. The experimental data of MIDAS data format is unpacked to
{\tt oaEvent} format.
\item Simulation: While Geant4  plays the central role in simulation, any other package can be cooperated in generating
and simulating events. The simulated data of each proton are collected into a proton-bunch event.
The detector responses are applied afterwards.
\item Reconstruction: 
A global reconstruction includes the signal track finding and fitting,  and subsequent analysis. This 
generates simpler event data with the format provided by {\tt oaAnalysis} package.  
\end{itemize}
Details are described in the following sections. 

Significant improvements have been made to the software in recent years, and
the physics content of the software is essentially independent between the
experiments, given the differences between 
COMET and T2K ND280.
Many of these improvements
are now being fed back into the ND280 software.

To ensure that software developed for ICEDUST is done in a consistent way
across the collaboration, naming and coding conventions have been
defined and followed.

\subsection{Data Formats}
The key strength of the approach used is the ability to treat experimental data on such an
equal footing as the simulated data.  This is achieved in two ways:
\begin{itemize}
\item An unpacking mechanism which converts the raw MIDAS data into offline root files.
\item A wrapping package which can provide a semi-transparent method to process raw data.
\end{itemize}

The core event structure of ICEDUST is provided by {\tt oaEvent} package. 
Data in the the {\tt oaEvent} format is built by a hierarchical structure of objects 
from the {\tt TNamed} class of  ROOT, 
that provides a stronger memory management policy and avoids the need 
for global pointers to access the output data hierarchy. 
Collections of data are stored in a similar manner with the map class of Standard Template Library (STL) of C++.

The description of the geometry is stored alongside the data, either in the
form of a hash-tag pointing to a particular archived geometry which is
automatically retrieved as needed, or else as a persisted ROOT object.  The
ROOT format uses the various TGeo classes which implement all geometry needs
such as navigation, mass calculations, material descriptions and visualisation.
This means that all packages throughout the framework use a common geometry
description as well as providing an easy book-keeping mechanism.

\subsection{Simulation}
The simulation of COMET has been sub-divided into smaller tasks.  The standard
simulation chain for production Monte-Carlo data is:
\begin{enumerate}
\item Simulate production target.  This is done separately so we can use
	packages that include different hadron production models.
\item Particle tracking in Geant4 (SimG4). Geant4 has a highly optimised
	tracking algorithm as well as many well tested experiment-based physics
	models.  This package tracks particles from the production target to
	the various detectors and produces the simulated energy deposits.
\item Detector Response Simulation.  Energy deposits produced by SimG4 are
	converted into realistic detector outputs such as ECAL crystal
	waveforms or CyDet wire hits.  Various detector effects such as finite
	resolution, cross-talk and random noise can be added here.
\item Rare-process selection.  Occasionally we may wish to focus a study on one
	of the rarer signal or background processes.  Since this would normally
	require the simulation of a lot of unimportant processes, a package is
	being developed to merge hits from rarer processes together to
	artificially increase their statistics.
\end{enumerate}

\begin{figure}[h!]
\begin{center}
\includegraphics[width=0.8\textwidth]{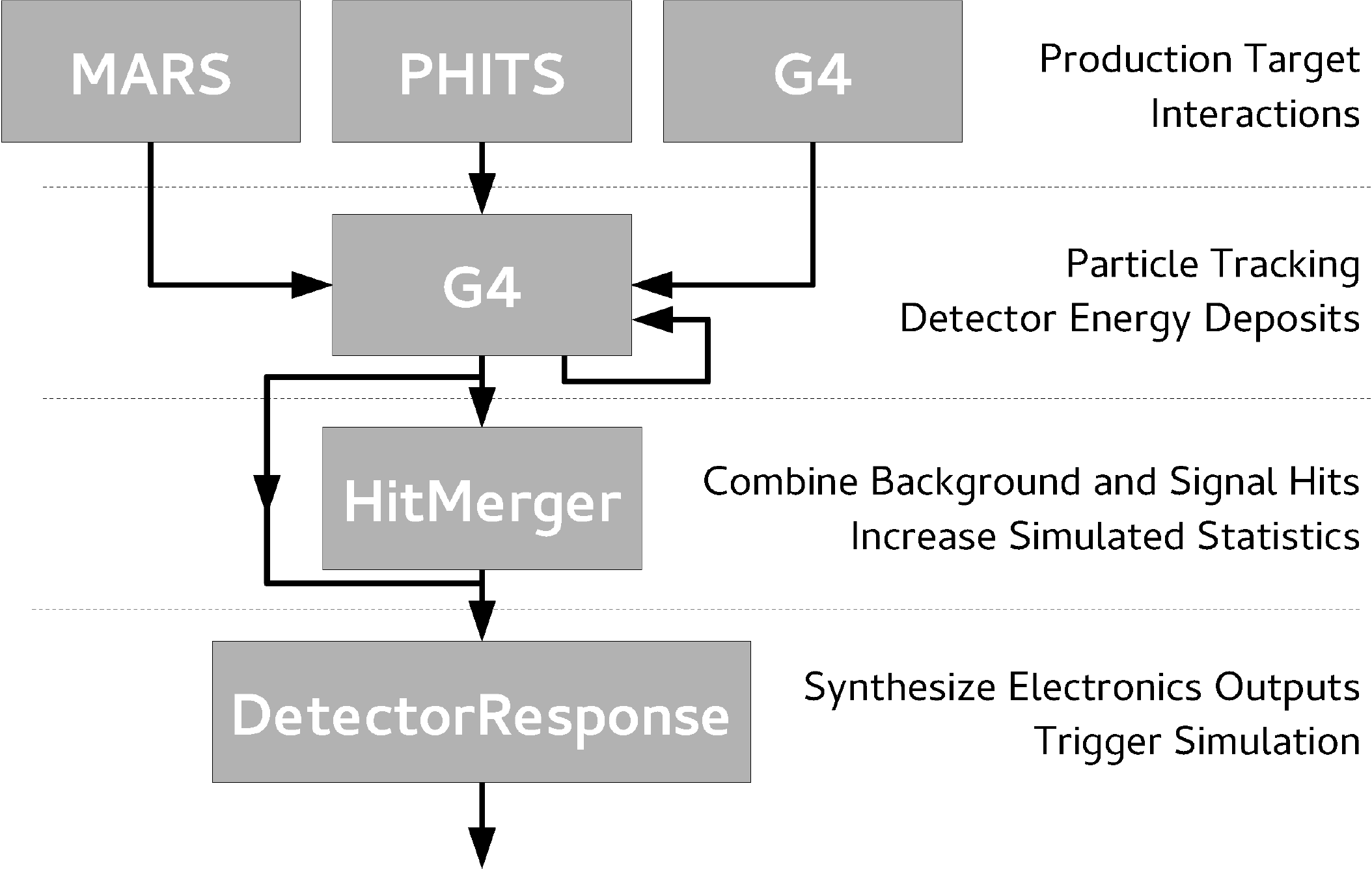}
\caption{
The package structure of simulation of ICEDUST.
}
\label{fig:ICEDUST:sim}
\end{center}
\end{figure}

Up to truth information, steps 1 and 2 can be done by any of the various
external Monte Carlo packages that have been incorporated into ICEDUST.  These
include Geant4~\cite{Agostinelli:2002hh}, MARS15~\cite{Mokhov:2007sz}, PHITS~\cite{Niita:2006zz} and
FLUKA~\cite{Battistoni:2007zzb,Ferrari:2005zk}.
\Cref{fig:ICEDUST:sim} shows the implementation of this simulation procedure in ICEDUST.

\subsubsection{Geometry definition}
SimG4 is the package tasked with running Geant4 tracking and producing geometry
files that all other packages use through the ROOT format described above.
This task is made non-trivial by the necessity for a highly detailed
description of the geometry in order to check all possible sources of
backgrounds combined with COMET's staged approach meaning the experiment is
likely to change quite drastically throughout its lifetime.

\begin{figure}[h!]
\begin{center}
\includegraphics[width=0.8\textwidth]{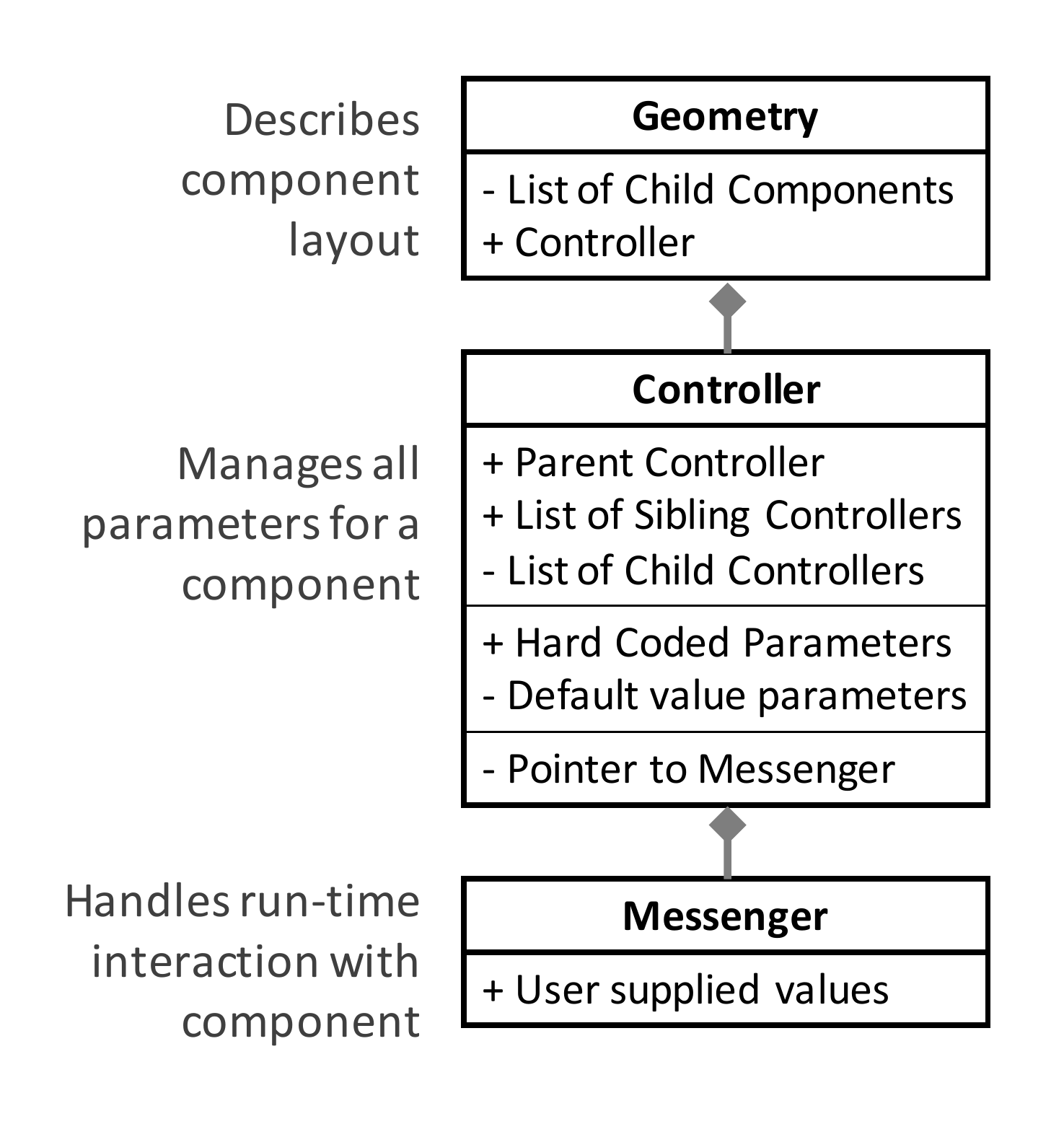}
\caption{Class structure of the Geometry-Controller-Messenger model.}
\label{fig:SimG4:ClassStruct}
\end{center}
\end{figure}

In practice, a user writes the geometry in C++ code, then defines all the
parameters in a Geant4 macro which is processed by a ``messenger'' and ``controller'',
as shown in \cref{fig:SimG4:ClassStruct}.
These parameters can be defined using complex expressions involving other parameters, which provides a flexible way for building the geometry.

\subsubsection{Custom Geant4 processes}
Custom modelling of physics processes has been developed within the
Geant4-based simulation to provide models that match experimental
data and models that include the latest theoretical updates.
In particular, modelling for negative muons stopping in 
aluminium
has been implemented into ICEDUST.
A custom class for physic process is defined by inheriting {\tt G4HadronicInteraction} class,
to describe the muon bound or capture process in an aluminium atom. The 
AlCap data and DIO spectrum estimation~\cite{czarnecki11} are implemented in the class.
This implementation of the modelling should be robust to
future Geant4 updates, but also allows new spectra to be included
easily as they become available.

\subsubsection{Hadron production models}
There is a large variation in the pion and muon yield predicted by different
hadron production codes.  It is therefore essential that the simulation has the
ability to use different hadron production codes.  Currently, simulations have
been done using FLUKA~\cite{Battistoni:2007zzb}, Geant4, MARS and PHITS and
these codes have all been integrated into the ICEDUST framework. Ensuring the
consistency of the geometry requires careful consideration when using FLUKA and
PHITS as they do not have native support for the ROOT geometry.

Since these packages contain very detailed, experimentally-supported hadron interaction models,
they are particularly useful for studying and simulating the pion production target and running shielding calculations.

\subsubsection{Refining simulation against experimental data}
In order to achieve the single event sensitivity of Phase-I, it will be
necessary to fully understand all 
sources of
particles that could mimic an electron
produced by muon to electron conversion.  This requires accurate simulation of
the experimental apparatus as well as understanding the production mechanisms
of rare processes that produce signal-like electrons and ensuring that these are 
well modeled
in the simulation.

It is important to make use of Phase-I to characterise the beam line and thus
understand the transport characteristics of the curved solenoid channel in
order to understand the background rates and validate the simulation.  This can be done using a relatively
simple detector to make flux measurements in the Phase-I beam line and by
varying the magnetic fields and placing absorbers in the beam.  Preliminary
studies to show what sort of measurements can be done with Phase-I is presented
in~\cite{Kurup:2013lsa}.  These results show that it is possible to alter the
composition of the beam, by making simple changes to the magnetic field or
using absorbers, and therefore understand the transport properties of the beam
line better and will provide a way to ensure the simulation accurately models the experiment.

Important information will be provided by other experiments (e.g.\ AlCap)
and beam tests (e.g. calorimeter resolution tests).  The information from these
will be fed back to improve the physics and detector descriptions in the offline
software.

\subsection{Reconstruction}
The aim of the reconstruction software is to take a collection of hits stored
in a ROOT file, either from simulation or experimental data, and produce a
collection of reconstructed objects such as tracks and clusters.  This requires
track or cluster finding and fitting code specific to each detector system.

GENFIT~\cite{Hoppner:2009af} is integrated into ICEDUST, which provides:
\begin{itemize}
\item integration with SimG4;
\item the ability to run on experimental data;
\item integration with analysis codes;
\item a simple interface to validate geometries and magnetic fields.
\end{itemize}
One important requirement of the ICEDUST framework is to provide a full audit
trail that allows 
an exact recreation of an analysis plot.
This
requires persisting the precise process used by the reconstruction to create
the reconstruction objects.  This will obviously be different depending on
which code is used, e.g. for track reconstruction.  Thus, the minimal set of
information required to fully specify the reconstruction process needs to be
defined for each code.  This can then be persisted in a database and used to
tag the simulated data that is produced.

\subsection{MC Data Production and Distribution}
With each major software release, there will also be the generation of
large-scale MC data.  This will be needed to debug the offline software as well
as improving the experimental details of Phase-I, reconstruction algorithms and
analysis code.  In addition to producing data that will mimic the data from
Phase-I, specific background modes will be simulated so that very rare
processes can be studied in a resource-efficient way.  The first of these MC production
runs has been done and was used to debug the software and provide estimates
of the computing resources required for future MC data production.  These data sets will
be used to develop reconstruction algorithms and analysis code and will provide estimates
of the resources needed to process data from Phase-I, where the data rate is currently estimated to be 7~Tb/day
for continuous running.

Grid computing resources will be used to distribute data.  A
similar production and distribution plan to that of the ND280 experiment will
be used.

 % works for both Part 1 and Part 2
\section{Physics Sensitivity and Background estimation}\label{ch:SensitivityBackgrounds} % Although we' ve removed "Backgrounds" from the heading for Part 1

COMET will operate in CyDet mode to search for \mue conversion in Phase-I.  The single event sensitivity (SES) is determined
for a given number of stopped muons, as described below.
The different sources of backgrounds are
identified in \cref{tb:backgrounds}. 

\begin{table}[tbh!]
    \caption{A list of potential backgrounds for the search for the \muec conversion at the COMET experiment. The items with $*$ would not produce 100 MeV/$c$ electrons but noise hits in the CyDet.}
  \begin{center}
    \begin{tabular}{lll}
    \hline \hline
       \multicolumn{3}{l}{} \\
       \multicolumn{3}{l}{Intrinsic physics backgrounds} \cr
    \hline
    1 &Muon decays in orbit (DIO)
    & Bound muons decay in a muonic atom \\ %\hline
    2 &Radiative muon capture (external)
    & $\mu^- A \rightarrow \nu_\mu A' \gamma$, \\
    & &followed by $\gamma \rightarrow e^- e^+$ \\ %\hline
    3 & Radiative muon capture (internal)
    & $\mu^- A \rightarrow \nu_\mu e^{+} e^{-} A'$, \\ %\hline
    4$^{*}$ & Neutron emission & $\mu^- A \rightarrow \nu_\mu A' n$, \\
    & after muon capture & and neutrons produce $e^-$ \\ %\hline
    5$^{*}$ & Charged particle emission & $\mu^- A \rightarrow \nu_\mu A' p$ (or $d$ or $\alpha$), \\
    & after muon capture & followed by charged particles produce $e^-$ \\
        \multicolumn{3}{l}{} \\
        \multicolumn{3}{l}{Beam related prompt/delayed backgrounds} \cr
    \hline
    6 & Radiative pion capture (external)
    & $\pi^- A \rightarrow \gamma A'$, $\gamma \rightarrow e^- e^+$  \\ %\hline
    7 & Radiative pion capture (internal)
    & $\pi^- A \rightarrow e^{+} e^{-} A'$ \\ %\hline
    8 & Beam electrons & $e^-$ scattering off a muon stopping target \\ %\hline
    9 &  Muon decay in flight & $\mu^-$ decays in flight to produce $e^-$ \\ %\hline
    10 & Pion decay in flight & $\pi^-$ decays in flight to produce $e^-$ \\ %\hline
    11 & Neutron-induced backgrounds & neutrons hit material to produce $e^-$ \\ %\hline
    12 & $\overline{p}$-induced backgrounds & $\overline{p}$ hits material to produce $e^-$ \\
           \multicolumn{3}{l}{} \\
           \multicolumn{3}{l}{Other backgrounds} \cr
    \hline
    14 & Cosmic ray-induced backgrounds & \\ %\hline
    15 & Room neutron-induced backgrounds & \\ %\hline
    16 & False tracking & \\ \hline\hline
\end{tabular}
\end{center}
    \label{tb:backgrounds}
\end{table}

\subsection{Signal Sensitivity}
\label{sec:SignalSensitivity}

 The signal acceptance  is determined by the geometrical acceptance of the CyDet, the track quality cuts and the acceptances of momentum and time windows of measurements.

\paragraph{Geometrical acceptance}\label{sec:geometricalacceptance}

\begin{figure}[htb!]
 \begin{center}
 \includegraphics[width=0.45\textwidth]{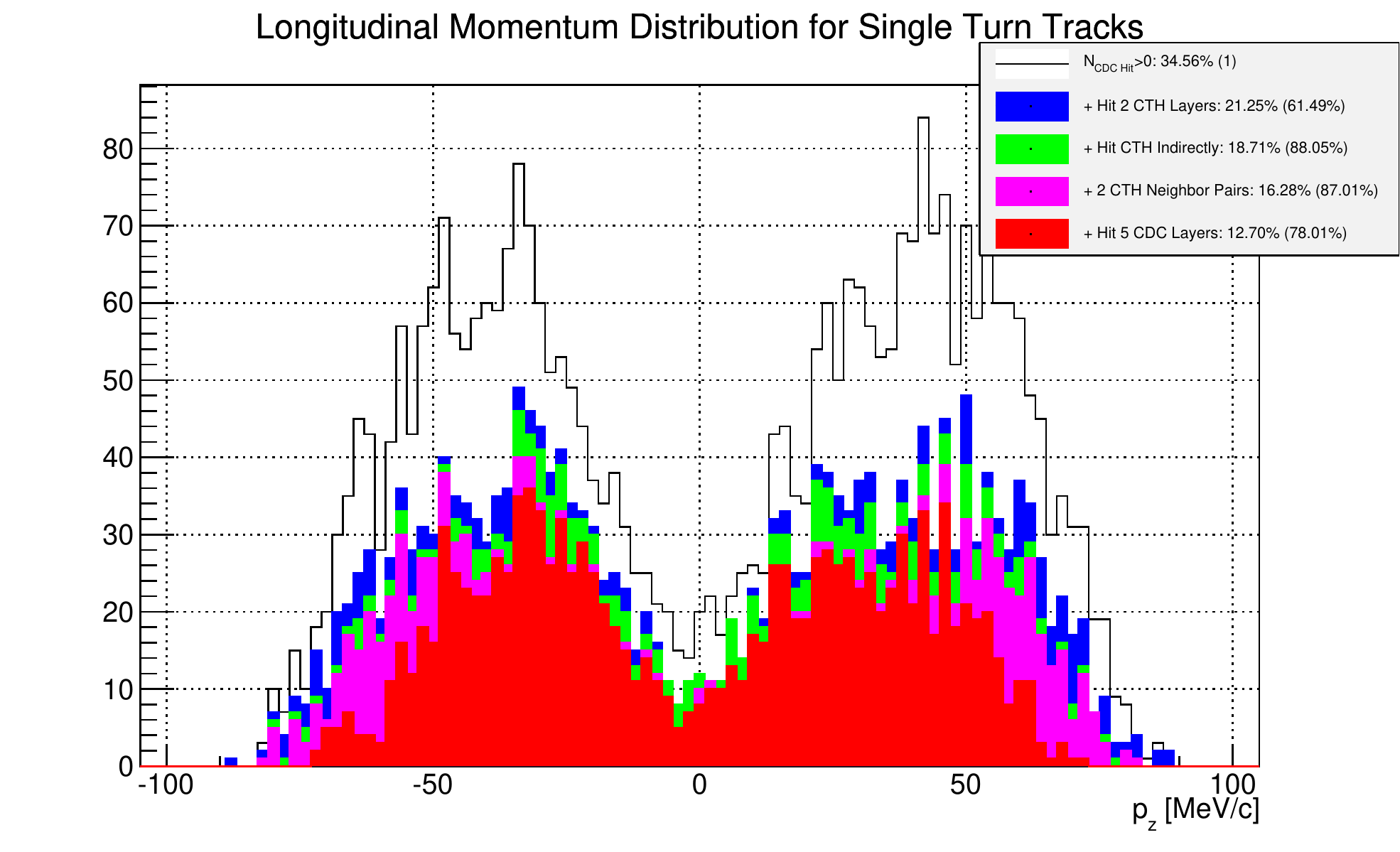}
 \includegraphics[width=0.45\textwidth]{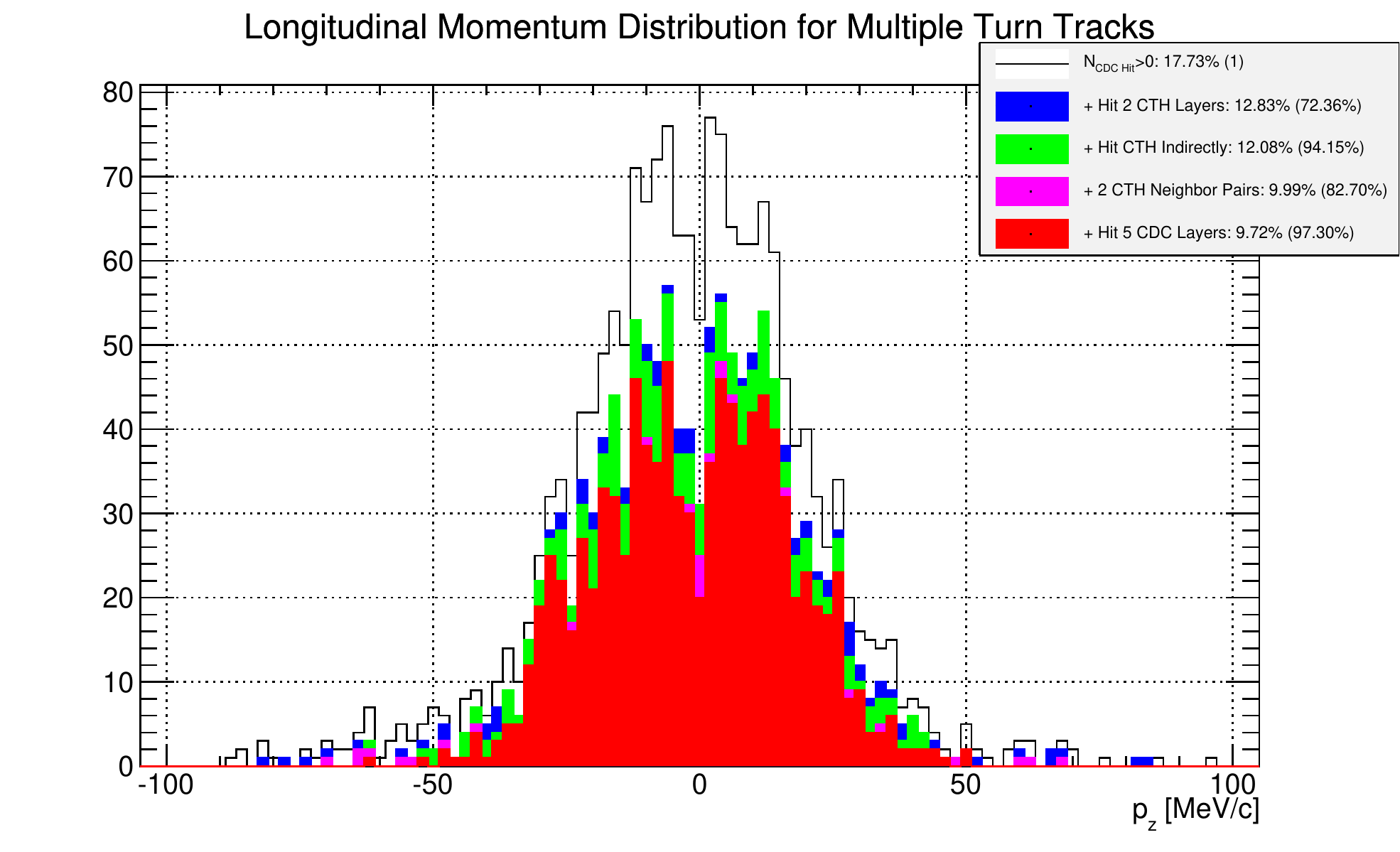}
 \end{center}
 \caption{%
Left: Distribution of longitudinal momentum ($P_Z$) of single turn tracks. Blue histogram is the tracks with two-fold coincidence of CTH, and magenta histogram is those with four-fold coincidence of CTH. Right: the same for multiple turn tracks.}
 \label{fig:ptacceptance}
\end{figure}

The geometrical acceptance is determined by the dimensions and positions of both the CDC and the CTH systems and the configuration of the magnetic field.  \Cref{fig:ptacceptance} shows the  longitudinal momenta ($P_L$) distributions for the tracks which enter the CDC (open histogram), and in addition those which make two-fold (blue histogram) and four-fold (magenta histogram) coincidence in the CTH.  
The tracks are generated isotropically at the muon stopping target. The acceptance of tracks reaching the CTH after a single turn (shown in \cref{fig:ptacceptance} (left) and after multiple turns (shown in \cref{fig:ptacceptance} (right)) are 0.21 and 0.13, respectively.
If a coincidence of hits in the CTH is required, the acceptance is further reduced. For a four-fold coincidence  the acceptance of single turn tracks and multiple turn tracks becomes 0.16 and 0.10 respectively, giving an overall value of 0.26.

\paragraph{Track quality cuts}\label{sec:trackqualitycuts}

In the tracking, the following requirements are made
 to ensure that only high quality tracks are considered:
\begin{itemize}
\item tracks must reach the 5th sense layer ({\tt NL5}),
\item at least one whole turn in the CDC is required ({\tt NHIT}),
\item the number of degrees of freedom must be greater than 30 ({\tt NDF30}),
\item the normalised $\chi^2$ must be less than two ({\tt $\chi^2$}), and
\item hits are required in more than three consecutive layers at both the entrance and exit points of the tracks ({\tt CL3}).
\end{itemize}
The breakdown of the tracking quality cuts are given separately for single-turn tracks and multiple-turn tracks in \cref{tbl:trackingbreakdown}.

\begin{table}[tbh!]
	%\vspace{3mm}
  \caption{Breakdown of the tracking quality cuts, together with the geometrical acceptance, separately for single turn and multiple turn tracks. The acceptance is normalised to all the signal tracks generated and emitted isotropically from the muon stopping target.}
	\begin{center}
		\begin{tabular}{lccc}\hline\hline
		& single turn tracks & multiple-turn tracks & single + multiple \cr\hline
		Geometrical & 0.16 & 0.10 & 0.26 \cr
		{\tt NL5} & 0.78 & 0.98 \cr
		{\tt NHIT} + {\tt NDF30} + {\tt $\chi^2$} + {\tt CL3} & 0.91 & 0.73  & \cr\hline
		total & 0.11 & 0.072 & 0.18 \cr\hline\hline
		\end{tabular}
	\label{tbl:trackingbreakdown}
	\end{center}
\end{table}

\begin{figure}[hbt!]
 \begin{center}
 \includegraphics[width=0.47\textwidth]{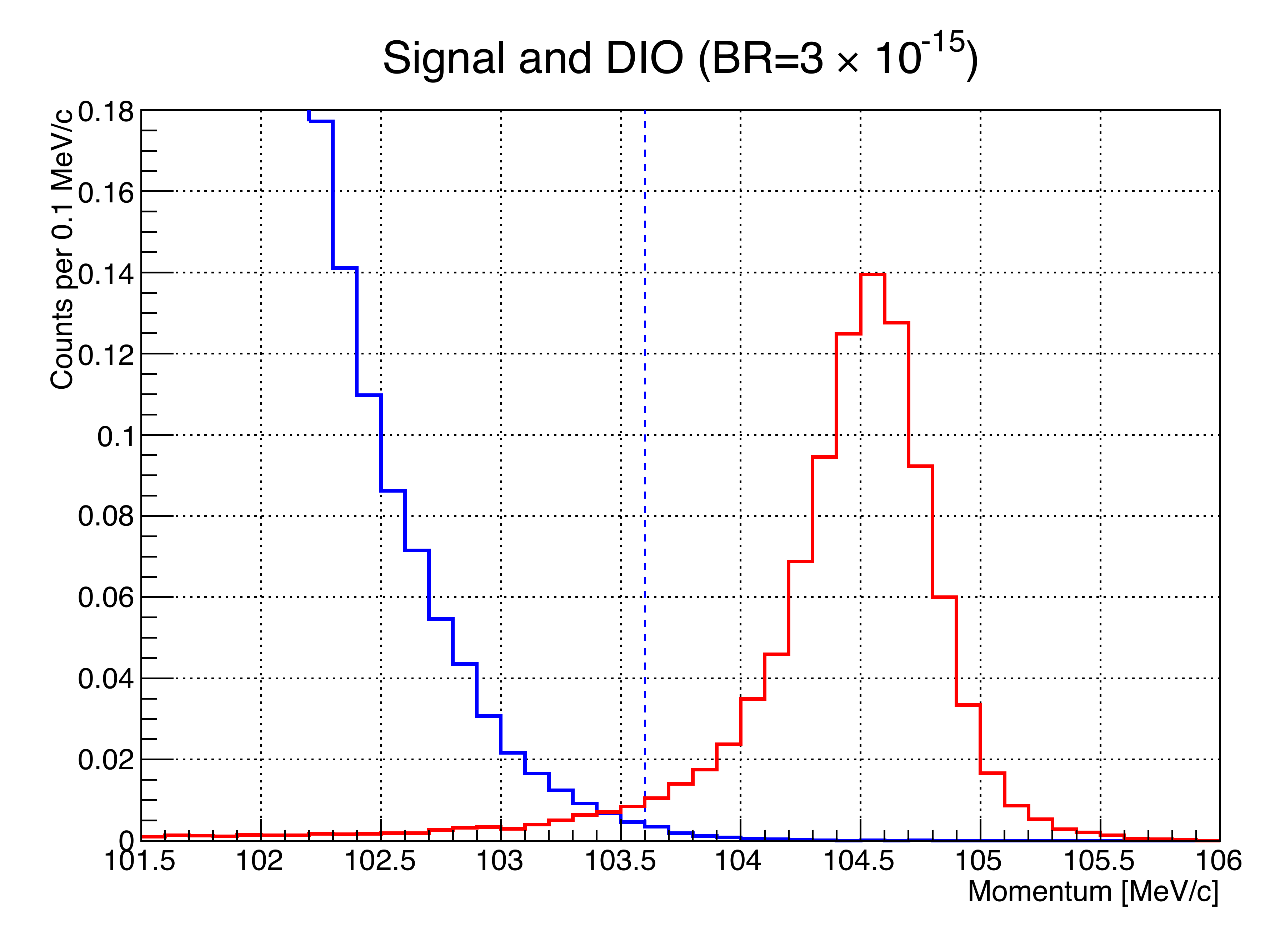}
 \includegraphics[width=0.51\textwidth]{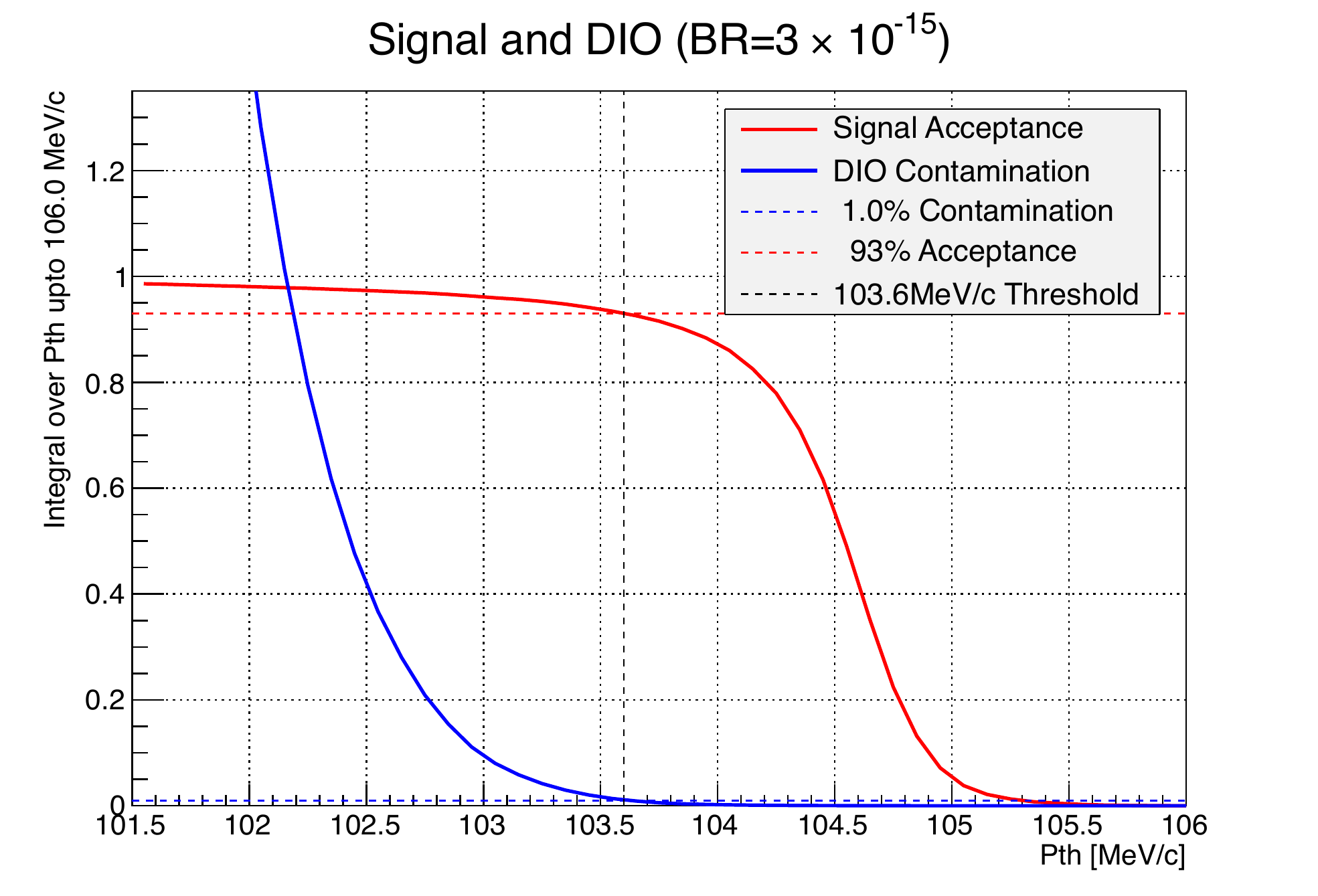}
 \end{center}
 \caption{Left: The momentum distributions for the reconstructed \mue
   conversion signals and reconstructed DIO events.  The vertical scale is
   normalised such that the integral of the signal curve is equal to one event.
   This assumes a branching ratio of $B(\mu N\rightarrow eN) = 3.1 \times
   10^{-15}$.  Right: The integrated fractions of the \mue conversion signals
   and DIO events as a function of the lower bound of the integration range. The
   momentum window for signals is selected to be from 103.6 MeV/$c$ to 106
   MeV/$c$, yielding a signal acceptance of 0.93.}
 \label{fig:signaldiomomentum}
\end{figure}

\paragraph{Signal momentum window}

A momentum cut is used to reduce contaminations from background events such as DIO electrons.
\Cref{fig:signaldiomomentum} shows the simulated momentum spectra for the \mue
conversion signal events and DIO electrons.
In \cref{fig:signaldiomomentum}, the vertical scale is normalised so that
the integral of the signal event curve is one event at a branching
ratio of $B(\mu N\rightarrow eN) = 3.1 \times 10^{-15}$. Using a momentum window of $103.6~{\rm MeV}/c < P_{e} < 106.0~{\rm MeV}/c$, as  shown in
\cref{fig:signaldiomomentum} results in a signal acceptance of $\varepsilon_{\rm mom} = 0.93$ is obtained for a SES of
$3.1 \times 10^{-15}$. An estimate of the contamination from DIO electrons is presented in
\cref{sec:decayinorbitbackground}.

\paragraph{Signal time window}\label{sec:signaltimewindow}
Muons stopped in  aluminium  have a mean lifetime of
864~ns and so \mue conversion electrons are detected between the
proton pulses to avoid the beam-related backgrounds.  The time window  is currently chosen to start at 700 ~ns after the
prompt timing but will be subsequently %As discussed in \cref{sec:backgrounds},
 optimized.

\begin{figure}[htp]
\begin{center}
\subfigure[][Efficiency versus the start time of the time window with the stop time fixed at 1170~ns.]{
	\includegraphics[width=0.45\textwidth]{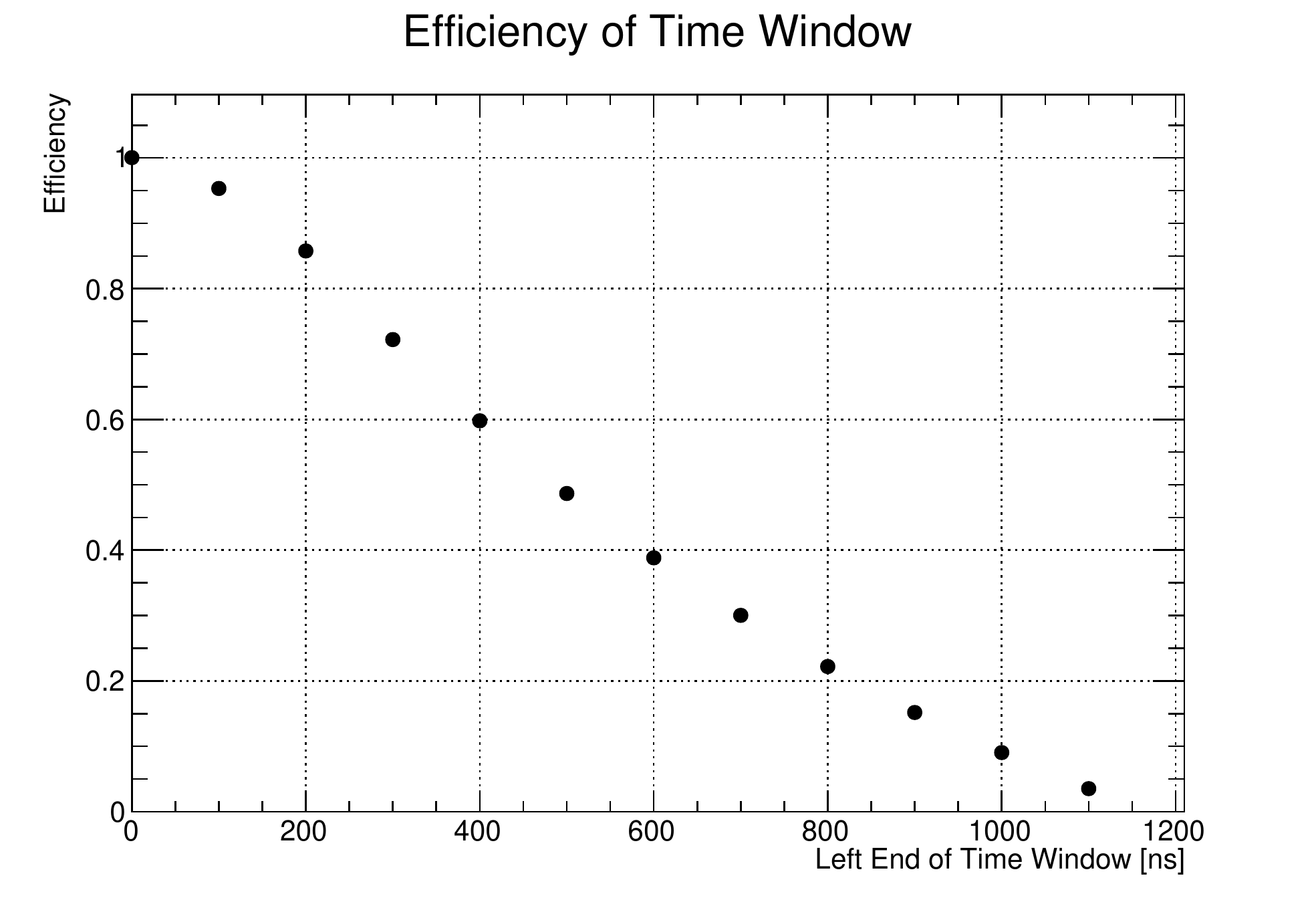}
	\label{fig:cutmomentumtime:left}
}
\subfigure[][Efficiency versus the stop time of the time window with the start time fixed at 700~ns.]{
	\includegraphics[width=0.45\textwidth]{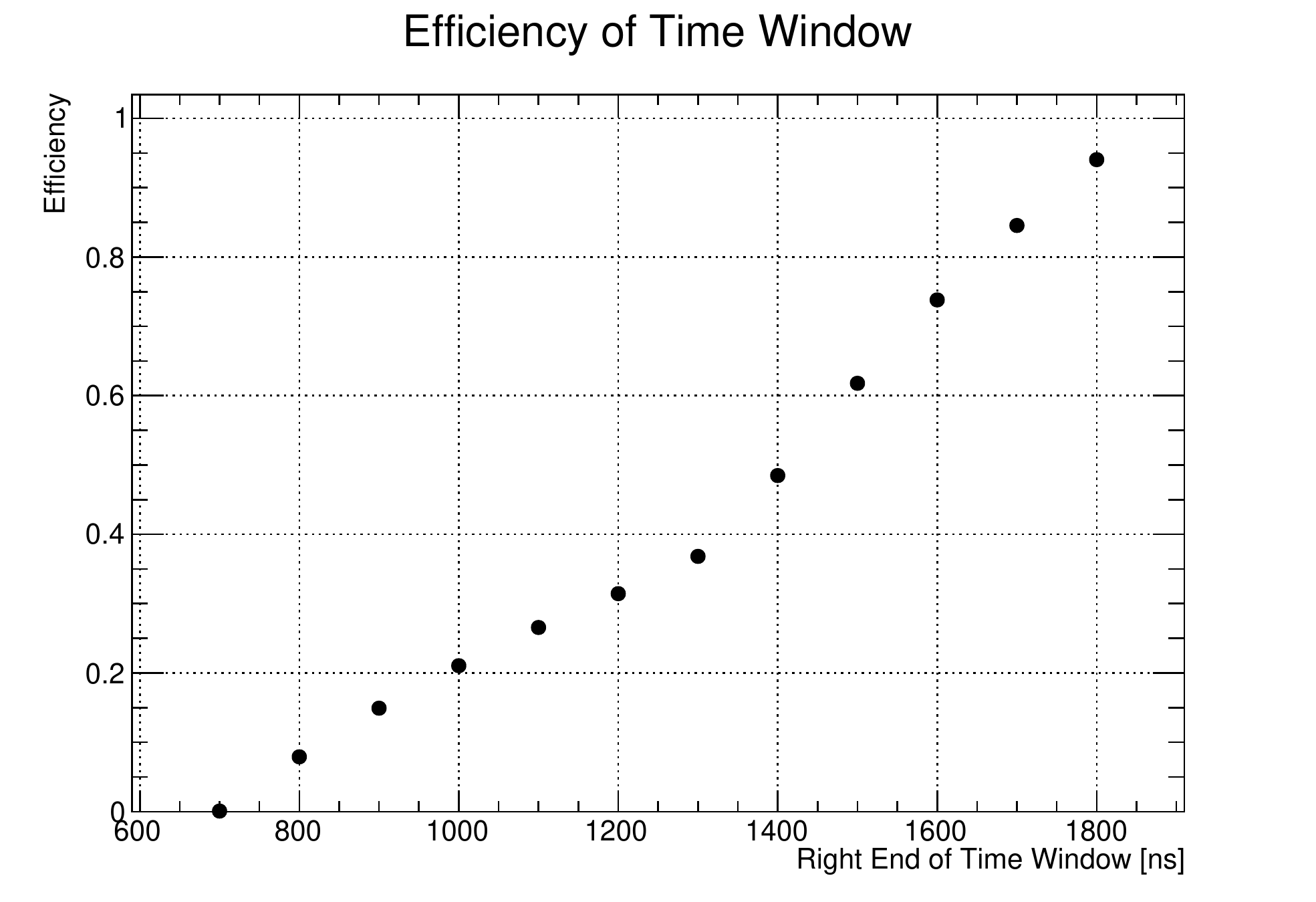}
	\label{fig:cutmomentumtime:right}
}~
\end{center}
\caption{Acceptances of the time window of measurement of the time window as a function of (a) start time and (b) stop time. The width of the proton pulses of 100~ns is included. The periodic time structure is considered with bunch separation time $T_\text{sep}=1170$~ns.}
 \label{fig:cutmomentumtime}
\end{figure}

The acceptance due to the time window cut is shown in \cref{fig:cutmomentumtime} for a varying start time $T_{1}$ and a fixed stop time of $T_2=1170$~ns (left) and for a fixed $T_1=$ 700~ns  and a variable $T_{2}$ (right).  Both assume a pulse separation $T_\text{sep}$ of 1170~ns.  Currently, the baseline design is that $T_1=700$~ns and $t_2=1170$~ns, and $T_\text{sep}$ is 1170~ns.  The signal acceptance resulting from the time window is $\varepsilon_{\rm time}=0.30$.

\paragraph{Net signal acceptance \& single event sensitivity}

The SES is given by:
\begin{eqnarray}
B(\mu^- + \mbox{Al} \rightarrow e^- + \mbox{Al}) & = &
\frac{1}{N_\mu \cdot f_{\mbox{\scriptsize cap}} \cdot f_{\mbox{\scriptsize gnd}} \cdot A_{\mbox{\scriptsize $\mu$-$e$}}},\label{eq:singleevent}
\end{eqnarray}
where $N_\mu$ is the number of muons stopped in the target.
The fraction of captured muons to total muons on target $f_{\mbox{\scriptsize cap}}=0.61$ is taken, while
the fraction of \mue conversion to the ground state in the final state of $f_{\mbox{\scriptsize gnd}}=0.9$ is
taken~\cite{Oset:1993iw}.
$A_{\mbox{\scriptsize $\mu$-$e$}}=0.041$ is the net signal acceptance.
The contributing factors to the overall acceptance are shown in \cref{tb:SignalAcceptance}.
\begin{table}[tbh!]
      \caption{Factors contributing to the \mue conversion signal acceptance value.} \label{tb:SignalAcceptance}
  \begin{center}
    \begin{tabular}{lcc} \hline\hline
      Event selection                  & Value & Comments \cr\hline
       Online event selection efficiency               & 0.9\phantom{1}   & \cref{sec:trigger:algorithm} \cr%
             DAQ efficiency                   & 0.9\phantom{1}   &  \cr%\hline%\hline
      Track finding efficiency & 0.99 & \cref{sec:trackfinding}  \cr%
     Geometrical acceptance + Track quality cuts               & 0.18  & \cr
      Momentum window ($\varepsilon_{\rm mom}$)                  & 0.93  & 103.6~MeV/$c$ $<P_{e}<$106.0~MeV/$c$ \cr%\hline
      Timing window ($\varepsilon_{\rm time}$)                   & 0.3\phantom{1}   & 700~ns~$< t <$ 1170~ns \cr\hline
      Total                            & \phantom{1}0.041 & \cr\hline\hline
    \end{tabular}
  \end{center}
\end{table}
To achieve 
SES of $3 \times 10^{-15}$, %\textcolor{red}{
$N_{\mu}=1.5 \times 10^{16}$ is needed.
By using the muon yield per proton of $4.7 \times 10^{-4}$
a total number of protons on target (POT) of $3.2 \times 10^{19}$ is needed.
With a proton beam current of 0.4 $\mu$A, the measurement requires about 146 days although there are considerable uncertainties such as the pion production yield.

\subsection{Intrinsic Physics Backgrounds} \label{sec:IntrinsicPhysicsBackground}

Negative muons stopped in material  form a muonic atom and then cascade down to the 1$s$ orbit. From there the fate of the bound $\mu^-$ is dominated by two (Standard Model) allowed processes, muon decay in orbit (DIO), and nuclear muon capture (NMC).

\paragraph{Muon decays in orbit (DIO)}\label{sec:decayinorbitbackground}

In a free muon decay the electron momentum must be balanced against that of the neutrinos but in DIO the nuclear
recoil from the Michel decay allows the electron to carry much more energy. This causes the maximum energy of
the $e^{-}$ to exceed the end point energy of the free Michel decay at rest (52.8~MeV), extending it to the momentum range of the \mue conversion signal. The endpoint energy of DIO occurs when the neutrinos are produced at rest and  can be very close to the \mue conversion signal energy $E_{\mu e}$.

\begin{figure}[htb!]
  \begin{center}
  \includegraphics[width=0.45\textwidth]{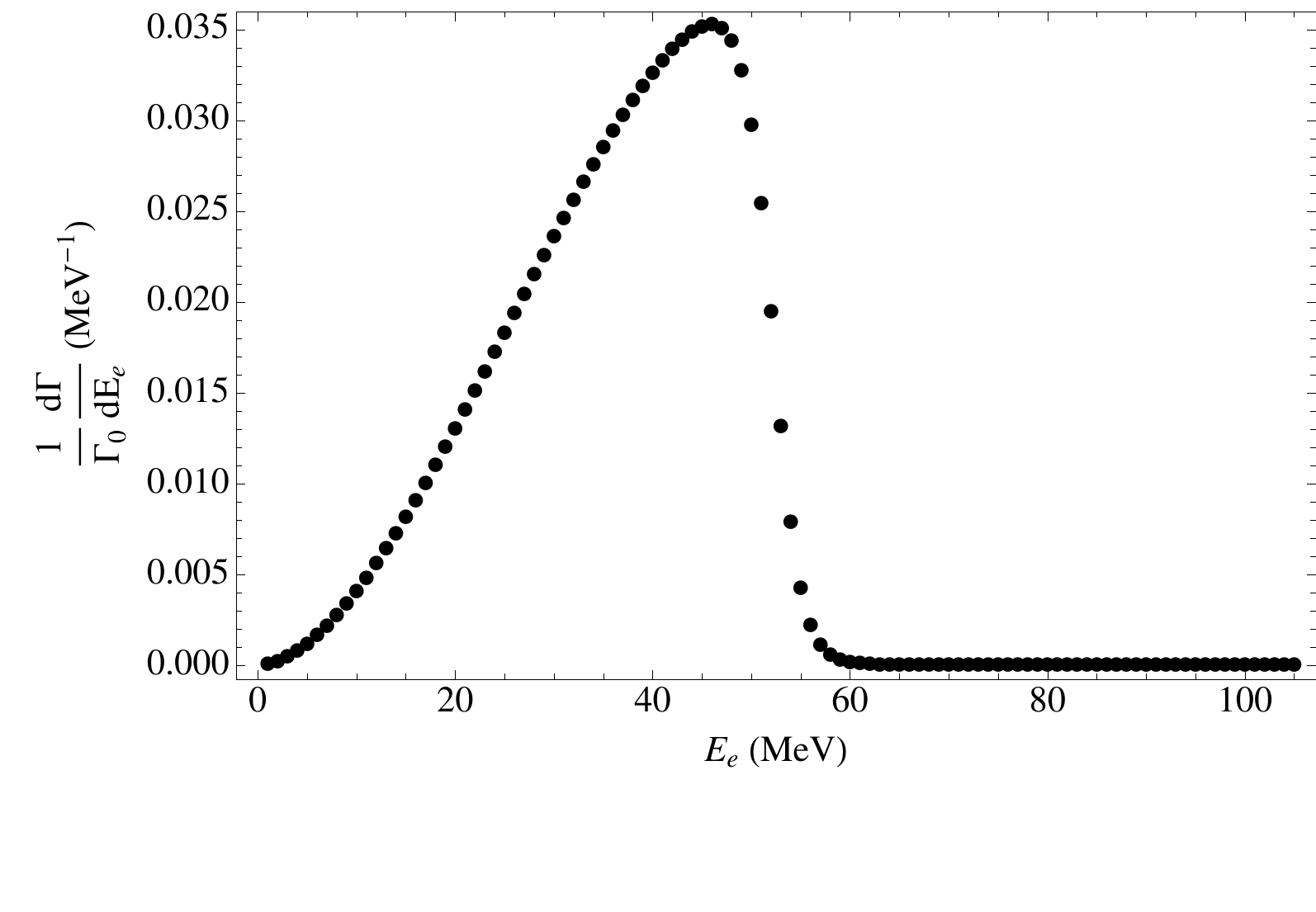}
  \includegraphics[width=0.45\textwidth]{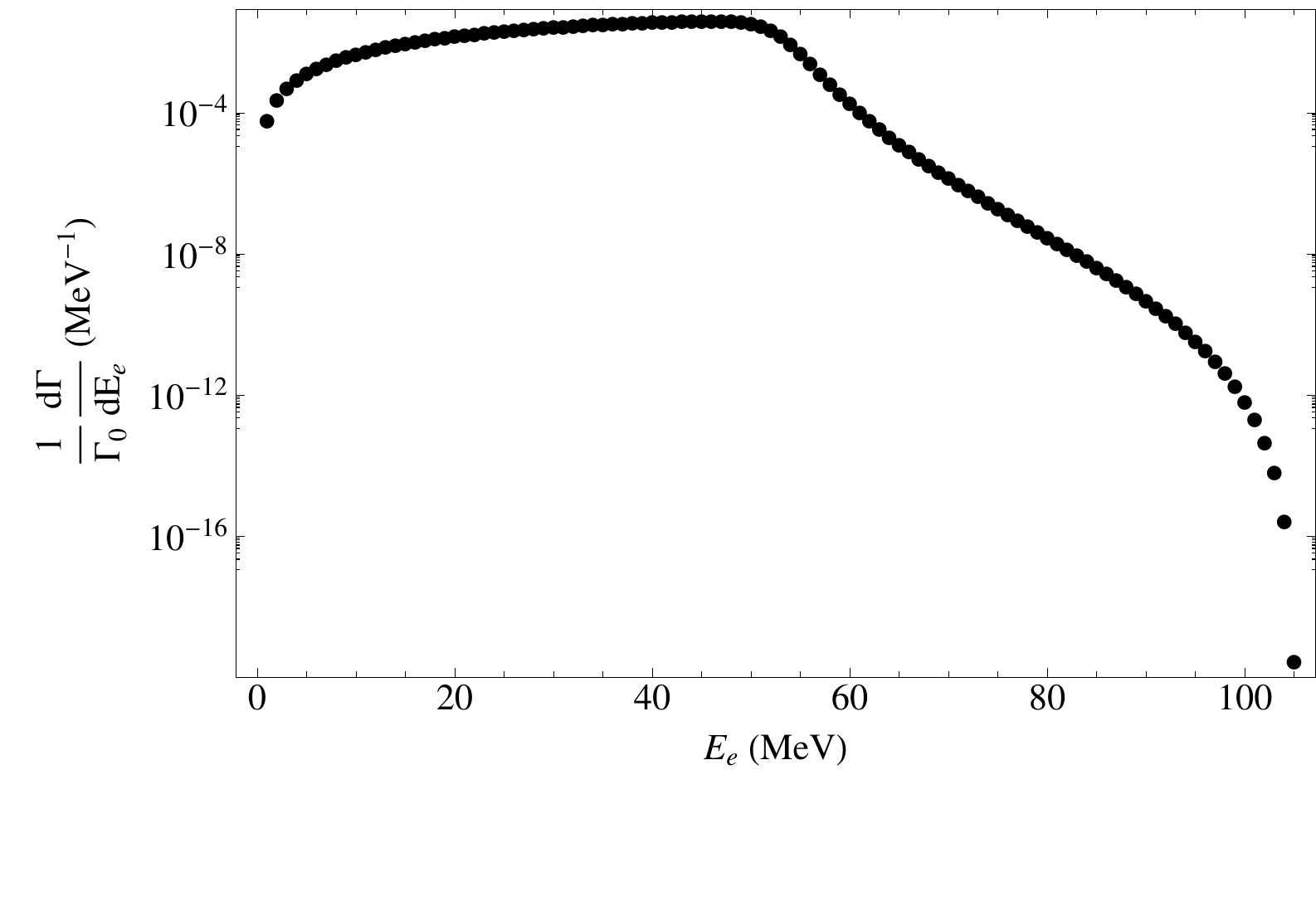}
  \end{center}
  \caption{DIO electron spectrum for aluminium. The left is linear scale and the right is a logarithmic scale. From Reference~\cite{czarnecki11}. }
  \label{fig:diospectrum}
\end {figure}

\begin{figure}[htb!]
  \begin{center}
  \includegraphics[width=0.8\textwidth]{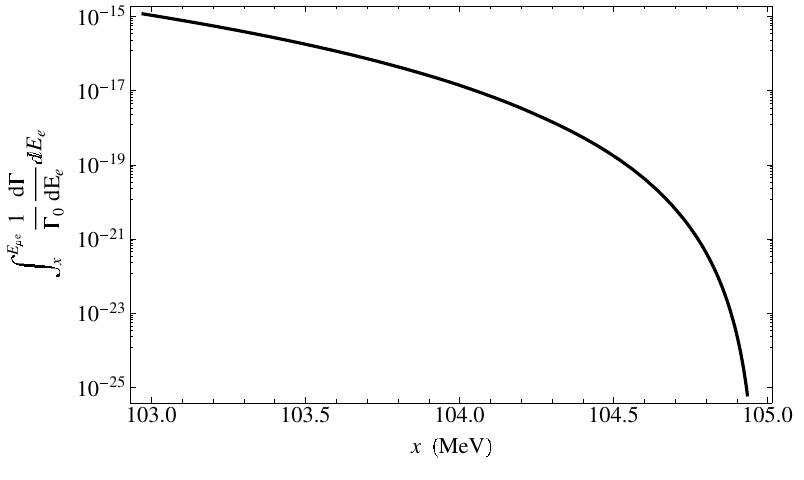}
  \end{center}
  \caption{Total rate of DIO for aluminium above the electron energy $x=E_e$\,(MeV) value, normalised by the free muon decay rate from Reference~\cite{czarnecki11}.}
  \label{fig:diototalrate}
\end {figure}

The momentum spectrum of DIO electrons for aluminium has been calculated based on
the model described in References~\cite{czarnecki11, Czarnecki:2014cxa, Szafron:2015mxa, Szafron:2015kja}.
\Cref{fig:diospectrum} shows the momentum spectrum of DIO electrons from aluminium and \cref{fig:diototalrate} shows the proportion of the aluminium DIO spectrum  with energy above $x$ (MeV)~\cite{czarnecki11}. Hence to reduce the DIO contribution down to $\mathcal{O}(10^{-16})$, the lower side of the momentum region for \mue conversion signals should be above about 103.6~MeV.

\begin{figure}[htb!]
  \begin{center}
  \includegraphics[trim={48pt 0 0 115pt},clip,width=0.8\textwidth]{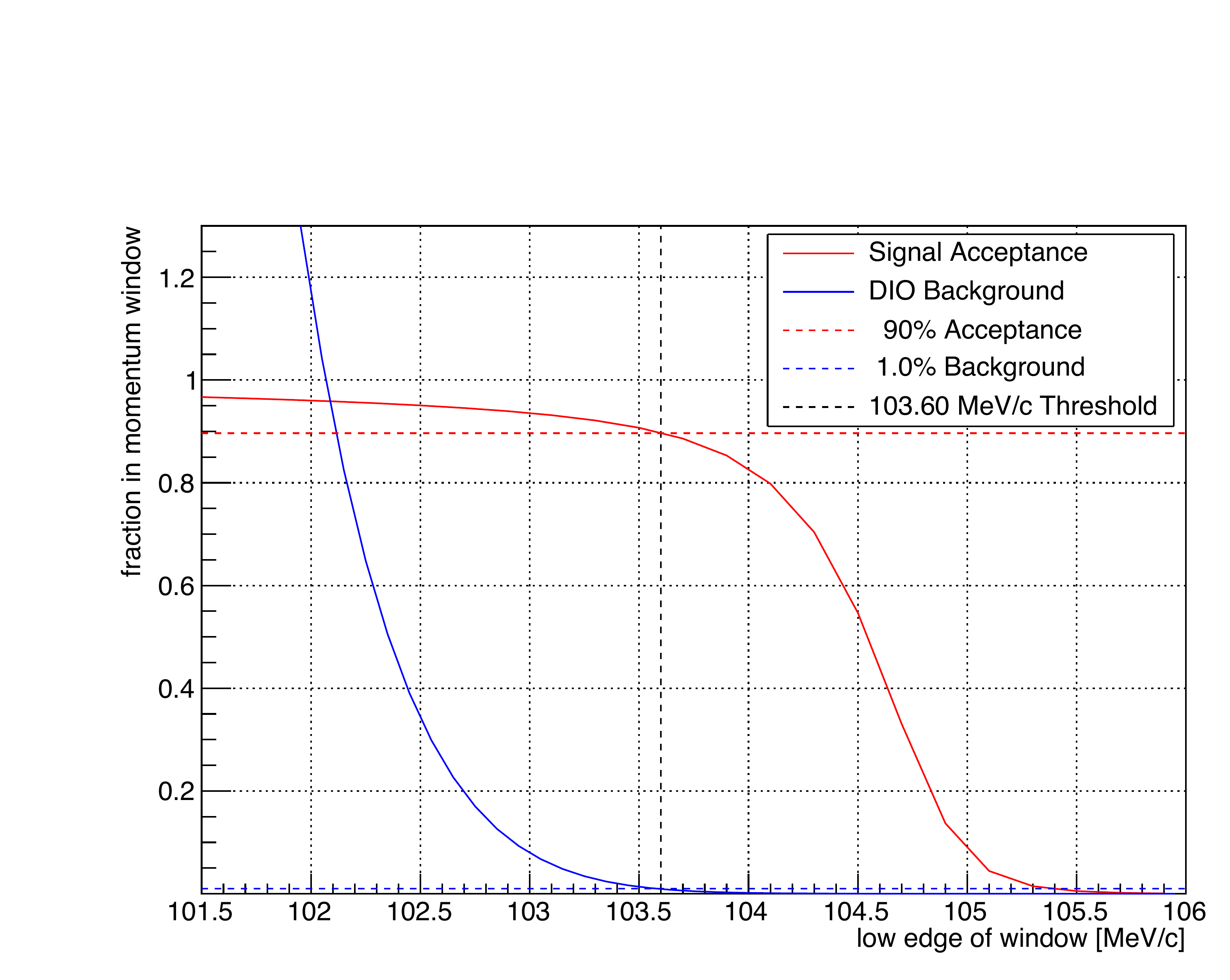}
  \end{center}
  \caption{Reconstructed DIO spectrum for aluminium, normalised to
    one single event of \mue conversion with $3 \times 10^{-15}$. The red
    line shows the integrated event rate above the energy given. The
    lower edge of the momentum window of the signal is set to be 103.6
    MeV.}
  \label{fig:dio-momentumwindow}
\end {figure}

In \cref{fig:dio-momentumwindow},
the reconstructed momentum spectrum of DIO electrons is shown (blue
line), normalised to the rate of a single \mue conversion event at a
branching fraction of $3 \times 10^{-15}$.
For a momentum window of $103.6~{\rm MeV}/c < P_{e} < 106~{\rm
  MeV}/c$ for the \mue conversion signals
the fraction of DIO electrons in the signal region is 0.01 events
for an SES of $3
\times 10^{-15}$.

\paragraph{Radiative muon capture (RMC)}

Radiative muon capture is given by

\begin{equation}
\mu^{-} + N(A,Z) \rightarrow \nu_{\mu} + N(A,Z-1) + \gamma.
\end{equation}

If the emitted $\gamma$-ray is followed by asymmetric $e^+ e^-$ conversion, or Compton scattering of the photon, it forms an important source of intrinsic background. This is referred to as ``external'' RMC. There is also ``internal'' conversion of the (virtual) photon:
\begin{equation}
\mu^{-} + N(A,Z) \rightarrow \nu_{\mu} + N(A,Z-1) + e^{+} + e^{-},
\end{equation}
which can make a similar contribution to the background  when the $e^+ e^-$ are asymmetric.
As is the case with DIO, energy measurement is the only means to combat RMC, hence understanding the   spectrum shape towards the endpoint is most important.

For the external process, the kinematic endpoint ($E^{end}_{rmc}$) of the emitted photon from RMC is given by,
\begin{equation}
E^\mathrm{end}_\text{rmc} \approx m_{\mu} - B_{\mu} - \Delta_{Z-1} - E_{recoil} = E_{\mu e} - \Delta_{Z-1},
\label{eq:rmc}
\end{equation}
where $m_{\mu}$ is the muon mass, $B_{\mu}$ is the muon binding energy in a muonic atom, and $E_{recoil}$ is the recoiling energy of the final nucleus. $\Delta_{Z-1}$ is the difference in the mass of the initial $(A,Z)$ and final $(A,Z-1)$ nucleus involved in RMC.   In aluminium the nuclear mass difference, $\Delta_{Z-1}= +3.12$~MeV, and the RMC photon endpoint energy is 101.85~MeV.

If the photon from RMC produces an electron by  Compton scattering, the maximum momentum of the emitted electron  is  $m_e/2 = 0.255$~MeV larger than the original photon momentum. For pair production the maximum momentum of the electron is about $m_e$ smaller than the original photon momentum. Therefore, Compton scattering is more important than external pair production or RMC with internal conversion.

The total background contribution from RMC for a single signal event
is given by
\begin{eqnarray}
N_\text{RMC} & = & N_\text{proton}
\times R_{\mu-\mathrm{stop}/p}
\times B_\text{RMC93}  \nonumber \\
& & \times P_{\gamma-e}
\times A_\text{geo}
\times A_\text{mom}
\times A_\text{time}
\times \varepsilon_\text{tracking}\,, \label{eq:rmcbackgrounds}
\end{eqnarray}
where
$N_\text{proton}$ is the total number of protons on the pion production target;
$R_{\mu-\mathrm{stop}/p}$ is the number of $\mu^{-}$ arriving at the muon stopping target per proton;
$B_\text{RMC93}$ is the branching ratio of RMC producing a photon with more than 93~MeV;
$P_{\gamma-e}$ is the probability of conversion of the RMC photon to an electron in the signal region;
$A_\text{geo}$ is the detector acceptance of the RMC-originated electrons in the signal region;
$A_\text{mom}$ and $A_\text{time}$ are the acceptances of momentum cut and timing cut, respectively; and $\varepsilon_{tracking}$ is the tracking efficiency.

\begin{figure}[hbt!]
 \begin{center}
   \includegraphics[width=0.8\textwidth]{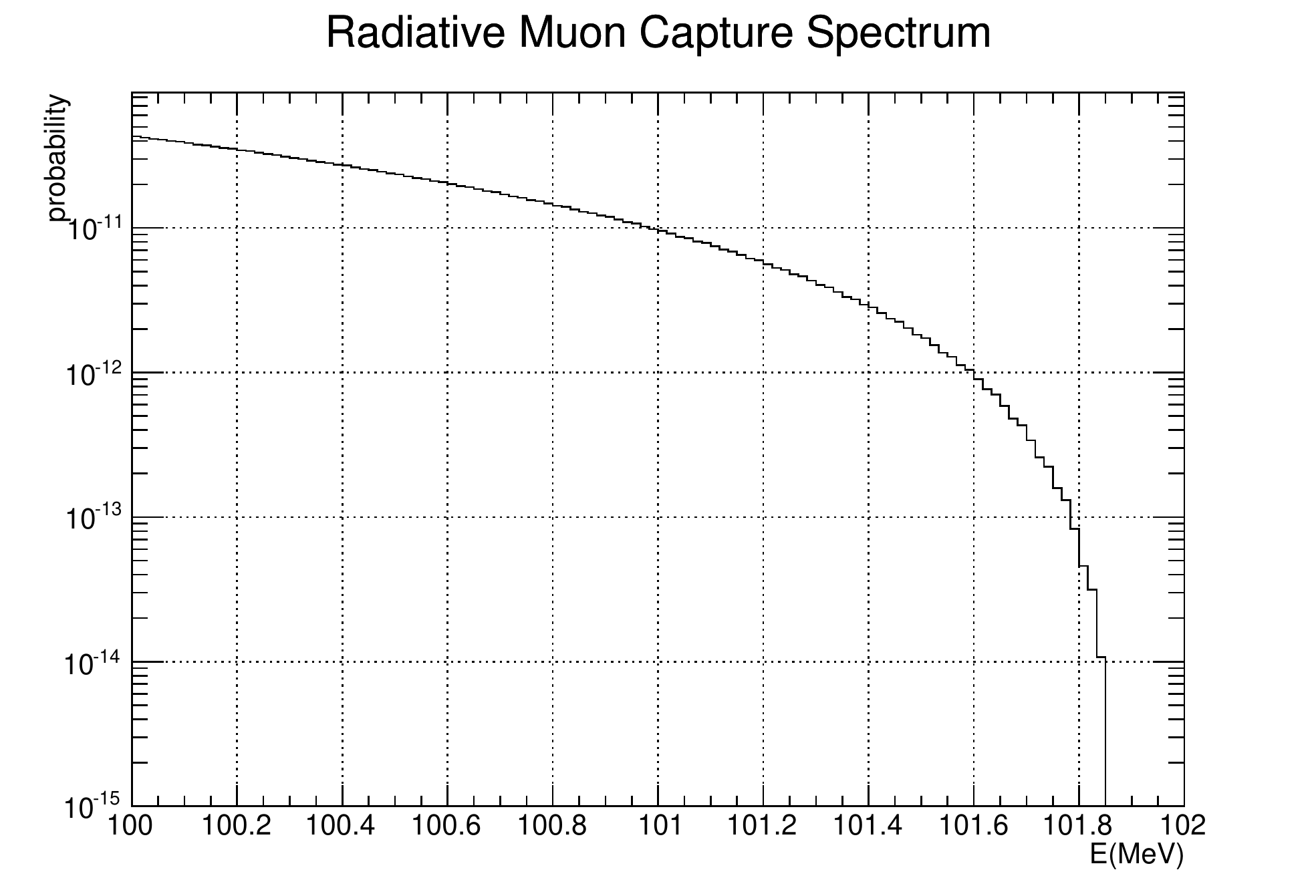}
 \end{center}
 \caption{
Extrapolated momentum distribution of photons from RMC, based on theoretical prediction~\cite{Hwang:1980rm}.
}
 \label{fig:rmc_spectrum}
\end{figure}

\begin{figure}[hbt!]
 \begin{center}
   \includegraphics[width=0.45\textwidth]{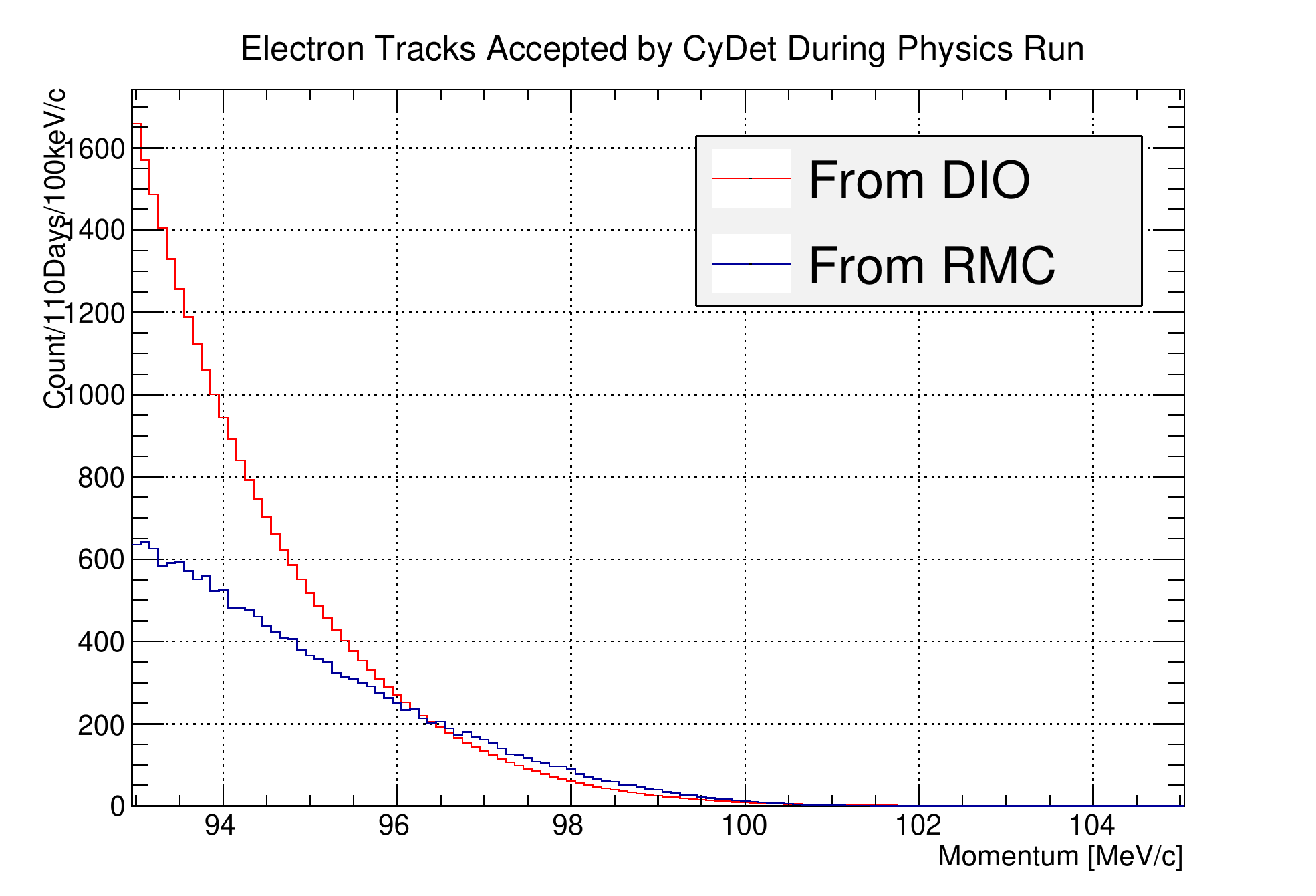}
   \includegraphics[width=0.45\textwidth]{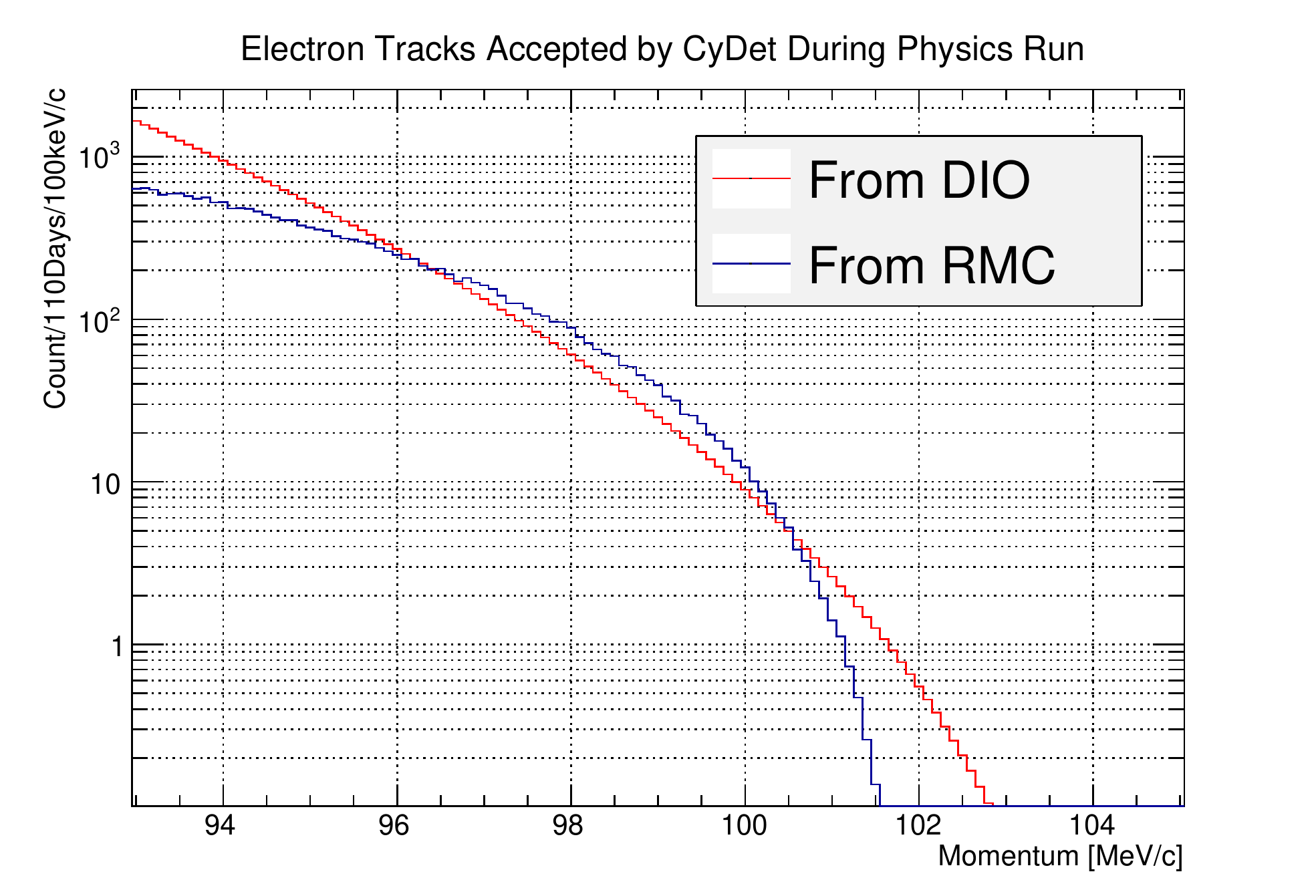}
 \end{center}
 \caption{
 Comparison of the rates and spectra between RMC and DIO, linear (left) and log (right) scales. }
 \label{fig:rmcVSdio}
\end{figure}

No experimental data of the photon spectrum from RMC on
aluminium near the endpoint is available and so
theoretical predictions
must be used to extrapolate to the endpoint. Following  Hwang~\cite{Hwang:1980rm}, the spectrum
based on  Hwang-Primakoff theory  is given by
\begin{equation}
R(x) = C (1-2x+2x^2)x(1-x)^2,
\label{eq:rmc-theory}
\end{equation}
where $x=k/k_\text{max}$, $k$ is the photon energy and $C$ is a
constant. For the overall normalisation, $C$, we make use of the measured rates
of RMC on aluminium  from Reference~\cite{Measday:2001yr}.  The
result of this extrapolation is shown in
\cref{fig:rmc_spectrum}. From this, the probability
per muon capture of producing a photon with energy exceeding 93~MeV
is estimated to be $B_\text{RMC93}= 2.97 \times 10^{-7}$.
One aim of COMET Phase-I will be to measure the RMC photon spectrum on aluminium.
The prediction of the electron
spectrum resulting from RMC is shown in \cref{fig:rmcVSdio}, along with
the DIO spectrum. This gives the number of
RMC backgrounds, $N_\text{RMC}=0.0019$, in the momentum window of the signal

\subsection{Beam-Related Prompt Backgrounds}

The beam-induced prompt backgrounds arise from protons circulating in
the MR buckets between the intended beam pulses. They are
suppressed by the proton beam extinction factor, $R_\text{extinction}$, which, in this background estimation, is assumed to be $3 \times 10^{-11}$ from  recent experimental measurements,
as given in \cref{sec:accelerationtest}.

\paragraph{Radiative pion capture (RPC)}
Pions
contaminating the muon beam can be  captured by an aluminium nucleus in the target  to
form an excited state of the daughter nucleus. As with RMC, there are both the external and internal conversion mechanisms which can produce the background electron events.

 According to~\cite{amaro97}, the probability
of $\gamma$ emission has a very small $Z$ dependence, being about 2\%
for C, O, and Ca,  with the energy of
the $\gamma$ ranging from 50~MeV to 140~MeV. The overall shapes of the spectra are also very
similar and so the experimentally
obtained spectrum from Ca was used for the RPC simulation.

The number of RPC backgrounds is expressed as
\begin{eqnarray}
N_\text{RPC} & = & N_{\rm proton} \times R_\text{extinction}
\times R_{\pi-\mathrm{stop}/p}
\times \nonumber \\
& & B_\text{RPC}
\times P_{\gamma-e}
\times A_\text{geo}
\times A_\text{mom}
\times A_\text{time}
\times \varepsilon_\text{tracking}\,, \label{eq:rpc}
\end{eqnarray}
where
$N_\text{proton}$ is the total number of protons on the pion production target;
$R_\text{extinction}$ is the proton beam extinction factor;
$R_{\pi-\mathrm{stop}/p}$ is the number of $\pi^{-}$s arriving at the muon stopping target per proton;
$B_\text{RPC}$ is the branching ratio of radiative pion capture;
$P_{\gamma-e}$ is the probability of conversion of the RPC photon to an electron of 105 MeV/$c$;
$A_\text{geo}$ is the detector acceptance of the RPC-originated electrons of 105 MeV/$c$;
$A_\text{mom}$ and $A_\text{time}$ are the acceptances of momentum cut and timing cut, respectively; and $\varepsilon_\text{tracking}$ is the tracking efficiency.

With $3 \times 10^{19}$
protons on target, a total of $1.4 \times 10^{-3}$ background events
from the external conversion of radiative pion capture is
predicted. The contribution from internal conversion is about the same
and therefore, an expectation of $2.8
\times 10^{-3}$ RPC events is estimated with a proton beam extinction
factor of $3 \times 10^{-11}$.

\paragraph{Beam electrons, electrons from muon and pion decays in flight }

Electron contamination of the muon beam can arise from $\gamma$ conversion following $\pi^0$ decays and the decays of muons and pions in flight. For the decay electrons to have
$p_\text{total}>102$~MeV/$c$, the muon momentum ($p_\mu$) must
exceed 77~MeV/$c$ and the $\pi$ momentum must exceed 60~MeV/$c$.

From simulations,
 the total number of electrons with momenta
greater than 80~MeV/$c$ after the beam collimator is $R_{e\mathrm{-beam}/p} = 1.7 \times
10^{-5}$ per proton. As the electron also needs the transverse momentum, $P_T$, to be greater than 70 MeV/$c$ to reach the
CDC\@.
Out of 40,000 electrons in the simulation none  reached
the CDC, and therefore  an upper limit estimate of the  background from beam electrons is less than $3.8 \times 10^{-3}$.

\paragraph{Background induced by beam neutrons}

Background events could be induced by high energy beam
neutrons which pass through the muon beam line by
continuously reflecting from the inner sides of the beam duct. Simulations predict the average transit time of the neutrons which arrive
at the  stopping target is around 300~ns, with
far fewer arriving at the signal window start time of 700~ns.
Therefore, this background is regarded as a prompt background.

The dominant process to produce a
100~MeV electron is $\pi^0$ production from energetic neutrons,
followed by $\pi^0$ decay and photon conversion.

The prompt background rate $N_\text{neutron}$ can be
estimated by
\begin{eqnarray}
N_\text{neutron} & = & N_{\rm proton} \times R_\text{extinction}
\times R_{n/p}
\times R_{\pi^0/n}
\times R_{e/\pi^0}
\label{eq:nbg}
\end{eqnarray}
and the ICEDUST simulation yields $1 \times 10^{-9}$, so the neutron background through
$\pi^0$s is expected to be negligible.

\begin{figure}[htb!]
\begin{center}
\includegraphics[width=\textwidth]{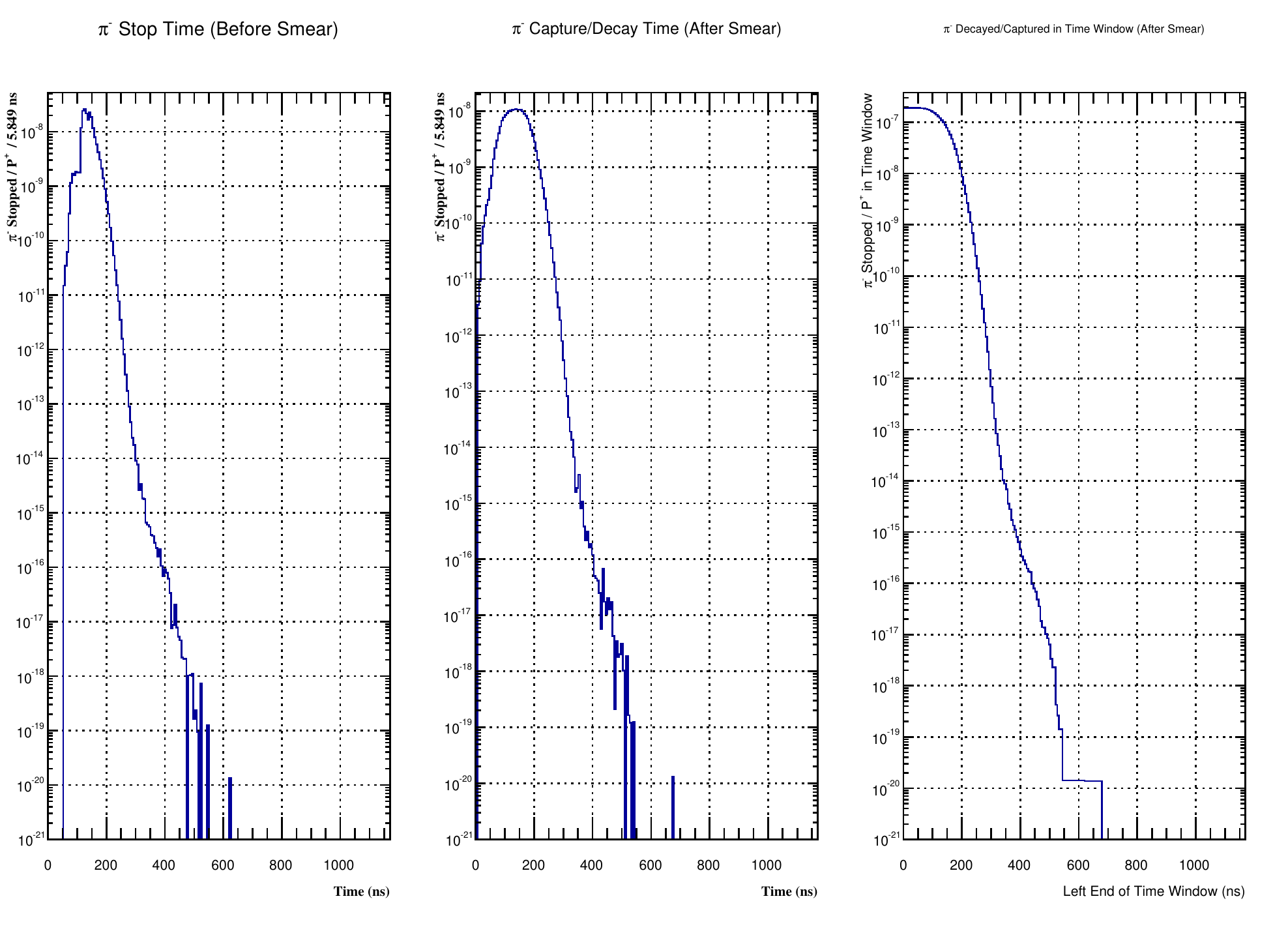}
\end{center}
\caption{(Left) Pion
  arrival times when the incoming proton is at time $0$. \textrm{(Middle)} Pion arrival times when
  the incoming proton pulse is assumed to have a 100~ns square time distribution.
  (Right) Pion survival rate as a function of the
  starting time of the time window of measurement.}
\label{fig:BG-PionSurvive}
\end{figure}

\subsection{Beam-Related Delayed Backgrounds}
The beam-related delayed backgrounds arise from slowly-arriving particles in the
muon beam line. In general they traverse through the solenoids with a small pitch angle in their
helical trajectories (namely a small $P_L$), and thereby arrive late.
They have specific time distribution to their arrival and they are suppressed by the use of the delayed time window for
measurement.
\paragraph{Beam-related delayed pion backgrounds}

It is critical to avoid delayed pions in the
beam, since pion decays can easily produce 100~MeV/$c$ electrons.

\Cref{fig:BG-PionSurvive} (Left) shows the distribution of pion arrival times relative to the time of the primary proton, and \cref{fig:BG-PionSurvive} (Middle) is the distribution when the primary proton time profile is assumed to be a square pulse of 100~ns duration. \Cref{fig:BG-PionSurvive} (Right) shows the integrated pion survival rate as a function of the starting time of the signal window. When the starting time is set to 700 ns,
a pion survival rate of $10^{-21}$
pions/proton is achieved, which is lower than the proton
extinction rate $R_{\rm extinction}=3\times 10^{-11}$. Therefore, the
beam-related delayed backgrounds are expected to be below the level of the
prompt backgrounds. The best signal time window will be
determined after some initial running to measure the time
distribution of pion arrival.

\paragraph{Antiproton-induced delayed backgrounds}

Low energy antiprotons produced in the proton target  can pass through the muon beam line and annihilate on materials in the detector region, producing other energetic particles, leading directly or indirectly to 100~MeV/$c$ electrons. These
antiprotons have very low kinetic energy and low velocity and are therefore
 not suppressed by the delayed time window of measurement.

Two ways to suppress the antiproton-induced backgrounds are
\begin{itemize}
\item Decreasing the proton beam energy
\item Adding a thin absorber material  in the muon beam line.
\end{itemize}

The choice of 8~GeV proton energy
 is specifically to reduce the production rate of antiprotons,
with keeping reasonably high rates of pion production. %(although it is
However  the yield of antiprotons in the backward direction from an 8~GeV
proton beam is not known, so an estimate has been made from predictions made by the MECO experiment 
using MARS. This yields a rate of 
antiproton production per proton
of $4 \times 10^{-5}$ is obtained.

Consideration is also being given to installing two absorber foils in
the muon beam line. One  would be a 500~\micro{}m thick
titanium foil with a diameter of 360~mm, placed at the entrance
of the curved muon-transport solenoid. It would also serve as a
vacuum window to separate
the muon beam line (in vacuum) and the proton beam line.  The thickness of the titanium
is sufficient to maintain a pressure difference of one atmosphere. 
The second absorber foil would be 500~\micro{}m thick titanium
with a diameter of 255~mm, and  placed at the front part of the
Bridge Solenoid (BS). It would also serve as a vacuum window between the muon
beam line and the detector region, which contains helium at atmospheric pressure.

From simulations, antiproton background contributions of $1.2 \times 10^{-3} (3.5 \times 10^{-3})$ are obtained for measurement time windows from 700 (500) ns to 1170 ns.

\subsection{Cosmic Ray-Induced Backgrounds}\label{sec:cosmicraybackground}

Cosmic ray-induced backgrounds are one of the most important
backgrounds. They can be divided into two categories:
\begin{itemize}
\item cosmic-ray muons that produce an electron which enters the detector, and
\vspace{-1mm}
\item cosmic-ray muons which enter the detector and are misidentified as an electron.
\end{itemize}
To veto and eliminate cosmic ray-induced backgrounds, the Cosmic Ray
Veto (CRV) system is installed to cover a large portion of the solid
angle around the Detector Solenoid (DS).
 The CRV detector works as a veto with very small inefficiency of $10^{-4}$, 
in an environment that has a large flux of
neutrons. The COMET Phase-I detector also has good
particle identification to discriminate electrons from cosmic-ray
muons. Signal tracks are required to hit the CTH, with the Cherenkov
counters serving to actively identify electrons.

In order to study cosmic ray-induced backgrounds, two kinds of
simulation studies are being considered. One is a general approach in
which cosmic rays are generated widely around the COMET experimental
hall. It is useful to examine the overall performance of the CRV and
characteristics of cosmic ray-induced backgrounds. The second is a
focused approach, in which some specific combinations of location and
direction are chosen and cosmic rays are generated in these areas. In
particular the second studies will be made for the locations where
the cosmic ray veto is weak or does not provide complete coverage.

 \texttt{Geant4} simulations are used to
estimate cosmic ray-induced backgrounds.
The data set of cosmic rays is based on a CERN input
file\footnote{More recently, \texttt{CORSIKA} simulations of cosmic rays above
  J-PARC have been obtained from the T2K experiment, but were not
  available in time for this study.} which contains about 23 million
$\mu^{\pm}$ events. A full air shower simulation code based on
\texttt{CORSIKA} was used.  23 million cosmic ray events were
generated over a $50 \times 50 \mathrm{m}^2$ plane.   Among this
sample, there were no events containing electrons of about 100~MeV in the
CDC without being detectable by the CRV. In one event,  a
cosmic ray muon produced a shower and one of the shower electrons
scattered off the BS and entered the CDC and hit the
CTH. However, the electron lost much of its energy  and it would not have been mistaken for a signal
electron.

Additionally in the DS area, the CDC  also serves as an
active volume to detect cosmic ray muons, complementing the CRV. Overall a
net veto inefficiency to identify cosmic rays and/or an associated
shower should be much better than $10^{-4}$.

Additional simulations in the BS area have found 35 events  in which an electron of $85\sim110$~MeV/$c$ reaches the CDC from 100 million generated cosmic ray events. None of them  met the CTH trigger requirements and the track quality
cuts. Nevertheless an additional veto system close to the BS is under consideration. 
With the veto system at BS placed, 
an upper limit of the cosmic background contribution is
obtained to be $\le 0.01$ for the COMET Phase-I physics run.

\subsection{Summary of Background Estimations}

\Cref{tab:SummaryBackgrounds} shows a summary of the estimated
backgrounds. The total estimated background is about 0.032 events for
a single event sensitivity of $3 \times 10^{-15}$ with a proton
extinction factor of $3 \times 10^{-11}$. If the proton extinction
factor is improved, the expected background events will be further
reduced.

\begin{table}[htb!]
  \caption{Summary of the estimated background events for a
    single-event sensitivity of $3 \times 10^{-15}$ in COMET Phase-I
    with a proton extinction factor of $3 \times 10^{-11}$.
  }
  \label{tab:SummaryBackgrounds}
  \centering
    \smallskip
    \begin{tabular}{llr} \hline\hline
      Type & Background & Estimated events \cr \hline
      Physics & Muon decay in orbit & 0.01 \cr %\hline
      & Radiative muon capture & 0.0019 \cr %\hline
      & Neutron emission after muon capture & $<0.001$ \cr %\hline
      & Charged particle emission after muon capture & $<0.001$ \\ \hline
      Prompt Beam & \quad * Beam electrons &  \cr
      & \quad * Muon decay in flight  & \cr
      & \quad * Pion decay in flight  &  \cr
      & \quad * Other beam particles  &  \cr %\hline
      & All (*) Combined & $\le 0.0038$  \cr
      & Radiative pion capture &  0.0028 \cr %\hline
      & Neutrons  & $ \sim 10^{-9}$ \cr \hline
      Delayed Beam & Beam electrons & $\sim 0$ \cr %\hline
      & Muon decay in flight  & $\sim 0$ \cr %\hline
      & Pion decay in flight  &  $\sim 0$ \cr %\hline
      & Radiative pion capture  & $\sim 0$ \cr %\hline
      & Antiproton-induced backgrounds & 0.0012 \cr \hline
      Others & Cosmic rays$^\dagger$ & $<0.01$ \cr \hline
      Total & & 0.032 \cr\hline\hline
    \end{tabular}
    \\ \footnotesize{${\dagger}$ This estimate is currently limited by computing
    resources.}
\end{table}

\section{Run programs}

The COMET Phase-I experiment is the search for the \mue conversion, but
at the same time, it is an intermediate stage before the Phase-II experiment. 
Two special experimental runs will be carried out during the Phase-I experiment,
beam measurement and background assessment, which will be the dedicated measurements
aiming for the preparation of Phase-II experiment, and for the better understanding of
\mue conversion data of the Phase-I experiment
Both run programs will use StrECAL detector with augmented configurations. 

\subsection{Beam Measurement Programs}\label{ch:BeamMeasurement}

There is no measurement on the backward pion production rate 
with 8~GeV protons. The simulation study of various hadron production  codes 
such as \texttt{MARS} and \texttt{Geant4}
\texttt{QGSP(BERT/BIC)} estimate more than two times different rate, as described in
\cref{sec:different_hadroncode}.
In order to understand the pion production rate, in the COMET Phase-I experiment, 
it is planned to measure the muons, pions,
antiprotons and electrons in the beam,
with the StrECAL placed downstream of the muon transport.

During the beam measurement, 
(1) the momentum and profile of beam, and (2) the beam timing should be measured. 
For the momentum and profile measurement, 
the track reconstruction and   the particle identification (PID)  are necessary. 
This also requires lower hit rate and thus lower beam power. 
For the beam timing measurement, 
the goal is to reproduce the timing distribution of particles after the initial beam pulse,
which does not require track reconstruction and
momentum information.
This measurement can be made without the
Detector Solenoid and Straw detector, therefore
beam can be operated at full power.

\subsubsection{Particle Identification   by StrECAL}

While there is no detector dedicated to PID in
COMET, a special StrECAL configuration may be designed to optimize the PID capability, 
by placing a scintillating fibre (Sci-Fi) detector at the end of the muon beam line. 
In this configuration, $\dv{E}{x}$ and $E/p$ can be provided by the ECAL, and 
the time of particle flight (TOF) between ECAL and the scintillating fibre (Sci-Fi) detector 
can be used for PID. 
The pulse shape analysis of the ECAL signal will also be investigated. 

A prototype Sci-Fi detector with 1~mm-square scintillating fibres
and MPPC readouts was successfully tested to obtain the beam profile
during the StrECAL test in KEK.
However, smaller fibre such as 250-$\mu$m-fibre would be preferred 
to minimize secondary particle production. 
Recent measurements have with 250-$\mu$m-square fibres
achieved  a satisfactory timing resolution of 500~ps for electrons and 200~ps for muons.

\begin{figure}[htb!]
    \begin{center}
    \subfigure[][]{
    \includegraphics[width=0.45\textwidth]{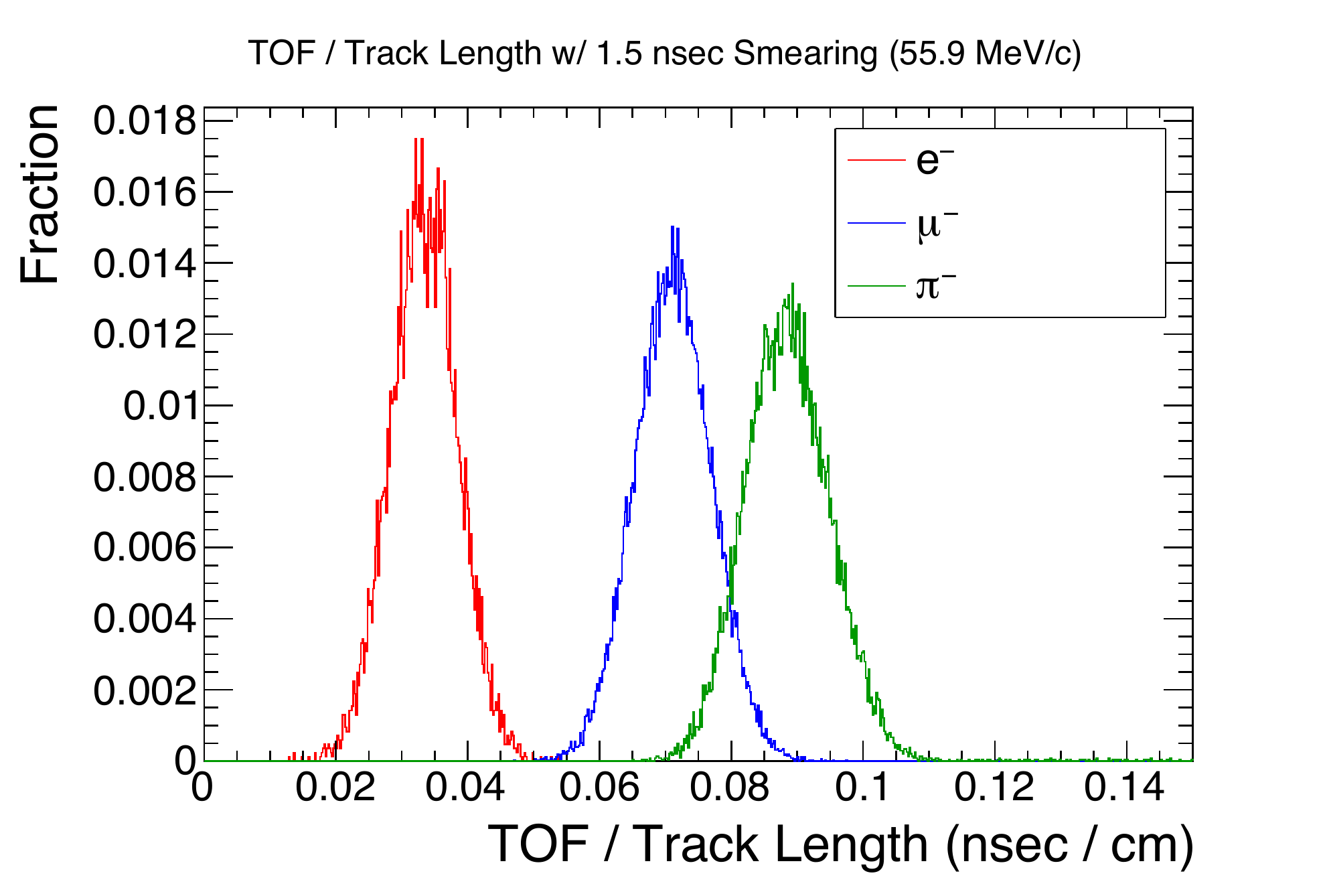}
    }
    \subfigure[][]{
    \includegraphics[width=0.45\textwidth]{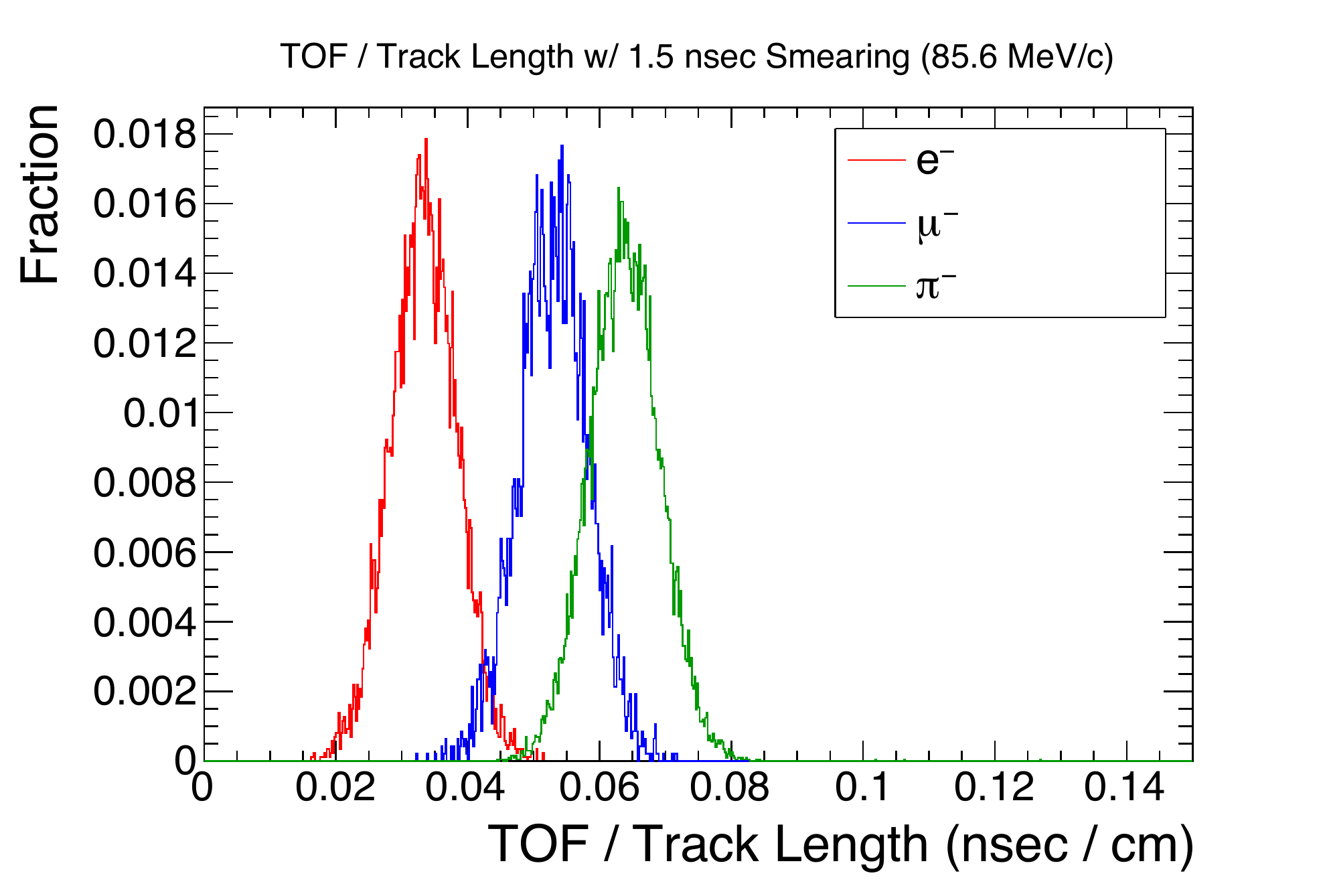}
    }
    \subfigure[][]{
    \includegraphics[width=0.45\textwidth]{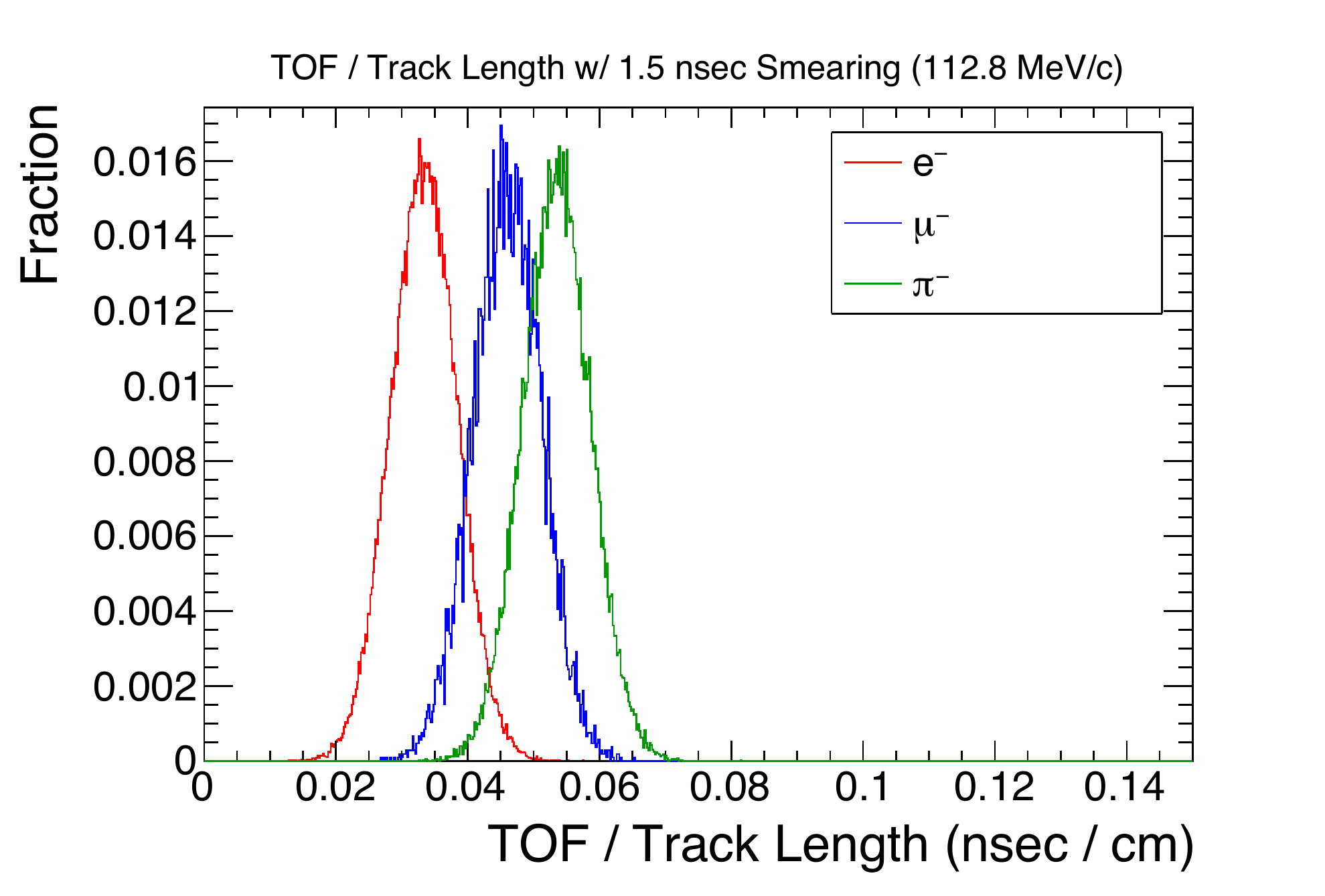}
    }
    \end{center}
    \caption{A simulation on TOF performance. The distributions of  TOF divide by the track length (therefore, inverse of velocity) are shown. The vertical scale is arbitrary. Particle momentum of simulation are 
       (a) 55.9 MeV/c, (b) 85.6 MeV/c, and (c) 112.8 MeV/c. 
       The red, blue and green lines correspond to e$^{-}$, $\mu^{-}$, and $\pi^{-}$, respectively. }%
    \label{fig:StrECAL_TOF_MC}%
\end{figure}

\paragraph{TOF performance} 
A dedicated simulation study was performed to evaluate the TOF performance.
\Cref{fig:StrECAL_TOF_MC} shows the 
 TOF distributions between the Sci-Fi detector and ECAL
of  e$^{-}$, $\mu^{-}$, and $\pi^{-}$, 
for three different momenta, 55.9, 85.6 and 112.8 MeV/$c$.
A TOF measurement accuracy of 1.5~ns for is assumed.
It is evident that 
electrons can be easily distinguished  from $\mu^{-},\pi^{-}$  at  55.9 MeV/$c$,
while such discrimination is not complete between $\mu^{-}$ and $\pi^{-}$.
In the  higher momentum region above 100 MeV/$c$,
even electron discrimination is not possible.
Therefore, PID performance is not sufficient only with TOF. 

\paragraph{PID using ECAL signal shape}
Another PID method using the different signal shape of ECAL for different particles is examined. 
A test on pulse shape measurement of LYSO crystal was carried out
by using the
intense
$e^{\pm}/\mu^{\pm}/\pi^{\pm}$ beams in PSI. 
The beam momentum was set to 115 MeV/$c$, and  varied by 
placing Lucite degraders of varying thicknesses.

\begin{figure}[htb!]
    \begin{center}
    \includegraphics[width=0.8\textwidth]{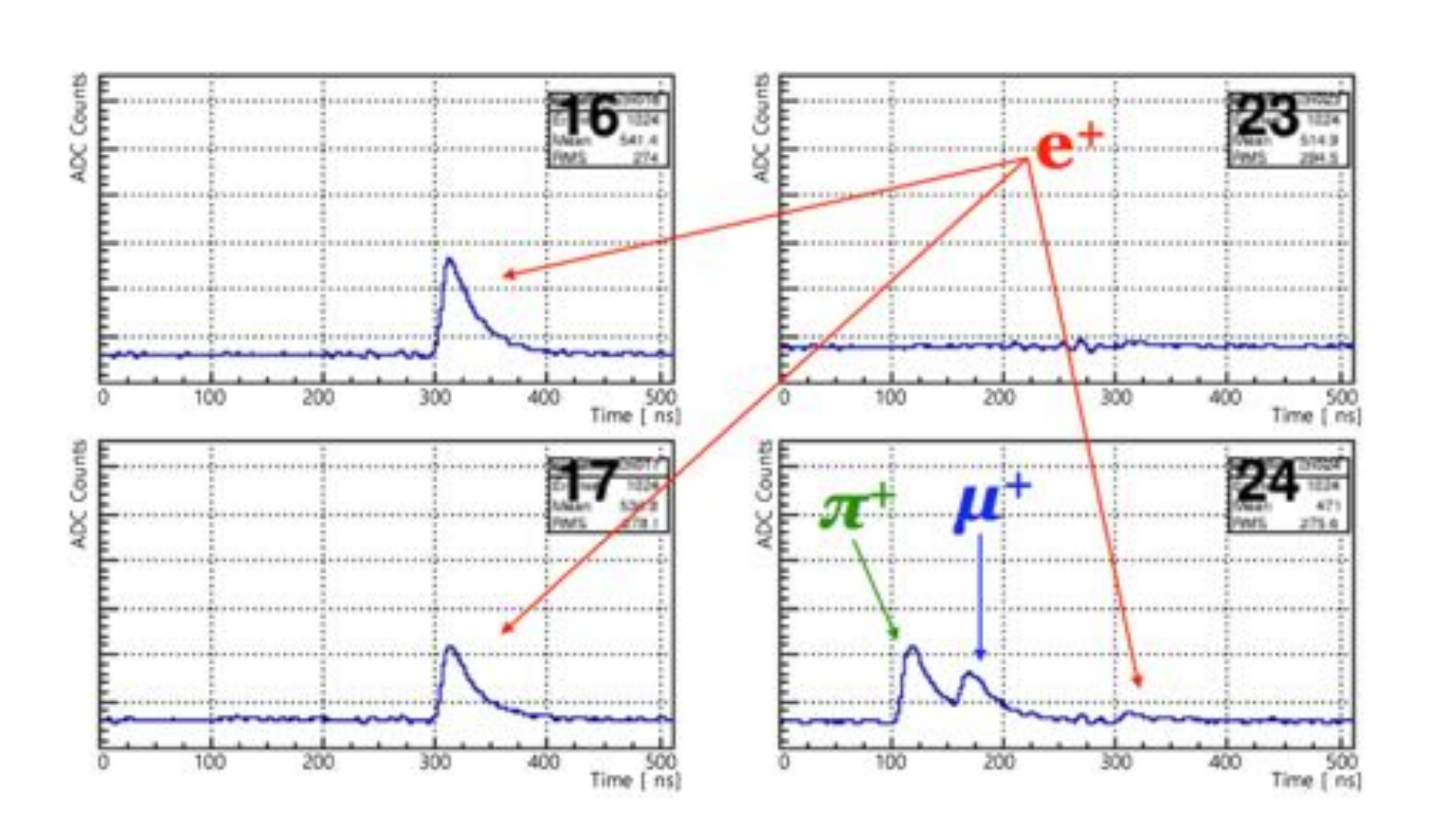}\\
    \includegraphics[width=0.8\textwidth]{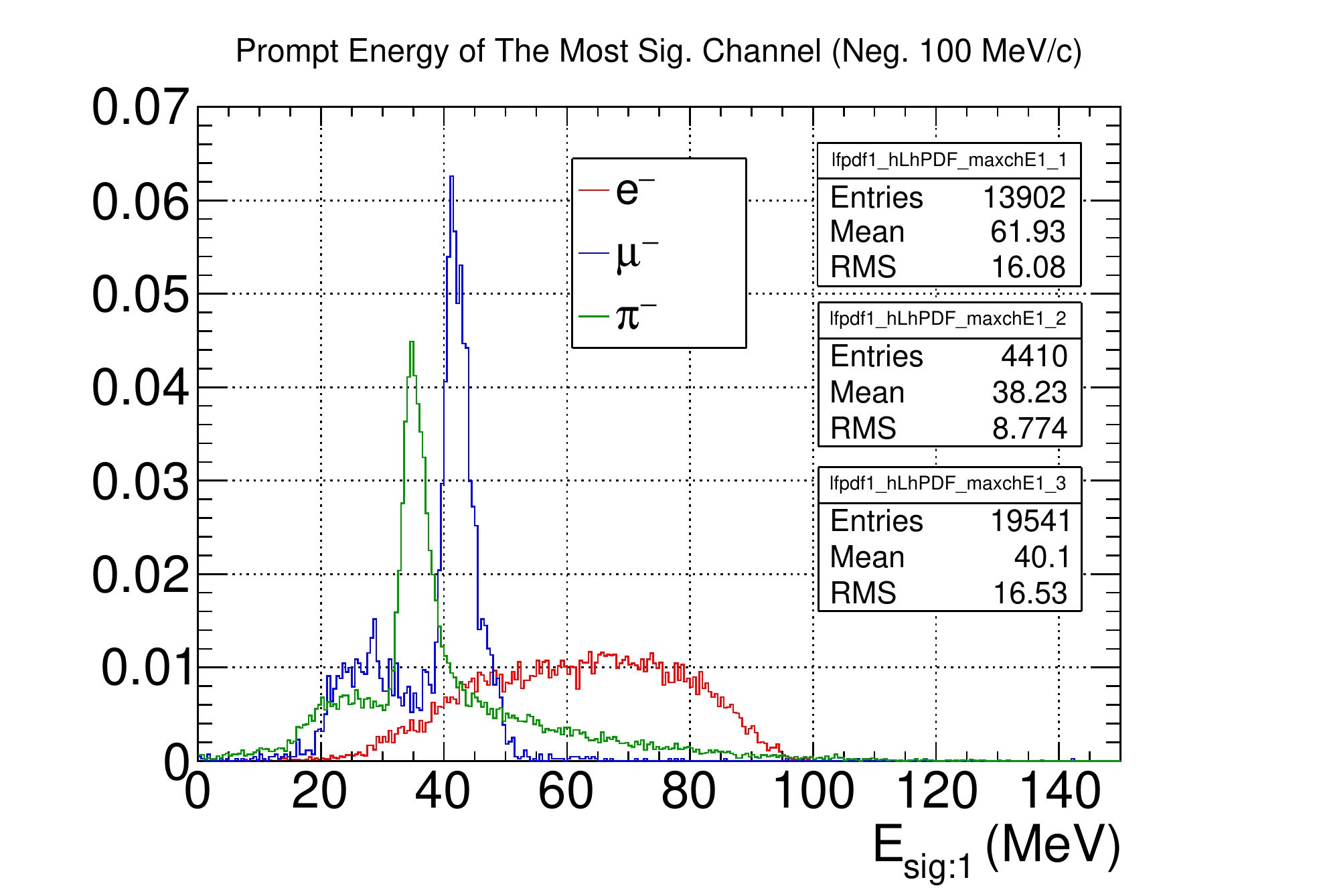}%
    \end{center}
    \caption{Performance of PID using ECAL signal shape.
     (Top) Observed decay chain of $\pi^{+}-\mu^{+}-e^{+}$,
     (Bottom) Distribution of the prompt energy deposition of negative particles }%
    \label{fig:StrECAL_ECALalone_output}%
\end{figure}

For  positive particles (e$^{+}$, $\mu^{+}$ and $\pi^{+}$),
 the decay chain of $\pi^{+}-\mu^{+}-\mathrm{e}^{+}$ 
in the crystal
is observed 
from the wave form measurement, 
as shown in \cref{fig:StrECAL_ECALalone_output} (Top).
This shows that a reasonable PID will be possible by combining the 
energy deposition information and this decay chain measurement. 
The decay chain measurement is not possible 
in case of negative particles which undergoes nuclear capture process. 
Instead, 
a weak particle discrimination will be possible by using 
the fraction of prompt energy deposition to the total energy deposition, 
which arise from the different nuclear capture processes. 
 \Cref{fig:StrECAL_ECALalone_output} (Bottom) shows the
distribution of prompt energy deposit for three negative particles
with 100 MeV/$c$. 
While a visible difference is observed between
$\mu^-$ and $\pi^-$,
 the difference is not clear
since the nuclear capture process is complicated
and not always same for every events.
Although, this demonstrates the feasibility of PID using ECAL only. 

\begin{figure}[htb!]
    \begin{center}
    \subfigure[][]{
      \includegraphics[trim={0 217pt 277pt 0},clip,width=0.45\textwidth]{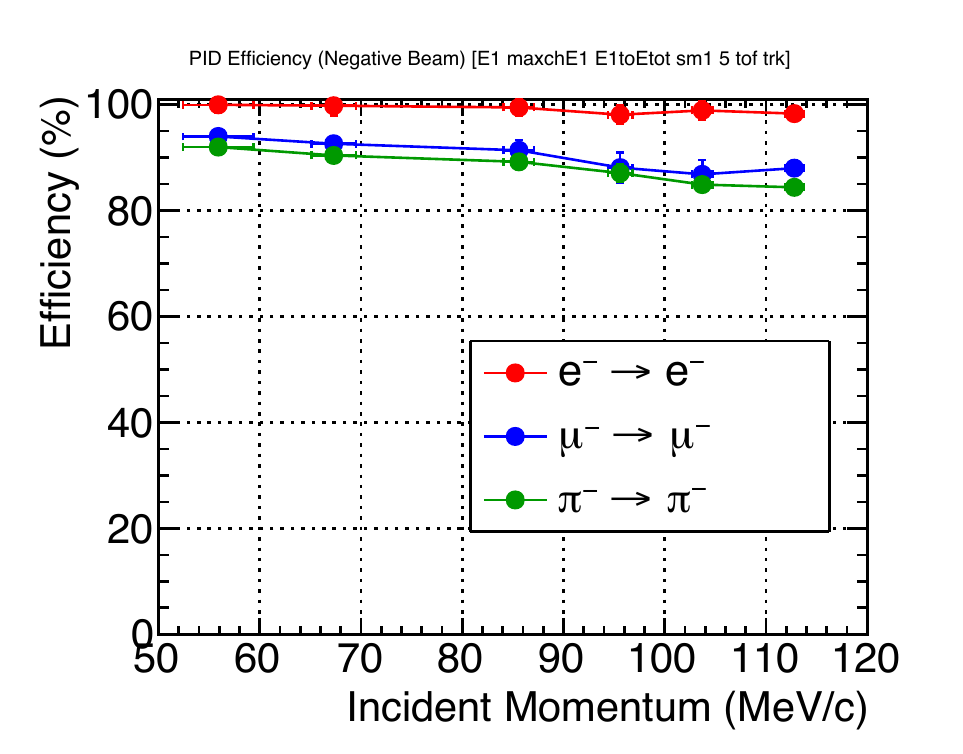}%
    }
    \subfigure[][]{
      \includegraphics[trim={0 217pt 277pt 0},clip,width=0.45\textwidth]{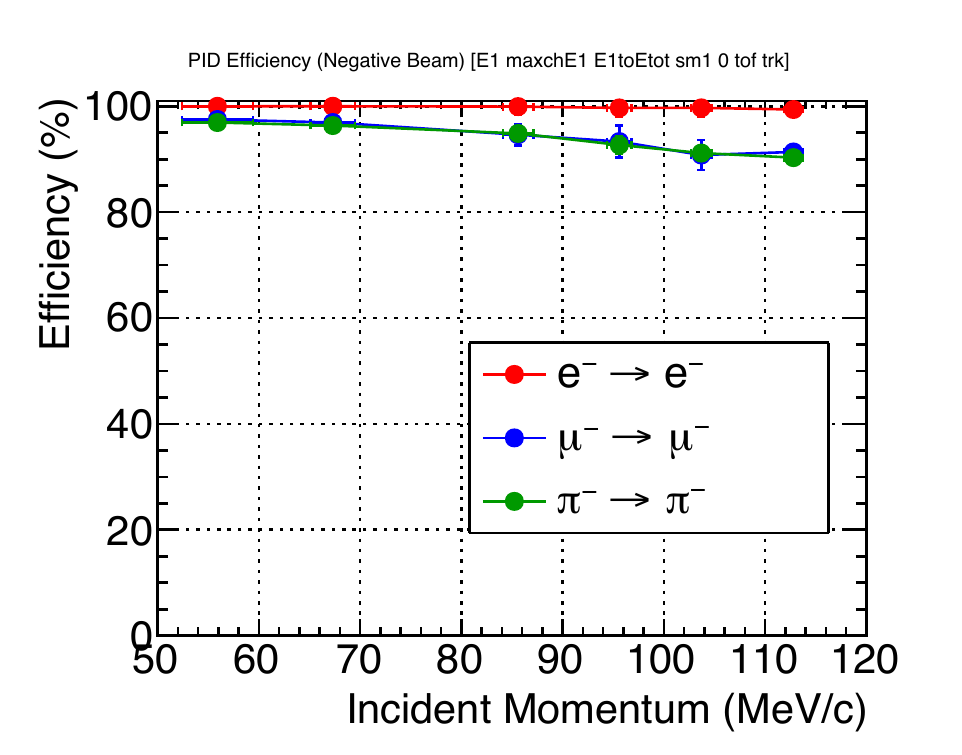}
    }
    \subfigure[][]{
      \includegraphics[trim={0 217pt 277pt 0},clip,width=0.45\textwidth]{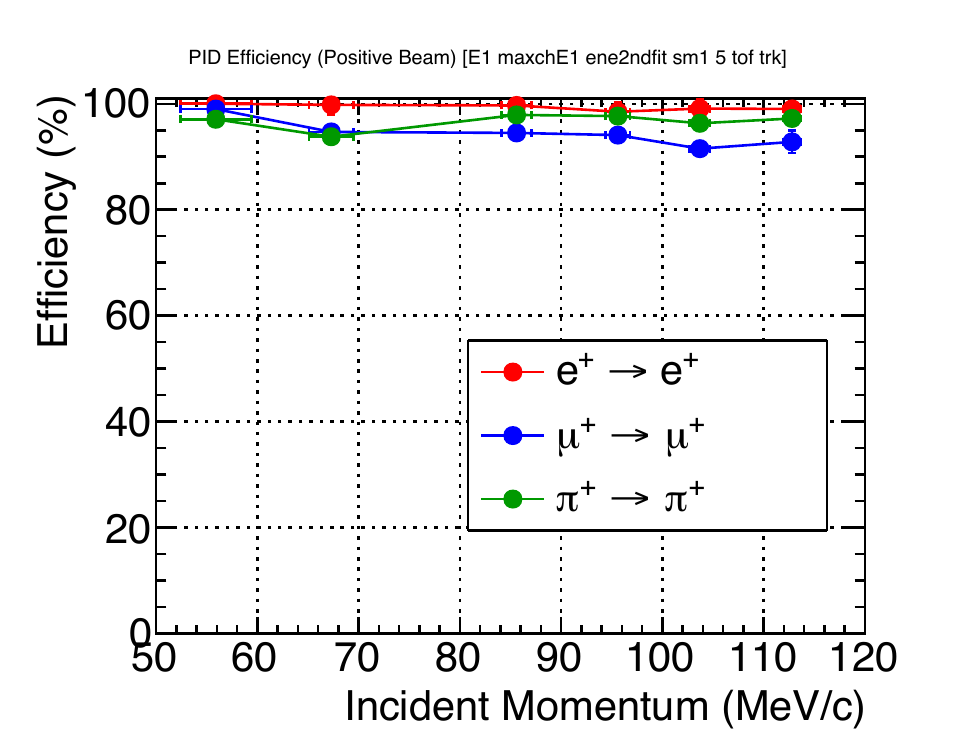}
    }
    \end{center}
    \caption{PID efficiencies by combining TOF and ECAL signal shape discriminator.
    (a) Negative particle cases with expected  TOF  timing resolution 1.5 ns,
    (b) Negative particle cases with expected  TOF  timing resolution 1.0 ns,
    (c) Positive particle cases with expected TOF timing resolution 1.5 ns.
      }
    \label{fig:StrECAL_PID_combine}%
\end{figure}

\paragraph{PID by combining TOF and ECAL signal shape}
In order to improve the PID performance both in low and high momentum ranges, 
a maximum likelihood analysis combining the TOF and ECAL signal shape discriminator has been studied  with a  simulation data.
Results of the PID efficiency estimation of are shown in \cref{fig:StrECAL_PID_combine}.
It is clear that the PID performance can be improved by combining the TOF and ECAL signal shape discriminator. 
In case of negative particle, it is clear that PID efficiency as high as 90 \% for all momentum range can be achieved when TOF timing resolution 
is 1.0 ns. 
The PID efficiency is still above 85 \% when TOF timing resolution is 1.5 ns.
In case of positive particle, the PID efficiency is better than negative particle, 
as the performance of PID by ECAL signal shape is better in the positive particle cases. 
In conclusion, the beam measurement in COMET Phase-I experiment will be possible by configuring 
StrECAL with TOF detector, 
and by using the combination analysis of TOF and ECAL signal shape discriminator.

\subsubsection{Rate capability consideration of StrECAL}
In the beam measurement program, not only the 
momentum and profile of beam, 
but the timing structure of beam will be measured. 
The rate capability of StrECAL detector is investigated with ICEDUST in order to understand the 
feasibility of measuring timing structure of beam. 

It is clear that the 3.2 kW beam of COMET Phase-I experiment is too strong 
for the the momentum and profile measurement. 
One option is to reduce the beam power, which is clearly not ideal.
Alternative option is 
to decrease its sensitivity in the StrECAL central region of the beam. 
A ``beam blocker'' and ``HV masking'' have been investigated.

\paragraph{Beam blocker}
Since the muon transport line for Phase-I is not long enough
and has only 90\Deg bending,
rate of transported beam to the detector section will be very high.
As most  beam particles are in the central region,
a beam blocker in that region of the beam can reduce the rate when operating at the Phase-I beam power.
A 2\,cm thick Tungsten disk in front of StrECAL is enough to reduce the hit rate down to 20 \%.

\paragraph{High voltage masking}
To reduce the hit rate further 
without reducing the beam power,
a partial turning off of the HV, (``HV masking''),
which result the reduction of sensitivity of StrECAL detector,
can be considered.

In the Straw detector configuration, 8 channels of the high voltage 
are controlled at the same time, which is called HV unit. 
From the simulation study with ICEDUST, it is found that 
HV masking (i.e. turning off) of three units (i.e. 24 channels) 
can reduce the hit rate around factor of 100, when the tracking performance degradation is not significant. 
The hit rate distribution versus momentum did not change by this three unit HV masking method. 

In case of ECAL, the hit rate does not change by HV masking. 
The beam blocker reduces the hit rate in the central part of the beam  and the Straw detector,
however, the reduction of hit rate in the ECAL is uniformly distributed over the ECAL. 
Therefore, reducing the beam power may be still required  for the beam measurements in the COMET Phase-I experiment.

\subsubsection{Plan for the Beam Measurement}
\label{sec:beam_measurement_program}

In case of the beam momentum and profile measurement, 
the beam power reduction by a factor of 1000 is necessary, 
as the tracking reconstruction is required by StrECAL. 
This  beam power reduction corresponds to $2 \times 10^{9}$ protons on target per second, 
and consequently 10 kHz hit rate in StrECAL. 
With this hit rate, the measurement less than an hour will be enough for the beam measurement. 
When including the positive particle measurements, a few days of data taking will be sufficient 
for the momentum and profile measurements.

For the beam timing measurement,
 track reconstruction is not required and hence the
Detector Solenoid  and the Straw detector  can be turned off
during this measurement.
PID will rely on the ECAL signal shape discriminator only. 
When the Detector Solenoid is off, 
the hit distribution in the ECAL is almost uniformly distributed, 
and the momentum spectrum above 40 MeV/$c$ is essentially unaffected.
Two classes of beam timing measurement will be made, prompt
and delayed.

The beam measurement programme is summarized in \cref{tab:beam_measurement_summary}. The programme is estimated to take two to three weeks.
\begin{table}[tbh!]
  \caption{Summary of beam measurement programme}
  \begin{center}
  \begin{tabular}{lcc}
    \hline\hline
  {} & {Momentum measurement} & {Timing measurement}\\
    \hline
  {Detector} & {StrECAL + SciFi} & {ECAL signal shape discriminator} \\
  {Beam mode} & {Normal SX} & {Bunched SX} \\
  {Detector solenoid} & {On} & {Off} \\
  {Beam suppression} & {Beam blocker, HV masking} & {No beam suppression} \\
  {PID} & {Full PID} & {Easy PID} \\
  {Beam power} & {1/1000} & {1/100 (prompt), Full power (delayed)} \\
     \hline\hline
  \end{tabular}
  \label{tab:beam_measurement_summary}
  \end{center}
\end{table}

\subsection{Background Assessment Programs}\label{ch:BackgroundAssessment}

The background processes of \mue conversion in aluminium have never been measured in most cases. 
The background 
estimations described in \cref{ch:SensitivityBackgrounds} are based on simulation.
The measurement of
background sources is one of the most important goals for COMET Phase-I, for 
understanding the Phase-I data and for preparing Phase-II experiment. 
Examples of such measurements are summarized in \cref{tb:backgroundassessment}.
\begin{table}[tbh!]
    \caption{Examples of potential backgrounds for the search for \mue conversion. The
    COMET Phase-I experiment will measure most of the background sources, which have not been
    measured in the past, with sufficient accuracy.}
  \begin{center}
    \begin{tabular}{lll}
    \hline \hline
       Intrinsic Physics backgrounds & Status & Plan  \cr
    \hline
    Muon decays in orbit (DIO)
    & endpoint not measured & by Phase-I  \\ %\hline
    Radiative muon capture
    & endpoint not measured & by Phase-I  \\
    Neutron emission & not measured & by AlCap  \\
    Charged particle emission & measured (AlCap) & by AlCap  \\[10pt]
    \hline %\hline
        Beam related backgrounds & Status & Plan \cr
    \hline
    Radiative pion capture
    &   \\ %\hline
    Beam electrons & not measured & by Phase-I \\ %\hline
    Muon decay in flight & not measured & by Phase-I \\ %\hline
    Pion decay in flight &  not measured & by Phase-I \\ %\hline
    Neutron-induced backgrounds &  not measured & by Phase-I \\ %\hline
    $\overline{p}$-induced backgrounds &  not known & by Phase-I \\[10pt]
    \hline %\hline
           Other backgrounds & Status & Plan \cr
    \hline
    Cosmic-ray-induced backgrounds & & by cosmic runs \\ %\hline
    Ambient neutron-induced backgrounds & & \\ \hline\hline
\end{tabular}
\end{center}
    \label{tb:backgroundassessment}
\end{table}

\paragraph{Muon decay in orbit}\label{sec:assessmentdio}

The electron momentum  spectrum of Muon decay in orbit (DIO) 
in the high-momentum region near the endpoint have not been measured. In the COMET
Phase-I experiment, CyDet will be used to measure the DIO electron spectrum precisely
with a momentum resolution of around 200~keV$/c$.

\paragraph{Cosmic-ray-induced background}\label{sec:assessmentcosmic}

Cosmic rays are potentially a significant source of backgrounds. 
In order to understand the impact of cosmic ray, 
The dedicated cosmic ray run will be performed using CyDet and  CTH triggers when there is no primary beam.
The duration of this measurement will be similar or longer than the beam time of \mue conversion physics run. 
The cosmic ray run will be done before the physics run to understand 
the necessity of additional configuration of the detector and trigger for 
further suppression of cosmic ray backgrounds.

\paragraph{Radiative muon capture}\label{sc:radiateivemuoncapture}

There is no measurement of radiative muon capture (RMC) with the
photon energy in the region of the endpoint for aluminium.
As the endpoint is only 3.06~MeV lower than the \mue conversion signal, 
the RMC background  measurement requires 1~MeV or better energy resolution.
The CyDet with a gold foil photon converter will be able to measure the 
photon of 100~MeV and above energy, with around 200~keV resolution. 
The partial branching ratio of RMC on $^{27}_{13}$Al, for a photon energy of greater than 100~MeV, is about $1.6 \times 10^{-9}$. 
Assuming a conversion efficiency of about 1.41\% with a 100~$\mu$m gold foil 
together with the simulated signal acceptance of 6.7\% and the expected muon yield of COMET Phase-I of 
$N_{\mu} = 1.2 \times 10^9$\,/sec, the running time of 10 days would accumulate about 1000 events.
The running time to accumulate 1000 events would be about 10 days.
A special trigger condition for this measurement must also be devised, which could be produced either from the CDC hits directly or by adding additional plastic scintillator.

\paragraph{Proton emission after muon capture}\label{sec:protonabsorber}

The maximum muon beam intensity  in the COMET Phase-I experiment will
be limited by the hit occupancy of the CDC. Protons,
emitted after nuclear muon capture, namely $\mu^{-} N \rightarrow N' p \nu_{\mu}$, were expected to be one of the major contributors
to the CDC hit rate, but measurements from the AlCap experiment have 
indicated that this is not the case.

\begin{figure}[tbh!]
 \begin{center}
 \includegraphics[width=0.6\textwidth]{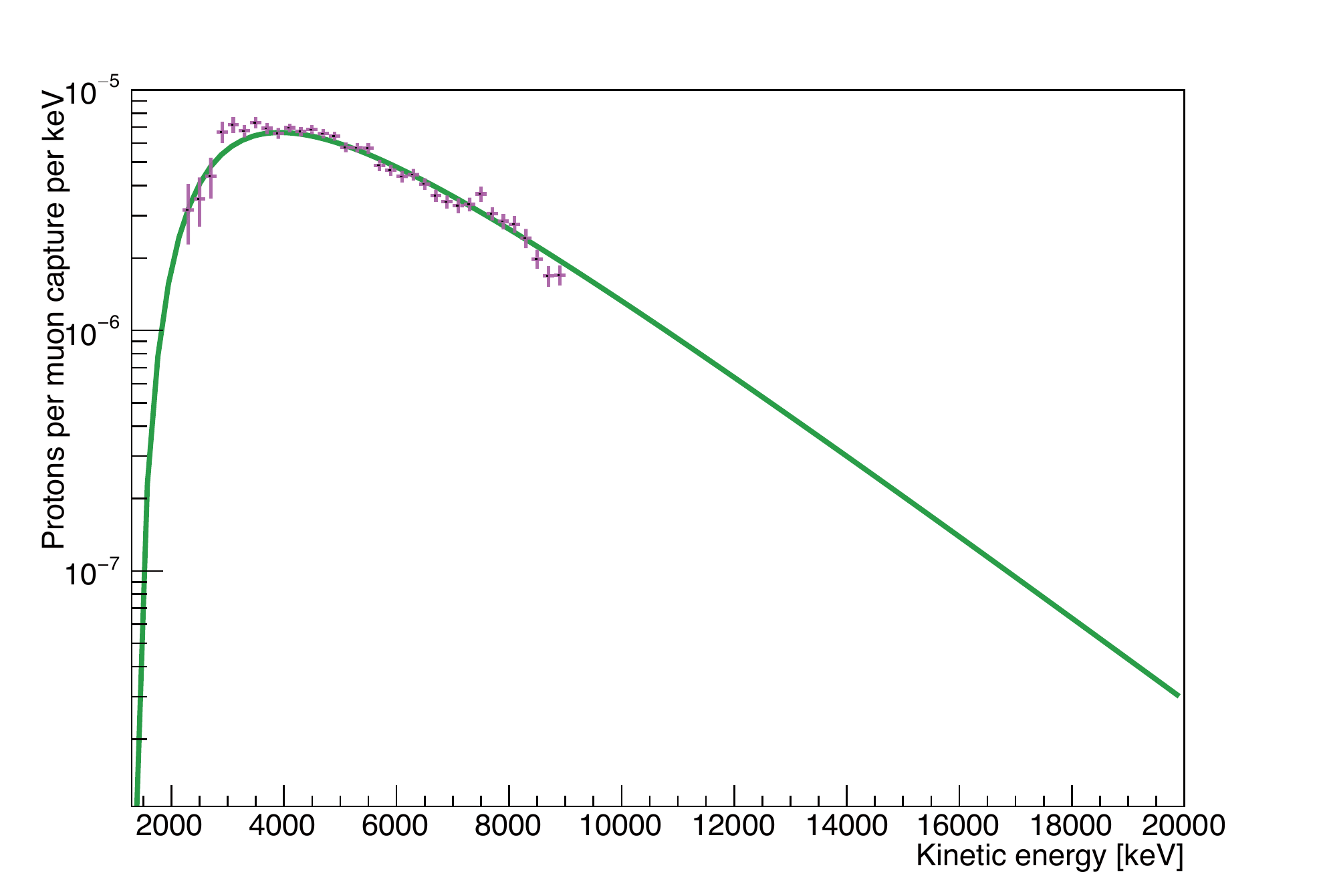}
 \end{center}
 \caption{Energy spectrum of protons emitted after nuclear muon capture on aluminium. Data points are
 from the AlCap measurement in 2013, the solid line shows the fitted function.}
 \label{fig:alcap-proton-spec}
\end{figure}

The energy spectrum of protons emitted after negative muon capture in
aluminium target has been measured in AlCap. 
A preliminary analysis shows
that the proton emission probability per muon capture in
the energy range  4 to 8 MeV is 0.017.  Fitting the measured
spectrum and extrapolating the fitted function gives a total emission
rate per muon capture of 0.035. The proton spectrum peaks at around 3.7~MeV,
then decreases exponentially with a decay constant of 2.5~MeV as shown in
\cref{fig:alcap-proton-spec}.
With a Monte Carlo simulation using this distribution the hit rate on a single cell of the CyDet is
estimated to be 1.4~kHz,  which is low enough for normal operation
of the CDC.
In the COMET Phase-I experiment, the rate and spectrum of proton
emission after muon nuclear capture will be measured with greater statistics than the AlCap results.

\section{Summary and Prospect}\label{sec:summary}

Details of the technical design of COMET Phase-I are described. 
This first stage of the COMET programme will provide an opportunity to fully understand the novel superconducting pion production system and muon beam line, with its charge-and-momentum selecting dipole fields which are superimposed on the curved solenoids which form the pion and muon transport section---a design that is unique to COMET amongst intense pulsed muon beam facilities.

The research programme for Phase-I encompasses both a search for \mue conversion with a sensitivity that is about 100 times better than the current limit, and a dedicated detector set-up which will allow us to make comprehensive measurements of the muon beam. Detailed rate and timing studies and other measurements from Phase-I will help us understand the backgrounds to the \mue conversion measurement. These will be crucial as COMET prepares to move to Phase-II, which is to improve the sensitivity by 
another two orders of magnitude.

The challenges to building and running this high-background rare-decay search experiment are addressed, including: proton and muon beam dynamics; the superconducting magnet systems; high-rate data-acquisition systems; operation in harsh radiation environments; software and computing systems that can meet the demands of the experiment.

The COMET Collaboration believes that rapid execution of Phase-I, which will consist of data taking in numerous different configurations of the beam line and detector systems, to be followed by the deployment of Phase-II soon after, is the most reliable path to a high-sensitivity search for \mue conversion. The programme has the potential to result in a paradigm-shifting discovery, which could lead to a entirely new field opening up of multiple measurements of different charged-lepton flavour violating processes---a new era of discovery in particle physics.

\section*{Acknowledgements}

We thank KEK and J-PARC, Japan for their support of infrastructure
and the operation of COMET.
This work is supported in part by:
Japan Society for the Promotion of Science (JSPS) KAKENHI Grant Nos. 25000004 and 18H05231; JSPS KAKENHI Grant No.JP17H06135;
Belarusian Republican Foundation for Fundamental Research Grant F18R-006;
National Natural Science Foundation of China (NSFC) under Contracts
No. 11335009 and 11475208; Research program of the Institute of High
Energy Physics (IHEP) under Contract No. Y3545111U2; the State Key
Laboratory of Particle Detection and Electronics of IHEP, China, under
Contract No.H929420BTD;
Supercomputer funding in Sun Yat-Sen University, China;
National Institute of Nuclear Physics and Particle Physics (IN2P3), France;
Shota Rustaveli National Science Foundation of Georgia (SRNSFG), grant No. DI-18-293;
Deutsche Forschungsgemeinschaft  grant  STO 876/7-1 of Germany;
Joint Institute for Nuclear Research (JINR), project COMET \#1134;
Institute for Basic Science (IBS) of Republic of Korea under Project
No. IBS-R017-D1-2018-a00;
Ministry of Education and Science of the Russian Federation and by the Russian Fund for Basic Research grants: 17-02-01073, 18-52-00004;
Science and Technology Facilities Council, United Kingdom; JSPS
London Short Term Predoctoral Fellowship program, Daiwa Anglo-Japanese
Foundation Small Grant; and Royal Society International Joint Projects
Grant.

Crucial computing support from all partners is gratefully
acknowledged, in particular from CC-IN2P3, France; GridPP, United
Kingdom; and Yandex Data Factory, Russia, which also contributed
expertise on Machine Learning methods.

\bibliography{COMET-TDR}{}
\bibliographystyle{ptephy}

\end{document}